\documentclass{IEEEtran}
\usepackage[T1]{fontenc}
\usepackage{cite}
\usepackage{amsmath,amssymb,amsfonts}
\usepackage{graphicx}
\graphicspath{{./}{../}}
\usepackage{float}
\usepackage{longtable}
\usepackage{textcomp,nicefrac}
\usepackage{bm}
\usepackage{microtype}
\DeclareFontShape{T1}{ptm}{m}{scit}{<->ssub * ptm/m/it}{}
\newtheorem{proposition}{Proposition}

\begin{document}
\setlength{\emergencystretch}{5em}
\title{Prior-Based Multi-Voltage Threshold Sampling as a Structured Inverse Problem}

\author{Ao Qiu,
  and Qingguo Xie
  \thanks{This work was supported in part by the National Natural Science Foundation of China, under Grant 625B2079, 61927801, 62250002, and 62050288. (\textit{Corresponding author: Qingguo Xie})}
  \thanks{Ao Qiu is with the Department of Biomedical Engineering, Huazhong University of Science and Technology, Wuhan, 430074 China (e-mail: aqiu@hust.edu.cn).}
  \thanks{Qingguo Xie is with the Department of Biomedical Engineering, Huazhong University of Science and Technology, Wuhan, 430074 China; Wuhan National Laboratory for Optoelectronics, Wuhan, 430074 China; and the Department of Electronic Engineering and Information Science, University of Science and Technology of China, Hefei, 230026 China (e-mail: qgxie@hust.edu.cn).}}

\maketitle
\bstctlcite{IEEEexample:BSTcontrol}

% =========================================================================
%  ABSTRACT
% =========================================================================
\begin{abstract}
  Prior-based Multi-Voltage Threshold (MVT) sampling reconstructs pulse parameters from sparse threshold-crossing times rather than full waveforms, making parameter recovery inherently a model-dependent inverse problem. However, prior-based MVT has lacked a formal mathematical statement, leaving identifiability, stochastic error propagation, and threshold design without a unified theoretical foundation. We formalize prior-based MVT for strictly unimodal pulse families as a structured inverse problem. On that foundation, we develop the first unified theory of prior-based MVT, comprising deterministic identifiability conditions, a stochastic timing-error model with leading-order mismatch bias, and a nuisance-profiled threshold-design theory centered on an effective-information equation for robust single-event and partial-trigger multi-event operation. We instantiate the framework for the bi-exponential pulse model, derive executable design recipes, and validate the resulting predictions on a 10{,}000-pulse $^{22}$Na/LYSO/SiPM dataset. The experiments confirm that the framework yields useful threshold designs in the photopeak regime while also revealing the regime boundary at which partial triggering and model mismatch limit the predictive power of Fisher-guided optimization. These results provide the first unified mathematical foundation for prior-based MVT and recast it from an empirical threshold heuristic as a principled inferential framework.
\end{abstract}


% =========================================================================
%  I. INTRODUCTION
% =========================================================================
\section{Introduction}
\IEEEPARstart{M}{ulti-Voltage} Threshold (MVT) sampling replaces dense waveform acquisition with sparse threshold-crossing times, thereby reducing the need for fast ADCs while simplifying front-end electronics and data throughput~\cite{xie_new_2005}. All-digital implementations of this strategy have since been demonstrated from FPGA-only digitizers and PET detector modules to dedicated brain and clinical PET platforms~\cite{xi_fpga-only_2013,xie_implementation_2013,fang_development_2024,zhang_performance_2025}. This paper concerns prior-based MVT, in which a pulse family is built into the observation model. Once such a prior is imposed, the threshold crossings cease to be merely hardware timing marks. In its standard form, prior-based MVT interprets sparse crossings through the pulse model to reconstruct the underlying pulse or its governing parameters.  Parameter recovery is therefore an inverse problem on paired-crossing observations.

Most prior studies have instead addressed one slice of the problem at a time. Representative examples include heuristic threshold selection~\cite{deng2017threshold} and task-specific estimators for energy or timing extraction~\cite{deng2013empirical,deng2015quadratic,xu2020neural}. Other work has focused on engineering refinements such as baseline restoration and automatic threshold calibration~\cite{xie2010baseline,fang2021atc}. Still other papers pursue compact digitizer variants~\cite{chu2023singleline}, application-specific detector studies~\cite{cheng2024dualended}, or model-centered pulse-recovery refinements~\cite{ling2024novel}. Yet these advances still do not define, in a unified way, the ordered crossing set itself as a mathematical observation map. Recoverability, uncertainty, model mismatch, and design therefore remain treated as separate engineering questions rather than consequences of one formal object. Without such a statement, it is difficult to say when a sparse crossing set uniquely determines pulse parameters, how timing perturbations propagate through reconstruction, or when Fisher-guided threshold design should be expected to predict estimation performance.

In this paper, prior-based MVT is recast for strictly unimodal pulse families, the minimal regime in which paired crossings define a fixed observation geometry, as a structured inverse problem. That recasting exposes the paired-crossing observation geometry and turns deterministic recovery, stochastic timing uncertainty, mismatch bias, and threshold design into coupled aspects of one theory. The resulting framework yields global identifiability conditions, a timing-error model with leading-order mismatch bias, and a nuisance-profiled design theory organized around effective information. The contribution is therefore foundational rather than heuristic: it provides a mathematical framework in which prior-based MVT becomes an analyzable inference problem.

We instantiate the general framework with a bi-exponential pulse model, derive operational design recipes, and evaluate them on measured $^{22}$Na/LYSO/SiPM data. The experiments serve both as validation of the theory and as a boundary test of its predictive regime: they show that Fisher-guided threshold design is useful in the photopeak regime while identifying where partial triggering and model mismatch erode its predictive accuracy. The paper moves from general formulation to model instantiation to experimental audit. Sections~\ref{sec:deterministic}--\ref{sec:general_opt} develop the general framework, Section~\ref{sec:optimal} instantiates it for the bi-exponential model, and Sections~\ref{sec:validation}--\ref{sec:discussion} provide implementation recipes and validation in the single-event and multi-event regimes.

% =========================================================================
%  II. DETERMINISTIC FOUNDATIONS
% =========================================================================
\section{Deterministic Identifiability}
\label{sec:deterministic}

Before examining statistical perturbations, we must first establish the deterministic topological conditions under which a discrete set of timestamps unambiguously encodes the continuous physical pulse parameters. If the deterministic mapping is not uniquely invertible, the underlying inverse problem is fundamentally non-identifiable in the sense of Hadamard~\cite{hadamard1923lectures,kabanikhin2008definitions}. In modern inverse-problems language, the relevant failure is loss of uniqueness or stability of the inverse map~\cite{clason2020regularization,bertero2021inverseimaging}; at the deterministic level, this appears here as non-uniqueness, local singularity, or discontinuity with respect to measurement perturbations.

\subsection{Pulse Model and the Observation Cone}
\label{sec:model}

\textbf{Physical Setup:} In a scintillation detector, an incident gamma photon deposits energy in a crystal, producing a burst of optical scintillation photons. These photons are detected by a photosensor (e.g., a Silicon Photomultiplier) and converted into a continuous voltage waveform, optionally conditioned by analog front-end electronics. The resulting macroscopic pulse rises rapidly during the initial photon avalanche, reaches a peak voltage proportional to the deposited energy, and subsequently decays as the crystal luminescence subsides~\cite{lecoq2016scintillation,gundacker2020silicon}. The MVT hardware records only the discrete time instants at which this waveform crosses predetermined voltage levels; the fundamental inverse problem is to reconstruct the underlying physical parameters from these sparse crossing timestamps~\cite{xie_new_2005,xi_fpga-only_2013}.

Let the continuous ideal pulse model be denoted by $f(t;\bm{\theta})$, parameterized by a physical state vector $\bm{\theta}=[\theta_1,\dots,\theta_M]^T \in \Theta \subset \mathbb{R}^M$ (typically encoding pulse amplitude, absolute arrival time, and intrinsic shape constants), where $\Theta$ is assumed to be an \emph{open, connected} subset of $\mathbb{R}^M$.

\emph{Assumption} (Single-event isolation): Each acquisition window is assumed to contain one dominant scintillation pulse, so that overlapping events, severe pile-up, or secondary afterpulses do not create additional local maxima comparable to the main pulse. This is the physical regime in which the strictly unimodal pulse model adopted below is intended to operate. If it fails, the waveform is no longer strictly unimodal and the paired-crossing observation cone must be replaced by a multimodal formulation with variable crossing multiplicity.

To represent such isolated signals, we require $f$ to be at least $C^2$-smooth in an open neighborhood of every threshold crossing, to be strictly unimodal in time $t$ (possessing exactly one local---and hence global---maximum), and, after subtracting any deterministic baseline offset, to satisfy an asymptotic decay condition. The first derivatives are needed for the transversality and implicit-differentiation arguments below, while the second derivatives are needed later for Taylor expansions and curvature-based robustness bounds.

In the baseline-free convention used throughout this general development, the asymptotic decay condition reduces to $\lim_{t \to \pm\infty} f(t;\bm{\theta}) = 0$ for all $\bm{\theta} \in \Theta$ (or, for causal pulses, $f(t;\bm{\theta}) = 0$ for $t < t_0$ and $f(t;\bm{\theta}) \to 0$ as $t \to +\infty$). Models with an explicit baseline parameter, such as the bi-exponential instantiation in Section~\ref{sec:biexp_model}, are handled by applying the same theory to the baseline-subtracted signal component and by requiring all thresholds to lie strictly above the baseline. Under this convention, every threshold level lying strictly between the baseline and the peak is crossed exactly twice, once on the rising branch and once on the falling branch.

For each admissible physical state $\bm{\theta}$, strict unimodality guarantees a unique peak time $t_p(\bm{\theta})$ and a well-defined peak amplitude $p(\bm{\theta}) \triangleq f(t_p(\bm{\theta});\bm{\theta})$. Assume, in the baseline-free convention of this general section, that the detection hardware defines an ordered set of $N$ strictly positive and strictly monotonic physical reference thresholds: $0 < V_1 < V_2 < \dots < V_N \equiv V_{\max}$. To ensure every threshold line intersects the waveform on both monotone branches, we restrict the parameter domain to the dynamically feasible state space:
\begin{equation}
  \Theta_V = \{ \bm{\theta} \in \Theta \mid p(\bm{\theta}) > V_{\max} \}.
\end{equation}
This excludes events whose peaks do not reach $V_{\max}$, since they would produce incomplete crossing data and fall outside the fixed-dimensional observation model.

\emph{Assumption} (Static calibrated thresholds): The hardware reference levels $\{V_n\}_{n=1}^{N}$ are treated as deterministic, event-independent voltages. Comparator offsets, hysteresis, and slow thermal drift are assumed negligible over the acquisition interval or absorbed by prior calibration, so that every recorded crossing is governed by the same physical threshold ladder. This is reasonable for laboratory-grade threshold generators and short acquisition runs; if these effects are not negligible, the effective threshold values must be promoted to additional nuisance parameters.

For each threshold level $V_n < p(\bm{\theta})$, strict unimodality yields exactly two crossings, denoted $t_{r,n}$ on the rising branch and $t_{f,n}$ on the falling branch. Collecting these paired crossings gives $K \triangleq 2N$ observations per event and the observation vector $\mathbf{t} \triangleq [t_{r,1}, \ldots, t_{r,N}, t_{f,N}, \ldots, t_{f,1}]^T \in \mathbb{R}^K$.

For later matrix formulas, we use two equivalent indexing conventions. The threshold index $n \in \{1,\dots,N\}$ labels the physical voltage levels $V_n$, while the ordered-crossing index $k \in \{1,\dots,K\}$ labels the components of $\mathbf{t}$ via
\begin{equation}
  \begin{aligned}
    t_k &=
    \begin{cases}
      t_{r,k}, & 1 \le k \le N, \\
      t_{f,2N+1-k}, & N+1 \le k \le 2N,
    \end{cases}
    \\
    \ell(k) &=
    \begin{cases}
      k, & 1 \le k \le N, \\
      2N+1-k, & N+1 \le k \le 2N.
    \end{cases}
  \end{aligned}
  \label{eq:crossing_index_map}
\end{equation}
Thus the $k$-th ordered crossing is generated by threshold $V_{\ell(k)}$ and satisfies $f(t_k;\bm{\theta}) = V_{\ell(k)}$. In later threshold-indexed formulas, the same ladder may still be written generically as $\{V_k\}_{k=1}^{N}$; whenever an ordered-crossing index is already in play, we write $V_{\ell(k)}$.

Monotonicity on each branch together with the threshold ordering implies
\begin{equation}
  \Omega_K = \left\{ \mathbf{t} \in \mathbb{R}^K \,\middle|\,
  t_{r,1} < \dots < t_{r,N} < t_{f,N} < \dots < t_{f,1} \right\}.
  \label{eq:obs_vector}
\end{equation}
Hence the deterministic observation map is $S:\Theta_V \to \Omega_K$, $S(\bm{\theta}) = \mathbf{t}(\bm{\theta})$.

The precise crossing instants are implicitly defined by the geometric level-set constraint. For each threshold $n \in \{1,\dots,N\}$, the rising-edge and falling-edge crossings satisfy $f(t_{r,n};\bm{\theta}) - V_n = 0$ and $f(t_{f,n};\bm{\theta}) - V_n = 0$, respectively. Because $\bm{\theta} \in \Theta_V$ enforces $p(\bm{\theta}) > V_n$, every crossing is pushed away from the stationary peak onto a strictly monotone branch.

\emph{Assumption} (No waveform plateau): The pulse model $f(t;\bm{\theta})$ does not possess any constant-voltage segment at any voltage level in the interval $(0, p(\bm{\theta}))$. Combined with the strict unimodality requirement, this ensures that every threshold crossing lies on a strictly monotone branch, so that the \emph{transversality condition} $\partial_t f(t;\bm{\theta}) \neq 0$ is satisfied at every crossing point. This assumption holds automatically for analytic pulse models (e.g., sums of real exponentials) and excludes only pathological cases such as hard-clipped or digitally saturated waveforms.

Define the level-set constraint function for the ordered crossing index $k \in \{1,\dots,K\}$ by $F_k(t_k, \bm{\theta}) \triangleq f(t_k;\bm{\theta}) - V_{\ell(k)}$. Applying the Implicit Function Theorem to the identity $F_k(t_k(\bm{\theta}),\bm{\theta}) = 0$, and noting that the transversality condition ensures $\partial F_k/\partial t_k = \partial_t f \neq 0$, guarantees that each crossing time $t_k(\bm{\theta})$ is a $C^2$-smooth function of $\bm{\theta}$ in a neighborhood of any admissible parameter. Because these branchwise crossings are uniquely defined for every admissible state, the local functions $t_k(\bm{\theta})$ are precisely the coordinate functions of $S$. Since smoothness is a local property, $S$ is therefore $C^2$ on $\Theta_V$. The global injectivity and properness properties needed later are addressed in the next subsection. Explicitly, differentiating $F_k(t_k(\bm{\theta}),\bm{\theta}) = 0$ with respect to $\theta_j$ gives $\partial_t f(t_k;\bm{\theta})\,\partial t_k/\partial\theta_j + \partial f(t_k;\bm{\theta})/\partial\theta_j = 0$. Solving for the temporal projection yields the Jacobian matrix $\mathbf{J} \in \mathbb{R}^{K \times M}$ (Appendix~\ref{app:jacobian}):
\begin{equation}
  J_{kj} = \frac{\partial t_k}{\partial \theta_j} = -\frac{\partial f/\partial\theta_j}{\partial_t f(t_k;\bm{\theta})}.
  \label{eq:jacobian}
\end{equation}
Physically, Eq.~\eqref{eq:jacobian} embodies a fundamental kinematic relationship: the temporal shift ($dt_k$) induced by a parameter perturbation ($d\theta_j$) is inversely proportional to the local temporal slew rate ($\partial_t f$). Steeper edges geometrically compress parameter variations into smaller temporal footprints.

\subsection{The Global Embedding Theorem}
\label{sec:hadamard}

\textbf{Theorem (Global embedding criterion).} Let $\Theta_{\mathrm{adm}} \subseteq \Theta_V$ be a physically admissible $C^2$ parameter domain, and let $S:\Theta_{\mathrm{adm}} \to \Omega_K$ be the deterministic observation map. If $S$ is injective, satisfies $\operatorname{rank}(\mathbf{J}(\bm{\theta})) = M$ for every $\bm{\theta} \in \Theta_{\mathrm{adm}}$, and is proper on $\Theta_{\mathrm{adm}}$, then $S$ is a $C^2$ embedding of $\Theta_{\mathrm{adm}}$ into $\Omega_K$; if $\Theta_{\mathrm{adm}}$ is compact, then this embedding is closed~\cite{lee2012smooth,mukherjee2015approximation}. Consequently, the deterministic inverse problem is globally identifiable on $\Theta_{\mathrm{adm}}$, and the inverse map $S^{-1}:S(\Theta_{\mathrm{adm}}) \to \Theta_{\mathrm{adm}}$ is continuous.

This theorem is the differential-topological criterion by which the Hadamard viewpoint of uniqueness and continuous deterministic recovery is implemented in the present setting. Although the hardware is typically overdetermined ($K = 2N > M$), overdetermination alone does not guarantee a globally well-posed inverse: one must also exclude macroscopic state degeneracy and parameter-escape paths that remain observationally bounded.

\textbf{Physical reading.} The three hypotheses exclude the three deterministic failure modes of MVT inversion: injectivity rules out distinct physical states producing the same timestamp vector, immersion rules out locally invisible parameter directions, and properness rules out divergent admissible sequences that would otherwise remain confined to a compact region of observation space.

For clarity, the required hypotheses are:

1) \textbf{Global Immersion ($\operatorname{rank}(\mathbf{J})=M$):} Prevents the loss of local measurement sensitivity across all dimensions of $\Theta_V$.

2) \textbf{Global Injectivity ($S(\bm{\theta}_a) = S(\bm{\theta}_b) \implies \bm{\theta}_a = \bm{\theta}_b$):} Structurally eliminates macroscopic state degeneracy.

3) \textbf{Proper Map Condition:} On the chosen admissible domain, the pre-image of every compact subset in $\Omega_K$ must be compact. Equivalently, every divergent admissible sequence must eventually leave the pre-image of any compact observation region.

The theorem itself is model-agnostic. The model-dependent task is to verify these hypotheses on the admissible domain relevant to a given pulse family and prior model. For the bi-exponential pulse analyzed in Section~\ref{sec:optimal}, immersion and injectivity are verified on the admissible open domain, whereas the embedding claim used for reconstruction is made on the physically bounded compact prior domain $\bar{\Theta}_V^{\mathrm{bi}}$; Appendix~\ref{app:proper} explains why properness is asserted on that compact domain rather than on the unbounded algebraic parameter space.

Sections~\ref{sec:stochastic}--\ref{sec:general_opt} build the stochastic estimation, misspecification, and threshold-design theory on top of this deterministic criterion, and Section~\ref{sec:optimal} later verifies the required hypotheses for the bi-exponential family.

% =========================================================================
%  III. STOCHASTIC MODEL AND CORRELATION LIMITS
% =========================================================================
\section{Stochastic Noise Model}
\label{sec:stochastic}

\subsection{Signal-Dependent Noise via Campbell's Theorem}
\label{sec:noise}

The continuous macroscopic voltage discussed in Section~\ref{sec:deterministic} represents the expected waveform $\mu_y(t) \triangleq \mathbb{E}[y(t)]$. In physical reality, the instantaneous signal $y(t)$ emerges from the superposition of discrete microscopic photoelectron avalanches, governed by an inhomogeneous Poisson point process with instantaneous photon intensity $\nu(t)$ (photons per unit time)~\cite{lane1984the}. Let $h_e(t)$ represent the effective single-photoelectron voltage impulse response at the readout output, carrying the physical dimension of Volts $[\mathrm{V}]$~\cite{seifert2009simulation,gundacker2020silicon}.

\emph{Assumption} (Linear superposition and stationary single-photoelectron response): Over the operating range considered, the sensor and front-end are approximated as sufficiently linear and time-invariant that the macroscopic voltage can be represented as a superposition of translated copies of a common effective single-photoelectron voltage impulse response $h_e(t)$. This is a standard first-order description for direct-readout scintillation pulses operated below severe saturation; departures such as strong SiPM gain compression, recharge distortion, or afterpulsing are treated later as waveform misspecification rather than as part of the well-specified stochastic model.

To connect this discrete shot-noise picture with the continuous macroscopic model $\mu_y(t) = f(t;\bm{\theta})$, we use Appendix~\ref{app:poisson} in two steps: Campbell's theorem~\cite{kingman1992poisson} gives exact first- and second-moment convolution formulas for the filtered Poisson process, and a slowly varying SPE-window approximation then reduces those formulas to the first-order affine crossing-variance law Eq.~\eqref{eq:poisson_voltage}.

\emph{Assumption} (Short single-photoelectron window / slowly varying intensity): The effective duration $\tau_e$ of $h_e(t)$ is short relative to the timescale on which the photon intensity $\nu(t)$ varies. In the regime considered here, this means $\tau_e/\tau_d \ll 1$ on the decay branch, while $\tau_e/\tau_r = \mathcal{O}(1)$ may occur on the fastest part of the rising edge. Representative numerical scales are recorded in Appendix~\ref{app:poisson}; accordingly, Eq.~\eqref{eq:poisson_voltage} is used here as a controlled first-order surrogate rather than as an exact microscopic variance law, particularly at the lowest thresholds.

Combining Campbell's exact moment formulas with the narrow-SPE approximation in Appendix~\ref{app:poisson} gives the first-order relation $\sigma_p^2(t) \approx \kappa \mu_y(t)$ for the intrinsic Poisson contribution, where $\kappa \triangleq \int h_e^2(x)\,dx \big/ \int h_e(x)\,dx$ carries the physical dimension $[\mathrm{V}]$ and represents an \emph{effective single-photoelectron voltage amplitude} of the readout chain. Evaluating this relation at a deterministic mean crossing time $t_c$ satisfying $\mu_y(t_c)=V$ yields a signal-dependent contribution $\kappa V$ in the baseline-free convention; adding the stationary thermal noise floor $\sigma_{\mathrm{th}}^2$ $[\mathrm{V}^2]$~\cite{johnson1928thermal,nyquist1928thermal} then gives the first-order affine analog voltage model used below:
\begin{equation}
  \sigma_{V}^2(V) = \sigma_{\mathrm{th}}^2 + \kappa V.
  \label{eq:poisson_voltage}
\end{equation}

\emph{Surrogate status of the stochastic model.} The Campbell-based affine voltage law above, the later diagonal crossing-covariance approximation, and the pseudo-Gaussian treatment of TDC quantization are used as tractable first-order analytical surrogates. Their role is to expose the closed-form Fisher geometry and stationarity conditions; they are not asserted to be an exact microscopic model of a SiPM/comparator/TDC chain. Residual colored noise, off-diagonal crossing covariance, clock-locked quantization structure, and unmodeled nonlinear front-end effects are treated in this manuscript as part of the model-misspecification channel $\varepsilon(t)$ and are empirically audited through the amplitude bias-to-variance diagnostic $\rho_{\mathrm{bias}}^{(A)}$ in Section~\ref{sec:delta_validation}.

\emph{Assumption} (Noise-source independence): The thermal noise $\sigma_{\mathrm{th}}^2$, the Poisson shot noise $\kappa V$, and the TDC quantization noise $\sigma_{\mathrm{TDC}}^2$ (introduced in Section~\ref{sec:bandwidth} below) arise from physically distinct and uncoupled mechanisms---electronic thermal fluctuations, photon counting statistics, and digital clock quantization, respectively---and are therefore treated as mutually statistically independent. This independence justifies the additive voltage-domain variance composition in Eq.~\eqref{eq:poisson_voltage}; after the slope projection of Section~\ref{sec:bandwidth}, the same decomposition carries over to the temporal variance model Eq.~\eqref{eq:variance} below.

When the model explicitly includes a deterministic baseline $b$ (as in the bi-exponential instantiation of Section~\ref{sec:biexp_model}), the Poisson term becomes $\kappa(V - b)$, since only the net signal portion $V - b$ arises from photon statistics (see Section~\ref{sec:biexp_stochastic}).

\subsection{Error Propagation and Covariance Diagnostics}
\label{sec:bandwidth}

To map the vertical voltage variance $\sigma_{V}^2$ onto the horizontal time axis, we employ a first-order geometric Taylor expansion around the crossing. A voltage fluctuation $\Delta V$ projects into a timing jitter $\Delta t$ inversely proportional to the localized waveform slope: $\Delta t \approx \Delta V / |\partial_t f(t_k)|$.

\emph{Assumption} (Small-jitter local linearization): The combined analog and quantization perturbation is small enough that each threshold crossing remains within a neighborhood where the local inverse map of $f(t;\bm{\theta}) = V$ is well-approximated by its linear term. Equivalently, the local slew rate $\partial_t f(t;\bm{\theta})$ does not vary appreciably across the induced timing jitter, so higher-order curvature corrections to $\Delta t \approx \Delta V / |\partial_t f(t_k)|$ are negligible.

For the diagonal-covariance approximation underlying the closed-form Gaussian Maximum Likelihood Estimator (MLE), any pair of crossings that is represented by separate scalar variances should have negligible cross-covariance; under the Gaussian approximation adopted below, this is equivalent to treating those crossings as approximately independent. This requirement applies both to neighboring thresholds on the same monotonic branch and, when a single threshold pair is modeled by two scalar denominators as in Section~\ref{sec:pair_general}, to the rising and falling crossings generated by that same threshold. Any physical readout chain inherently behaves as a finite-bandwidth low-pass filter, conventionally characterized by its $-3\,\mathrm{dB}$ analog bandwidth $B$ $[\mathrm{Hz}]$ (the frequency at which the transmitted power falls to one half, equivalently the amplitude to $1/\sqrt{2}$, of its low-frequency value). Because the output autocorrelation function is the Fourier transform of the filtered noise power spectrum, a spectrum whose effective width is $O(B)$ induces a characteristic autocorrelation timescale $\tau_c = O(1/B)$. The precise constant depends on the effective transfer function and on the chosen definition of correlation time; below we use the conservative engineering proxy $\tau_c \approx 1/(2B)$~\cite{meyer2002noise,chen1982minimum,anthonys2021jitter}. Thresholds whose crossings occur too close in time on the same monotonic branch may therefore fall within the same autocorrelation window and inherit correlated macroscopic noise fluctuations.

To rigorously preserve the diagonal-theory Fisher Information additivity while maintaining the strict ordering of the observation cone $\Omega_K$, thresholds must maintain sufficient vertical clearance. For adjacent thresholds $V_i, V_j$ with same-branch crossing times $t_i, t_j$, defining the supremum of the local slew rate $L_{ij}(\bm{\theta}) \triangleq \sup_{t \in [t_i, t_j]} |\partial_t f(t;\bm{\theta})|$, the temporal independence condition ($|t_i - t_j| \ge \tau_c$) translates via the Mean Value Theorem into a conservative sufficient diagonal-covariance design condition:
\begin{equation}
  |V_i - V_j| \ge L_{ij}(\bm{\theta}) \cdot \tau_c.
  \label{eq:autocorrelation}
\end{equation}
Equation~\eqref{eq:autocorrelation} should be read as a conservative decorrelation diagnostic for the diagonal surrogate, not as a theorem that same-branch crossings become statistically independent once the inequality is met. In a real readout chain, crossing-time perturbations are functionals of the same colored analog noise waveform, and any non-negligible off-diagonal terms should be retained in a full covariance matrix $\bm{\Sigma}$ when quantitative accuracy is required. The diagonal form is retained here because it yields the additive threshold-wise Fisher structure and closed-form stationarity equations; residual correlation error is included in the practical misspecification budget assessed in Section~\ref{sec:delta_validation}.
This condition controls same-branch cross-threshold correlations only. If one also wishes to use a diagonal covariance for the two crossings generated by a single threshold $V_n$, the pair gap must separately satisfy $t_{f,n}-t_{r,n} \ge \tau_c$; otherwise that pair should be retained as a correlated $2 \times 2$ covariance block even when Eq.~\eqref{eq:autocorrelation} holds. Because $L_{ij}(\bm{\theta})$ depends on the unknown event parameters, practical hardware design uses a design-point or worst-case upper bound over the intended operating range; Section~\ref{sec:partial_trigger} later introduces the corresponding worst-case quantity $L_{ij}^{\max}$ for multi-event design.
Under this bandwidth-aware diagonal approximation, the total independent temporal variance $\sigma_k^2$ for the $k$-th ordered crossing correctly aggregates the slope-projected analog variance and the intrinsic digital quantization jitter of the TDC, $\sigma_{\mathrm{TDC}}^2 = \mathrm{LSB}^2/12$ $[\mathrm{s}^2]$, where $\mathrm{LSB}$ denotes the TDC least-significant-bit time resolution (timing bin width; see Appendix~\ref{app:quantization_error}). Here, $t_{k,\mathrm{true}}$ denotes the exact physical crossing instant in the absence of stochastic jitter (i.e., the deterministic solution of $f(t;\bm{\theta}) = V_{\ell(k)}$):
\begin{equation}
  \sigma_k^2 = \sigma_{\mathrm{TDC}}^2 + \frac{\sigma_{\mathrm{th}}^2 + \kappa V_{\ell(k)}}{\bigl[\partial_t f(t_{k,\mathrm{true}};\bm{\theta})\bigr]^2}.
  \label{eq:variance}
\end{equation}
Equation~\eqref{eq:variance} is written in the baseline-free convention of Section~\ref{sec:model}; for models with an explicit deterministic baseline, the signal-dependent Poisson numerator is replaced by $\sigma_{\mathrm{th}}^2 + \kappa(V_{\ell(k)}-b)$.

% =========================================================================
%  IV. PARAMETER ESTIMATION AND MANIFOLD OPTIMIZATION
% =========================================================================
\section{Parameter Estimation}
\label{sec:estimation}

\subsection{Weighted Maximum Likelihood Formulation}
\label{sec:mle}

Under the Central Limit Theorem applied to high-photon-count filtered Poisson processes, together with the diagonal-covariance and pseudo-Gaussian quantization approximations stated below, the observation vector is modeled approximately as $\mathbf{t} = S(\bm{\theta}) + \bm{\epsilon}$, with $\bm{\epsilon}\sim\mathcal{N}(\mathbf{0},\bm{\Sigma})$, where $\mathcal{N}$ denotes the multivariate normal distribution, and diagonal covariance $\bm{\Sigma}=\operatorname{diag}(\sigma_1^2,\dots,\sigma_K^2)$.

\emph{Assumption} (Gaussian regime): The photon count per scintillation event is sufficiently large that the filtered Poisson process at each crossing can be well-approximated by a Gaussian random variable~\cite{hero1991timing}. Quantitatively, this requires the mean number of photoelectrons contributing to the signal at each threshold level $V$ to satisfy $N_{\mathrm{pe}}(V) \gg 1$. For typical scintillation detectors at 511\,keV, $N_{\mathrm{pe}} \sim \mathcal{O}(10^3$--$10^4)$, typically satisfying this condition~\cite{thompson2013measurement,mao2013crystal}; for low-energy events ($\lesssim 50$\,keV), the Gaussian approximation may degrade and non-Gaussian estimators should be considered.

\emph{Assumption} (Diagonal covariance): The off-diagonal elements of $\bm{\Sigma}$ are negligible for every pair of crossings that is treated as separate scalar observations in the Gaussian likelihood. A sufficient set of conditions is that Eq.~\eqref{eq:autocorrelation} holds for adjacent thresholds on each monotonic branch and, for every threshold pair modeled diagonally, the within-pair separation satisfies $t_{f,n}-t_{r,n} \ge \tau_c$.

\emph{Scope of this approximation.}  Violating the diagonal-covariance condition does not invalidate the deterministic embedding, the implicit-crossing Jacobian, or identifiability of the MVT forward map.  It changes the stochastic weighting model.  With correlated threshold errors, the Gaussian likelihood and Gauss--Newton estimator remain well defined after replacing the diagonal covariance by a full covariance matrix $\bm{\Sigma}$ and using $\mathbf{W}=\bm{\Sigma}^{-1}$.  The subsequent bias-projection and Schur-complement arguments also remain structurally valid with this full precision matrix.  What is special to the diagonal/independent approximation is the simple per-crossing denominator form in Eq.~\eqref{eq:variance}, the additive per-threshold Fisher decomposition, and the closed-form scalar threshold-design rules derived from those diagonal weights.  Thus the diagonal theory should be read as a tractable closed-form limit; when adjacent same-branch crossings or same-threshold rising/falling pairs are measurably correlated, the same framework should be evaluated with empirical or model-based off-diagonal covariance terms.

\emph{Assumption} (Pseudo-Gaussian quantization): The underlying TDC quantization error is modeled as uniform with variance $\sigma_{\mathrm{TDC}}^2 = \mathrm{LSB}^2/12$, as derived in Appendix~\ref{app:quantization_error} and consistent with the classical high-resolution quantization model~\cite{bennett1948spectra}. This requires that the true sub-bin crossing phase carry no systematic bias or clock-locked structure across a TDC bin, so midpoint assignment is an adequate first-order description. Modeling that uniform term as an additive zero-mean Gaussian noise is an additional approximation, justified when the TDC least-significant-bit resolution is small relative to the analog timing jitter so that the total timing error remains well-approximated by a Gaussian. If the LSB is comparable to the analog jitter, the exact convolution of analog jitter with a bounded uniform quantization law should replace the pseudo-Gaussian surrogate; the variance-only form is retained here only for the closed-form Fisher and design equations.

Under these approximations, the Gaussian log-likelihood takes the form~\cite{kay1993estimation}:
\begin{equation}
  \ln L(\bm{\theta}\mid\mathbf{t})
  = -\frac{1}{2}\sum_{k=1}^{K}
  \left[\frac{\bigl(t_k - t_k(\bm{\theta})\bigr)^2}{\sigma_k^2}
    + \ln(2\pi\sigma_k^2)\right].
  \label{eq:loglik}
\end{equation}
This likelihood retains parameter dependence through both the mean map $t_k(\bm{\theta})$ and, in principle, the variance terms $\sigma_k^2(\bm{\theta})$.

In standard scintillation dynamics (the High-SNR regime), covariance variations contribute only a higher-order correction to the Fisher Information. Appendix~\ref{app:slepian} shows that, under the uniform noise scaling $\sigma_{\mathrm{th}}^2,\kappa,\sigma_{\mathrm{TDC}}^2 = O(\epsilon^2)$, the mean-shift term in the Slepian-Bangs formula scales as $O(\epsilon^{-2})$ whereas the covariance term scales as $O(1)$~\cite{slepian1958some,abeida2019slepian,el2024full}. The covariance contribution is therefore smaller by a factor $O(\epsilon^2)$ and is asymptotically negligible. Treating $\sigma_k^2$ as locally constant when differentiating the likelihood therefore incurs only a higher-order error.

Consequently, within this locally constant-covariance high-SNR approximation, maximizing the Gaussian log-likelihood is equivalent up to $\bm{\theta}$-independent normalization terms to minimizing a geometrically weighted non-linear least-squares objective $\mathcal{E}_{\mathrm{WLS}}$:
\begin{equation}
  \mathcal{E}_{\mathrm{WLS}}(\bm{\theta})
  = \frac{1}{2}\,\Delta\mathbf{t}(\bm{\theta})^T\,\mathbf{W}\,\Delta\mathbf{t}(\bm{\theta})
  \label{eq:cost}
\end{equation}
where $\Delta\mathbf{t}(\bm{\theta})\triangleq\mathbf{t}-\mathbf{t}(\bm{\theta})\in\mathbb{R}^K$ is the residual error, and the precision matrix is:
\begin{equation}
  \mathbf{W} \triangleq \bm{\Sigma}^{-1} = \operatorname{diag}\!\left(\frac{1}{\sigma_1^2},\,\dots,\,\frac{1}{\sigma_K^2}\right).
  \label{eq:weight_matrix}
\end{equation}
\emph{Physical Interpretation:} $\mathbf{W}$ operates as an intrinsic \emph{precision-weighting mechanism}: it suppresses timestamps residing on shallow, highly stochastic waveform tails, dynamically concentrating the estimation algorithm's focus onto steep, low-noise geometric features.

\subsection{Gauss--Newton Solver}
\label{sec:gauss_newton}

To navigate the non-Euclidean parameter manifold $\Theta_V$ toward $\nabla_{\bm{\theta}}\mathcal{E}_{\mathrm{WLS}}=\mathbf{0}$, we project the nonlinear mapping onto its local linear tangent space around the current estimate $\bm{\theta}^{(\ell)}$ following the Gauss--Newton approximation~\cite{nocedal2006numerical}:
\begin{equation}
  \mathbf{t}(\bm{\theta})
  \approx \mathbf{t}\!\left(\bm{\theta}^{(\ell)}\right)
  + \mathbf{J}\!\left(\bm{\theta}^{(\ell)}\right)
  \bigl(\bm{\theta}-\bm{\theta}^{(\ell)}\bigr).
  \label{eq:gn_linearize}
\end{equation}
Define the current residual and parameter increment by $\Delta\mathbf{t}^{(\ell)} \triangleq \mathbf{t}-\mathbf{t}(\bm{\theta}^{(\ell)})$ and $\Delta\bm{\theta} \triangleq \bm{\theta}-\bm{\theta}^{(\ell)}$. Differentiating the linearized cost then yields the Gauss-Newton normal equations:
\begin{equation}
  \underbrace{\mathbf{J}^T\mathbf{W}\mathbf{J}}_{\displaystyle\approx\,\mathbf{H}}
  \;\Delta\bm{\theta}
  = \mathbf{J}^T\mathbf{W}\,\Delta\mathbf{t}^{(\ell)},
  \qquad
  \bm{\theta}^{(\ell+1)}=\bm{\theta}^{(\ell)}+\Delta\bm{\theta}.
  \label{eq:gauss_newton}
\end{equation}
By replacing the exact Hessian of $\mathcal{E}_{\mathrm{WLS}}$ with the Gauss-Newton proxy $\mathbf{H}=\mathbf{J}^T\mathbf{W}\mathbf{J}$, the algorithm avoids explicit second-order tensor derivatives, which is attractive for high-throughput or FPGA-oriented implementations. Because $\mathbf{W} \succ 0$ and $\mathbf{H}$ is positive semidefinite by construction, $\mathbf{H}$ is positive definite at an iterate whenever $\mathbf{J}(\bm{\theta}^{(\ell)})$ has full column rank. The global immersion condition guarantees this rank property on the admissible manifold $\Theta_V$, but the standard local Gauss--Newton convergence statement additionally assumes that $S$ is $C^2$ with locally Lipschitz Jacobian and that the iterates remain in a neighborhood where the linearization is accurate. Under those regularity conditions the method enjoys the usual local Gauss--Newton convergence guarantees; in the zero-residual limit the rate becomes quadratic. If an unconstrained step leaves $\Theta_V$ or $\mathbf{H}$ becomes ill-conditioned, damping or line-search safeguards may be required.

% =========================================================================
%  V. CRLB AND SYSTEMATIC BIAS
% =========================================================================
\section{Performance Bounds and Model Misspecification}
\label{sec:misspecification}

The preceding sections established the statistical estimator and its convergence properties under the assumption that the analytical model $f(t;\bm{\theta})$ faithfully represents the true physical waveform. A central question is how estimator performance degrades when the model deviates from reality. In practice, no analytical model is perfect---real detectors exhibit SiPM gain saturation, baseline pile-up, optical cross-talk, and other distortions~\cite{rosado2015characterization,gola2014sipm,couce2008parametrization}. This section quantifies the resulting performance degradation.

When the assumed analytical model $f(t;\bm{\theta})$ accurately captures the underlying physics and the high-SNR approximation of Appendix~\ref{app:slepian} is valid, the covariance of an unbiased estimator is, to leading order, bounded by the model-based Cram\'{e}r-Rao Lower Bound (CRLB). In that regime, the covariance-derivative term in the Slepian-Bangs formula is asymptotically negligible, and the Fisher Information Matrix (FIM) is correspondingly approximated by its geometric projection: $\mathcal{I}(\bm{\theta}) \approx \mathbf{J}^T\mathbf{W}\mathbf{J}$.

In laboratory reality, however, idealized analytical models inevitably deviate from physical waveforms:
\begin{equation}
  y_{\mathrm{true}}(t) = f(t; \bm{\theta}_0) + \varepsilon(t).
\end{equation}
The residual $\varepsilon(t)$ encapsulates unmodeled deterministic physics: Silicon Photomultiplier (SiPM) gain saturation, non-linear baseline pile-up, or optical cross-talk. Unlike quantum noise, these \emph{deterministic deviations} cannot be smoothed away by ensemble averaging over infinitely many events.
Throughout this section, expectations are taken over the remaining zero-mean stochastic timing noise only; the mismatch field $\varepsilon(t)$ is treated as a fixed deterministic perturbation.

\emph{Assumption} (Perturbative and locally flat mismatch near each crossing): The misspecification remains small and slowly varying in a neighborhood of every ideal ordered crossing $t_k$, specifically $|\varepsilon(t_k)| \ll |\partial_t f(t_k;\bm{\theta}_0)|\,\Delta t_{\mathrm{char}}$ and $|\partial_t \varepsilon(t_k)| \ll |\partial_t f(t_k;\bm{\theta}_0)|$, where $\Delta t_{\mathrm{char}}$ is a characteristic temporal width of the pulse (e.g., the full-width at half-maximum). These inequalities keep the perturbation within the local linearization regime around $\bm{\theta}_0$ and justify neglecting higher-order terms in the crossing perturbation.

\textbf{Geometric Projection of Mismatch:} Under this local regime, the vertical voltage deviation is linearly projected into a deterministic horizontal temporal shift. Writing the perturbed crossing equation as $f(t_k + \Delta t_{\mathrm{bias},k};\bm{\theta}_0) + \varepsilon(t_k + \Delta t_{\mathrm{bias},k}) = V_{\ell(k)}$, where $t_k$ denotes the ideal $k$-th ordered crossing satisfying $f(t_k;\bm{\theta}_0)=V_{\ell(k)}$, and expanding to first order gives $\bigl[\partial_t f(t_k;\bm{\theta}_0) + \partial_t \varepsilon(t_k)\bigr]\,\Delta t_{\mathrm{bias},k} + \varepsilon(t_k) \approx 0$. Under the local-flatness condition $|\partial_t \varepsilon(t_k)| \ll |\partial_t f(t_k;\bm{\theta}_0)|$, this reduces to the elementwise temporal bias (Appendix~\ref{app:mismatch_proof}):
\begin{equation}
  \Delta t_{\mathrm{bias},k} \approx -\frac{\varepsilon(t_k)}{\partial_t f(t_k;\bm{\theta}_0)}.
  \label{eq:dt_bias_scalar}
\end{equation}
Assembling all $K$ crossings into a vector,
\begin{equation}
  \Delta\mathbf{t}_{\mathrm{bias}} \;\triangleq\; [\Delta t_{\mathrm{bias},1},\;\dots,\;\Delta t_{\mathrm{bias},K}]^T \;\in\;\mathbb{R}^K.
  \label{eq:dt_bias_vec}
\end{equation}

As analytically derived via the expectation of the perturbed Gauss-Newton score equation in Appendix~\ref{app:mismatch_proof}, and under the same locally constant-covariance approximation used in Section~\ref{sec:mle}, the Quasi-Maximum Likelihood Estimator absorbs these geometric temporal distortions, projecting them directly onto the parameter manifold to form a leading-order systematic bias vector $\bm{\beta} \triangleq \mathbb{E}[\hat{\bm{\theta}}] - \bm{\theta}_0$:
\begin{equation}
  \bm{\beta} \approx \bigl(\mathbf{J}^T \mathbf{W} \mathbf{J}\bigr)^{-1} \mathbf{J}^T \mathbf{W} \Delta\mathbf{t}_{\mathrm{bias}}.
  \label{eq:bias_vector}
\end{equation}

To convert the bias expression into a local MSE surrogate bound, we make two standard local assumptions.

\emph{Assumption} (Locally constant bias): The systematic bias vector $\bm{\beta}$ is approximately independent of $\bm{\theta}$ in a neighborhood of $\bm{\theta}_0$, i.e., $\partial\bm{\beta}/\partial\bm{\theta} \approx \mathbf{0}$. This holds when the model misspecification $\varepsilon(t)$ varies smoothly and slowly across the relevant crossing neighborhoods, so that the geometric projection Eq.~\eqref{eq:bias_vector} remains nearly constant under small parameter perturbations.

\emph{Assumption} (Model-based FIM validity): The FIM $\mathcal{I}(\bm{\theta}_0) = \mathbf{J}^T\mathbf{W}\mathbf{J}$ is computed from the \emph{assumed} model $f$, not from the true waveform $f + \varepsilon$. This substitution is valid to leading order when the mismatch remains uniformly small over the crossing neighborhoods relevant to the estimator, so that the model-based score function is close to the true score. For large misspecification, the usual sandwich covariance $\mathcal{I}^{-1}\mathcal{I}_{\mathrm{true}}\mathcal{I}^{-1}$, where $\mathcal{I}_{\mathrm{true}}$ denotes the true score covariance, would be required; that regime is not pursued here.

Under these assumptions, the matrix identity $\mathrm{MSE}(\hat{\bm{\theta}})=\mathrm{Cov}(\hat{\bm{\theta}})+\bm{\beta}\bm{\beta}^T$, where $\mathrm{Cov}(\hat{\bm{\theta}})$ is the covariance about $\mathbb{E}[\hat{\bm{\theta}}]$, together with the model-based CRLB for the covariance component yields the following leading-order local model-based misspecified MSE surrogate bound in the Loewner partial order (i.e., $\mathbf{A}\succeq\mathbf{B}$ means $\mathbf{A}-\mathbf{B}$ is positive semidefinite):
\begin{equation}
  \mathrm{MSE}(\hat{\bm{\theta}}) \succeq \mathcal{I}^{-1}(\bm{\theta}_0) + \bm{\beta}\bm{\beta}^T.
  \label{eq:mse_bound}
\end{equation}
This surrogate bound is intended within the same small-misspecification, locally constant-covariance regime that underlies Eq.~\eqref{eq:bias_vector}; outside that regime the sandwich covariance based on $\mathcal{I}_{\mathrm{true}}$ is the relevant object.

\emph{Physical Consequence:} For baseline-free pulse models whose signal component starts from zero at onset, progressively pushing thresholds toward the baseline ($V \to 0$; equivalently $V \to b^+$ for explicit-baseline models) drives the rising-edge crossing into a region of steep temporal slope, thereby reducing the analog timing variance. However, the paired falling-edge crossing simultaneously moves into the far tail where $\partial_t f \to 0$, so the temporal mismatch projection $\Delta t_{\mathrm{bias},k}\propto \varepsilon/\partial_t f$ becomes increasingly ill-conditioned and can diverge unless the mismatch decays commensurately with the slope. The attainable error is therefore governed by a structural trade-off between stochastic precision and deterministic misspecification.

% =========================================================================
%  VI. PAIRED SAMPLING AND EFFECTIVE INFORMATION
% =========================================================================
\section{Paired-Crossing Information Structure}
\label{sec:paired}

Having established both the well-specified Cram\'{e}r-Rao bound and the misspecified bias-variance decomposition for the full parameter vector, we now address the central practical question: how much information does each hardware threshold contribute specifically to the timing parameter of interest, after accounting for the unavoidable uncertainty in nuisance parameters such as energy amplitude?

\textbf{Physical Motivation:} MVT sampling imposes a geometric coupling: each admissible hardware threshold $V$ with $b < V < p(\bm{\theta})$ generates a rising/falling observation pair $(t_r, t_f)$. In many timing experiments the parameter of interest is a specific scalar quantity (e.g., the absolute photon arrival time $\vartheta$), while the remaining parameters (such as the deposited-energy amplitude) are treated as nuisance variables. Because these quantities are estimated jointly from the same coupled pair, nuisance uncertainty propagates directly into the attainable precision of the target estimate.

\emph{Assumption} (Single-pair diagonal covariance): Throughout this section and Appendix~\ref{app:schur}, the diagonal single-pair formulas are used only for admissible thresholds $V$ whose rising and falling crossings satisfy $t_f(V;\bm{\theta}) - t_r(V;\bm{\theta}) \ge \tau_c$, so the within-pair covariance can be approximated as diagonal. If this separation condition fails, the same two observations must be treated with their full $2 \times 2$ covariance block.

\subsection{Single-Pair Fisher Information}
\label{sec:pair_general}

To explicitly quantify this leakage, we define the pairwise waveform sensitivities of the $j$-th physical parameter at the two edges:
\begin{equation}
  g_{j,r} \triangleq \frac{\partial f(t_r;\bm{\theta})}{\partial \theta_j},
  \qquad
  g_{j,f} \triangleq \frac{\partial f(t_f;\bm{\theta})}{\partial \theta_j}.
  \label{eq:pair_sensitivity}
\end{equation}
Aggregating the Poisson variance, thermal floor, and the slope-projected TDC jitter, we define the composite noise denominators for each edge (see Eq.~\eqref{eq:pair_denominators} in Appendix~\ref{app:schur}):
\begin{equation}
  \begin{aligned}
    D_r & \triangleq \sigma_{\mathrm{th}}^2 + \kappa (V-b) + \sigma_{\mathrm{TDC}}^2 \bigl[\partial_t f(t_r;\bm{\theta})\bigr]^2, \\
    D_f & \triangleq \sigma_{\mathrm{th}}^2 + \kappa (V-b) + \sigma_{\mathrm{TDC}}^2 \bigl[\partial_t f(t_f;\bm{\theta})\bigr]^2.
  \end{aligned}
\end{equation}
Here $b$ denotes any deterministic baseline offset; in the baseline-free convention of the general framework one simply sets $b=0$. Thus the Poisson shot-noise term depends on the signal component $V-b$ at the crossing, consistent with Eq.~\eqref{eq:poisson_voltage}.
By direct matrix multiplication (Appendix~\ref{app:schur}), the slope factors from the Jacobian cancel against the slope-projected analog-noise terms in the weights. The resulting diagonal single-pair contribution to the FIM is:
\begin{equation}
  \mathcal{I}_{jl}^{(\mathrm{pair})}(V)
  = \frac{g_{j,r}\,g_{l,r}}{D_r} + \frac{g_{j,f}\,g_{l,f}}{D_f}.
  \label{eq:pair_fim}
\end{equation}
Equation~\eqref{eq:pair_fim} is therefore the diagonal closed-form expression for a single threshold pair.

\subsection{Effective Information and Nuisance Projection}
\label{sec:master_eq}

In PET timing applications, the primary physical observable is the photon arrival time, while all other parameters (particularly the pulse amplitude $A$, which encodes the deposited energy) serve as nuisance variables: they must be jointly estimated from the same data but are not the quantity of primary physical interest~\cite{hero1991a,ruizgonzalez2016joint,mohammaddjafari2004on}. \emph{Notation:} $\vartheta$ denotes the generic scalar target parameter, and the framework applies identically to any component of $\bm{\theta}$ or a scalar functional thereof (cf.~Section~\ref{sec:biexp_paired}).

To rigorously isolate the effective information belonging to $\vartheta$, we partition the parameter vector $\bm{\theta}=[\bm{\eta}^T,\vartheta]^T$, separating $\vartheta$ from the nuisance vector $\bm{\eta}\in\mathbb{R}^{M-1}$. Under this partition, the $M\times M$ single-pair FIM from Eq.~\eqref{eq:pair_fim} decomposes into the block structure:
\begin{equation}
  \mathcal{I}^{(\mathrm{pair})}(V)
  = \begin{bmatrix}
    \bm{\mathcal{I}}_{\bm{\eta}\bm{\eta}}   & \bm{\mathcal{I}}_{\bm{\eta}\vartheta} \\[4pt]
    \bm{\mathcal{I}}_{\bm{\eta}\vartheta}^T & \mathcal{I}_{\vartheta\vartheta}
  \end{bmatrix},
  \label{eq:pair_fim_block}
\end{equation}
Defining the nuisance-edge sensitivity vectors $\mathbf{g}_{\bm{\eta},r} \triangleq [g_{1,r},\,\dots,\allowbreak\,g_{M-1,r}]^T$, $\mathbf{g}_{\bm{\eta},f} \triangleq [g_{1,f},\,\dots,\allowbreak\,g_{M-1,f}]^T \in \mathbb{R}^{M-1}$, each sub-block is constructed explicitly from Eq.~\eqref{eq:pair_fim}:
\begin{equation}
  \bm{\mathcal{I}}_{\bm{\eta}\bm{\eta}}
  = \frac{\mathbf{g}_{\bm{\eta},r}\,\mathbf{g}_{\bm{\eta},r}^T}{D_r}
  + \frac{\mathbf{g}_{\bm{\eta},f}\,\mathbf{g}_{\bm{\eta},f}^T}{D_f}
  \;\in\;\mathbb{R}^{(M-1)\times(M-1)},
  \label{eq:I_eta_eta}
\end{equation}
\begin{equation}
  \bm{\mathcal{I}}_{\bm{\eta}\vartheta}
  = \frac{g_{\vartheta,r}\,\mathbf{g}_{\bm{\eta},r}}{D_r}
  + \frac{g_{\vartheta,f}\,\mathbf{g}_{\bm{\eta},f}}{D_f}
  \;\in\;\mathbb{R}^{(M-1)\times 1},
  \label{eq:I_eta_tau}
\end{equation}
\begin{equation}
  \mathcal{I}_{\vartheta\vartheta}
  = \frac{g_{\vartheta,r}^{2}}{D_r}
  + \frac{g_{\vartheta,f}^{2}}{D_f}
  \;\in\;\mathbb{R},
  \label{eq:I_tau_tau}
\end{equation}
where $g_{\vartheta,r} \triangleq \partial f(t_r;\bm{\theta})/\partial\vartheta$ and $g_{\vartheta,f} \triangleq \partial f(t_f;\bm{\theta})/\partial\vartheta$ are the target-parameter sensitivities at the two edges.

By profiling out the nuisance uncertainty through the generalized Schur complement, we define the single-pair \emph{generalized nuisance-profiled information score} for the target parameter $\vartheta$ (Appendix~\ref{app:schur}) as:
\begin{equation}
  \begin{aligned}
    \Delta\mathcal{I}_{\mathrm{eff}}^{(\vartheta)}(V)
     & \;\triangleq\; \mathcal{I}_{\vartheta\vartheta}
    - \bm{\mathcal{I}}_{\bm{\eta}\vartheta}^T\,
    \bm{\mathcal{I}}_{\bm{\eta}\bm{\eta}}^{+}\,
    \bm{\mathcal{I}}_{\bm{\eta}\vartheta}              \\
     & = \min_{\bm{\gamma}\in\mathbb{R}^{M-1}}
    \left[
      \begin{aligned}
        \mathcal{I}_{\vartheta\vartheta}
         & - 2\,\bm{\gamma}^{T}\bm{\mathcal{I}}_{\bm{\eta}\vartheta}         \\
         & + \bm{\gamma}^{T}\bm{\mathcal{I}}_{\bm{\eta}\bm{\eta}}\bm{\gamma}
      \end{aligned}
      \right].
  \end{aligned}
  \label{eq:schur_general}
\end{equation}
Here $(\cdot)^{+}$ denotes the Moore--Penrose inverse. The minimum exists because
$\bm{\mathcal{I}}_{\bm{\eta}\vartheta}$ lies in the range of
$\bm{\mathcal{I}}_{\bm{\eta}\bm{\eta}}$: indeed,
$\bm{\mathcal{I}}_{\bm{\eta}\vartheta}$ is a weighted linear combination of
$\mathbf{g}_{\bm{\eta},r}$ and $\mathbf{g}_{\bm{\eta},f}$, while
$\operatorname{range}(\bm{\mathcal{I}}_{\bm{\eta}\bm{\eta}})
  = \operatorname{span}\{\mathbf{g}_{\bm{\eta},r},\mathbf{g}_{\bm{\eta},f}\}$.
Equivalently, the minimizer set is the affine space
$\bm{\gamma}\in
  \bm{\mathcal{I}}_{\bm{\eta}\bm{\eta}}^{+}\bm{\mathcal{I}}_{\bm{\eta}\vartheta}
  + \ker(\bm{\mathcal{I}}_{\bm{\eta}\bm{\eta}})$.

\emph{Remark (Rank condition and single-pair degeneracy).}
The single-pair nuisance block $\bm{\mathcal{I}}_{\bm{\eta}\bm{\eta}}$ in Eq.~\eqref{eq:I_eta_eta} is the sum of two rank-one matrices and therefore has $\operatorname{rank}(\bm{\mathcal{I}}_{\bm{\eta}\bm{\eta}}) \le 2$. When $M - 1 > 2$ (e.g., the five-parameter bi-exponential model), a single threshold pair cannot identify all nuisance parameters in the ordinary inverse sense; nevertheless, the generalized Schur complement~\eqref{eq:schur_general} remains well-defined as an algebraic profiled-information score. In fact, if $M-1 \ge 2$ and the two nuisance sensitivity vectors $\mathbf{g}_{\bm{\eta},r}$ and $\mathbf{g}_{\bm{\eta},f}$ are linearly independent, one can choose $\bm{\gamma}$ so that $\bm{\gamma}^{T}\mathbf{g}_{\bm{\eta},r} = g_{\vartheta,r}$ and $\bm{\gamma}^{T}\mathbf{g}_{\bm{\eta},f} = g_{\vartheta,f}$, which drives the minimum in Eq.~\eqref{eq:schur_general} to zero. Thus, in the general $M$-parameter setting, a standalone finite target bound cannot be inferred from a single pair; after aggregation, one must verify both strict positivity of the total nuisance block and positive profiled target information. If $\bm{\mathcal{I}}_{\bm{\eta}\bm{\eta}}$ is itself invertible, Eq.~\eqref{eq:schur_general} reduces to the ordinary Schur complement and recovers the usual profiled CRLB.

\emph{Assumption} (Global nuisance and target identifiability): Throughout Sections~\ref{sec:master_eq}--\ref{sec:general_opt}, we require both (i) the total nuisance block $\bm{\mathcal{I}}_{\bm{\eta}\bm{\eta}}^{\mathrm{tot}}$ to be strictly positive definite ($\bm{\mathcal{I}}_{\bm{\eta}\bm{\eta}}^{\mathrm{tot}} \succ 0$), and (ii) the scalar profiled information for the target to be strictly positive, $\Delta\mathcal{I}_{\mathrm{eff}}^{(\vartheta),\mathrm{tot}} > 0$.  The first condition makes the ordinary Schur complement well-defined and, by rank counting, requires $N \ge \lceil (M-1)/2 \rceil$ threshold pairs at generically distinct voltages.  The second condition is the one that makes the CRLB/MSE bound for $\vartheta$ finite; equivalently, the full Fisher matrix must identify the target direction after nuisance profiling.  A necessary generic rank count is $2N \ge M$.  Thus, in the five-parameter bi-exponential amplitude-target problem, two threshold pairs can make the $4\times4$ nuisance block full rank, but at least three generic active pairs---or direct verification that $\Delta\mathcal{I}_{\mathrm{eff}}^{(A)} > 0$---are needed for a finite amplitude bound.

Accordingly, Eq.~\eqref{eq:schur_general} should be interpreted as the isolated single-pair profiled-information score entering the joint variational analysis below, not as a standalone finite MSE bound for the general $M$-parameter problem.

\subsection{Superadditivity of Joint Threshold Design}
\label{sec:synergy}

Equation~\eqref{eq:schur_general} characterizes the isolated profiled contribution of a single threshold pair. We now consider the joint information from $N$ threshold pairs at voltages $V_1,\dots,V_N$.

\emph{Assumption} (Inter-pair stochastic independence): When using the additive diagonal-information form, the bandwidth-separation condition Eq.~\eqref{eq:autocorrelation} is enforced across distinct thresholds, so crossing-time errors from different threshold pairs are approximately independent. This assumption concerns inter-pair additivity only; if the rising and falling crossings within one threshold pair remain correlated, that pair should be handled with its own $2 \times 2$ covariance block rather than Eq.~\eqref{eq:pair_fim}. Under this condition the total FIM is the sum of the per-pair FIM blocks.

Write $\Delta\mathcal{I}_{\mathrm{eff}}^{(\vartheta),(n)}(V_n) \triangleq \Delta\mathcal{I}_{\mathrm{eff}}^{(\vartheta)}(V_n)$ for the single-pair generalized profiled-information score (Eq.~\eqref{eq:schur_general}) evaluated at the $n$-th threshold, and let the \emph{total} effective information from joint estimation over all $N$ pairs be
\begin{equation}
  \Delta\mathcal{I}_{\mathrm{eff}}^{(\vartheta),\mathrm{tot}}
  \;\triangleq\;
  \mathcal{I}_{\vartheta\vartheta}^{\mathrm{tot}}
  - (\bm{\mathcal{I}}_{\bm{\eta}\vartheta}^{\mathrm{tot}})^T\,
  (\bm{\mathcal{I}}_{\bm{\eta}\bm{\eta}}^{\mathrm{tot}})^{-1}\,
  \bm{\mathcal{I}}_{\bm{\eta}\vartheta}^{\mathrm{tot}},
  \label{eq:eff_info_tot}
\end{equation}
where, writing $\bm{\mathcal{I}}_{\bm{\eta}\bm{\eta}}^{(n)}$, $\bm{\mathcal{I}}_{\bm{\eta}\vartheta}^{(n)}$, and $\mathcal{I}_{\vartheta\vartheta}^{(n)}$ for the sub-blocks (Eqs.~\eqref{eq:I_eta_eta}--\eqref{eq:I_tau_tau}) evaluated at $V = V_n$, the summed FIM blocks are
\begin{equation}
  \begin{aligned}
    \mathcal{I}_{\vartheta\vartheta}^{\mathrm{tot}}      & = \textstyle\sum_{n=1}^{N} \mathcal{I}_{\vartheta\vartheta}^{(n)},      \\
    \bm{\mathcal{I}}_{\bm{\eta}\vartheta}^{\mathrm{tot}} & = \textstyle\sum_{n=1}^{N} \bm{\mathcal{I}}_{\bm{\eta}\vartheta}^{(n)}, \\
    \bm{\mathcal{I}}_{\bm{\eta}\bm{\eta}}^{\mathrm{tot}} & = \textstyle\sum_{n=1}^{N} \bm{\mathcal{I}}_{\bm{\eta}\bm{\eta}}^{(n)}
  \end{aligned}
  \label{eq:summed_blocks}
\end{equation}
(additivity holds by the inter-pair independence assumption above).

A key geometric insight emerges when multiple independent pairs are aggregated. If one were to mathematically eliminate nuisance effects sequentially, pair by pair, the total effective information would merely be the linear sum $\sum \Delta\mathcal{I}_{\mathrm{eff}}^{(\vartheta),(n)}$.

However, this linear assumption is generally invalid as an information-combination rule. \textbf{Physical Phase-Space Interpretation:} Each threshold pair interrogates the amplitude-time covariance at a different vertical elevation along the pulse envelope. Governed by disparate local slew rates, these sensitivity vectors $\mathbf{g}$ point in geometrically diverse directions in the parameter phase space, creating non-collinear uncertainty ellipses.

When aggregated by the global joint estimator, the geometric intersection of these rotated covariance structures can uncouple the amplitude-time degeneracy more efficiently than any single pair could. Mathematically, this synergistic rotation is captured by the superadditivity of generalized Schur complements, proven in Appendix~\ref{app:synergy} through the common variational representation of the per-pair and global effective information. Defining the corresponding leading-order target-bias proxy from all $N$ pairs as
\begin{equation}
  \beta_{\vartheta,\mathrm{tot}} \;\triangleq\; \bigl[(\mathbf{J}_{\mathrm{tot}}^T \mathbf{W}_{\mathrm{tot}} \mathbf{J}_{\mathrm{tot}})^{-1}\,\mathbf{J}_{\mathrm{tot}}^T \mathbf{W}_{\mathrm{tot}}\,\Delta\mathbf{t}_{\mathrm{bias}}^{\mathrm{tot}}\bigr]_{\vartheta},
  \label{eq:b_tau_tot}
\end{equation}
where $\mathbf{J}_{\mathrm{tot}}$, $\mathbf{W}_{\mathrm{tot}}$, and $\Delta\mathbf{t}_{\mathrm{bias}}^{\mathrm{tot}}$ are the vertically stacked Jacobian, block-diagonal weight matrix, and concatenated temporal bias vector over all $N$ threshold pairs, the corresponding leading-order model-based MSE proxy for $\vartheta$ from all $N$ pairs is defined as
\begin{align}
  \mathcal{B}_{\vartheta,\min}^{\mathrm{tot}}
   & \triangleq
  \bigl[\Delta\mathcal{I}_{\mathrm{eff}}^{(\vartheta),\mathrm{tot}}\bigr]^{-1} + \beta_{\vartheta,\mathrm{tot}}^{2} \nonumber \\
   & \le
  \left[ \sum_{n=1}^{N} \Delta\mathcal{I}_{\mathrm{eff}}^{(\vartheta),(n)}(V_n) \right]^{-1} + \beta_{\vartheta,\mathrm{tot}}^{2}.
  \label{eq:synergy_bound}
\end{align}
This shows that structurally heterogeneous threshold deployments need not behave as merely additive data collection; they can yield a superadditive information gain that suppresses parameter cross-coupling and lowers the corresponding leading-order model-based error proxy.

\emph{Remark (Equality condition).}
Equality in Eq.~\eqref{eq:synergy_bound} holds if and only if the global minimizer $\bm{\gamma}^{*} = \bm{A}_{\mathrm{tot}}^{-1}\,\mathbf{u}_{\mathrm{tot}}$ (Appendix~\ref{app:synergy}, Eq.~\eqref{eq:autonum:F8}) simultaneously minimizes every per-pair objective $\psi_n(\bm{\gamma})$ (Eq.~\eqref{eq:autonum:F3}). In the singular case this means
\begin{equation}
  \bm{\gamma}^{*}
  \in
  \bigcap_{n=1}^{N}\left(\bm{A}_n^{+}\mathbf{u}_n + \ker(\bm{A}_n)\right),
  \label{eq:synergy_equality_condition}
\end{equation}
not merely that a particular Moore--Penrose representative $\bm{A}_n^{+}\mathbf{u}_n$ matches across pairs. When each per-pair nuisance block is invertible ($\bm{A}_n \succ 0$), Eq.~\eqref{eq:synergy_equality_condition} reduces to $\bm{\gamma}_n^{*} \triangleq \bm{A}_n^{-1}\,\mathbf{u}_n = \bm{\gamma}^{*}$ for all $n$---i.e., all threshold pairs share an identical optimal nuisance coupling direction. Physically, this means that every pair's $(\vartheta,\bm{\eta})$ sensitivity vectors are equally aligned: the nuisance information provided by distinct pairs is perfectly collinear, leaving no geometric diversity to exploit. Strict superadditivity occurs precisely when the intersection condition in Eq.~\eqref{eq:synergy_equality_condition} fails.

\emph{Remark (Low-rank pairs).}
Because Eq.~\eqref{eq:schur_general} is stated directly in generalized-Schur-complement form, no separate low-rank exception is needed in the superadditivity bound~\eqref{eq:synergy_bound}. When a per-pair nuisance block happens to be invertible, Eq.~\eqref{eq:schur_general} reduces to the ordinary Schur complement. When it is singular, the generalized value remains the correct per-pair effective information, while the joint quantity $\Delta\mathcal{I}_{\mathrm{eff}}^{(\vartheta),\mathrm{tot}}$ still uses the ordinary inverse because the global nuisance identifiability assumption enforces $\bm{\mathcal{I}}_{\bm{\eta}\bm{\eta}}^{\mathrm{tot}} \succ 0$.

% =========================================================================
%  VII. GENERAL OPTIMAL THRESHOLD CONFIGURATION
% =========================================================================
\section{Optimal Threshold Design}
\label{sec:general_opt}

\subsection{Problem Formulation}
\label{sec:design_problem}

The superadditivity analysis of the preceding section established the leading-order model-based MSE proxy for the target parameter $\vartheta$ from $N$ threshold pairs (Eq.~\eqref{eq:synergy_bound}):
\begin{equation}
  \mathcal{B}_{\vartheta,\min}^{\mathrm{tot}}(\{V_n\})
  \;=\;
  \bigl[\Delta\mathcal{I}_{\mathrm{eff}}^{(\vartheta),\mathrm{tot}}(\{V_n\})\bigr]^{-1}
  + \beta_{\vartheta,\mathrm{tot}}^{2}(\{V_n\}).
  \label{eq:mse_objective}
\end{equation}
Both the effective Fisher Information $\Delta\mathcal{I}_{\mathrm{eff}}^{(\vartheta),\mathrm{tot}}$ and the systematic bias $\beta_{\vartheta,\mathrm{tot}}$ depend on the threshold configuration $\{V_1, \dots, V_N\}$ through the waveform geometry, the noise structure, and the model mismatch function $\varepsilon(t)$. Within the standing diagonal-covariance, transversality, and identifiability assumptions, these dependencies are fully determined by the pulse model $f(t;\bm{\theta})$ and the physical noise parameters $(\sigma_{\mathrm{th}}^2, \kappa, \sigma_{\mathrm{TDC}}^2)$.

\emph{Assumption} (Single-event design point): Throughout Sections~\ref{sec:design_problem}--\ref{sec:opt_derivation}, the threshold optimization is conditioned on a fixed representative parameter vector $\bm{\theta}$. Within this diagonal-covariance, transversality, and global-identifiability regime, the threshold dependence is fully determined by the pulse model, noise parameters, and mismatch field. Sections~\ref{sec:robustness} and \ref{sec:partial_trigger} then relax this single-design-point setting to parameter neighborhoods and event populations.

From an engineering design perspective, the central question is: \emph{given a fixed number $N$ of hardware comparators, what explicit threshold voltage assignment $\{V_1^*, \ldots, V_N^*\}$ minimizes the leading-order model-based MSE proxy for the target parameter $\vartheta$?} Answering this question transforms the abstract information-theoretic result into a concrete hardware prescription.

\textbf{Formal diagonal-covariance optimization problem.}
In the closed-form diagonal-information limit, the general optimal threshold design problem is:
\begin{equation}
  \boxed{\begin{aligned}
      \{V_n^*\}_{n=1}^N & = \arg\min_{\{V_n\}_{n=1}^{N}}\;
      \mathcal{B}_{\vartheta,\min}^{\mathrm{tot}}(\{V_n\})                \\[4pt]
      \text{s.t.}\quad
                        & b < V_1 < V_2 < \cdots < V_N < p(\bm{\theta}),  \\
                        & t_f(V_n;\bm{\theta}) - t_r(V_n;\bm{\theta}) \ge \tau_c,
      \quad n=1,\ldots,N,                                                   \\
                        & |V_i - V_j| \ge L_{ij}(\bm{\theta})\cdot\tau_c,
      \;\;\forall\;(i,j)\;\text{adjacent},
    \end{aligned}}
  \label{eq:general_opt_problem}
\end{equation}
where $b$ is the deterministic baseline offset (set to $0$ in the baseline-free convention), $p(\bm{\theta})$ is the pulse peak amplitude (Section~\ref{sec:model}), $L_{ij}(\bm{\theta})$ is the supremum slew rate between thresholds $i$ and $j$ (Section~\ref{sec:bandwidth}), and $\tau_c \approx 1/(2B)$ is the noise autocorrelation time. The pair-gap condition is the within-pair analogue of the same-branch spacing condition: the former keeps the rising/falling crossings generated by a single threshold far enough apart to justify a diagonal pair covariance, while the latter controls correlations between neighboring thresholds on a common branch. In practice, these parameter-dependent conditions are enforced through a design-point or worst-case bound over the intended operating range when one wants the diagonal closed form to be quantitatively valid. If either type of correlation is retained, the same optimization objective should instead be evaluated with a full covariance model rather than with additive per-threshold weights.

\textbf{Structural observations.}
Two features of the objective function are essential:
\begin{enumerate}
  \item \emph{Non-separability:} Unlike a simple sum of per-pair costs, $\mathcal{B}_{\vartheta,\min}^{\mathrm{tot}}$ is a \emph{non-separable} function of all $N$ thresholds through the matrix Schur complement in $\Delta\mathcal{I}_{\mathrm{eff}}^{(\vartheta),\mathrm{tot}}$ (Eq.~\eqref{eq:eff_info_tot}). The nuisance-parameter information matrix $\bm{\mathcal{I}}_{\bm{\eta}\bm{\eta}}^{\mathrm{tot}}$ couples all threshold pairs, so that modifying a single threshold alters the effective information contribution of every other threshold.
  \item \emph{Bias--variance tension:} The two terms in $\mathcal{B}_{\vartheta,\min}^{\mathrm{tot}}$ generically create an engineering trade-off. Lower thresholds often reduce analog timing variance by probing steeper waveform regions, but they can simultaneously amplify the bias contribution $\beta_{\vartheta,\mathrm{tot}}^2$ from model mismatch near the pulse onset. Higher thresholds generally encounter gentler slopes and increased shot-noise influence, yet can provide complementary geometric directions in the parameter phase space that improve nuisance decoupling through superadditivity. The optimal configuration resolves this tension at the marginal-return balance point.
\end{enumerate}
\subsection{Optimality Conditions}
\label{sec:opt_derivation}

We now derive the first-order optimality conditions for Problem~\eqref{eq:general_opt_problem} by differentiating $\mathcal{B}_{\vartheta,\min}^{\mathrm{tot}}$ with respect to each threshold $V_n$ in the interior of the feasible region.

\emph{Assumption} (Interior optimum): The optimal thresholds $\{V_n^*\}$ lie strictly in the interior of the feasible set defined by Problem~\eqref{eq:general_opt_problem}, i.e., no active inequality constraints. When a threshold sits at the boundary (e.g., at the minimum same-branch spacing or pair-gap limit imposed by the diagonal-covariance conditions), the stationarity conditions below must be replaced by Karush-Kuhn-Tucker (KKT) conditions with appropriate complementary slackness.

\emph{Assumption} (Smooth dependence on $V_n$): The crossing times $t_{r}(V_n;\bm{\theta})$ and $t_f(V_n;\bm{\theta})$, hence all FIM blocks and bias terms, depend smoothly ($C^2$) on each threshold $V_n$. This is guaranteed by the transversality condition $\partial_t f \neq 0$ at every crossing, via the Implicit Function Theorem.

\textbf{The nuisance coupling vector.}
A central quantity governing the inter-threshold coupling is the \emph{nuisance coupling vector}:
\begin{equation}
  \bm{\gamma}
  \;\triangleq\;
  \bigl(\bm{\mathcal{I}}_{\bm{\eta}\bm{\eta}}^{\mathrm{tot}}\bigr)^{-1}\,
  \bm{\mathcal{I}}_{\bm{\eta}\vartheta}^{\mathrm{tot}}
  \;\in\;\mathbb{R}^{M-1},
  \label{eq:nuisance_coupling_vector}
\end{equation}
which represents the direction in nuisance-parameter space most correlated with the target parameter $\vartheta$ under the global joint estimator. Geometrically, $\bm{\gamma}$ is the coefficient vector of the weighted least-squares projection of the $\vartheta$-column of the Jacobian onto the subspace spanned by the nuisance columns.

\textbf{Nuisance-projected effective sensitivities.}
For each threshold $V_n$, define the \emph{nuisance-projected sensitivities} at the rising and falling edges:
\begin{equation}
  \begin{aligned}
    \tilde{g}_{r}^{(n)}
     & \;\triangleq\;
    g_{\vartheta,r}(V_n) - \bm{\gamma}^{T}\,\mathbf{g}_{\bm{\eta},r}(V_n), \\
    \tilde{g}_{f}^{(n)}
     & \;\triangleq\;
    g_{\vartheta,f}(V_n) - \bm{\gamma}^{T}\,\mathbf{g}_{\bm{\eta},f}(V_n).
  \end{aligned}
  \label{eq:projected_sensitivity}
\end{equation}
These quantities measure the ``residual'' waveform sensitivity of the target parameter at each edge after geometrically subtracting the component attributable to nuisance parameters. A large $|\tilde{g}|$ signifies that the corresponding edge carries substantial timing information that is fundamentally \emph{orthogonal} to the nuisance directions.

\textbf{Nuisance-projected per-pair Fisher contribution.}
Using these projected sensitivities, define the nuisance-projected Fisher contribution from the $n$-th threshold pair:
\begin{equation}
  \Phi_{\vartheta}^{(n)}(V_n;\,\bm{\gamma})
  \;\triangleq\;
  \frac{[\tilde{g}_{r}^{(n)}]^{2}}{D_r(V_n)}
  + \frac{[\tilde{g}_{f}^{(n)}]^{2}}{D_f(V_n)},
  \label{eq:projected_fisher}
\end{equation}
where $D_r$, $D_f$ are the composite noise denominators defined in Section~\ref{sec:pair_general}.

\textbf{Global decomposition of effective information.}
A fundamental algebraic identity links the total effective information to the per-pair projected contributions.

\begin{proposition}[Fisher Decomposition]
  \label{prop:fisher_decomp}
  When $\bm{\gamma}$ is evaluated at the global coupling vector $\bm{\gamma} = (\bm{\mathcal{I}}_{\bm{\eta}\bm{\eta}}^{\mathrm{tot}})^{-1}\bm{\mathcal{I}}_{\bm{\eta}\vartheta}^{\mathrm{tot}}$, the total effective Fisher Information decomposes exactly as:
  \begin{equation}
    \Delta\mathcal{I}_{\mathrm{eff}}^{(\vartheta),\mathrm{tot}}
    = \sum_{n=1}^{N}
    \Phi_{\vartheta}^{(n)}(V_n;\,\bm{\gamma}).
    \label{eq:fisher_decomposition}
  \end{equation}
\end{proposition}

\noindent\emph{Proof.}\;
Expanding $\sum_n \Phi_{\vartheta}^{(n)}$ using Eqs.~\eqref{eq:projected_sensitivity}--\eqref{eq:projected_fisher} and the FIM block definitions Eqs.~\eqref{eq:I_eta_eta}--\eqref{eq:I_tau_tau}, and recollecting terms via the additivity relations Eq.~\eqref{eq:summed_blocks}:
\begin{align}
  \sum_{n=1}^{N} \Phi_{\vartheta}^{(n)}
   & = \mathcal{I}_{\vartheta\vartheta}^{\mathrm{tot}}
  - 2\,\bm{\gamma}^{T}\,\bm{\mathcal{I}}_{\bm{\eta}\vartheta}^{\mathrm{tot}}
  + \bm{\gamma}^{T}\,
  \bm{\mathcal{I}}_{\bm{\eta}\bm{\eta}}^{\mathrm{tot}}\,
  \bm{\gamma}.
  \label{eq:decomp_expanded}
\end{align}
Substituting $\bm{\gamma} = (\bm{\mathcal{I}}_{\bm{\eta}\bm{\eta}}^{\mathrm{tot}})^{-1}\,\bm{\mathcal{I}}_{\bm{\eta}\vartheta}^{\mathrm{tot}}$ into the last two terms collapses them to $-(\bm{\mathcal{I}}_{\bm{\eta}\vartheta}^{\mathrm{tot}})^{T}(\bm{\mathcal{I}}_{\bm{\eta}\bm{\eta}}^{\mathrm{tot}})^{-1}\bm{\mathcal{I}}_{\bm{\eta}\vartheta}^{\mathrm{tot}}$, recovering the Schur complement definition Eq.~\eqref{eq:eff_info_tot}. \hfill$\square$

\emph{Physical interpretation:}
The total effective information is \emph{exactly} the sum of per-pair projected contributions $\Phi_{\vartheta}^{(n)}$, evaluated at the \emph{global} nuisance coupling vector~$\bm{\gamma}$. The Schur complement's apparent non-separability is resolved upon introduction of $\bm{\gamma}$; this decomposition is a mathematical identity, not an approximation.

\emph{Connection to variational superadditivity.}
The decomposition also provides a constructive proof of the superadditivity inequality (Section~\ref{sec:synergy}). For each pair viewed in isolation, its generalized effective information admits the variational representation
$$
  \Delta\mathcal{I}_{\mathrm{eff}}^{(\vartheta),(n)} = \min_{\bm{\gamma}} \Phi_\vartheta^{(n)}(V_n;\bm{\gamma}),
$$
where the minimizer set is
$\bm{\gamma} \in (\bm{\mathcal{I}}_{\bm{\eta}\bm{\eta}}^{(n)})^{+}\bm{\mathcal{I}}_{\bm{\eta}\vartheta}^{(n)} + \ker(\bm{\mathcal{I}}_{\bm{\eta}\bm{\eta}}^{(n)})$.
Since $\Phi_\vartheta^{(n)}$ is a convex quadratic in $\bm{\gamma}$ (its Hessian is $2\,\bm{\mathcal{I}}_{\bm{\eta}\bm{\eta}}^{(n)} \succeq 0$), any choice of $\bm{\gamma}$ yields
\begin{equation}
  \Phi_\vartheta^{(n)}(V_n;\bm{\gamma})
  \;\ge\;
  \Delta\mathcal{I}_{\mathrm{eff}}^{(\vartheta),(n)},
  \quad \forall\;\bm{\gamma}.
  \label{eq:synergy_mechanism}
\end{equation}
Summing over $n$ and evaluating at $\bm{\gamma} = \bm{\gamma}^*$ immediately yields $\Delta\mathcal{I}_{\mathrm{eff}}^{(\vartheta),\mathrm{tot}} \ge \sum_n \Delta\mathcal{I}_{\mathrm{eff}}^{(\vartheta),(n)}$. The synergy arises because the global coupling $\bm{\gamma}^*$ is a compromise among all pairs that generically cannot lie in every per-pair minimizer set simultaneously, thereby leaving a non-negative residual projected contribution in each pair.

\textbf{Gradient of $\mathcal{B}_{\vartheta,\min}^{\mathrm{tot}}$ with respect to $V_n$.}
Differentiating the objective Eq.~\eqref{eq:mse_objective}:
\begin{equation}
  \frac{\partial\mathcal{B}_{\vartheta,\min}^{\mathrm{tot}}}{\partial V_n}
  = -\frac{1}{\bigl[\Delta\mathcal{I}_{\mathrm{eff}}^{(\vartheta),\mathrm{tot}}\bigr]^{2}}\,
  \frac{\partial\Delta\mathcal{I}_{\mathrm{eff}}^{(\vartheta),\mathrm{tot}}}{\partial V_n}
  + 2\,\beta_{\vartheta,\mathrm{tot}}\,
  \frac{\partial \beta_{\vartheta,\mathrm{tot}}}{\partial V_n}.
  \label{eq:mse_gradient}
\end{equation}
The gradient of the effective information is obtained by differentiating the Schur complement Eq.~\eqref{eq:eff_info_tot} with respect to $V_n$. Since only the $n$-th pair's FIM blocks depend on $V_n$, applying the standard matrix derivative identity $\partial_{V_n}[\mathbf{c}^{T}\mathbf{M}^{-1}\mathbf{c}] = 2(\partial_{V_n}\mathbf{c})^{T}\mathbf{M}^{-1}\mathbf{c} - \mathbf{c}^{T}\mathbf{M}^{-1}(\partial_{V_n}\mathbf{M})\mathbf{M}^{-1}\mathbf{c}$ (with $\mathbf{c} = \bm{\mathcal{I}}_{\bm{\eta}\vartheta}^{\mathrm{tot}}$, $\mathbf{M} = \bm{\mathcal{I}}_{\bm{\eta}\bm{\eta}}^{\mathrm{tot}}$) yields:
\begin{align}
  \frac{\partial\Delta\mathcal{I}_{\mathrm{eff}}^{(\vartheta),\mathrm{tot}}}{\partial V_n}
   & = \frac{\partial \mathcal{I}_{\vartheta\vartheta}^{(n)}}{\partial V_n}
  - 2\left(\frac{\partial \bm{\mathcal{I}}_{\bm{\eta}\vartheta}^{(n)}}{\partial V_n}\right)^{\!T}\bm{\gamma}
  + \bm{\gamma}^{T}\,
  \frac{\partial \bm{\mathcal{I}}_{\bm{\eta}\bm{\eta}}^{(n)}}{\partial V_n}\,
  \bm{\gamma}.
  \label{eq:eff_info_gradient_expanded}
\end{align}
Comparing with the definition of $\Phi_{\vartheta}^{(n)}$ (Eqs.~\eqref{eq:projected_sensitivity}--\eqref{eq:projected_fisher}), the three terms in Eq.~\eqref{eq:eff_info_gradient_expanded} are precisely the derivative of the nuisance-projected per-pair Fisher contribution with $\bm{\gamma}$ held constant:
\begin{equation}
  \frac{\partial\Delta\mathcal{I}_{\mathrm{eff}}^{(\vartheta),\mathrm{tot}}}{\partial V_n}
  = \frac{d\,\Phi_{\vartheta}^{(n)}(V;\,\bm{\gamma})}{dV}
  \bigg|_{\substack{V = V_n \\ \bm{\gamma}\,=\,\mathrm{const}}}.
  \label{eq:eff_info_gradient}
\end{equation}
Physically, Eq.~\eqref{eq:eff_info_gradient} states that each threshold's marginal contribution to the global effective information is locally determined by the rate of change of its own nuisance-projected Fisher contribution, evaluated against the globally determined nuisance coupling backdrop.

\textbf{General stationarity system.}
Setting $\partial\mathcal{B}_{\vartheta,\min}^{\mathrm{tot}}/\partial V_n = 0$ for each interior threshold and substituting Eqs.~\eqref{eq:mse_gradient}--\eqref{eq:eff_info_gradient} yields the system of $N$ coupled stationarity equations:
\begin{equation}
  \boxed{
    \begin{gathered}
      \frac{d\,\Phi_{\vartheta}^{(n)}(V;\,\bm{\gamma}^*)}{dV}
      \bigg|_{V=V_n^*} \\
      = 2\bigl[\Delta\mathcal{I}_{\mathrm{eff}}^{(\vartheta),\mathrm{tot}}\bigr]^{2}\,
      \beta_{\vartheta,\mathrm{tot}}\,
      \frac{\partial \beta_{\vartheta,\mathrm{tot}}}{\partial V_n}\bigg|_{V_n = V_n^*}
    \end{gathered}}
  \label{eq:general_stationarity}
\end{equation}
for $n = 1, \ldots, N$. Here $\bm{\gamma}^*$ is the global coupling vector of Eq.~\eqref{eq:nuisance_coupling_vector}, evaluated at $\{V_n^*\}$.

\emph{Structural interpretation:}
The left-hand side of Eq.~\eqref{eq:general_stationarity} is the marginal gain in nuisance-projected Fisher Information from increasing the $n$-th threshold. The right-hand side is the marginal cost in squared systematic bias, amplified by the square of the total effective information. The optimal threshold $V_n^*$ sits at the exact voltage where these two marginal rates balance.

\textbf{The well-specified case ($\varepsilon = 0$).}
When the pulse model is perfectly specified, the systematic bias identically vanishes ($\beta_{\vartheta,\mathrm{tot}} = 0$) and the right-hand side of Eq.~\eqref{eq:general_stationarity} is zero. The stationarity system simplifies to:
\begin{equation}
  \frac{d\,\Phi_{\vartheta}^{(n)}(V;\,\bm{\gamma}^*)}{dV}
  \bigg|_{V=V_n^*}
  = 0,
  \quad n = 1, \dots, N.
  \label{eq:bias_free_stationarity}
\end{equation}
In this regime, each optimal threshold satisfies the stationary condition for its own nuisance-projected Fisher contribution $\Phi_{\vartheta}^{(n)}(V;\bm{\gamma}^*)$, subject to the globally determined coupling structure~$\bm{\gamma}^*$. It is a local maximizer only if the corresponding second derivative with respect to $V$ is negative.
\textbf{Fixed-point structure and self-consistency.}
Equations~\eqref{eq:general_stationarity} and \eqref{eq:nuisance_coupling_vector} jointly define a self-consistent fixed-point system:
\begin{enumerate}
  \item Given $\bm{\gamma}$, the thresholds $\{V_n^*\}$ are updated by solving the stationarity system (Eq.~\eqref{eq:general_stationarity} or \eqref{eq:bias_free_stationarity}) together with the ordering and spacing constraints from Problem~\eqref{eq:general_opt_problem}.
  \item Given $\{V_n^*\}$, the coupling vector updates:
        \begin{equation}
          \bm{\gamma}^*
          = \left(\sum_{n=1}^{N}
          \bm{\mathcal{I}}_{\bm{\eta}\bm{\eta}}^{(n)}(V_n^*)\right)^{\!-1}
          \sum_{n=1}^{N}
          \bm{\mathcal{I}}_{\bm{\eta}\vartheta}^{(n)}(V_n^*).
          \label{eq:fixed_point_lambda}
        \end{equation}
\end{enumerate}
      This structure naturally suggests an \emph{alternating optimization algorithm}~\cite{si2015off,kharfati2026block}: initialize $\bm{\gamma}^{(0)}$ from a heuristic threshold placement, solve the threshold-update subproblem with $\bm{\gamma}$ frozen, update $\bm{\gamma}^{(1)}$ from Eq.~\eqref{eq:fixed_point_lambda}, and iterate to convergence. Only in the bias-free interior approximation, where $\beta_{\vartheta,\mathrm{tot}} = 0$ and the feasibility constraints are inactive, does the threshold-update subproblem reduce to $N$ independent one-dimensional stationary searches.

\textbf{Leading-order model-based MSE proxy.}
Substituting the optimal thresholds $\{V_n^*\}$ into the objective, the general leading-order model-based MSE proxy for the target parameter $\vartheta$ is:
\begin{equation}
  \boxed{
    \mathcal{B}_{\vartheta}^{*}
    = \left[\sum_{n=1}^{N}
      \Phi_{\vartheta}^{(n)}(V_n^*;\,\bm{\gamma}^*)\right]^{-1}
    + \beta_{\vartheta,\mathrm{tot}}^{2}(\{V_n^*\}).}
  \label{eq:min_mse_general}
\end{equation}
This expression is the corresponding leading-order model-based MSE proxy for an MVT system operating with $N$ threshold pairs under the specified pulse model, noise structure, and regularity assumptions; it is not intended as a claim about architectures outside those assumptions.

\subsection{Robustness under Parameter Perturbation}
\label{sec:robustness}

The optimal thresholds $\{V_n^*\}$ derived in Section~\ref{sec:opt_derivation} are computed at a specific \emph{design-point} parameter vector $\bm{\theta}_0$, representing, e.g., the nominal pulse shape at a particular energy (such as the 511\,keV photopeak).  In practice, the true parameter vector $\bm{\theta}$ varies from event to event and across operating conditions, while the hardware thresholds remain fixed at their design-point values.  This subsection quantifies how the MSE performance floor degrades when the actual parameter departs from $\bm{\theta}_0$ by a perturbation $\Delta\bm{\theta}$.

\textbf{Setup.}
Let $\{V_n^*\}$ and $\bm{\gamma}^*$ denote the optimal thresholds and nuisance coupling vector obtained by solving the stationarity system~\eqref{eq:general_stationarity}--\eqref{eq:fixed_point_lambda} at $\bm{\theta} = \bm{\theta}_0$.  Define the effective Fisher Information and leading-order MSE proxy evaluated at the \emph{fixed} thresholds $\{V_n^*\}$ but at an arbitrary parameter point~$\bm{\theta}$:
\begin{align}
  \mathcal{I}_{\mathrm{eff}}(\bm{\theta})
   & \;\triangleq\;
  \sum_{n=1}^{N}
  \Phi_{\vartheta}^{(n)}\!\bigl(V_n^*;\,\bm{\gamma}(\bm{\theta}),\,\bm{\theta}\bigr),
  \label{eq:Ieff_of_theta} \\
  \mathcal{B}_{\vartheta}(\bm{\theta})
   & \;\triangleq\;
  \bigl[\mathcal{I}_{\mathrm{eff}}(\bm{\theta})\bigr]^{-1}
  + \beta_{\vartheta,\mathrm{tot}}^{2}(\bm{\theta}),
  \label{eq:mse_of_theta}
\end{align}
where $\bm{\gamma}(\bm{\theta}) = \bigl(\bm{\mathcal{I}}_{\bm{\eta}\bm{\eta}}^{\mathrm{tot}}(\bm{\theta})\bigr)^{-1}\bm{\mathcal{I}}_{\bm{\eta}\vartheta}^{\mathrm{tot}}(\bm{\theta})$ is the nuisance coupling vector re-evaluated at $\bm{\theta}$ with thresholds held fixed, and $\Phi_{\vartheta}^{(n)}$ and $\beta_{\vartheta,\mathrm{tot}}$ depend on $\bm{\theta}$ through the waveform sensitivities $g_{j,r/f}(V_n^*;\bm{\theta})$ and the noise denominators $D_{r/f}(V_n^*;\bm{\theta})$.
At the design point, $\mathcal{B}_{\vartheta}(\bm{\theta}_0) = \mathcal{B}_{\vartheta}^*$.

\textbf{Envelope theorem: elimination of indirect dependence through $\bm{\gamma}$.}
A key simplification arises from the stationarity of $\mathcal{I}_{\mathrm{eff}}$ with respect to $\bm{\gamma}$ at the global optimum.

\emph{Assumption} (Envelope theorem regularity): The nuisance coupling vector $\bm{\gamma}^*(\bm{\theta})$ is a $C^1$ function of $\bm{\theta}$ in a neighborhood of $\bm{\theta}_0$. This is guaranteed by the Implicit Function Theorem applied to the stationarity condition $\partial\mathcal{I}_{\mathrm{eff}}/\partial\bm{\gamma} = 0$, provided $\bm{\mathcal{I}}_{\bm{\eta}\bm{\eta}}^{\mathrm{tot}}(\bm{\theta})$ remains strictly positive definite in a neighborhood of $\bm{\theta}_0$ (i.e., the global nuisance identifiability assumption holds uniformly near the design point).

From the expanded form Eq.~\eqref{eq:decomp_expanded}:
\begin{equation}
  \frac{\partial\,\mathcal{I}_{\mathrm{eff}}}{\partial\bm{\gamma}}
  = -2\,\bm{\mathcal{I}}_{\bm{\eta}\vartheta}^{\mathrm{tot}}
  + 2\,\bm{\mathcal{I}}_{\bm{\eta}\bm{\eta}}^{\mathrm{tot}}\,\bm{\gamma},
\end{equation}
which vanishes identically at $\bm{\gamma} = \bm{\gamma}^* = (\bm{\mathcal{I}}_{\bm{\eta}\bm{\eta}}^{\mathrm{tot}})^{-1}\bm{\mathcal{I}}_{\bm{\eta}\vartheta}^{\mathrm{tot}}$. By the envelope theorem, the \emph{total} derivative of $\mathcal{I}_{\mathrm{eff}}$ with respect to any physical parameter $\theta_j$ equals its \emph{partial} derivative with $\bm{\gamma}$ held constant:
\begin{equation}
  \frac{d\,\mathcal{I}_{\mathrm{eff}}}{d\theta_j}\bigg|_{\bm{\theta}_0}
  = \sum_{n=1}^{N}
  \frac{\partial\,\Phi_{\vartheta}^{(n)}(V_n^*;\,\bm{\gamma}^*,\,\bm{\theta})}
  {\partial\theta_j}\bigg|_{\bm{\theta}_0},
  \label{eq:envelope_sensitivity}
\end{equation}
where the right-hand side involves only the \emph{explicit} $\bm{\theta}$-dependence of $\Phi_{\vartheta}^{(n)}$ through the waveform sensitivities and noise denominators, with $\bm{\gamma}^*$ frozen at its design-point value.  This eliminates the need to differentiate the implicitly defined coupling vector, substantially simplifying the perturbation analysis.

\textbf{Parameter sensitivity vector.}
Define the \emph{Fisher sensitivity vector} $\mathbf{S} \in \mathbb{R}^M$ with components
\begin{equation}
  S_j
  \;\triangleq\;
  \frac{d\,\mathcal{I}_{\mathrm{eff}}}{d\theta_j}\bigg|_{\bm{\theta}_0}
  = \sum_{n=1}^{N}
  \frac{\partial\,\Phi_{\vartheta}^{(n)}}{\partial\theta_j}\bigg|_{\bm{\theta}_0,\,\bm{\gamma}^*},
  \quad j = 1, \ldots, M,
  \label{eq:sensitivity_vector}
\end{equation}
and the \emph{bias sensitivity vector} $\mathbf{s}_\beta \in \mathbb{R}^M$ with components
\begin{equation}
  s_{\beta,j}
  \;\triangleq\;
  \frac{\partial\,\beta_{\vartheta,\mathrm{tot}}}{\partial\theta_j}\bigg|_{\bm{\theta}_0}.
  \label{eq:bias_sensitivity_vector}
\end{equation}

\textbf{First-order MSE perturbation.}
Expanding $\mathcal{B}_{\vartheta}(\bm{\theta}_0 + \Delta\bm{\theta})$ to first order (see Appendix~\ref{app:perturbation} for the full derivation):
\begin{equation}
  \boxed{
    \Delta\mathcal{B}_{\vartheta}^{(1)}
    = -\frac{\mathbf{S}^{T}\Delta\bm{\theta}}
    {\mathcal{I}_{\mathrm{eff}}^{*\,2}}
    + 2\,\beta_{\vartheta}^{*}\,\mathbf{s}_\beta^{T}\Delta\bm{\theta},}
  \label{eq:mse_first_order}
\end{equation}
where $\mathcal{I}_{\mathrm{eff}}^{*} \equiv \mathcal{I}_{\mathrm{eff}}(\bm{\theta}_0)$ and $\beta_{\vartheta}^{*} \equiv \beta_{\vartheta,\mathrm{tot}}(\bm{\theta}_0)$.  In the well-specified case ($\beta_{\vartheta}^{*} = 0$), the second term vanishes and the MSE shift is governed entirely by how the effective Fisher Information changes with $\bm{\theta}$.  To first order, the sign of $\mathbf{S}^{T}\Delta\bm{\theta}$ determines whether the perturbation improves ($\mathbf{S}^{T}\Delta\bm{\theta} > 0$) or degrades ($\mathbf{S}^{T}\Delta\bm{\theta} < 0$) the design-point proxy.

\textbf{Second-order MSE perturbation.}
Including second-order terms requires care: while the first-order envelope theorem allows $\bm{\gamma}$ to be held fixed, the \emph{second-order} total derivative of $\mathcal{I}_{\mathrm{eff}}$ with respect to $\bm{\theta}$ acquires a correction from the implicit variation of $\bm{\gamma}^*(\bm{\theta})$. By the second-order envelope theorem (proved in Appendix~\ref{app:perturbation}, Step~M.0), the true Hessian of $\mathcal{I}_{\mathrm{eff}}$ with respect to $\bm{\theta}$ is
\begin{align}
  \mathbf{H}_{\mathcal{I}}
  \;\triangleq\;
  \frac{d^2\,\mathcal{I}_{\mathrm{eff}}}{d\bm{\theta}^2}\bigg|_{\bm{\theta}_0}
   & = \frac{\partial^2\mathcal{I}_{\mathrm{eff}}}{\partial\bm{\theta}^2}\bigg|_{\bm{\gamma}^*} \notag          \\
   & \quad - \left(\frac{\partial^2\mathcal{I}_{\mathrm{eff}}}{\partial\bm{\theta}\,\partial\bm{\gamma}}\right)
  \left(\frac{\partial^2\mathcal{I}_{\mathrm{eff}}}{\partial\bm{\gamma}^2}\right)^{\!-1}
  \left(\frac{\partial^2\mathcal{I}_{\mathrm{eff}}}{\partial\bm{\gamma}\,\partial\bm{\theta}}\right),
  \label{eq:true_hessian}
\end{align}
where all partials are evaluated at $(\bm{\theta}_0,\bm{\gamma}^*)$.  The first term is the ``na\"ive'' Hessian with $\bm{\gamma}$ frozen; the second term corrects for the implicit response $d\bm{\gamma}^*/d\bm{\theta}$.  From the quadratic structure Eq.~\eqref{eq:decomp_expanded}, $\partial^2\mathcal{I}_{\mathrm{eff}}/\partial\bm{\gamma}^2 = 2\,\bm{\mathcal{I}}_{\bm{\eta}\bm{\eta}}^{\mathrm{tot}}$, which is positive definite under the global rank condition, ensuring the correction is well-defined.  Similarly, let $\mathbf{H}_\beta$ denote the Hessian of $\beta_{\vartheta,\mathrm{tot}}$ with respect to $\bm{\theta}$.  The full second-order expansion is:
\begin{equation}
  \boxed{
    \begin{aligned}
      \Delta\mathcal{B}_{\vartheta}^{(2)}
       & = \Delta\mathcal{B}_{\vartheta}^{(1)}
      + \frac{(\mathbf{S}^{T}\Delta\bm{\theta})^{2}}
        {\mathcal{I}_{\mathrm{eff}}^{*\,3}}
      - \frac{\Delta\bm{\theta}^{T}\mathbf{H}_{\mathcal{I}}\,\Delta\bm{\theta}}
        {2\,\mathcal{I}_{\mathrm{eff}}^{*\,2}} \\
       & \quad
         + (\mathbf{s}_\beta^{T}\Delta\bm{\theta})^{2}
      + \beta_{\vartheta}^{*}\,\Delta\bm{\theta}^{T}\mathbf{H}_{\beta}\,\Delta\bm{\theta}.
    \end{aligned}}
  \label{eq:mse_second_order}
\end{equation}
The three additional terms have transparent physical origins: (i)~$(\mathbf{S}^{T}\Delta\bm{\theta})^{2}/\mathcal{I}_{\mathrm{eff}}^{*3}$ is the convexity of $1/\mathcal{I}$ and is always non-negative, representing the accelerating degradation of the CRB as information is lost; (ii)~the Hessian term $\mathbf{H}_{\mathcal{I}}$ (Eq.~\eqref{eq:true_hessian}) captures the curvature of the Fisher landscape including the implicit readjustment of the nuisance coupling; (iii)~the bias terms account for the quadratic growth of systematic error with parameter perturbation.

\textbf{Worst-case MSE degradation over a parameter uncertainty region.}
In hardware design, the parameter uncertainty is typically bounded componentwise: $|\Delta\theta_j| \le \epsilon_j$ for $j = 1, \ldots, M$, defining a hyper-rectangular uncertainty region $\mathcal{R}(\bm{\epsilon})$.  The exact first-order worst-case MSE degradation over this region is the support function of that box applied to the linear form Eq.~\eqref{eq:mse_first_order} (Appendix~\ref{app:perturbation}, Step~M.5):
\begin{equation}
  \begin{aligned}
    \sup_{\Delta\bm{\theta}\,\in\,\mathcal{R}(\bm{\epsilon})}
    \Delta\mathcal{B}_{\vartheta}^{(1)}
     & = \sum_{j=1}^{M}
    \left|
    -\frac{S_j}{\mathcal{I}_{\mathrm{eff}}^{*\,2}}
    + 2\,\beta_{\vartheta}^{*}\,s_{\beta,j}
    \right|\epsilon_j \\
     & \le \frac{1}{\mathcal{I}_{\mathrm{eff}}^{*\,2}}
    \sum_{j=1}^{M} |S_j|\,\epsilon_j
    + 2\,|\beta_{\vartheta}^{*}|
    \sum_{j=1}^{M} |s_{\beta,j}|\,\epsilon_j.
  \end{aligned}
  \label{eq:worst_case_first_order}
\end{equation}
The equality is the exact first-order worst-case value; the inequality is the conservative decoupled upper bound obtained from the triangle inequality. In the well-specified case $\beta_{\vartheta}^{*}=0$, the two coincide.

\emph{Relative robustness metric.}
Normalizing by the design-point MSE:
\begin{equation}
  \rho(\bm{\epsilon})
  \;\triangleq\;
  \frac{\sup_{\Delta\bm{\theta}\,\in\,\mathcal{R}(\bm{\epsilon})}
    \Delta\mathcal{B}_{\vartheta}^{(1)}}
  {\mathcal{B}_{\vartheta}^{*}},
  \label{eq:robustness_metric}
\end{equation}
which quantifies the fractional degradation.  A threshold design is considered robust when $\rho(\bm{\epsilon}) \ll 1$ over the anticipated parameter range.

\emph{Physical interpretation.}
The sensitivity vector $\mathbf{S}$ reveals which physical parameters most strongly affect
the timing precision at the design point.  Large $|S_j|$ identifies fragile directions in parameter space where the optimal threshold placement is sensitive, while $S_j \approx 0$ marks robust directions.  This provides direct guidance for system design: if timing performance is particularly sensitive to, say, the decay time constant $\tau_d$, this signals that either (i)~the detector should be characterized more precisely in that parameter, or (ii)~the threshold design should be regularized toward robustness against $\tau_d$ variations, potentially at a modest cost to design-point optimality.

\emph{Remark (Design conditioning).}
A subtler form of fragility arises not from parameter perturbation but from \emph{threshold configuration}: the single-parameter design objective (Eq.~\eqref{eq:projected_fisher}) can steer all thresholds toward a small number of support voltages (classical optimal-design theory permits at most $\lceil M/2\rceil$ support points for an $M$-parameter model).  The resulting threshold concentration may render the full Fisher matrix $\mathbf{I}$ near-singular, making joint parameter estimation numerically ill-conditioned even though the scalar CRLB bound for the target parameter remains small.  This ``design--reconstruction gap'' motivates the use of dimensionless joint criteria such as normalized D-optimality, which simultaneously controls all eigenvalues of the scaled Fisher matrix and thereby enforces design-point identifiability while materially improving numerical feasibility; the empirical consequences are documented in the experimental validation (Section~\ref{sec:doptimal_mvt}).

\subsection{Multi-Event Extension with Partial Triggering}
\label{sec:partial_trigger}

The preceding subsections developed the optimal threshold design framework under the full-triggering assumption $\bm{\theta} \in \Theta_V$, which requires the pulse peak to exceed all $N$ hardware thresholds.  In this subsection, we relax this restriction to address a critical engineering scenario: the recovery of \emph{low-energy} detection events whose pulse amplitude is insufficient to trigger the complete comparator chain.

\subsubsection{Physical Motivation}

In PET imaging, a significant fraction of recorded events arise from inter-crystal scattering (ICS) or Compton interactions in which the incident 511\,keV gamma photon deposits only a portion of its energy in a given scintillation crystal.  The resulting pulse amplitude $A$ is proportionally reduced, and the peak voltage $p(\bm{\theta}) = A\cdot h_{\mathrm{peak}} + b$ may fall below one or more of the higher hardware thresholds.  These events carry valuable spatial and temporal information for image reconstruction---particularly in high-resolution detector arrays where ICS recovery improves spatial resolution~\cite{fu2016recovery,lee2020recovery,comanor1996algorithms,ollinger1995detector}---yet they produce an \emph{incomplete} set of threshold crossings: only the lower thresholds, for which $V_n < p(\bm{\theta})$, generate valid rising- and falling-edge timestamps.

Formally, the full-triggering domain $\Theta_V = \{\bm{\theta} \in \Theta \mid p(\bm{\theta}) > V_{\max}\}$ introduced in Section~\ref{sec:model} is replaced by the \emph{partial-triggering regime} in which $p(\bm{\theta})$ may lie between adjacent thresholds.  To handle this, we define the \emph{active threshold set} for a given event with physical parameters $\bm{\theta}$:
\begin{equation}
  \mathcal{A}(\bm{\theta})
  \;\triangleq\;
  \bigl\{n \in \{1,\ldots,N\} : V_n < p(\bm{\theta})\bigr\},
  \label{eq:active_set}
\end{equation}
with cardinality $K_{\mathrm{act}}(\bm{\theta}) \triangleq |\mathcal{A}(\bm{\theta})|$.  When $K_{\mathrm{act}} = N$, the event lies in $\Theta_V$ and all thresholds are triggered; when $K_{\mathrm{act}} < N$, the event is \emph{partially triggering}, and only $2K_{\mathrm{act}}$ crossing times are observed.

\emph{Assumption} (Partial-triggering transversality): For every active threshold $n \in \mathcal{A}(\bm{\theta})$, the transversality condition $\partial_t f(t_{r,n};\bm{\theta}) \neq 0$ and $\partial_t f(t_{f,n};\bm{\theta}) \neq 0$ holds at the rising and falling crossings.  This is guaranteed because $V_n < p(\bm{\theta})$ implies each active crossing lies on a strictly monotone branch of the pulse, away from the stationary peak.  Under this condition, the same implicit-function argument used in Section~\ref{sec:hadamard} ensures that, on any region of parameter space where the active set is fixed, the corresponding restricted forward map $S_{\mathcal{A}}$ into $\Omega_{2K_{\mathrm{act}}}$ remains well-defined and smooth. Across boundaries where some threshold satisfies $V_n = p(\bm{\theta})$ and the active set changes, the partial-triggering map is only piecewise smooth and must be treated stratum by stratum.

\subsubsection{Leading-Order MSE Proxy under Partial Triggering}

For an event with parameters $\bm{\theta}$ and active set $\mathcal{A}(\bm{\theta})$, the FIM is constructed exclusively from the $K_{\mathrm{act}}(\bm{\theta})$ active threshold pairs.  The active FIM blocks (restricting the summation in Eq.~\eqref{eq:summed_blocks} to $\mathcal{A}$) are:
\begin{equation}
  \begin{aligned}
    \mathcal{I}_{\vartheta\vartheta}^{\mathrm{act}}(\bm{\theta})      & = \sum_{n \in \mathcal{A}(\bm{\theta})} \mathcal{I}_{\vartheta\vartheta}^{(n)}(\bm{\theta}),      \\
    \bm{\mathcal{I}}_{\bm{\eta}\vartheta}^{\mathrm{act}}(\bm{\theta}) & = \sum_{n \in \mathcal{A}(\bm{\theta})} \bm{\mathcal{I}}_{\bm{\eta}\vartheta}^{(n)}(\bm{\theta}), \\
    \bm{\mathcal{I}}_{\bm{\eta}\bm{\eta}}^{\mathrm{act}}(\bm{\theta}) & = \sum_{n \in \mathcal{A}(\bm{\theta})} \bm{\mathcal{I}}_{\bm{\eta}\bm{\eta}}^{(n)}(\bm{\theta}),
  \end{aligned}
  \label{eq:active_blocks}
\end{equation}
where each per-pair block $\bm{\mathcal{I}}^{(n)}$ is defined by Eqs.~\eqref{eq:I_eta_eta}--\eqref{eq:I_tau_tau} evaluated at threshold $V_n$ with parameter $\bm{\theta}$.

\emph{Assumption} (Active nuisance and target identifiability): The active nuisance block satisfies $\bm{\mathcal{I}}_{\bm{\eta}\bm{\eta}}^{\mathrm{act}}(\bm{\theta}) \succ 0$, and the active profiled target information satisfies $\Delta\mathcal{I}_{\mathrm{eff}}^{(\vartheta),\mathrm{act}}(\bm{\theta}) > 0$.  Since each threshold pair contributes a rank-$\le 2$ matrix (Sec.~\ref{sec:master_eq}), nuisance-block invertibility requires at least $K_{\mathrm{act}}(\bm{\theta}) \ge \lceil(M-1)/2\rceil$ active pairs placed at generically distinct voltages.  A finite scalar-target proxy additionally requires the target direction to remain identifiable after nuisance profiling; a necessary generic rank count is $2K_{\mathrm{act}}(\bm{\theta}) \ge M$.  For the five-parameter bi-exponential amplitude-target model ($M = 5$), two active pairs can make the $4\times4$ nuisance block full rank, but three generic active pairs are needed to make the full five-parameter Fisher matrix full rank and to obtain $\Delta\mathcal{I}_{\mathrm{eff}}^{(A),\mathrm{act}} > 0$.  When these conditions fail, the Moore--Penrose generalized score remains algebraically computable, but it should not be interpreted as a standalone finite MSE proxy; the generic singular-pair interpretation is discussed in Appendix~\ref{app:schur}, Section~E.4, and the bi-exponential amplitude--baseline special case is worked out in Appendix~\ref{app:baseline_penalty}.

Under these active-identifiability conditions, the effective Fisher Information for the target parameter $\vartheta$ from the active pairs is (cf.\ Eq.~\eqref{eq:eff_info_tot}):
\begin{equation}
  \Delta\mathcal{I}_{\mathrm{eff}}^{(\vartheta),\mathrm{act}}(\bm{\theta})
  \;\triangleq\;
  \mathcal{I}_{\vartheta\vartheta}^{\mathrm{act}}
  - (\bm{\mathcal{I}}_{\bm{\eta}\vartheta}^{\mathrm{act}})^T\,
  (\bm{\mathcal{I}}_{\bm{\eta}\bm{\eta}}^{\mathrm{act}})^{-1}\,
  \bm{\mathcal{I}}_{\bm{\eta}\vartheta}^{\mathrm{act}}.
  \label{eq:eff_info_active}
\end{equation}

The nuisance coupling vector restricted to the active pairs is:
\begin{equation}
  \bm{\gamma}^{\mathrm{act}}(\bm{\theta})
  \;=\;
  \bigl(\bm{\mathcal{I}}_{\bm{\eta}\bm{\eta}}^{\mathrm{act}}(\bm{\theta})\bigr)^{-1}\,
  \bm{\mathcal{I}}_{\bm{\eta}\vartheta}^{\mathrm{act}}(\bm{\theta}).
  \label{eq:lambda_active}
\end{equation}

\emph{Corollary} (Active Fisher decomposition): The nuisance-projected Fisher decomposition of Proposition~1 in Section~\ref{sec:opt_derivation} holds verbatim with the summation restricted to the active set:
\begin{equation}
  \Delta\mathcal{I}_{\mathrm{eff}}^{(\vartheta),\mathrm{act}}(\bm{\theta})
  = \sum_{n \in \mathcal{A}(\bm{\theta})}
  \Phi_{\vartheta}^{(n)}\bigl(V_n;\,\bm{\gamma}^{\mathrm{act}}(\bm{\theta})\bigr),
  \label{eq:active_decomp}
\end{equation}
where $\Phi_{\vartheta}^{(n)}$ is the per-pair nuisance-projected Fisher contribution defined in Eq.~\eqref{eq:projected_fisher}.  The proof is identical---the algebraic identity underlying the decomposition depends only on the block structure of the FIM, not on the number of summands---and is recorded for completeness in Appendix~\ref{app:partial_trigger} (Step~N.1).

A crucial structural difference from the full-triggering case is that the coupling vector $\bm{\gamma}^{\mathrm{act}}$ now depends on the event parameters $\bm{\theta}$ not only through the FIM block values but also through the \emph{composition} of the active set $\mathcal{A}(\bm{\theta})$.  Two events with different amplitudes generally activate different threshold subsets, yielding distinct coupling vectors and hence distinct per-pair contributions $\Phi_{\vartheta}^{(n)}$.

Similarly, the systematic timing bias restricted to the active crossings is:
\begin{equation}
  \beta_{\vartheta,\mathrm{act}}(\bm{\theta})
  \;\triangleq\;
  \bigl[(\mathbf{J}_{\mathrm{act}}^T \mathbf{W}_{\mathrm{act}}\,\mathbf{J}_{\mathrm{act}})^{-1}\,
    \mathbf{J}_{\mathrm{act}}^T \mathbf{W}_{\mathrm{act}}\,
    \Delta\mathbf{t}_{\mathrm{bias}}^{\mathrm{act}}\bigr]_{\vartheta},
  \label{eq:bias_active}
\end{equation}
where $\mathbf{J}_{\mathrm{act}} \in \mathbb{R}^{2K_{\mathrm{act}} \times M}$, $\mathbf{W}_{\mathrm{act}} \in \mathbb{R}^{2K_{\mathrm{act}} \times 2K_{\mathrm{act}}}$, and $\Delta\mathbf{t}_{\mathrm{bias}}^{\mathrm{act}} \in \mathbb{R}^{2K_{\mathrm{act}}}$ are the stacked Jacobian, diagonal weight matrix, and temporal bias vector restricted to the $2K_{\mathrm{act}}$ active crossings (cf.\ Eq.~\eqref{eq:b_tau_tot}).

Combining the CRB and bias contributions, the corresponding leading-order model-based MSE proxy for the target parameter $\vartheta$ from a partially triggering event is:
\begin{equation}
  \resizebox{\columnwidth}{!}{$\displaystyle
      \boxed{
        \mathcal{B}_{\vartheta,\min}(\{V_n\};\,\bm{\theta})
        \;=\;
        \left[\sum_{n \in \mathcal{A}(\bm{\theta})}
          \Phi_{\vartheta}^{(n)}\bigl(V_n;\,\bm{\gamma}^{\mathrm{act}}(\bm{\theta})\bigr)\right]^{-1}
        \!+ \beta_{\vartheta,\mathrm{act}}^{2}(\bm{\theta}).}$}%
  \label{eq:partial_mse}
\end{equation}

\emph{Remark} (Reduction to the full-triggering case): When $p(\bm{\theta}) > V_N$, the active set is $\mathcal{A} = \{1,\ldots,N\}$, the coupling vector $\bm{\gamma}^{\mathrm{act}} = \bm{\gamma}$ (Eq.~\eqref{eq:nuisance_coupling_vector}), and Eq.~\eqref{eq:partial_mse} reduces identically to the full-triggering leading-order proxy Eq.~\eqref{eq:mse_objective}.

\subsubsection{Population-Weighted Optimisation}

In a realistic detection system, the incoming event population spans a range of physical parameters $\bm{\theta}$.  Different events activate different subsets of thresholds, and the hardware designer must choose a \emph{single} threshold configuration $\{V_1,\ldots,V_N\}$ that serves the entire event population.  We now formulate the optimal threshold design as a multi-event optimization problem.

\textbf{Setup.}
Consider $P$ representative events with known (or estimated) parameter vectors $\bm{\theta}^{(1)},\ldots,\bm{\theta}^{(P)}$ and associated importance weights $w_1,\ldots,w_P > 0$ satisfying $\sum_{p=1}^{P} w_p = 1$.  Here each $w_p$ encodes both the relative frequency of the event type in the detector's energy spectrum and the physicist's priority for that event class.  For instance, in a PET system, $w_p$ may be proportional to the energy-dependent detection efficiency times a user-defined utility factor that up-weights scatter events for ICS recovery.  The active threshold set for the $p$-th event is:
\begin{equation}
  \mathcal{A}_p
  \;\triangleq\;
  \mathcal{A}(\bm{\theta}^{(p)})
  = \bigl\{n : V_n < p(\bm{\theta}^{(p)})\bigr\}.
  \label{eq:active_set_p}
\end{equation}

\textbf{Formal diagonal-covariance optimization problem.}
Let $V_{\mathrm{HW,max}}$ denote the finite upper threshold admitted by the threshold-DAC/readout hardware.  In the diagonal-covariance closed-form approximation, the multi-event optimal threshold design problem is:
\begin{equation}
  \resizebox{\columnwidth}{!}{$\displaystyle
      \boxed{\begin{aligned}
          \{V_n^*\}_{n=1}^N
           & = \arg\min_{\{V_n\}_{n=1}^{N}}\;
          \mathcal{C}(\{V_n\})
          \;\triangleq\;
          \sum_{p=1}^{P} w_p\,
          \mathcal{B}_{\vartheta,\min}\bigl(\{V_n\};\,\bm{\theta}^{(p)}\bigr) \\[4pt]
          \text{s.t.}\quad
           & \max_{1 \le p \le P} b^{(p)} < V_1 < V_2 < \cdots < V_N,         \\
           & V_N \le V_{\mathrm{HW,max}},                                      \\
           & |V_i - V_j| \ge L_{ij}^{\max}\cdot\tau_c,
          \;\;\forall\;(i,j)\;\text{adjacent},                       \\
           & t_f(V_n;\bm{\theta}^{(p)}) - t_r(V_n;\bm{\theta}^{(p)}) \ge \tau_c,
             \quad n \in \mathcal{A}_p,\; p=1,\ldots,P,                       \\
           & K_p(\mathbf{V}) \ge K_{\min}^{(\vartheta)},\quad
             \Delta\mathcal{I}_{\mathrm{eff}}^{(\vartheta),\mathrm{act},(p)}(\mathbf{V}) > 0,
          \;\;p=1,\ldots,P,
        \end{aligned}}$}%
  \label{eq:multi_event_opt}
\end{equation}
where $L_{ij}^{\max} \triangleq \max_{1 \le p \le P} L_{ij}(\bm{\theta}^{(p)})$ is the worst-case slew rate over all events, $K_p(\mathbf{V}) \triangleq |\{n: V_n < p(\bm{\theta}^{(p)})\}|$ is the active-pair count induced by the candidate thresholds, and $\mathcal{B}_{\vartheta,\min}(\{V_n\};\bm{\theta}^{(p)})$ is given by Eq.~\eqref{eq:partial_mse}.  The pair-gap constraint is the within-pair analogue of the worst-case same-branch spacing condition and is imposed only on active thresholds whose two crossings are represented by diagonal single-pair weights. The feasibility constant $K_{\min}^{(\vartheta)}$ is the target-specific rank-count floor; for the five-parameter bi-exponential amplitude target, $K_{\min}^{(A)}=3$.  The explicit positivity constraint on $\Delta\mathcal{I}_{\mathrm{eff}}^{(\vartheta),\mathrm{act},(p)}$ excludes representative events whose active-pair proxy would be infinite. If either the inter-threshold or within-pair separation conditions are intentionally relaxed, the same multi-event objective should be evaluated with event-dependent full covariance blocks instead of the additive diagonal approximation. In the common baseline-free convention, the lower-bound constraint reduces to $0 < V_1$.

Two structural differences from the single-event optimization (Eq.~\eqref{eq:general_opt_problem}) are essential:
\begin{enumerate}
  \item \emph{No full-triggering upper-bound constraint on $V_N$}: The full-triggering requirement $V_N < p(\bm{\theta})$ is removed because partial triggering is explicitly permitted; the finite hardware bound $V_N \le V_{\mathrm{HW,max}}$ remains.  Thresholds may exceed the peak amplitude of some events, which simply excludes those thresholds from the active set of the corresponding event.
  \item \emph{Event-dependent coupling}: Each event $p$ carries its own active set $\mathcal{A}_p$, its own nuisance coupling vector $\bm{\gamma}_p^{\mathrm{act}} \triangleq \bm{\gamma}^{\mathrm{act}}(\bm{\theta}^{(p)})$, and its own effective information $\Delta\mathcal{I}_{\mathrm{eff}}^{(\vartheta),\mathrm{act},(p)}$.  The threshold optimization must simultaneously balance these $P$ distinct information landscapes.
\end{enumerate}

\emph{Assumption} (Interior optimum): As in Section~\ref{sec:opt_derivation}, we assume the optimal thresholds lie in the interior of the feasible set.  Additionally, we assume that at the optimum, no pulse peak $p(\bm{\theta}^{(p)})$ coincides exactly with any threshold voltage $V_n^*$---i.e., the active sets are locally constant.  This regularity condition ensures that $\mathcal{C}$ is differentiable with respect to each $V_n$ at the optimum.  (If a pulse peak exactly equals a threshold, the MSE has a discontinuous derivative arising from the discrete change in $\mathcal{A}_p$; this boundary case is addressed via directional derivatives in Appendix~\ref{app:partial_trigger}, Step~N.2.)

\emph{Remark} (Nonsmooth active-set boundaries): More precisely, $\mathcal{C}$ is a piecewise-smooth objective whose smooth strata are indexed by the active-set pattern $\{\mathcal{A}_p\}_{p=1}^{P}$. At boundaries $V_n=p(\bm{\theta}^{(p)})$, the classical gradient may fail to exist and the appropriate first-order object is a directional derivative, or equivalently the Clarke subdifferential generated by the adjacent active-set strata. The stationarity equations below are therefore stratum-wise interior conditions. The algorithmic recipes in Section~\ref{sec:recipe_multi} update thresholds with fixed active sets and then explicitly monitor active-set changes.

\textbf{Gradient of the multi-event cost.}
On any local stratum where the active sets $\{\mathcal{A}_p\}$ are fixed, differentiating $\mathcal{C}$ with respect to $V_n$ shows that only events for which threshold $n$ is active contribute (for $n \notin \mathcal{A}_p$ and away from the boundary $V_n = p(\bm{\theta}^{(p)})$, the proxy $\mathcal{B}_{\vartheta,\min}(\cdot\,;\bm{\theta}^{(p)})$ is locally independent of $V_n$):
\begin{equation}
  \frac{\partial\,\mathcal{C}}{\partial V_n}
  = \sum_{p:\,n\in\mathcal{A}_p}
  w_p\,
  \frac{\partial\,\mathcal{B}_{\vartheta,\min}(\{V_n\};\,\bm{\theta}^{(p)})}{\partial V_n}.
  \label{eq:multi_event_gradient}
\end{equation}
The per-event derivative is obtained by applying the envelope theorem (the same stationarity-at-$\bm{\gamma}$ argument as in Section~\ref{sec:robustness}, Eq.~\eqref{eq:envelope_sensitivity}) to the active Fisher decomposition Eq.~\eqref{eq:active_decomp}, yielding (Appendix~\ref{app:partial_trigger}, Steps~N.2--\eqref{eq:autonum:N3}):
\begin{multline}
  \frac{\partial\,\mathcal{B}_{\vartheta,\min}^{(p)}}{\partial V_n}
  = -\frac{1}{\bigl[\Delta\mathcal{I}_{\mathrm{eff}}^{(\vartheta),\mathrm{act},(p)}\bigr]^{2}}\,
  \frac{d\,\Phi_{\vartheta}^{(n)}(V;\,\bm{\gamma}_p^{\mathrm{act}})}{dV}\bigg|_{V=V_n}\\
  + 2\,\beta_{\vartheta,\mathrm{act}}^{(p)}\,
  \frac{\partial\,\beta_{\vartheta,\mathrm{act}}^{(p)}}{\partial V_n},
  \label{eq:per_event_derivative}
\end{multline}
where the superscript $(p)$ denotes evaluation at $\bm{\theta}^{(p)}$ with coupling vector $\bm{\gamma}_p^{\mathrm{act}}$.

\textbf{Multi-event stationarity system.}
Setting $\partial\mathcal{C}/\partial V_n = 0$ for each $n = 1,\ldots,N$ and substituting Eq.~\eqref{eq:per_event_derivative} into Eq.~\eqref{eq:multi_event_gradient}:
\begin{equation}
  \resizebox{\columnwidth}{!}{$\displaystyle
      \boxed{
        \sum_{p:\,n\in\mathcal{A}_p}
        w_p\left[
          \frac{1}{\bigl[\Delta\mathcal{I}_{\mathrm{eff}}^{(\vartheta),\mathrm{act},(p)}\bigr]^{2}}\,
          \frac{d\,\Phi_{\vartheta}^{(n)}(V;\,\bm{\gamma}_p^{\mathrm{act}})}{dV}\bigg|_{V_n}
          - 2\,\beta_{\vartheta,\mathrm{act}}^{(p)}\,
          \frac{\partial\,\beta_{\vartheta,\mathrm{act}}^{(p)}}{\partial V_n}
          \right] = 0}$}%
  \label{eq:multi_event_stationarity}
\end{equation}
for $n = 1, \ldots, N$.  This is a system of $N$ coupled nonlinear equations in the $N$ threshold voltages, generalizing the single-event stationarity Eq.~\eqref{eq:general_stationarity}.

Each stationarity equation is a \emph{weighted sum} of marginal Fisher-information gains minus marginal bias costs, accumulated across all events that trigger threshold $n$.  The weighting factor $1/[\Delta\mathcal{I}_{\mathrm{eff}}^{(\vartheta),\mathrm{act},(p)}]^{2}$ acts as an automatic \emph{relevance weight}: events with lower effective information (i.e., higher MSE) exert a proportionally stronger pull on the threshold placement, reflecting the convexity of $1/\mathcal{I}$.

\emph{Well-specified simplification} ($\varepsilon \equiv 0$).
In the absence of model misspecification, the bias terms vanish and the stationarity system reduces to:
\begin{multline}
  \sum_{p:\,n\in\mathcal{A}_p}
  \frac{w_p}{\bigl[\Delta\mathcal{I}_{\mathrm{eff}}^{(\vartheta),\mathrm{act},(p)}\bigr]^{2}}\,
  \frac{d\,\Phi_{\vartheta}^{(n)}(V;\,\bm{\gamma}_p^{\mathrm{act}})}{dV}\bigg|_{V_n}\\
  = 0,
  \quad n = 1,\ldots,N.
  \label{eq:multi_event_bias_free}
\end{multline}

\textbf{Self-consistent fixed-point iteration.}
The multi-event problem involves $P$ coupled nuisance coupling vectors $\{\bm{\gamma}_p^{\mathrm{act}}\}_{p=1}^P$, each depending on the threshold configuration through the active set:
\begin{equation}
  \begin{split}
    \bm{\gamma}_p^{\mathrm{act}}
     & = \left(\sum_{n \in \mathcal{A}_p}
    \bm{\mathcal{I}}_{\bm{\eta}\bm{\eta}}^{(n)}(\bm{\theta}^{(p)})\right)^{\!-1} \\
     & \quad \sum_{n \in \mathcal{A}_p}
    \bm{\mathcal{I}}_{\bm{\eta}\vartheta}^{(n)}(\bm{\theta}^{(p)}),
    \quad p = 1,\ldots,P.
  \end{split}
  \label{eq:multi_event_lambda}
\end{equation}
The alternating optimization of Section~\ref{sec:opt_derivation}, Eq.~\eqref{eq:fixed_point_lambda}, extends naturally:
\begin{enumerate}
  \item \emph{$\bm{\gamma}$-update}: Given current thresholds $\{V_n\}$, compute $\mathcal{A}_p$ via Eq.~\eqref{eq:active_set_p} and update $\bm{\gamma}_p^{\mathrm{act}}$ via Eq.~\eqref{eq:multi_event_lambda} for each event $p = 1,\ldots,P$.
  \item \emph{$V$-update}: Given the fixed coupling vectors $\{\bm{\gamma}_p^{\mathrm{act}}\}$, solve (or iterate toward a solution of) the $N$ stationarity equations~\eqref{eq:multi_event_stationarity} for the updated thresholds $\{V_n\}$.
\end{enumerate}

\emph{Remark} (Reduction to the single-event, full-triggering case): Setting $P = 1$ and $\mathcal{A}_1 = \{1,\ldots,N\}$ recovers the single-event stationarity system Eq.~\eqref{eq:general_stationarity} exactly, confirming that the multi-event formulation is a proper generalization.

\emph{Remark} (Threshold allocation trade-off):
The structure of the stationarity system~\eqref{eq:multi_event_stationarity} reveals a fundamental resource-allocation principle.  Lower thresholds ($V_n$ small) are \emph{shared resources}: they belong to the active set $\mathcal{A}_p$ of every event regardless of amplitude, so their placement is influenced by the entire event population.  Upper thresholds ($V_n$ large) are \emph{dedicated resources}: they are active only for high-energy events and their placement is governed predominantly by those events.  The optimal design allocates lower thresholds to maximize information for the worst-served (lowest-energy) events---which exert the largest automatic relevance weight $1/[\Delta\mathcal{I}]^2$---while the upper thresholds are fine-tuned for photopeak events.  This creates a natural spectral partitioning of the threshold array that mirrors the detection system's energy spectrum.

The following section instantiates the general design framework---including the robustness analysis (Section~\ref{sec:robustness}) and the multi-event partial-triggering optimization developed above---for the concrete bi-exponential pulse model.

% =========================================================================
%  VIII. BI-EXPONENTIAL INSTANTIATION AND COMPREHENSIVE ANALYSIS
% =========================================================================
\section{Bi-Exponential Instantiation}
\label{sec:optimal}

The preceding sections developed a general, pulse-shape-agnostic theoretical framework for MVT sampling.  We now translate that abstract machinery into concrete, verifiable results by selecting a specific analytical pulse model and systematically instantiating every layer of the theory---from deterministic identifiability through stochastic modeling, parameter estimation, performance bounds, paired information extraction, and ultimately optimal threshold design.

\subsection{Pulse Model}
\label{sec:biexp_model}

\textbf{Physical motivation.}
In inorganic scintillation crystals (e.g., LYSO:Ce, BGO), the optical photon emission process following a gamma-ray energy deposit is well described by the competition between two characteristic relaxation mechanisms: (i) the rapid initial population of luminescent states, governed by a \emph{rise time constant} $\tau_r$, and (ii) the subsequent slow de-excitation, governed by a \emph{decay time constant} $\tau_d \gg \tau_r$.  The superposition of these two exponential processes yields a strictly unimodal macroscopic voltage pulse.  Although more elaborate multi-component or non-exponential models exist (e.g., incorporating afterglow or non-radiative channels), the bi-exponential model strikes an effective balance: it captures the dominant first-order temporal dynamics with only a small number of free parameters, making it amenable to real-time FPGA-based fitting while remaining analytically tractable~\cite{kimble2002scintillation,mao2013crystal,brunner2017bgo,okajima1982characteristics}.

In the fitted voltage model below, $\tau_r$ and $\tau_d$ should be read as \emph{effective macroscopic} time constants of the measured readout waveform.  They absorb convolution with the scintillation emission profile, the SiPM single-photoelectron response, and any analog front-end shaping; they are therefore not identical to microscopic material constants unless those additional responses are negligible or separately deconvolved.

\emph{Assumption} (Effective bi-exponential macromodel): Over the threshold range used for reconstruction, the measured single-pulse waveform is assumed to be dominated by one effective rise component and one effective decay component after baseline subtraction.  This is physically reasonable for isolated LYSO/SiPM pulses in direct readout because the combined scintillation emission, SiPM response, and oscilloscope bandwidth produce a smooth single-peaked voltage waveform.  It does not assert that the crystal or sensor has exactly two microscopic kinetic channels; secondary luminescence components, saturation, afterpulsing, or shaping tails are represented by the deterministic mismatch field $\varepsilon(t)$ and are diagnosed experimentally in Section~\ref{sec:delta_validation}.

\textbf{Mathematical definition.}
The full bi-exponential pulse model, including an explicit voltage baseline offset $b$ $[\mathrm{V}]$, reads:
\begin{equation}
  f(t;\bm{\theta}) = A\!\left(e^{-\frac{t-t_0}{\tau_d}}-e^{-\frac{t-t_0}{\tau_r}}\right)\cdot u(t-t_0) + b,
  \label{eq:biexp}
\end{equation}
where the complete parameter vector is
\begin{equation}
  \bm{\theta} = [A,\; t_0,\; \tau_d,\; \tau_r,\; b]^T \in \mathbb{R}^5,
  \label{eq:biexp_theta}
\end{equation}
and each component carries the following physical meaning:
\begin{itemize}
  \item $A > 0$ $[\mathrm{V}]$: the waveform scale amplitude, proportional to the total deposited gamma-ray energy; the observable peak voltage is $A h_{\mathrm{peak}} + b$, not $A$ itself;
  \item $t_0$ $[\mathrm{s}]$: the absolute photon arrival time (the timestamp of the scintillation event onset);
  \item $\tau_d > 0$ $[\mathrm{s}]$: the effective macroscopic decay time constant, dominated by the slow luminescence relaxation and readout shaping;
  \item $\tau_r > 0$ $[\mathrm{s}]$: the effective macroscopic rise time constant, governing the leading-edge buildup of the measured voltage waveform, with the physical constraint $\tau_d > \tau_r$;
  \item $b$ $[\mathrm{V}]$: the DC voltage baseline offset of the analog front-end, representing the quiescent output level in the absence of any scintillation signal.
\end{itemize}
The Heaviside unit step function $u(\cdot)$ ($u(s)=1$ for $s>0$, $u(s)=0$ for $s\le 0$) enforces physical causality: the signal contribution is identically zero before the photon arrives at $t_0$.  Although $u(\cdot)$ introduces a derivative discontinuity at $t = t_0$, all threshold crossings occur strictly at $t > t_0$ (since $V_k > b = f(t_0;\bm{\theta})$), where $f$ reduces to a sum of smooth exponentials and is $C^{\infty}$.  Thus the $C^2$ regularity required in Section~\ref{sec:model} is satisfied at every crossing point.

\textbf{Normalized shape function.}
Defining the dimensionless shape kernel $h(x) \triangleq e^{-x/\tau_d} - e^{-x/\tau_r}$ for $x > 0$, the pulse factors as $f(t;\bm{\theta}) = A\,h(t - t_0) + b$.  The function $h(x)$ rises from $h(0^+) = 0$, attains its unique peak $h_{\mathrm{peak}} \triangleq \max_{x>0} h(x)$ at the normalized time
\begin{equation}
  x_p = \frac{\tau_r\,\tau_d}{\tau_d - \tau_r}\ln\!\left(\frac{\tau_d}{\tau_r}\right),
  \label{eq:biexp_xp}
\end{equation}
obtained by solving $h'(x_p) = 0$ (see Appendix~\ref{app:peak_time}), and then decays monotonically to zero.

\textbf{Threshold crossing structure.}
At each hardware threshold $V_n$, the crossing condition $f(t;\bm{\theta}) = V_n$ reduces to $A\,h(x) = V_n - b$, or equivalently $h(x) = (V_n - b)/A$, where $x \triangleq t - t_0$. This scalar equation yields two roots: $x_{r,n} < x_p$ on the rising branch and $x_{f,n} > x_p$ on the falling branch. The physically admissible parameter domain is:
\begin{equation}
  \begin{aligned}
    \Theta_V^{\mathrm{bi}} = \bigl\{ \bm{\theta} \mid\;
     & A > (V_{\max} - b)/h_{\mathrm{peak}},                         \\
     & \tau_d > \tau_r > 0,\; t_0 \in \mathbb{R},\; b < V_1 \bigr\}.
  \end{aligned}
  \label{eq:biexp_domain}
\end{equation}
The condition $b < V_1$ ensures that all thresholds lie above the baseline, i.e., the net signal $V_n - b > 0$ at every threshold level.

For the topological embedding claim used by the inverse-problem theory, we work on a physically bounded compact prior subdomain
\begin{equation}
  \begin{aligned}
    \bar{\Theta}_V^{\mathrm{bi}}
    = \bigl\{\bm{\theta}\in\Theta_V^{\mathrm{bi}} \;\big|\;&
    A_- \le A \le A_+,\quad
    t_0^- \le t_0 \le t_0^+,                                      \\
    & \tau_d^- \le \tau_d \le \tau_d^+,
    \quad \tau_r^- \le \tau_r \le \tau_r^+,                       \\
    & \tau_d-\tau_r \ge \Delta_\tau,
    \quad b_- \le b \le b_+ < V_1,                                \\
    & A h_{\mathrm{peak}}(\tau_d,\tau_r)+b \ge V_{\max}+\Delta_p
    \bigr\},
  \end{aligned}
  \label{eq:biexp_compact_prior_domain}
\end{equation}
with fixed margins $A_->0$, $\Delta_\tau>0$, and $\Delta_p>0$. These bounds encode ordinary hardware and material priors: finite SiPM and oscilloscope dynamic range, finite threshold-DAC span, bounded baseline motion, and bounded effective crystal/readout kinetics. They also exclude purely algebraic coupled escape modes that have no counterpart in a calibrated detector, such as simultaneous $A,\tau_d\to\infty$ with $b\to-\infty$ arranged to keep a finite set of threshold crossings nearly fixed.

\subsection{Identifiability Analysis}
\label{sec:biexp_embedding}

We now verify the ingredients needed for the closed-embedding statement used in reconstruction. Immersion and injectivity are algebraic properties of the five-parameter bi-exponential family on the admissible open domain $\Theta_V^{\mathrm{bi}}$. Properness, however, is asserted on the compact physical prior domain $\bar{\Theta}_V^{\mathrm{bi}}$ in Eq.~\eqref{eq:biexp_compact_prior_domain}; Appendix~\ref{app:proper} explains why this compact-domain formulation is the mathematically stable version relevant to real hardware.

\textbf{A rigorous five-parameter reduction.}
Introduce the post-onset exponential coordinates
\begin{equation}
  \lambda_d \triangleq \frac{1}{\tau_d},
  \qquad
  \lambda_r \triangleq \frac{1}{\tau_r},
  \qquad
  D \triangleq A e^{\lambda_d t_0},
  \qquad
  E \triangleq A e^{\lambda_r t_0},
  \label{eq:biexp_reparam}
\end{equation}
for which $0 < \lambda_d < \lambda_r$ and $D,E > 0$. Since every threshold crossing satisfies $t_k > t_0$, the post-onset branch of Eq.~\eqref{eq:biexp} can be written exactly as
\begin{equation}
  f(t;\bm{\theta}) = b + D e^{-\lambda_d t} - E e^{-\lambda_r t},
  \qquad t > t_0.
  \label{eq:biexp_post_onset}
\end{equation}
For later estimation and Fisher-information calculations, the parameter sensitivities at a crossing $t_k = t_0 + x_k$ remain
\begin{equation}
  \begin{aligned}
    \frac{\partial f}{\partial A}      & = h(x_k), \qquad
    \frac{\partial f}{\partial t_0} = -A\,h'(x_k),                                    \\
    \frac{\partial f}{\partial \tau_d} & = A\,\frac{x_k}{\tau_d^2}\,e^{-x_k/\tau_d},  \\
    \frac{\partial f}{\partial \tau_r} & = -A\,\frac{x_k}{\tau_r^2}\,e^{-x_k/\tau_r}, \\
    \frac{\partial f}{\partial b}      & = 1,
  \end{aligned}
  \label{eq:biexp_partials}
\end{equation}
The change of variables $\Psi:(A,t_0,\tau_d,\tau_r,b) \mapsto (D,E,\lambda_d,\lambda_r,b)$ is a smooth diffeomorphism onto its image, with inverse
\begin{equation}
  \begin{aligned}
    \tau_d & = \frac{1}{\lambda_d},
           & \tau_r                                  & = \frac{1}{\lambda_r},                         \\
    t_0    & = \frac{\ln(E/D)}{\lambda_r-\lambda_d},
           & A                                       & = D e^{-\lambda_d t_0} = E e^{-\lambda_r t_0}.
  \end{aligned}
  \label{eq:biexp_reparam_inverse}
\end{equation}

The crucial observation is that the five-function family
\begin{equation}
  \mathcal{F}_5(\lambda_d,\lambda_r)
  = \left\{1,\; e^{-\lambda_d t},\; t e^{-\lambda_d t},\right.
  \left.e^{-\lambda_r t},\; t e^{-\lambda_r t}\right\}
  \label{eq:biexp_ect_family}
\end{equation}
is an extended complete Chebyshev system on every interval of post-onset times. This conclusion follows from the classical theory of real exponential-polynomial Chebyshev systems with prescribed multiplicities; the Wronskian below verifies the specific two-node confluent case needed here. Its Wronskian is
\begin{equation}
  W_{\mathcal{F}_5}(t)
  = \lambda_d^2 \lambda_r^2 (\lambda_r-\lambda_d)^4 e^{-2(\lambda_d+\lambda_r)t}
  > 0,
  \label{eq:biexp_ect_wronskian}
\end{equation}
as proved in Appendix~\ref{app:rank}. Consequently, every nonzero linear combination of the functions in Eq.~\eqref{eq:biexp_ect_family} has at most four zeros, counted with multiplicity.

\textbf{1) Global Immersion ($\operatorname{rank}(\mathbf{J}) = 5$):}
Let $\delta\bm{\theta}$ be a tangent perturbation and denote its image under the coordinate change by $\delta\bm{\omega} = (\delta D,\delta E,\delta \lambda_d,\delta \lambda_r,\delta b) = d\Psi(\bm{\theta})\,\delta\bm{\theta}$. If $\mathbf{J}\,\delta\bm{\theta} = \mathbf{0}$, then the first-order shift of every crossing time vanishes. Differentiating the level-set equation $f(t_k;\bm{\theta}) = V_{\ell(k)}$ with $dt_k = 0$ gives
\begin{equation}
  \begin{aligned}
    \delta b
     & + \delta D\,e^{-\lambda_d t_k}
    - \delta E\,e^{-\lambda_r t_k}                   \\
     & - D\,\delta\lambda_d\, t_k e^{-\lambda_d t_k}
    + E\,\delta\lambda_r\, t_k e^{-\lambda_r t_k}
    = 0
  \end{aligned}
  \label{eq:biexp_immersion_zero}
\end{equation}
for every crossing $t_k$. The left-hand side is a linear combination of the five functions in Eq.~\eqref{eq:biexp_ect_family}. Because the observation cone $\Omega_K$ enforces $K = 2N$ distinct crossing times and $N \ge 3$ gives $K \ge 6$, Eq.~\eqref{eq:biexp_immersion_zero} yields a nonzero member of $\operatorname{span}\mathcal{F}_5$ vanishing at at least six distinct points, which is impossible by Eq.~\eqref{eq:biexp_ect_wronskian}. Therefore $\delta\bm{\omega} = \mathbf{0}$, hence $\delta\bm{\theta} = \mathbf{0}$, and $\operatorname{rank}(\mathbf{J}) = 5$ everywhere on $\Theta_V^{\mathrm{bi}}$.

\textbf{Minimum threshold count requirement:} Since the parameter dimension is $M = 5$ and each threshold generates $2$ observations, the counting condition $K = 2N \ge M$ requires $N \ge 3$. The argument above shows that three distinct thresholds are not merely necessary: they are already sufficient for full-rank immersion.

\textbf{2) Global Injectivity:}
Assume $S(\bm{\theta}) = S(\tilde{\bm{\theta}})$ for two admissible parameter vectors. Let the common crossing times be $s_1,\dots,s_K$ with $K = 2N \ge 6$. Because every threshold lies strictly above both baselines, each common crossing time satisfies $s_i > t_0$ and $s_i > \tilde t_0$, so both pulses admit the post-onset representations
\begin{equation}
  \begin{aligned}
    f_{\bm{\theta}}(t)         & = b + D e^{-\lambda_d t} - E e^{-\lambda_r t},                                  \\
    f_{\tilde{\bm{\theta}}}(t) & = \tilde b + \tilde D e^{-\tilde\lambda_d t} - \tilde E e^{-\tilde\lambda_r t}.
  \end{aligned}
  \label{eq:biexp_two_reps}
\end{equation}
Their difference
\begin{equation}
  g(t)
  = (b-\tilde b)
  + D e^{-\lambda_d t}
  - E e^{-\lambda_r t}
  - \tilde D e^{-\tilde\lambda_d t}
  + \tilde E e^{-\tilde\lambda_r t}
  \label{eq:biexp_injectivity_difference}
\end{equation}
vanishes at all $K \ge 6$ common crossings. After merging repeated exponents, $g$ becomes a linear combination of at most five functions of the form $1,e^{-\lambda_1 t},\dots,e^{-\lambda_{m_\lambda} t}$ with distinct positive exponents $\lambda_j$, where $m_\lambda$ is the local count of distinct positive exponents. This family is again a Chebyshev system (Appendix~\ref{app:rank}), so any nonzero member can have at most $m_\lambda \le 4$ zeros. Hence $g \equiv 0$.

Linear independence of distinct real exponentials implies $b = \tilde b$ and the exponent multisets coincide: $\{\lambda_d,\lambda_r\} = \{\tilde\lambda_d,\tilde\lambda_r\}$. Since both parameter vectors satisfy $0 < \lambda_d < \lambda_r$ and $0 < \tilde\lambda_d < \tilde\lambda_r$, we obtain $\lambda_d = \tilde\lambda_d$ and $\lambda_r = \tilde\lambda_r$, after which Eq.~\eqref{eq:biexp_injectivity_difference} reduces to $(D-\tilde D)e^{-\lambda_d t} - (E-\tilde E)e^{-\lambda_r t} \equiv 0$, forcing $D = \tilde D$ and $E = \tilde E$. Inverting Eq.~\eqref{eq:biexp_reparam_inverse} yields $\bm{\theta} = \tilde{\bm{\theta}}$. Therefore the forward map is globally injective.

\textbf{3) Properness on the compact physical prior domain and asymptotic information degeneracy:}
Appendix~\ref{app:proper} records the compact-domain properness argument for the forward map restricted to $\bar{\Theta}_V^{\mathrm{bi}}$. We do not claim properness on the unbounded algebraic domain $\Theta_V^{\mathrm{bi}}$ without physical bounds: coupled limits can be constructed in which several parameters diverge together while a finite set of crossings remains bounded. In real detector operation these directions are excluded by finite dynamic range, threshold span, baseline calibration, and bounded effective pulse kinetics. On $\bar{\Theta}_V^{\mathrm{bi}}$, properness follows from compactness and continuity, while the immersion and injectivity arguments above provide the local differential and global uniqueness ingredients.

Together with the immersion and injectivity results above, this establishes the forward map $S|_{\bar{\Theta}_V^{\mathrm{bi}}}$ as a \emph{closed embedding} on the physically bounded prior domain used for reconstruction, completing the identifiability analysis for the five-parameter bi-exponential model in the experimentally relevant regime.

\emph{Physical consequence (asymptotic information degeneracy):} If the hardware records \emph{only} rising edges, large-amplitude pulses compress the rising timestamps toward $t_0$, weakening the Fisher Information for $A$ and simultaneously reducing sensitivity to $\tau_d$ (which governs the tail). Thus a rising-edge-only architecture is fragile near the high-amplitude edge of the physical prior; retaining both edges is the robust choice when one seeks a well-conditioned embedding over the intended operating range.

\subsection{Noise Model}
\label{sec:biexp_stochastic}

Applying the compound noise model of Section~\ref{sec:stochastic} to the bi-exponential pulse, the total analog voltage variance at the $k$-th ordered crossing is $\sigma_V^2(V_{\ell(k)}) = \sigma_{\mathrm{th}}^2 + \kappa(V_{\ell(k)} - b)$. Here the Poisson shot noise term is $\kappa(V_{\ell(k)} - b)$ rather than $\kappa V_{\ell(k)}$ because only the net signal component $V_{\ell(k)} - b = A\,h(x_k)$ originates from photoelectron statistics; the baseline $b$ is a deterministic DC offset of the front-end amplifier and carries no shot noise.

The projected temporal variance at each crossing follows from Eq.~\eqref{eq:variance}:
\begin{equation}
  \sigma_k^2 = \sigma_{\mathrm{TDC}}^2 + \frac{\sigma_{\mathrm{th}}^2 + \kappa (V_{\ell(k)} - b)}{A^2 [h'(x_k)]^2}.
  \label{eq:biexp_variance}
\end{equation}

\textbf{Bandwidth-correlation diagnostic.}
The autocorrelation limit Eq.~\eqref{eq:autocorrelation} is a sufficient condition for the diagonal-covariance approximation: if adjacent thresholds $V_i$, $V_j$ on the same slope branch satisfy $|V_i - V_j| \ge L_{ij}\,\tau_c$, their crossing-time errors can be treated as approximately independent.  For a single threshold pair treated by the diagonal closed form Eq.~\eqref{eq:pair_fim}, one additionally needs the pair gap $t_f(V)-t_r(V) \ge \tau_c$; this is the condition most likely to fail near the pulse peak, where the pair gap collapses. It is not a requirement for threshold sampling itself.  If either condition is not met, the MVT observations remain valid, but the off-diagonal covariance terms between the corresponding crossings should be retained rather than discarded.  For the bi-exponential pulse, the supremum slew rate on the rising edge near $t_0$ is $L_{\mathrm{max}}^{(r)} = A(\lambda_r - \lambda_d)$ (attained as $x \to 0^+$, where $\lambda_d = 1/\tau_d$, $\lambda_r = 1/\tau_r$). On the falling edge, the exact maximum slope magnitude occurs at $x=2x_p>x_p$ and is strictly smaller than $A\lambda_d$; asymptotically, the tail obeys $|\partial_t f| \sim A\lambda_d e^{-x/\tau_d}$ because it is dominated by the slow decay exponential.  Using the global rising-edge supremum gives the conservative worst-case spacing diagnostic
\begin{equation}
  \Delta V \ge A\!\left(\frac{1}{\tau_r} - \frac{1}{\tau_d}\right) \cdot \tau_c.
  \label{eq:biexp_bandwidth}
\end{equation}

\textbf{Physical implication:} Equation~\eqref{eq:biexp_bandwidth} is homogeneous in the pulse amplitude, so the useful quantity is the normalized spacing
\begin{equation}
  \frac{\Delta V}{A}
  \ge
  \left(\frac{1}{\tau_r} - \frac{1}{\tau_d}\right)\tau_c .
  \label{eq:biexp_bandwidth_normalized}
\end{equation}
No numerical value of $A$ is needed to obtain the coefficient multiplying $A$.  For the experimental setup used in Section~\ref{sec:exp_setup}, the oscilloscope analog bandwidth is $B = 16\,\mathrm{GHz}$, so the engineering proxy $\tau_c \approx 1/(2B)$ gives $\tau_c \approx 31\,\mathrm{ps}$.  Using the \emph{effective macroscopic} rise constant fitted to the voltage waveform, $\tau_{r,\mathrm{eff}} \simeq 0.5\,\mathrm{ns}$, together with $\tau_d \simeq 40\,\mathrm{ns}$ therefore gives
\begin{equation}
  \frac{\Delta V}{A}
  \gtrsim
  \left(2.000 - 0.025\right) \times 0.031 \simeq 0.062 .
\end{equation}
Thus $\Delta V \gtrsim 0.062A$ is the per-amplitude, worst-case global-slope diagnostic under the stated $16\,\mathrm{GHz}$ bandwidth assumption: for a $100\,\mathrm{mV}$ pulse it corresponds to about $6.2\,\mathrm{mV}$, while for a $1\,\mathrm{V}$ pulse it corresponds to about $62\,\mathrm{mV}$.  This number is intentionally very conservative because it uses the global rising-edge supremum $L_{\max}^{(r)} = A(\lambda_r - \lambda_d)$, attained only in the limit $x \to 0^+$ immediately after threshold emergence.  On the actual rising branch $0 < x < x_p$, the local slew is
\begin{equation}
  \partial_t f(x) = A\!\left(\lambda_r e^{-\lambda_r x} - \lambda_d e^{-\lambda_d x}\right),
\end{equation}
and, since $h''(x)<0$ for $0<x<x_p$, it decreases monotonically from that supremum to $0$ at the peak; hence for any finite threshold interval away from the immediate onset, the relevant local bound $L_{ij}$ is strictly smaller, often substantially smaller, than $A(\lambda_r - \lambda_d)$.  This calculation should not be performed with the intrinsic scintillation rise time of LYSO, $\tau_{r,\mathrm{scint}} \approx 70\,\mathrm{ps}$, because that microscopic kinetic constant is not the same parameter as the fitted macroscopic $\tau_r$ in Eq.~\eqref{eq:biexp}.  The latter absorbs convolution with the single-photoelectron response $h_e(t)$ and front-end shaping, and is therefore typically slower.  A much larger value such as $\tau_c \simeq 1\,\mathrm{ns}$ would correspond instead to an effective bandwidth of only about $500\,\mathrm{MHz}$, and should therefore be used only if an independently justified \emph{effective} noise bandwidth, rather than the instrument's nominal analog bandwidth, is intended.  Under the present experimental assumptions, the diagnostic no longer implies an unrealistically large threshold spacing; it should be read as an upper-envelope warning for the earliest and steepest part of the rising edge, not as a representative spacing requirement across the full threshold range.  Practical designs should therefore prefer the local-slope estimate $L_{ij}$ on the actual threshold interval, an empirically measured correlation time, or a full covariance model when adjacent same-branch crossings or near-peak rising/falling pair members are correlated.

\subsection{Estimator Specification}
\label{sec:biexp_estimation}

With $M = 5$ parameters and $K = 2N$ observations, the weighted least-squares objective Eq.~\eqref{eq:cost} becomes:
\begin{equation}
  \mathcal{E}_{\mathrm{WLS}}(\bm{\theta}) = \frac{1}{2}\sum_{k=1}^{2N} \frac{[t_k^{\mathrm{obs}} - t_k(\bm{\theta})]^2}{\sigma_k^2(\bm{\theta})},
  \label{eq:biexp_cost}
\end{equation}
where each model-predicted crossing time $t_k(\bm{\theta})$ is the implicit solution of $A\,h(t_k - t_0) + b = V_{\ell(k)}$.

The Gauss-Newton iteration Eq.~\eqref{eq:gauss_newton} operates on the $5 \times 5$ approximate Hessian $\mathbf{H} = \mathbf{J}^T \mathbf{W} \mathbf{J}$, with the $K \times 5$ Jacobian $\mathbf{J}$ specified by Eqs.~\eqref{eq:jacobian} and \eqref{eq:biexp_partials}, and the diagonal precision matrix $\mathbf{W}$ from Eq.~\eqref{eq:weight_matrix}.  The update step is:
\begin{equation}
  \Delta\bm{\theta} = (\mathbf{J}^T \mathbf{W} \mathbf{J})^{-1}\,\mathbf{J}^T \mathbf{W}\,\Delta\mathbf{t}^{(\ell)}.
  \label{eq:biexp_gn_update}
\end{equation}

\textbf{Convergence considerations.}
The global immersion condition ($\operatorname{rank}(\mathbf{J}) = 5$, verified in Section~\ref{sec:biexp_embedding}) guarantees that $\mathbf{H}$ is strictly positive definite, ensuring local convergence (quadratic in the zero-residual limit) of the Gauss-Newton iterates.  However, enlarging the parameter space from $2$ to $5$ dimensions increases the curvature of the model manifold.  In practice, the shape parameters $\tau_d$ and $\tau_r$ introduce non-negligible second-order effects, and a Levenberg-Marquardt damping term $\mu\,\mathbf{I}$ may be beneficial during the initial iterations to prevent overshooting in the $(\tau_d, \tau_r)$ subspace.

\textbf{Baseline parameter $b$.}
The baseline $b$ enters the crossing condition linearly: $A\,h(x_k) = V_{\ell(k)} - b$. This means the Jacobian column for $b$ is simply $J_{k,b} = -1/[\partial_t f(t_k)] = -1/[A\,h'(x_k)]$. The amplitude column is $J_{k,A} = -h(x_k)/[A\,h'(x_k)]$, so for a single threshold pair the two columns are proportional because $h(x_r)=h(x_f)=(V-b)/A$. In a multi-threshold system with at least two distinct threshold levels, however, the corresponding $h(x_k)$ values differ across threshold pairs, breaking this proportionality and making the amplitude-baseline coupling generically non-degenerate.

\subsection{Performance Bounds}
\label{sec:biexp_crlb}

\textbf{Well-specified bound.}
Under the High-SNR Slepian-Bangs reduction (Appendix~\ref{app:slepian}), the $5 \times 5$ Fisher Information Matrix for the bi-exponential model is $\mathcal{I}(\bm{\theta}) \approx \mathbf{J}^T \mathbf{W} \mathbf{J}$, with entries computed from Eq.~\eqref{eq:biexp_partials}.  The Cram\'{e}r-Rao Lower Bound for each parameter $\theta_j$ is:
\begin{equation}
  \mathrm{Var}(\hat{\theta}_j) \ge [\mathcal{I}^{-1}(\bm{\theta})]_{jj}.
  \label{eq:biexp_crlb}
\end{equation}

\textbf{Misspecification analysis.}
When physical distortions $\varepsilon(t)$ (SiPM saturation, pile-up, afterpulsing) perturb the true waveform away from the ideal Eq.~\eqref{eq:biexp}, the leading-order systematic bias vector from Eq.~\eqref{eq:bias_vector} is approximated by:
\begin{equation}
  \bm{\beta} \approx (\mathbf{J}^T \mathbf{W} \mathbf{J})^{-1}\,\mathbf{J}^T \mathbf{W}\,\Delta\mathbf{t}_{\mathrm{bias}} \;\in\; \mathbb{R}^5,
  \label{eq:biexp_bias}
\end{equation}
where $\Delta t_{\mathrm{bias},k} \approx -\varepsilon(t_k)/\partial_t f(t_k;\bm{\theta}_0)$ as before.  The bias now has components along all five parameter axes: $\bm{\beta} = [\beta_A,\; \beta_{t_0},\; \beta_{\tau_d},\; \beta_{\tau_r},\; \beta_b]^T$.

The overarching leading-order misspecified MSE surrogate bound from Eq.~\eqref{eq:mse_bound} becomes the $5 \times 5$ matrix inequality:
\begin{equation}
  \mathrm{MSE}(\hat{\bm{\theta}}) \succeq \mathcal{I}^{-1}(\bm{\theta}_0) + \bm{\beta}\bm{\beta}^T.
  \label{eq:biexp_mse}
\end{equation}

\emph{Physical consequence for the baseline parameter:} The inclusion of $b$ as an explicit free parameter absorbs a significant portion of low-frequency model mismatch that would otherwise manifest as correlated biases in $A$ and $t_0$.  In particular, slow baseline wander---previously a dominant source of deterministic bias---is now captured within the statistical framework.  However, the cost is an enlarged Fisher matrix whose inversion yields increased marginal variances for all parameters due to inter-parameter coupling.

\subsection{Paired Information Structure}
\label{sec:biexp_paired}

For the five-parameter model, the pairwise waveform sensitivity vector Eq.~\eqref{eq:pair_sensitivity} at threshold $V$ has five components per edge:
\begin{equation}
  \begin{aligned}
    g_{A,k}      & = h(x_k) = (V - b)/A,                \\
    g_{t_0,k}    & = -A\,h'(x_k),                       \\
    g_{\tau_d,k} & = A\,x_k\,e^{-x_k/\tau_d}/\tau_d^2,  \\
    g_{\tau_r,k} & = -A\,x_k\,e^{-x_k/\tau_r}/\tau_r^2, \\
    g_{b,k}      & = 1.
  \end{aligned}
  \label{eq:biexp_sensitivity}
\end{equation}

The single-pair FIM Eq.~\eqref{eq:pair_fim} is now a $5 \times 5$ matrix:
\begin{equation}
  \mathcal{I}_{jl}^{(\mathrm{pair})}(V) = \frac{g_{j,r}\,g_{l,r}}{D_r} + \frac{g_{j,f}\,g_{l,f}}{D_f},
  \label{eq:biexp_pair_fim}
\end{equation}
with composite noise denominators $D_r$, $D_f$ as defined in Section~\ref{sec:pair_general}.

\textbf{Single-pair profiled score for $t_0$.}
To isolate the timing information, we partition $\bm{\theta} = [\bm{\eta}^T,\, t_0]^T$ with nuisance vector $\bm{\eta} = [A,\, \tau_d,\, \tau_r,\, b]^T \in \mathbb{R}^4$.  Applying the generalized Schur complement Eq.~\eqref{eq:schur_general}:
\begin{equation}
  \Delta\mathcal{I}_{\mathrm{eff}}^{(t_0)}(V) = \mathcal{I}_{t_0 t_0} - \bm{\mathcal{I}}_{\bm{\eta} t_0}^T\,\bm{\mathcal{I}}_{\bm{\eta}\bm{\eta}}^{+}\,\bm{\mathcal{I}}_{\bm{\eta} t_0},
  \label{eq:biexp_eff_t0}
\end{equation}
For the full five-parameter model this isolated single-pair profiled score is generically zero, because two edge observations cannot profile out four nuisance degrees of freedom. Accordingly, a finite timing bound arises only after multiple threshold pairs are aggregated as in Section~\ref{sec:synergy}. Appendix~\ref{app:biexp_eff} derives the corresponding nondegenerate two-parameter timing special case Eq.~\eqref{eq:eff_linear}, which serves as the timing-specialized reference case before reintroducing the full nuisance set.

\textbf{Effective information for $A$.}
Similarly, we can isolate the amplitude information by partitioning $\bm{\theta} = [\bm{\eta}'^T,\, A]^T$ with $\bm{\eta}' = [t_0,\, \tau_d,\, \tau_r,\, b]^T$:
\begin{equation}
  \Delta\mathcal{I}_{\mathrm{eff}}^{(A)}(V) = \mathcal{I}_{AA} - \bm{\mathcal{I}}_{\bm{\eta}' A}^T\,\bm{\mathcal{I}}_{\bm{\eta}'\bm{\eta}'}^{+}\,\bm{\mathcal{I}}_{\bm{\eta}' A}.
  \label{eq:biexp_eff_A}
\end{equation}
For a \emph{single} threshold pair this generalized profiled score is identically zero: because both edges correspond to the same threshold voltage, $g_{A,k} = h(x_k) = (V-b)/A = [(V-b)/A]g_{b,k}$ for $k \in \{r,f\}$, so the amplitude sensitivity lies exactly in the baseline-nuisance span.  Appendix~\ref{app:baseline_penalty} works out this $A$--$b$ degeneracy explicitly in the $\{A,t_0,b\}$ submodel.  A nonzero amplitude bound is therefore a genuinely multi-pair effect and emerges only after sufficiently many distinct threshold pairs are aggregated to make the stacked nuisance block identifiable.

\textbf{Impact of additional nuisance parameters.}
Compared to the two-parameter model ($\bm{\theta} = [A, t_0]^T$), the enlarged nuisance space now includes $\tau_d$, $\tau_r$, and $b$.  By the Schur complement structure, the effective information for $A$ cannot increase and is generically reduced:
\begin{equation}
  \Delta\mathcal{I}_{\mathrm{eff}}^{(A),\mathrm{tot}}\big|_{M=5} \le \Delta\mathcal{I}_{\mathrm{eff}}^{(A),\mathrm{tot}}\big|_{M=2},
  \label{eq:info_reduction}
\end{equation}
once enough threshold pairs are aggregated that the nuisance blocks are interpreted by the ordinary or generalized Schur complement in the usual way.  Thus profiling out additional uncertain nuisance parameters through the Schur complement can only decrease (never increase) the effective information for the target parameter.  The magnitude of this degradation depends on how strongly $t_0$, $\tau_d$, $\tau_r$, and $b$ couple to $A$ through the waveform sensitivities, with the shape-parameter and baseline couplings varying across threshold placements.

\textbf{Information superadditivity.}
The superadditivity bound Eq.~\eqref{eq:synergy_bound} generalizes directly: when $N$ threshold pairs are jointly estimated, the total effective information satisfies $\Delta\mathcal{I}_{\mathrm{eff}}^{(A),\mathrm{tot}} \ge \sum_{n=1}^N \Delta\mathcal{I}_{\mathrm{eff}}^{(A),(n)}$.  For the full five-parameter model the single-pair amplitude scores on the right-hand side are zero, so any strictly positive amplitude information is itself a collective multi-pair effect.  With five parameters, geometrically heterogeneous threshold placements can gain additional leverage because the timing, shape, and baseline nuisance directions ($t_0$, $\tau_d$, $\tau_r$, $b$) provide more axes along which differently oriented pairwise covariance structures can decorrelate.  This is the amplitude-target analogue of the superadditivity effect described in Section~\ref{sec:synergy}.

\subsection{Optimal Threshold Design for $A$}
\label{sec:opt_A}

We now instantiate the general optimal threshold design framework of Section~\ref{sec:general_opt} for the bi-exponential model with the amplitude $A$ as the target parameter. The timing-specialized case follows by interchanging the target and nuisance roles of $A$ and $t_0$ in the partition $\bm{\theta} = [\bm{\eta}^T,\,\vartheta]^T$ and re-evaluating the nuisance-projected quantities accordingly.

\textbf{Nuisance coupling instantiation.}
Partitioning $\bm{\theta} = [\bm{\eta}'^T,\, A]^T$ with nuisance vector $\bm{\eta}' = [t_0,\, \tau_d,\, \tau_r,\, b]^T \in \mathbb{R}^{4}$, the global nuisance coupling vector (Eq.~\eqref{eq:nuisance_coupling_vector}) becomes:
\begin{equation}
  \bm{\gamma}^*
  = \bigl(\bm{\mathcal{I}}_{\bm{\eta}'\bm{\eta}'}^{\mathrm{tot}}\bigr)^{-1}\,
  \bm{\mathcal{I}}_{\bm{\eta}' A}^{\mathrm{tot}}
  \;\in\;\mathbb{R}^{4},
  \label{eq:biexp_lambda}
\end{equation}
where the FIM blocks are constructed from the five-parameter sensitivities Eq.~\eqref{eq:biexp_sensitivity} summed over all $N$ pairs (Eq.~\eqref{eq:summed_blocks}).

\textbf{Nuisance-projected sensitivities.}
From Eq.~\eqref{eq:projected_sensitivity}, the projected sensitivities at each edge of the $n$-th threshold pair take the bi-exponential form:
\begin{equation}
  \begin{aligned}
    \tilde{g}_{r}^{(n)}
     & = g_{A,r}(V_n) - \bm{\gamma}^{*T}\,\mathbf{g}_{\bm{\eta}',r}(V_n)               \\
     & = h(x_r^{(n)})
    + \gamma_{t_0}^*\,A\,h'(x_r^{(n)})
    - \gamma_{\tau_d}^*\,\frac{A\,x_r^{(n)}\,e^{-x_r^{(n)}/\tau_d}}{\tau_d^2}         \\
     & \quad+ \gamma_{\tau_r}^*\,\frac{A\,x_r^{(n)}\,e^{-x_r^{(n)}/\tau_r}}{\tau_r^2}
    - \gamma_b^*,
  \end{aligned}
  \label{eq:biexp_proj_sens}
\end{equation}
where $\bm{\gamma}^* = [\gamma_{t_0}^*,\, \gamma_{\tau_d}^*,\, \gamma_{\tau_r}^*,\, \gamma_b^*]^T$ and $x_r^{(n)}$, $x_f^{(n)}$ are the rising and falling roots of $h(x) = (V_n - b)/A$ (the falling-edge expression $\tilde{g}_{f}^{(n)}$ is analogous with $x_r^{(n)} \to x_f^{(n)}$).

\textbf{Nuisance-projected Fisher contribution.}
The per-pair contribution from the $n$-th threshold (Eq.~\eqref{eq:projected_fisher}) instantiates as:
\begin{equation}
  \Phi_{A}^{(n)}(V_n;\,\bm{\gamma}^*)
  = \frac{[\tilde{g}_{r}^{(n)}]^{2}}{D_r(V_n)}
  + \frac{[\tilde{g}_{f}^{(n)}]^{2}}{D_f(V_n)},
  \label{eq:biexp_phi}
\end{equation}
with composite noise denominators $D_r = \sigma_{\mathrm{th}}^2 + \kappa(V_n - b) + \sigma_{\mathrm{TDC}}^2\,A^2[h'(x_r^{(n)})]^2$ and $D_f$ defined analogously.

\textbf{Stationarity equations.}
Applying the general stationarity system Eq.~\eqref{eq:general_stationarity} with $\vartheta = A$, the $N$ coupled optimality conditions for the bi-exponential model read:
\begin{equation}
  \begin{split}
     & \frac{d\,\Phi_{A}^{(n)}(V;\,\bm{\gamma}^*)}{dV}
    \bigg|_{V=V_n^*}                                                                 \\
     & \quad= 2\bigl[\Delta\mathcal{I}_{\mathrm{eff}}^{(A),\mathrm{tot}}\bigr]^{2}\,
    \beta_{A,\mathrm{tot}}\,
    \frac{\partial \beta_{A,\mathrm{tot}}}{\partial V_n}\bigg|_{V_n = V_n^*}
  \end{split}
  \label{eq:biexp_stationarity}
\end{equation}
for $n = 1, \ldots, N$. The gradient $d\Phi_{A}^{(n)}/dV$ is evaluated via the chain rule through the implicit crossing derivatives $dx_k/dV = 1/[A\,h'(x_k)]$ (Appendix~\ref{app:opt_threshold}).

\textbf{Well-specified case ($\varepsilon = 0$).}
When the pulse model is perfectly specified, the systematic amplitude bias vanishes ($\beta_{A,\mathrm{tot}} = 0$) and the general stationarity system Eq.~\eqref{eq:bias_free_stationarity} instantiates as:
\begin{equation}
  \frac{d\,\Phi_{A}^{(n)}(V;\,\bm{\gamma}^*)}{dV}
  \bigg|_{V=V_n^*}
  = 0,
  \quad n = 1, \dots, N.
  \label{eq:biexp_bias_free_stationarity}
\end{equation}
In this regime, each threshold satisfies its own one-dimensional stationary condition for $\Phi_{A}^{(n)}(V;\bm{\gamma}^*)$ with $\bm{\gamma}^*$ frozen.  Because $\bm{\gamma}^*$ is determined globally from all pairs, these stationarity conditions remain coupled; a given $V_n^*$ is a local maximizer only if the corresponding second derivative is negative.  Expanding the derivative using the chain rule through the crossing derivatives $dx_k/dV = 1/[A\,h'(x_k)]$, this yields:
\begin{equation}
  \sum_{k \in \{r,f\}}
  \frac{\tilde{g}_k^{(n)}}{D_k}
  \!\left(2\,\tilde{g}_k'
  - \frac{\tilde{g}_k^{(n)}\,D_k'}{D_k}\right)
  \bigg|_{V=V_n^*}
  = 0,
  \quad n = 1, \dots, N,
  \label{eq:biexp_bias_free_expanded}
\end{equation}
where $\tilde{g}_k'$ and $D_k'$ are the projected-sensitivity and noise-denominator derivatives defined in Eqs.~\eqref{eq:gtilde_deriv} and \eqref{eq:Dk_deriv} below.  Equation~\eqref{eq:biexp_bias_free_expanded} states that, absent model misspecification, the marginal gain in nuisance-projected amplitude sensitivity exactly balances the marginal compound noise penalty at each optimal threshold, with no bias-cost term.

\textbf{Step 1: Explicit stationarity equation.}
Retaining both model misspecification ($\varepsilon \neq 0$) and TDC quantization jitter ($\sigma_{\mathrm{TDC}} > 0$), we directly expand the full stationarity condition Eq.~\eqref{eq:biexp_stationarity}.  Applying the quotient rule to each edge's contribution $\tilde{g}_k^2/D_k$ and the chain rule through $dx_k/dV = 1/[A\,h'(x_k)]$ (Appendix~\ref{app:opt_threshold}), the composite noise-denominator derivative is
\begin{equation}
  D_k' \;\triangleq\; \frac{dD_k}{dV}
  = \kappa + 2\sigma_{\mathrm{TDC}}^2\,A\,h''(x_k),
  \quad k \in \{r,f\},
  \label{eq:Dk_deriv}
\end{equation}
where the Poisson-noise contribution $\kappa$ is augmented by the TDC jitter variation $2\sigma_{\mathrm{TDC}}^2 A\,h''(x_k)$ through the pulse curvature; and the projected-sensitivity derivative is
\begin{equation}
  \tilde{g}_k' \;\triangleq\; \frac{d\tilde{g}_k}{dV}
  = \frac{1}{A} + \frac{\mathcal{P}_k}{h'(x_k)},
  \label{eq:gtilde_deriv}
\end{equation}
with nuisance-projected amplitude-shape factor
\begin{align}
  \mathcal{P}_k & \triangleq \gamma_{t_0}^*\,h''(x_k)
  - \frac{\gamma_{\tau_d}^*}{\tau_d^2}\,e^{-x_k/\tau_d}\bigl(1 - x_k/\tau_d\bigr) \nonumber                \\
                & \quad\; + \frac{\gamma_{\tau_r}^*}{\tau_r^2}\,e^{-x_k/\tau_r}\bigl(1 - x_k/\tau_r\bigr).
  \label{eq:proj_curvature}
\end{align}
Substituting into Eq.~\eqref{eq:biexp_stationarity}, the general optimal threshold equation for the bi-exponential model reads (Appendix~\ref{app:opt_threshold}):
\begin{equation}
  \boxed{
    \begin{aligned}
       & \sum_{k \in \{r,f\}}
      \frac{\tilde{g}_k^{(n)}}{D_k}
      \!\left(2\,\tilde{g}_k'
      - \frac{\tilde{g}_k^{(n)}\,D_k'}{D_k}\right)
      \bigg|_{V=V_n^*}        \\
       & \quad=\;
      2\bigl[\Delta\mathcal{I}_{\mathrm{eff}}^{(A),\mathrm{tot}}\bigr]^{2}\,
      \beta_{A,\mathrm{tot}}\,
      \frac{\partial \beta_{A,\mathrm{tot}}}{\partial V_n}
      \bigg|_{V_n = V_n^*}
    \end{aligned}
  }
  \label{eq:opt_A_condition}
  \end{equation}
Here $D_k$ is the full composite noise denominator (cf.~Eq.~\eqref{eq:biexp_phi}) and $\beta_{A,\mathrm{tot}}$ is the amplitude component of the total systematic bias vector (Eq.~\eqref{eq:b_tau_tot}).
  All left-hand-side quantities are evaluated on the $n$-th threshold pair, i.e., with $x_k = x_k^{(n)}$ and $V = V_n^*$ for $k \in \{r,f\}$.

On the left-hand side, $2\tilde{g}_k'\tilde{g}_k/D_k$ is the marginal gain in nuisance-projected amplitude sensitivity, while $-\tilde{g}_k^2 D_k'/D_k^2$ is the marginal compound noise penalty from Poisson shot noise (via~$\kappa$) and TDC jitter (via~$\sigma_{\mathrm{TDC}}^2 h''$). The right-hand side is the marginal misspecification-bias cost weighted by the squared total effective information. The optimal $V_n^*$ sits at the voltage where these three marginal rates---sensitivity gain, noise growth, and bias penalty---reach mutual equilibrium. The $N$ thresholds satisfying Eq.~\eqref{eq:opt_A_condition} are solved jointly via the fixed-point scheme Eq.~\eqref{eq:fixed_point_lambda}, alternating per-pair threshold optimization with the global nuisance coupling update~$\bm{\gamma}^*$.

\textbf{Step 2: Model-based MVT amplitude-MSE proxy.}
Substituting the optimal threshold set $\{V_n^*\}$ and the converged coupling vector $\bm{\gamma}^*$ into the general performance formula Eq.~\eqref{eq:min_mse_general} with $\vartheta = A$, the corresponding leading-order model-based amplitude-MSE proxy from the MVT digitization stage is:
\begin{equation}
  \boxed{
    \mathcal{B}_{A}^{*}
    = \left[\sum_{n=1}^{N}
      \Phi_{A}^{(n)}(V_n^*;\,\bm{\gamma}^*)\right]^{-1}
    + \beta_{A,\mathrm{tot}}^{2}(\{V_n^*\}).}
  \label{eq:biexp_min_mse}
\end{equation}

\textbf{Step 3: System-level amplitude MSE.}
In a physical detector chain, the MVT digitization stage is preceded by stochastic frontend processes that also affect amplitude reconstruction, including photon-counting fluctuations, gain variation, and analog baseline perturbations not already absorbed by the local threshold-crossing model.  Denoting the aggregate upstream amplitude variance from these statistically independent contributions by $\sigma_{A,\mathrm{front}}^2$, the total single-detector amplitude MSE is:
\begin{equation}
  \mathrm{MSE}_{A,\mathrm{det}} = \mathcal{B}_{A}^{*} + \sigma_{A,\mathrm{front}}^2.
  \label{eq:total_det_mse}
\end{equation}
When $A$ is proportional to deposited energy, a convenient dimensionless summary is the FWHM relative amplitude resolution:
\begin{equation}
  \boxed{
    \mathcal{R}_{A,\mathrm{FWHM}}
    = 2\sqrt{2\ln 2}\cdot
    \frac{\sqrt{\mathrm{MSE}_{A,\mathrm{det}}}}{A}.}
  \label{eq:amp_resolution_fwhm}
\end{equation}
Equation~\eqref{eq:amp_resolution_fwhm} expresses the system-level relative amplitude resolution as a direct function of the optimal MVT amplitude floor $\mathcal{B}_{A}^{*}$ and the irreducible upstream contribution $\sigma_{A,\mathrm{front}}^2$.  Threshold optimization is therefore most consequential in the intermediate-SNR regime where the nuisance-projected MVT floor and the upstream amplitude fluctuations are comparable.

\subsection{Robustness Analysis}
\label{sec:biexp_robustness}

We now instantiate the general robustness framework of Section~\ref{sec:robustness} for the five-parameter bi-exponential model $\bm{\theta} = [A,\, t_0,\, \tau_d,\, \tau_r,\, b]^T$ with target parameter $\vartheta = A$.  All quantities below are evaluated at the design-point parameter vector $\bm{\theta}_0$ with optimal thresholds $\{V_n^*\}$ and converged coupling vector $\bm{\gamma}^*$.

\textbf{Fisher sensitivity vector.}
Applying the envelope theorem (Eq.~\eqref{eq:envelope_sensitivity}), the $j$-th component of the Fisher sensitivity vector $\mathbf{S} \in \mathbb{R}^5$ (Eq.~\eqref{eq:sensitivity_vector}) is:
\begin{equation}
  S_j
  = \sum_{n=1}^{N}
  \frac{\partial\,\Phi_{A}^{(n)}(V_n;\,\bm{\gamma}^*)}{\partial\theta_j}\bigg|_{\bm{\theta}_0},
  \quad j \in \{A,\, t_0,\, \tau_d,\, \tau_r,\, b\}.
  \label{eq:biexp_S_vector}
\end{equation}
Each per-pair derivative $\partial\Phi_{A}^{(n)}/\partial\theta_j$ is obtained by the chain rule through (i)~the crossing positions $x_r^{(n)}(\bm{\theta})$, $x_f^{(n)}(\bm{\theta})$ (implicitly defined by $A\,h(x) + b = V_n$), (ii)~the projected sensitivities $\tilde{g}_r^{(n)}$, $\tilde{g}_f^{(n)}$ (Eq.~\eqref{eq:biexp_proj_sens}), and (iii)~the composite noise denominators $D_r$, $D_f$.  The detailed derivation is given in Appendix~\ref{app:biexp_robustness} (Steps~O.1--\eqref{eq:autonum:O4}); here we state the results.

The implicit differentiation of the crossing positions (Appendix~\ref{app:biexp_robustness}, Step~O.1) yields:
\begin{equation}
  \begin{aligned}
    \frac{\partial x_r}{\partial A} = \frac{\partial x_f}{\partial A}
                                      & = -\frac{h(x_k)}{A\,h'(x_k)},                     \\
    \frac{\partial x_k}{\partial t_0} & = 0,                                              \\
    \frac{\partial x_k}{\partial \tau_d}
                                      & = -\frac{x_k\,e^{-x_k/\tau_d}/\tau_d^2}{h'(x_k)}, \\
    \frac{\partial x_k}{\partial \tau_r}
                                      & = \frac{x_k\,e^{-x_k/\tau_r}/\tau_r^2}{h'(x_k)},  \\
    \frac{\partial x_k}{\partial b}
                                      & = -\frac{1}{A\,h'(x_k)},
  \end{aligned}
  \label{eq:crossing_derivatives}
\end{equation}
where $k \in \{r, f\}$ and the sign of $h'(x_k)$ distinguishes the rising edge ($h'(x_r) > 0$) from the falling edge ($h'(x_f) < 0$).

Substituting into the chain rule for $\Phi_{A}^{(n)} = \tilde{g}_r^2/D_r + \tilde{g}_f^2/D_f$ (Eq.~\eqref{eq:biexp_phi}), each component of $\mathbf{S}$ takes the form (Appendix~\ref{app:biexp_robustness}, Step~O.4):
\begin{equation}
  S_j = \sum_{n=1}^{N} \sum_{k \in \{r,f\}}
  \frac{1}{D_k}\left[
    2\,\tilde{g}_k^{(n)}\,\frac{\partial\tilde{g}_k^{(n)}}{\partial\theta_j}
    - \frac{[\tilde{g}_k^{(n)}]^2}{D_k}\,\frac{\partial D_k}{\partial\theta_j}
    \right],
  \label{eq:biexp_S_expanded}
\end{equation}
where the partial derivatives $\partial\tilde{g}_k/\partial\theta_j$ and $\partial D_k/\partial\theta_j$ are expressed in terms of the crossing derivatives~\eqref{eq:crossing_derivatives} and the bi-exponential sensitivities~\eqref{eq:biexp_sensitivity}.

\textbf{Physical interpretation of each sensitivity component.}
\begin{itemize}
  \item $S_A$ (design-amplitude sensitivity): Varying the design-point amplitude changes both the crossing geometry and the composite noise denominators.  In the usual high-SNR regime, larger $A$ improves the precision of amplitude estimation, but the nuisance-projected information also depends on how the fixed threshold ladder samples the waveform.  The magnitude of $S_A$ therefore quantifies the sensitivity of the threshold design to energy-calibration errors.
  \item $S_{t_0}$ (arrival-time sensitivity): The effective information for $A$ is invariant under global time translation at fixed thresholds because the crossing geometry depends on $t-t_0$, not on absolute time.  Hence $S_{t_0} = 0$ identically, and the amplitude-oriented threshold design is intrinsically robust to arrival-time offsets.
  \item $S_{\tau_d}$ (decay-time sensitivity): Perturbations in $\tau_d$ shift the trailing-edge crossings and alter the nuisance coupling between amplitude and waveform shape.  Thresholds sampling the pulse tail are therefore most sensitive to errors in the decay constant.
  \item $S_{\tau_r}$ (rise-time sensitivity): Perturbations in $\tau_r$ modify the early rising-edge geometry, which directly affects how well amplitude can be separated from timing.  Designs that place several low thresholds near the leading edge can therefore exhibit substantial $\tau_r$ sensitivity.
  \item $S_b$ (baseline sensitivity): Baseline drift shifts all crossing levels through $A\,h(x)+b=V_n$, perturbing both the inferred normalized amplitudes $h(x_k)$ and the nuisance projection.  A stable baseline remains important because baseline errors reduce the usable dynamic range for amplitude estimation.
\end{itemize}

\textbf{Bias sensitivity vector.}
The bias sensitivity vector $\mathbf{s}_\beta \in \mathbb{R}^5$ (Eq.~\eqref{eq:bias_sensitivity_vector}) is:
\begin{equation}
  s_{\beta,j}
  = \frac{\partial\,\beta_{A,\mathrm{tot}}}{\partial\theta_j}\bigg|_{\bm{\theta}_0},
  \quad j \in \{A,\, t_0,\, \tau_d,\, \tau_r,\, b\}.
  \label{eq:biexp_sb_vector}
\end{equation}
Since $\beta_{A,\mathrm{tot}}$ depends on $\bm{\theta}$ through the Jacobian $\mathbf{J}$, weight matrix $\mathbf{W}$, and temporal bias vector $\Delta\mathbf{t}_{\mathrm{bias}}$ (Eq.~\eqref{eq:b_tau_tot}), the differentiation involves the chain rule through these three quantities.  The full expansion is given in Appendix~\ref{app:biexp_robustness} (Step~O.5).

\textbf{First-order MSE perturbation.}
Substituting the bi-exponential sensitivity vectors into Eq.~\eqref{eq:mse_first_order}, the first-order MSE shift for a parameter perturbation $\Delta\bm{\theta}$ about $\bm{\theta}_0$ is:
\begin{equation}
  \boxed{
    \Delta\mathcal{B}_{A}^{(1)}
    = -\frac{\sum_{j} S_j\,\Delta\theta_j}
    {\mathcal{I}_{\mathrm{eff}}^{*\,2}}
    + 2\,\beta_{A}^{*}\sum_{j} s_{\beta,j}\,\Delta\theta_j,}
  \label{eq:biexp_mse_first}
\end{equation}
where $j$ ranges over $\{A,t_0,\tau_d,\tau_r,b\}$ and $S_{t_0} = 0$ eliminates the arrival-time component from the first term.

\textbf{Hessian and second-order MSE perturbation.}
The true Hessian $\mathbf{H}_{\mathcal{I}} \in \mathbb{R}^{5 \times 5}$ (Eq.~\eqref{eq:true_hessian}) for the bi-exponential model includes the implicit-$\bm{\gamma}$ correction from the second-order envelope theorem.  From the expanded Fisher decomposition form Eq.~\eqref{eq:decomp_expanded}:
\begin{equation}
  \frac{\partial^2\mathcal{I}_{\mathrm{eff}}}{\partial\theta_j\,\partial\gamma_k}
  = -2\frac{\partial}{\partial\theta_j}\left[
    \bm{\mathcal{I}}_{\bm{\eta}\vartheta}^{\mathrm{tot}}
    - \bm{\mathcal{I}}_{\bm{\eta}\bm{\eta}}^{\mathrm{tot}}\,\bm{\gamma}
    \right]_k\bigg|_{\bm{\gamma}^*},
  \label{eq:biexp_cross_deriv}
\end{equation}
and $\partial^2\mathcal{I}_{\mathrm{eff}}/\partial\bm{\gamma}^2 = 2\,\bm{\mathcal{I}}_{\bm{\eta}\bm{\eta}}^{\mathrm{tot}}$ (from the quadratic structure).  The full $5 \times 5$ Hessian is assembled from these blocks in Appendix~\ref{app:biexp_robustness} (Steps~O.6--O.7) and takes the form:
\begin{equation}
  \mathbf{H}_{\mathcal{I}}
  = \frac{\partial^2\mathcal{I}_{\mathrm{eff}}}{\partial\bm{\theta}^2}\bigg|_{\bm{\gamma}^*}
  - \frac{1}{2}\left(\frac{\partial^2\mathcal{I}_{\mathrm{eff}}}{\partial\bm{\theta}\,\partial\bm{\gamma}}\right)
  \bigl(\bm{\mathcal{I}}_{\bm{\eta}\bm{\eta}}^{\mathrm{tot}}\bigr)^{-1}
  \left(\frac{\partial^2\mathcal{I}_{\mathrm{eff}}}{\partial\bm{\gamma}\,\partial\bm{\theta}}\right),
  \label{eq:biexp_hessian}
\end{equation}
where the positive-definiteness of $\bm{\mathcal{I}}_{\bm{\eta}\bm{\eta}}^{\mathrm{tot}}$ (guaranteed for $N \ge 2$ generically distinct thresholds) ensures the correction term is well-defined.
For the four-dimensional bi-exponential nuisance block, this invertibility is already generic once two distinct threshold pairs are aggregated; the amplitude-target design problem in the present section nevertheless continues to require $N \ge 3$ for positive profiled target information, as noted in Sections~\ref{sec:biexp_paired} and \ref{sec:biexp_partial}.

Substituting into Eq.~\eqref{eq:mse_second_order}, with the bias Hessian $\mathbf{H}_\beta \in \mathbb{R}^{5\times 5}$ defined analogously, the second-order MSE expansion for the bi-exponential model is:
\begin{equation}
  \boxed{
    \begin{aligned}
      \Delta\mathcal{B}_{A}^{(2)}
       & = \Delta\mathcal{B}_{A}^{(1)}
      + \frac{(\mathbf{S}^{T}\Delta\bm{\theta})^{2}}
        {\mathcal{I}_{\mathrm{eff}}^{*\,3}}
      - \frac{\Delta\bm{\theta}^{T}\mathbf{H}_{\mathcal{I}}\,\Delta\bm{\theta}}
        {2\,\mathcal{I}_{\mathrm{eff}}^{*\,2}} \\
       & \quad
         + (\mathbf{s}_\beta^{T}\Delta\bm{\theta})^{2}
      + \beta_{A}^{*}\,\Delta\bm{\theta}^{T}\mathbf{H}_{\beta}\,\Delta\bm{\theta}.
    \end{aligned}}
  \label{eq:biexp_mse_second}
\end{equation}

\textbf{Fragility ranking.}
For a given detector hardware configuration, the absolute magnitudes $|S_j|/\mathcal{I}_{\mathrm{eff}}^{*2}$ and $|s_{\beta,j}|$ (appropriately normalized by the characteristic scale of each parameter) rank the five physical parameters by their impact on amplitude performance.  In amplitude-oriented designs, the dominant fragility directions are typically the waveform-shape and baseline parameters $(\tau_d,\tau_r,b)$, while $t_0$ remains a null direction in the Fisher term because of time-translation invariance.  The exact ordering depends on how the chosen threshold ladder distributes sensitivity between the leading edge, the pulse peak, and the trailing tail.

\subsection{Multi-Event Extension}
\label{sec:biexp_partial}

We now instantiate the partial-triggering and multi-event optimization framework of Section~\ref{sec:partial_trigger} for the bi-exponential model with target parameter $\vartheta = A$.

\textbf{Active threshold set.}
For the bi-exponential model, the peak voltage is $p(\bm{\theta}) = A\cdot h_{\mathrm{peak}} + b$ (Appendix~\ref{app:peak_time}).  The active threshold set (Eq.~\eqref{eq:active_set}) for the $p$-th event with parameters $\bm{\theta}^{(p)} = [A^{(p)},\, t_0^{(p)},\, \tau_d^{(p)},\, \tau_r^{(p)},\, b^{(p)}]^T$ is:
\begin{equation}
  \mathcal{A}_p
  = \bigl\{n \in \{1,\ldots,N\} : V_n < A^{(p)}\,h_{\mathrm{peak}}^{(p)} + b^{(p)}\bigr\},
  \label{eq:biexp_active_set}
\end{equation}
where $h_{\mathrm{peak}}^{(p)} = h_{\mathrm{peak}}(\tau_d^{(p)}, \tau_r^{(p)})$ depends on the shape parameters of event $p$.  In the common case where the shape parameters are shared across events ($\tau_d^{(p)} = \tau_d$, $\tau_r^{(p)} = \tau_r$ for all $p$), the active set is determined solely by the amplitude and baseline: $\mathcal{A}_p = \{n : V_n < A^{(p)} h_{\mathrm{peak}} + b^{(p)}\}$.

\emph{Assumption} (Minimum active pairs): Each event used in the five-parameter amplitude-target design satisfies $K_p \triangleq |\mathcal{A}_p| \ge 3$ with the active pairs placed at distinct voltages, so that the active Fisher matrix generically attains full rank and $\Delta\mathcal{I}_{\mathrm{eff}}^{(A),\mathrm{act},(p)} > 0$ (cf.\ the active nuisance and target identifiability assumption in Section~\ref{sec:partial_trigger}).  The weaker condition $K_p \ge 2$ is sufficient only for the $4\times4$ active nuisance block to become generically full rank; by itself it gives at most four crossing observations and does not guarantee a finite amplitude CRLB in the five-parameter model.

\textbf{Event-specific nuisance coupling.}
For each event $p$, the nuisance coupling vector (Eq.~\eqref{eq:multi_event_lambda}) instantiates as:
\begin{equation}
  \bm{\gamma}_p^{\mathrm{act}}
  = \left(\sum_{n \in \mathcal{A}_p}
  \bm{\mathcal{I}}_{\bm{\eta}'\bm{\eta}'}^{(n)}(\bm{\theta}^{(p)})\right)^{\!-1}
  \sum_{n \in \mathcal{A}_p}
  \bm{\mathcal{I}}_{\bm{\eta}' A}^{(n)}(\bm{\theta}^{(p)})
  \;\in\;\mathbb{R}^{4},
  \label{eq:biexp_lambda_p}
\end{equation}
where $\bm{\eta}' = [t_0,\, \tau_d,\, \tau_r,\, b]^T$ and each FIM block is constructed from the five-parameter sensitivities Eq.~\eqref{eq:biexp_sensitivity} evaluated at $\bm{\theta}^{(p)}$.  The coupling vector $\bm{\gamma}_p^{\mathrm{act}}$ differs across events both because the FIM block values depend on $\bm{\theta}^{(p)}$ and because the summation range $\mathcal{A}_p$ varies with event amplitude.

\textbf{Multi-event stationarity system.}
Substituting the bi-exponential projected Fisher $\Phi_{A}^{(n)}$ (Eq.~\eqref{eq:biexp_phi}) and its derivative structure (Eqs.~\eqref{eq:Dk_deriv}--\eqref{eq:proj_curvature}) into the general multi-event stationarity Eq.~\eqref{eq:multi_event_stationarity} with $\vartheta = A$:
\begin{equation}
  \resizebox{\columnwidth}{!}{$\displaystyle
      \boxed{
        \begin{aligned}
           & \sum_{p:\,n\in\mathcal{A}_p}
          \frac{w_p}{\bigl[\Delta\mathcal{I}_{\mathrm{eff}}^{(A),\mathrm{act},(p)}\bigr]^{2}}
          \sum_{k \in \{r,f\}}
          \frac{\tilde{g}_k^{(n,p)}}{D_k^{(p)}}
          \!\left(2\,\tilde{g}_k'^{(n,p)}
          - \frac{\tilde{g}_k^{(n,p)}\,D_k'^{(p)}}{D_k^{(p)}}\right) \\
           & \quad=\;
          2\sum_{p:\,n\in\mathcal{A}_p}
          w_p\,
          \beta_{A,\mathrm{act}}^{(p)}\,
          \frac{\partial\,\beta_{A,\mathrm{act}}^{(p)}}{\partial V_n}
        \end{aligned}
      }$}%
  \label{eq:biexp_multi_stationarity}
\end{equation}
for $n = 1, \ldots, N$, where the superscript $(n,p)$ denotes evaluation at the $n$-th threshold with parameters $\bm{\theta}^{(p)}$ and coupling vector $\bm{\gamma}_p^{\mathrm{act}}$.  The quantities $\tilde{g}_k'^{(n,p)}$ and $D_k'^{(n,p)}$ are the projected-sensitivity and noise-denominator derivatives defined in Eqs.~\eqref{eq:gtilde_deriv}--\eqref{eq:Dk_deriv}, evaluated at the $p$-th event's parameters and coupling vector.

\emph{Remark} (Reduction to single-event bi-exponential): Setting $P = 1$ with $\mathcal{A}_1 = \{1,\ldots,N\}$ recovers the single-event optimal threshold condition Eq.~\eqref{eq:opt_A_condition} exactly.

\textbf{Physical interpretation.}
The multi-event stationarity system~\eqref{eq:biexp_multi_stationarity} reveals the threshold allocation principle anticipated in Section~\ref{sec:partial_trigger} in concrete bi-exponential terms:

\begin{itemize}
  \item \emph{Low-energy events} ($A^{(p)}$ small, $K_p \ll N$): These events trigger only the lowest thresholds.  Their effective information $\Delta\mathcal{I}_{\mathrm{eff}}^{(A),\mathrm{act},(p)}$ is small (fewer active pairs and lower signal-to-noise ratio), producing a large relevance weight $1/[\Delta\mathcal{I}]^2$ in the stationarity equation.  Consequently, the lowest thresholds are pulled toward placements that preserve amplitude discrimination for the weakest events.

  \item \emph{High-energy events} ($A^{(p)}$ large, $K_p = N$): These photopeak events trigger all thresholds and contribute to the stationarity equations at every threshold level.  However, their large effective information produces a small relevance weight, so their influence is attenuated relative to the low-energy events.  The upper thresholds, which are active \emph{only} for these high-energy events, are optimized primarily for photopeak amplitude performance via the single-event stationarity Eq.~\eqref{eq:opt_A_condition}.

  \item \emph{Spectral partitioning}: The natural consequence is a \emph{spectral partition} of the threshold array: lower thresholds serve as shared resources whose placement reflects the entire energy spectrum, while upper thresholds act as dedicated ``photopeak refinement'' levels.  This architecture introduces an energy-aware weighting that accounts for the detector's operational energy range.
\end{itemize}

\textbf{Algorithmic implementation.}
The alternating fixed-point iteration of Section~\ref{sec:partial_trigger} specializes as follows for the bi-exponential model:
\begin{enumerate}
  \item \emph{Initialization}: Set thresholds $\{V_n^{(0)}\}$ using the single-event ($P = 1$, photopeak) optimal solution from Section~\ref{sec:opt_A}.
  \item \emph{$\bm{\gamma}$-update}: For each event $p$, compute $\mathcal{A}_p$ from current $\{V_n\}$ via Eq.~\eqref{eq:biexp_active_set}, then update $\bm{\gamma}_p^{\mathrm{act}}$ via Eq.~\eqref{eq:biexp_lambda_p}.
  \item \emph{$V$-update}: Solve (or take a Newton step toward) the stationarity system~\eqref{eq:biexp_multi_stationarity} for updated $\{V_n\}$.
  \item Iterate steps~2--3 until convergence; monitor for discrete changes in the active sets $\mathcal{A}_p$.
\end{enumerate}

% =========================================================================
%  IX. PRACTICAL IMPLEMENTATION AND MODEL-ASSUMPTION VALIDATION
% =========================================================================
\section{Implementation and Model Validation}
\label{sec:validation}

The preceding sections developed a comprehensive analytical framework for optimal MVT threshold design, culminating in explicit stationarity equations (Eq.~\eqref{eq:opt_A_condition}) and a closed-form leading-order model-based amplitude-MSE proxy (Eq.~\eqref{eq:biexp_min_mse}) for the bi-exponential pulse model.  This section bridges the gap between abstract theory and hardware practice.  We first describe the experimental apparatus and the simulation-based MVT digitization methodology (Section~\ref{sec:exp_setup}), then distill the theoretical framework into self-contained, step-by-step engineering recipes for both single-event (Section~\ref{sec:recipe}) and multi-event (Section~\ref{sec:recipe_multi}) threshold optimization.  Finally, we apply the perturbative misspecification diagnostic developed in Section~\ref{sec:recipe_algorithm} to a 10{,}000-pulse dataset to assess whether the well-specified model assumption ($\varepsilon(t) \approx 0$) underlying the recipes is empirically justified for amplitude-oriented threshold design in the present detector configuration (Section~\ref{sec:delta_validation}).

Throughout this section, the bi-exponential parameters retain the meanings fixed in Section~\ref{sec:biexp_model}: $\tau_r$ and $\tau_d$ denote \emph{effective macroscopic} time constants of the measured voltage waveform, not bare microscopic material constants, and $A$ (hence $A_{\mathrm{design}}$ and $\{A^{(p)}\}$) denotes the waveform \emph{scale} amplitude, not the observable peak voltage, which is $A h_{\mathrm{peak}} + b$.

\subsection{Experimental Setup}
\label{sec:exp_setup}

\textbf{Radioactive source and detector module.}
A $^{22}$Na radioactive source emitting 511~keV annihilation photons and 1274~keV gamma rays was used to provide the excitation photons.  The source itself is therefore discrete in energy; the continuous 0--1274~keV spectrum discussed below is the \emph{deposited-energy} spectrum observed in the detector after Compton scattering in air and in the scintillator crystal.  The detector module comprises a lutetium-yttrium oxyorthosilicate (LYSO) scintillator coupled to a SiPM and read out without external preamplification.  The LYSO scintillator has dimensions of $3.9~\mathrm{mm} \times 3.9~\mathrm{mm} \times 20~\mathrm{mm}$.  It is optically coupled to a SiPM.  The SiPM output waveform---i.e., the macroscopic scintillation voltage pulse at the readout output---is digitized directly by an oscilloscope (50~GS/s real-time sampling rate, 16~GHz analog bandwidth) without any intermediate preamplifier stage.  Channel~1 is configured with a rising-edge trigger at a threshold of 100~mV.  Each acquisition captures a 1000~ns window at 50~GS/s, yielding 50{,}000 sample points per pulse.  A total of 10{,}000 individual scintillation pulses were recorded across ten sequential \texttt{.wfm} files (1{,}000 pulses per file).  The vertical scale is set to 200~mV/div and the horizontal scale to 100~ns/div.  This direct-readout configuration preserves the intrinsic SiPM pulse shape and avoids the additional pole-zero structure that would be introduced by a shaping amplifier~\cite{xu2015optimization,grodzicka2013energy,iwai2017background,cates2022low}.

\emph{Assumption} (Calibration-rate isolation and direct-readout linearity): The validation dataset is treated as a collection of isolated single-pulse events acquired at sufficiently modest count rate that pile-up within the 1000~ns window is rare and the pre-pulse baseline is locally stationary.  The SiPM and oscilloscope front-end are also assumed to operate in a moderate dynamic range where direct-readout gain compression is not the dominant effect.  These assumptions are reasonable for the triggered calibration runs used here; residual saturation, afterpulsing, or baseline motion are not dismissed, but enter the measured mismatch $\varepsilon(t)$ and are quantified in Section~\ref{sec:delta_validation}.

\textbf{MVT digitization via offline simulation.}
To emulate an 8-threshold MVT digitizer on the recorded waveforms, we retain the same single-extreme timestamp-selection convention used in the validation scripts.  Let $v_i$, $i = 1, 2, \ldots, m$, denote the oscilloscope samples, let $\Delta t_{\mathrm{samp}}$ denote the sampling interval, and let $V_j$, $j = 1, 2, \ldots, 8$, denote the threshold voltages.  We first form the piecewise-linear interpolant $\tilde v(t)$ through the sample pairs $(t_i,v_i)$ with $t_i=i\,\Delta t_{\mathrm{samp}}$.  Whenever $\tilde v(t)$ crosses threshold $V_j$ on interval $[t_i,t_{i+1}]$, the crossing time is estimated by linear interpolation~\cite{mog2004zero}:
\begin{equation}
  t_{\mathrm{cross}} = t_i + \frac{V_j - v_i}{v_{i+1} - v_i}\,\Delta t_{\mathrm{samp}},
  \label{eq:linear_interp_crossing}
\end{equation}
where $t_i = i\,\Delta t_{\mathrm{samp}}$ is the time of the $i$-th sample.  This linear interpolation provides a sub-sample-accurate estimate of each threshold-crossing instant, more closely approximating the behavior of an analog comparator followed by a TDC.  For a simple crossing with nonzero local slope, the resulting crossing-time truncation error is $\mathcal{O}\!\bigl((\partial_t^2 f/\partial_t f)\,\Delta t_{\mathrm{samp}}^2\bigr)$, i.e., second order in the sampling interval with a constant set by the local curvature-to-slope ratio; at 50~GS/s ($\Delta t_{\mathrm{samp}} = 20~\mathrm{ps}$) this contribution is negligible except arbitrarily close to slope-vanishing points.

\emph{Assumption} (Offline MVT emulation): The 50~GS/s piecewise-linear waveform is treated as a sufficiently dense surrogate for the analog comparator input, and the interpolated crossing time is treated as the timestamp that an ideal comparator/TDC chain would have reported.  This is reasonable because the sampling interval is much shorter than the fitted macroscopic pulse time constants and because the analysis avoids crossings at the stationary peak.  Comparator hysteresis, metastability, finite propagation delay, and clock-correlation effects are not represented by the offline emulation; if those effects are appreciable in hardware, the timestamp-generation rule and covariance model must be replaced by a hardware-calibrated model.

Let $\hat t_p$ denote the observed time of the waveform maximum.  For each threshold $V_j$, define the pre-peak and post-peak crossing sets
\begin{equation}
  \begin{aligned}
    \mathcal{T}_{j}^{(r)} &\triangleq \{t < \hat t_p : \tilde v(t)=V_j\}, \\
    \mathcal{T}_{j}^{(f)} &\triangleq \{t > \hat t_p : \tilde v(t)=V_j\}.
  \end{aligned}
  \label{eq:offline_crossing_sets}
\end{equation}
For a noise-free strictly unimodal pulse, both sets are singletons.  On measured waveforms, however, small fluctuations can create multiple roots, especially on the falling edge where the tail slope is shallow and late recrossings are common.  The offline pipeline used for the present validation therefore applies a simple single-extreme rule: among all pre-peak threshold-upcrossings it retains the earliest one, and among all post-peak threshold-downcrossings it retains the latest one.  Accordingly, the stored offline timestamps are
\begin{equation}
  \hat t_{r,j}
  =
  \min \mathcal{T}_{j}^{(r)},
  \qquad
  \hat t_{f,j}
  =
  \max \mathcal{T}_{j}^{(f)},
  \label{eq:offline_falling_timestamp_rule}
\end{equation}
whenever both crossing sets are nonempty.  Operationally, this is implemented by scanning all threshold-upcrossings and threshold-downcrossings of $\tilde v(t)$, storing the first rising crossing and the last falling crossing, and discarding that threshold if the selected falling time does not occur after the selected rising time.  This is exactly the timestamp-generation rule used in the current reconstruction scripts.  The resulting timestamp vector is then passed to the Gauss-Newton estimator of Section~\ref{sec:biexp_estimation} for waveform-parameter recovery.

\emph{Assumption} (Single-extreme crossing convention): Multiple threshold recrossings caused by high-frequency noise are compressed into one rising/falling pair by the earliest-rise/latest-fall rule.  This convention is a reproducible offline surrogate for a latched MVT readout and is reasonable when the underlying physical pulse remains unimodal and recrossings are small perturbations around the main branches.  It can introduce a systematic tail preference on noisy falling edges; that bias is part of the end-to-end validation rather than a property of the ideal theory.  If recrossings become frequent for a deployed comparator chain, the same reconstruction should be audited against alternative edge-selection rules.

\subsection{Single-Event Design Recipe}
\label{sec:recipe}

We now distill the optimal threshold design theory of Sections~\ref{sec:general_opt}--\ref{sec:optimal} into a self-contained, step-by-step engineering procedure for the 8-threshold MVT system described above.  This subsection is designed to be readable independently of the preceding theoretical development; all required formulas are restated explicitly.

\paragraph{Step-label convention.}
Recipe step labels are written in the form \texttt{XX-Yn}.  The first block identifies the recipe scope: \texttt{SE} for the single-event recipe and \texttt{ME} for the multi-event recipe.  The middle letter identifies the stage inside that recipe: \texttt{P} for preliminary computations, \texttt{A} for the main optimization algorithm, and \texttt{R} for the post-design robustness audit.  The trailing integer gives the local order within that stage.  Thus \texttt{SE-A7} means ``single-event algorithm, Step~7,'' while \texttt{ME-P2} means ``multi-event preliminaries, Step~2.''  This keeps the numbering local to each recipe block instead of mixing all steps into one section-wide sequence.

\subsubsection{Required Inputs}
\label{sec:recipe_inputs}

The engineer must supply the following quantities, organized by their physical origin:

{
  \begin{enumerate}
    \item \emph{Waveform shape parameters} (from full-waveform calibration, prior fits, or a separately justified deconvolution):
          \begin{itemize}
            \item $\tau_d$: effective macroscopic decay time constant of the fitted voltage waveform $[\mathrm{s}]$;
            \item $\tau_r$: effective macroscopic rise time constant of the fitted voltage waveform $[\mathrm{s}]$, with $\tau_d > \tau_r$.
          \end{itemize}
    \item \emph{Noise parameters} (from front-end characterization):
          \begin{itemize}
            \item $\sigma_{\mathrm{th}}^2$: thermal (signal-independent) voltage noise variance $[\mathrm{V}^2]$;
            \item $\kappa$: effective single-photoelectron voltage amplitude $[\mathrm{V}]$, defined as $\kappa = \int h_e^2(x)\,dx\,\big/\!\int h_e(x)\,dx$, where $h_e(x)$ is the single-photoelectron impulse response;
            \item $\sigma_{\mathrm{TDC}}^2$: TDC quantization jitter variance $[\mathrm{s}^2]$.
          \end{itemize}
    \item \emph{Design-point parameters} (from energy calibration):
          \begin{itemize}
            \item $A_{\mathrm{design}}$: bi-exponential waveform scale amplitude at the target operating energy (e.g., 511~keV photopeak) $[\mathrm{V}]$; the corresponding peak voltage is $A_{\mathrm{design}} h(x_p)+b$;
            \item $b$: DC baseline voltage of the front-end $[\mathrm{V}]$.
          \end{itemize}
    \item \emph{System parameters}:
          \begin{itemize}
            \item $N = 8$: number of hardware comparators (thresholds);
            \item $B$: analog bandwidth of the front-end $[\mathrm{Hz}]$;
            \item $\sigma_{A,\mathrm{front}}^2$ (optional): aggregate upstream amplitude variance used when converting the digitizer-only bound into a full-chain relative amplitude resolution.
          \end{itemize}
  \end{enumerate}
}

\subsubsection{Preliminary Computations}
\label{sec:recipe_prelim}

Before the optimization, compute the following auxiliary quantities.

\textbf{Step SE-P1: Normalized shape function and its derivative.}
Define the bi-exponential shape kernel (Eq.~\eqref{eq:biexp}):
\begin{equation}
  h(x) = e^{-x/\tau_d} - e^{-x/\tau_r}, \qquad x > 0,
  \label{eq:autonum:R1}
\end{equation}
with first derivative:
\begin{equation}
  h'(x) = -\frac{e^{-x/\tau_d}}{\tau_d} + \frac{e^{-x/\tau_r}}{\tau_r}.
  \label{eq:autonum:R2}
\end{equation}

\textbf{Step SE-P2: Peak time and peak voltage.}
The pulse reaches its maximum at the normalized time (Eq.~\eqref{eq:biexp_xp}):
\begin{equation}
  x_p = \frac{\tau_r\,\tau_d}{\tau_d - \tau_r}\,\ln\!\left(\frac{\tau_d}{\tau_r}\right).
  \label{eq:autonum:R3}
\end{equation}
The peak voltage of the physical pulse is:
\begin{equation}
  V_{\mathrm{peak}} = A_{\mathrm{design}} \cdot h(x_p) + b.
  \label{eq:autonum:R4}
\end{equation}
All thresholds must satisfy $b < V_n < V_{\mathrm{peak}}$ for $n = 1, \ldots, N$.

\textbf{Step SE-P3: Minimum threshold spacing (bandwidth diagnostic).}
If diagonal-covariance independence is enforced through the conservative global-slope surrogate of Eq.~\eqref{eq:biexp_bandwidth}, adjacent thresholds on the same waveform slope would require
\begin{equation}
  \Delta V_{\min} = A_{\mathrm{design}}\!\left(\frac{1}{\tau_r} - \frac{1}{\tau_d}\right) \cdot \frac{1}{2B}.
  \label{eq:autonum:R5}
\end{equation}
This quantity is a diagnostic for when adjacent same-branch crossings may fall within the same noise autocorrelation window and violate the statistical independence assumption underlying Fisher Information additivity (Section~\ref{sec:bandwidth}).  It does not by itself certify diagonality of the rising/falling pair produced by a single threshold; that pair must also satisfy $t_f(V_n)-t_r(V_n) \ge \tau_c$ if it is to be treated with diagonal pair weights.  In hardware recipes below, these correlation diagnostics should be replaced by a local-slew or empirically calibrated constraint whenever the global bound exceeds the usable voltage range.

\subsubsection{Threshold Optimisation Algorithm}
\label{sec:recipe_algorithm}

The following iterative algorithm computes the optimal threshold set $\{V_1^*, \ldots, V_N^*\}$ that minimizes the amplitude MSE for the target parameter $A$.  It implements the fixed-point iteration scheme of Eqs.~\eqref{eq:general_stationarity}--\eqref{eq:fixed_point_lambda}, specialized to the bi-exponential model under the following assumption.

\emph{Assumption} (Well-specified model): The bi-exponential pulse Eq.~\eqref{eq:biexp} is an accurate representation of the physical waveform, i.e., the model misspecification function $\varepsilon(t) \approx 0$.  Under this assumption, the systematic bias term $\beta_{A,\mathrm{tot}} = 0$ (Section~\ref{sec:biexp_crlb}), and the stationarity condition Eq.~\eqref{eq:opt_A_condition} reduces to the amplitude-specialized bias-free condition Eq.~\eqref{eq:biexp_bias_free_stationarity}:
\begin{equation}
  \frac{d\,\Phi_{A}^{(n)}(V;\,\bm{\gamma}^*)}{dV}\bigg|_{V = V_n^*} = 0, \quad n = 1, \ldots, N.
  \label{eq:autonum:R6}
\end{equation}
That is, each threshold is updated by imposing the stationary condition on its own nuisance-projected Fisher contribution, given the globally determined coupling vector $\bm{\gamma}^*$. It is a local maximizer only if the corresponding second derivative with respect to $V$ is negative.

\emph{Justification}: This assumption is most plausible when the SiPM output is digitized directly without a shaping amplifier (as in our experimental setup), because the dominant waveform distortions introduced by shaping networks (pole-zero tails, undershoot) are absent.  Residual distortions from SiPM nonlinear saturation and afterpulsing may still matter; Section~\ref{sec:delta_validation} shows that, with the present waveform model, they are not negligible for amplitude-oriented validation.  The recipe should therefore be read as a well-specified reference design rather than as a fully validated final prescription.

\emph{Perturbative error analysis and the direct bias projection.}
Let $\mathbf{V} = [V_1,\dots,V_N]^T$ denote the threshold vector, and define the idealized and misspecified MSE design objectives
\begin{align}
  \mathcal{B}_0^{(A)}(\mathbf{V})
   & \triangleq
  \bigl[\Delta\mathcal{I}_{\mathrm{eff}}^{(A),\mathrm{tot}}(\mathbf{V})\bigr]^{-1},
  \label{eq:autonum:R6a}    \\
  \mathcal{B}_{\varepsilon}^{(A)}(\mathbf{V})
   & \triangleq
  \mathcal{B}_0^{(A)}(\mathbf{V})
  + \beta_{A,\mathrm{tot}}(\mathbf{V};\varepsilon)^2,
  \label{eq:autonum:R6b}
\end{align}
with minimizers $\mathbf{V}^{(0)} = \arg\min \mathcal{B}_0^{(A)}$ and $\mathbf{V}^{(\varepsilon)} = \arg\min \mathcal{B}_{\varepsilon}^{(A)}$.  Evaluate the waveform mismatch at the crossing times induced by the ideal thresholds $\mathbf{V}^{(0)}$, and define the mean and maximum absolute mismatch amplitudes
\begin{equation}
  \bar{\varepsilon}
  \triangleq \frac{1}{K}\sum_{k=1}^{K} \bigl|\varepsilon(t_k^{(0)})\bigr|,
  \qquad
  \varepsilon_{\max}
  \triangleq \max_{1\le k\le K} \bigl|\varepsilon(t_k^{(0)})\bigr|,
  \label{eq:autonum:R6c}
\end{equation}
where $K = 2N$, $t_k^{(0)}$ are the $K$ crossing times at $\mathbf{V}^{(0)}$, and $\varepsilon_k \triangleq \varepsilon(t_k^{(0)})$ denotes the mismatch sampled at the $k$-th ideal crossing.  From the first-order bias projection $\Delta t_{\mathrm{bias},k} \approx -\varepsilon(t_k^{(0)})/\partial_t f(t_k^{(0)};\bm{\theta}_0)$ (Section~\ref{sec:misspecification}), the direct amplitude-bias projection is
\begin{equation}
  \beta_{A}^{\mathrm{direct}}
  \triangleq
  \mathbf{e}_A^T
  \bigl(\mathbf{J}^T\mathbf{W}\mathbf{J}\bigr)^{-1}
  \mathbf{J}^T\mathbf{W}\,
  \Delta\mathbf{t}_{\mathrm{bias}}
  \Bigg|_{\mathbf{V}=\mathbf{V}^{(0)}},
  \label{eq:beta_direct}
  \end{equation}
where $\mathbf{e}_A$ is the coordinate unit vector selecting the amplitude component and $\Delta\mathbf{t}_{\mathrm{bias}} \approx [-\varepsilon(t_1^{(0)})/\partial_t f(t_1^{(0)};\bm{\theta}_0),\,\ldots,\,-\varepsilon(t_K^{(0)})/\partial_t f(t_K^{(0)};\bm{\theta}_0)]^T$ is the element-wise timing bias vector.  This expression is the corresponding linearized Gauss--Newton projection of the crossing-point mismatch onto the amplitude parameter $A$ through the Fisher metric, accounting for both the signed voltage errors and their precision weights.

A conservative upper bound on $|\beta_{A}^{\mathrm{direct}}|$ follows by bounding $\|\Delta\mathbf{t}_{\mathrm{bias}}\|_2$.  Defining $m_{\min} \triangleq \min_{k}|\partial_t f(t_k^{(0)})|$ and the bias projection coefficient
\begin{equation}
  C_{\beta}
  \triangleq
  \left\|\mathbf{e}_A^T
  \bigl(\mathbf{J}^{T}\mathbf{W}\mathbf{J}\bigr)^{-1}
  \mathbf{J}^{T}\mathbf{W}\right\|_2
  \Bigg|_{\mathbf{V}=\mathbf{V}^{(0)}},
  \label{eq:autonum:R6e}
\end{equation}
we obtain---since $|\varepsilon_k| \le \varepsilon_{\max}$ implies $|\varepsilon_k|^2 \le |\varepsilon_k|\,\varepsilon_{\max}$ for each $k$, and summing gives $\sum_k |\varepsilon_k|^2 \le \varepsilon_{\max}\sum_k|\varepsilon_k| = K\bar{\varepsilon}\,\varepsilon_{\max}$---the bound
\begin{equation}
  \bigl|\beta_{A}^{\mathrm{direct}}\bigr|
  \le
  C_{\beta}
  \frac{\sqrt{K\,\bar{\varepsilon}\,\varepsilon_{\max}}}{m_{\min}},
  \qquad
  (\beta_{A}^{\mathrm{direct}})^2 = O(\bar{\varepsilon}\,\varepsilon_{\max}).
  \label{eq:autonum:R6f}
\end{equation}

Assume furthermore that the ideal design has a nondegenerate local minimum, with Hessian
\begin{equation}
  \mathbf{H}_0
  \triangleq
  \nabla_{\mathbf{V}}^2 \mathcal{B}_0^{(A)}(\mathbf{V}^{(0)}) \succ 0.
  \label{eq:autonum:R6g}
\end{equation}
Then the misspecified stationarity condition Eq.~\eqref{eq:general_stationarity} yields, by a first-order implicit-function expansion around $\mathbf{V}^{(0)}$,
\begin{equation}
  \begin{aligned}
    \mathbf{V}^{(\varepsilon)} - \mathbf{V}^{(0)}
     & = -\mathbf{H}_0^{-1}
    \nabla_{\mathbf{V}}
    \bigl[\beta_{A,\mathrm{tot}}(\mathbf{V};\varepsilon)^2\bigr]
    \Big|_{\mathbf{V}=\mathbf{V}^{(0)}}                                  \\
     & \quad + O\!\bigl((\bar{\varepsilon}\,\varepsilon_{\max})^2\bigr),
  \end{aligned}
  \label{eq:autonum:R6h}
\end{equation}
so the threshold perturbation is second-order in the waveform mismatch.  Consequently,
\begin{equation}
  \begin{aligned}
    \mathcal{B}_{\varepsilon}^{(A)}(\mathbf{V}^{(\varepsilon)})
    - \mathcal{B}_0^{(A)}(\mathbf{V}^{(0)})
     & = (\beta_{A}^{\mathrm{direct}})^2                                 \\
     & \quad + O\!\bigl((\bar{\varepsilon}\,\varepsilon_{\max})^2\bigr).
  \end{aligned}
  \label{eq:autonum:R6i}
\end{equation}
Equation~\eqref{eq:autonum:R6i} establishes two critical results.  First, the dominant gap between the true misspecified amplitude floor and the idealized amplitude floor is determined by second power of the direct bias projection $\beta_{A}^{\mathrm{direct}}$; re-optimizing the thresholds changes the leading-order proxy only at higher order.  Second, $\beta_{A}^{\mathrm{direct}}$ is computable from a single recorded pulse without any re-optimization: given an initial fit $\hat{\bm{\theta}}$ and the raw waveform $y_{\mathrm{data}}(t)$, one evaluates the crossing-point mismatches $\varepsilon_k$, assembles the Jacobian $\mathbf{J}$ and weights $\mathbf{W}$, and applies Eq.~\eqref{eq:beta_direct} directly.  The resulting ratio
\begin{equation}
  \rho_{\mathrm{bias}}^{(A)}
  \triangleq
  \frac{(\beta_{A}^{\mathrm{direct}})^2}{\mathcal{B}_0^{(A)}(\mathbf{V}^{(0)})}
  \label{eq:autonum:R6j}
\end{equation}
provides a practical engineering diagnostic of whether the $\varepsilon(t) = 0$ approximation is sufficiently accurate for threshold design: if $\rho_{\mathrm{bias}}^{(A)} \ll 1$, the bias is negligible relative to the variance floor and the idealized stationarity condition (Eq.~\eqref{eq:autonum:R6}) is adequate.

The algorithm proceeds as follows.

\textbf{Step SE-A1: Initialization.}
Choose an initial threshold set $\{V_n^{(0)}\}_{n=1}^{N}$. A simple ordering-preserving initialization is equally spaced thresholds:
\begin{equation}
  V_n^{(0)} = b + n \cdot \frac{V_{\mathrm{peak}} - b}{N + 1}, \quad n = 1, \ldots, N.
  \label{eq:autonum:R7}
\end{equation}

\textbf{Step SE-A2: Crossing root computation.}
For each threshold $V_n$, solve the level-set equation (Section~\ref{sec:biexp_model}):
\begin{equation}
  h(x) = \frac{V_n - b}{A_{\mathrm{design}}}
  \label{eq:autonum:R8}
\end{equation}
for the two roots $x_r^{(n)} < x_p$ (rising edge) and $x_f^{(n)} > x_p$ (falling edge).  This is a one-dimensional root-finding problem on each monotone branch of $h(x)$ and can be solved efficiently by bisection or Newton's method.

\textbf{Step SE-A3: Waveform sensitivities.}
At each crossing root $x_k$ ($k \in \{r, f\}$) of each threshold $n$, compute the five-component waveform sensitivity vector (Eqs.~\eqref{eq:biexp_partials}--\eqref{eq:biexp_sensitivity}):
\begin{equation}
  \begin{aligned}
    g_{A,k}^{(n)}      & = \frac{V_n - b}{A_{\mathrm{design}}},                                     \\
    g_{t_0,k}^{(n)}    & = -A_{\mathrm{design}}\,h'(x_k^{(n)}),                                     \\
    g_{\tau_d,k}^{(n)} & = A_{\mathrm{design}}\,\frac{x_k^{(n)}}{\tau_d^2}\,e^{-x_k^{(n)}/\tau_d},  \\
    g_{\tau_r,k}^{(n)} & = -A_{\mathrm{design}}\,\frac{x_k^{(n)}}{\tau_r^2}\,e^{-x_k^{(n)}/\tau_r}, \\
    g_{b,k}^{(n)}      & = 1.
  \end{aligned}
  \label{eq:autonum:R9}
\end{equation}

\textbf{Step SE-A4: Composite noise denominators.}
For each edge, compute the composite noise denominator (Eq.~\eqref{eq:biexp_phi}):
\begin{equation}
  D_k^{(n)} = \sigma_{\mathrm{th}}^2 + \kappa(V_n - b) + \sigma_{\mathrm{TDC}}^2\,A_{\mathrm{design}}^2\,[h'(x_k^{(n)})]^2.
  \label{eq:autonum:R10}
\end{equation}
This aggregates three independent noise sources: thermal noise ($\sigma_{\mathrm{th}}^2$), Poisson shot noise ($\kappa(V_n - b)$), and TDC jitter ($\sigma_{\mathrm{TDC}}^2 A_{\mathrm{design}}^2 [h']^2$).

\textbf{Step SE-A5: FIM block assembly.}
For each threshold pair $n$, construct the per-pair FIM blocks.  Partitioning the parameter vector as $\bm{\theta} = [\bm{\eta}'^T,\, A]^T$ with nuisance vector $\bm{\eta}' = [t_0,\, \tau_d,\, \tau_r,\, b]^T \in \mathbb{R}^4$ (Eq.~\eqref{eq:biexp_lambda}), the $4 \times 4$ nuisance--nuisance block and the $4 \times 1$ nuisance--target block are:
\begin{equation}
  \begin{aligned}
    \bm{\mathcal{I}}_{\bm{\eta}'\bm{\eta}'}^{(n)} & = \sum_{k \in \{r,f\}} \frac{\mathbf{g}_{\bm{\eta}',k}^{(n)}\,[\mathbf{g}_{\bm{\eta}',k}^{(n)}]^T}{D_k^{(n)}}, \\
    \bm{\mathcal{I}}_{\bm{\eta}' A}^{(n)}         & = \sum_{k \in \{r,f\}} \frac{\mathbf{g}_{\bm{\eta}',k}^{(n)}\,g_{A,k}^{(n)}}{D_k^{(n)}},
  \end{aligned}
  \label{eq:autonum:R11}
\end{equation}
where $\mathbf{g}_{\bm{\eta}',k}^{(n)} = [g_{t_0,k}^{(n)},\, g_{\tau_d,k}^{(n)},\, g_{\tau_r,k}^{(n)},\, g_{b,k}^{(n)}]^T$.

\textbf{Step SE-A6: Global nuisance coupling update ($\bm{\gamma}$-step).}
Sum the FIM blocks over all $N$ thresholds and compute the nuisance coupling vector (Eq.~\eqref{eq:fixed_point_lambda}):
\begin{equation}
  \bm{\gamma}^* = \left(\sum_{n=1}^{N} \bm{\mathcal{I}}_{\bm{\eta}'\bm{\eta}'}^{(n)}\right)^{\!-1} \sum_{n=1}^{N} \bm{\mathcal{I}}_{\bm{\eta}' A}^{(n)} \;\in\; \mathbb{R}^4.
  \label{eq:autonum:R12}
\end{equation}

\textbf{Step SE-A7: Nuisance-projected Fisher contribution ($V$-step).}
With $\bm{\gamma}^*$ fixed, compute the nuisance-projected sensitivities at each edge (Eq.~\eqref{eq:biexp_proj_sens}):
\begin{equation}
  \tilde{g}_k^{(n)} = g_{A,k}^{(n)} - [\bm{\gamma}^*]^T\,\mathbf{g}_{\bm{\eta}',k}^{(n)}, \quad k \in \{r, f\},
  \label{eq:autonum:R13}
\end{equation}
and form the per-pair nuisance-projected Fisher contribution (Eq.~\eqref{eq:biexp_phi}):
\begin{equation}
  \Phi_{A}^{(n)}(V_n;\,\bm{\gamma}^*) = \frac{[\tilde{g}_r^{(n)}]^2}{D_r^{(n)}} + \frac{[\tilde{g}_f^{(n)}]^2}{D_f^{(n)}}.
  \label{eq:autonum:R14}
\end{equation}
For each $n = 1, \ldots, N$, update $V_n$ by solving the one-dimensional maximization:
\begin{equation}
  V_n^{*} = \arg\max_{V \in (b,\, V_{\mathrm{peak}})} \;\Phi_{A}^{(n)}(V;\,\bm{\gamma}^*),
  \label{eq:autonum:R15}
\end{equation}
with the search interval clipped by the chosen same-branch spacing floor: the diagonal-covariance diagnostic $\Delta V_{\min}$ (Eq.~\eqref{eq:autonum:R5}) if one enforces independent crossings, or an empirical/coverage floor otherwise.  When the diagonal pair model Eq.~\eqref{eq:autonum:R14} is retained, the admissible interval must also exclude voltages for which the pair gap $t_f(V)-t_r(V)$ falls below $\tau_c$ at the design point; for the bi-exponential pulse this removes an upper neighborhood of $V_{\mathrm{peak}}$ where the rising and falling crossings coalesce.  If this within-pair separation condition is intentionally relaxed, the update should instead be evaluated with a $2\times2$ within-pair covariance block rather than the diagonal formula Eq.~\eqref{eq:autonum:R14}.  With $\bm{\gamma}^*$ fixed and the active constraints inactive, each update is a one-dimensional optimization that can be solved by golden-section search or Brent's method.

\textbf{Step SE-A8: Convergence check.}
Repeat Steps~SE-A2--SE-A7 (alternating between the $\bm{\gamma}$-update and the $V$-update) until the threshold values change by less than a prescribed tolerance:
\begin{equation}
  \max_{1 \le n \le N} \left|V_n^{(\ell+1)} - V_n^{(\ell)}\right| < \varepsilon_V.
  \label{eq:autonum:R16}
\end{equation}
A typical tolerance is $\varepsilon_V = 0.1~\mathrm{mV}$.  Convergence is typically achieved within 5--15 iterations.

\textbf{Step SE-A9: Performance evaluation.}
At convergence, the leading-order model-based amplitude-MSE proxy from the MVT digitization stage is (Eq.~\eqref{eq:biexp_min_mse}, with $\beta_{A,\mathrm{tot}} = 0$ under the well-specified assumption):
\begin{equation}
  \mathcal{B}_{A}^* = \left[\sum_{n=1}^{N} \Phi_{A}^{(n)}(V_n^*;\,\bm{\gamma}^*)\right]^{-1}.
  \label{eq:autonum:R17}
\end{equation}
If the upstream amplitude variance $\sigma_{A,\mathrm{front}}^2$ is known, the corresponding full-chain relative amplitude resolution is (Eq.~\eqref{eq:amp_resolution_fwhm}):
\begin{equation}
  \mathcal{R}_{A,\mathrm{FWHM}} = 2\sqrt{2\ln 2} \cdot \frac{\sqrt{\mathcal{B}_{A}^* + \sigma_{A,\mathrm{front}}^2}}{A_{\mathrm{design}}},
  \label{eq:autonum:R18}
\end{equation}
where $\sigma_{A,\mathrm{front}}^2$ is the aggregate upstream amplitude variance from the scintillator--SiPM frontend chain, independent of the MVT digitization stage.

\subsubsection{Summary}

For quick reference, the complete threshold design procedure is summarized below.

\begin{enumerate}
  \item \textbf{Characterize} your detector: measure or fit the effective macroscopic waveform parameters $\tau_r$ and $\tau_d$, together with $\sigma_{\mathrm{th}}^2$, $\kappa$, $\sigma_{\mathrm{TDC}}^2$, $B$, the scale-amplitude design point $A_{\mathrm{design}}$, $b$, and optionally $\sigma_{A,\mathrm{front}}^2$ for full-chain resolution reporting.
  \item \textbf{Compute} the peak voltage $V_{\mathrm{peak}}$ (Eq.~\eqref{eq:autonum:R3}--\eqref{eq:autonum:R4}), the bandwidth-derived diagnostic $\Delta V_{\min}$ (Eq.~\eqref{eq:autonum:R5}), and the spacing floor actually enforced in the design.
  \item \textbf{Initialize} 8 thresholds equally spaced in $[b, V_{\mathrm{peak}}]$ (Eq.~\eqref{eq:autonum:R7}).
  \item \textbf{Iterate} the fixed-point loop (Steps~SE-A2--SE-A7 above): for each threshold, find its two crossing roots on the pulse, compute the waveform sensitivities and noise, build the Fisher Information blocks, update the global nuisance coupling $\bm{\gamma}^*$, then re-optimize each threshold to maximize its nuisance-projected Fisher contribution.
  \item \textbf{Verify} that all threshold spacings satisfy the chosen same-branch spacing floor, that every threshold treated with diagonal pair weights also satisfies $t_f(V_n)-t_r(V_n) \ge \tau_c$, and that all thresholds lie within $(b, V_{\mathrm{peak}})$.
  \item \textbf{Evaluate} performance via Eqs.~\eqref{eq:autonum:R17}--\eqref{eq:autonum:R18}.
\end{enumerate}

\subsubsection{Post-Design Robustness Audit}
\label{sec:recipe_robust}

The single-event recipe above delivers the design-point-optimal ladder.  When the physical pulse parameters are expected to vary over a known operating range, Section~\ref{sec:biexp_robustness} provides a lightweight post-design screening and redesign loop that can be applied without re-deriving the threshold equations.

\textbf{Additional inputs.}
Specify anticipated componentwise uncertainty bounds for the five-parameter vector:
\begin{equation}
  |\Delta\theta_j| \le \epsilon_j,
  \quad
  \bm{\epsilon}
  = [\epsilon_A,\, \epsilon_{t_0},\, \epsilon_{\tau_d},\, \epsilon_{\tau_r},\, \epsilon_b]^T.
  \label{eq:autonum:R19}
\end{equation}
These bounds can be obtained from calibration drift, temperature coefficients, batch-to-batch detector variation, or direct empirical fitting statistics.

\textbf{Step SE-R1: Sensitivity evaluation at the converged design.}
Using the thresholds $\{V_n^*\}$ from Steps~SE-A1--SE-A9, compute the Fisher sensitivity vector $\mathbf{S}$ and the bias sensitivity vector $\mathbf{s}_{\beta}$ using the bi-exponential formulas of Section~\ref{sec:biexp_robustness}, namely Eqs.~\eqref{eq:biexp_S_vector}--\eqref{eq:biexp_S_expanded} and Eq.~\eqref{eq:biexp_sb_vector}.  In the well-specified case assumed by the base recipe, $\beta_A^* = 0$, so the first-order degradation is governed entirely by the Fisher sensitivity vector.

\textbf{Step SE-R2: First-order worst-case robustness screening.}
Over the hyper-rectangular uncertainty box~\eqref{eq:worst_case_first_order}, a conservative first-order upper bound on the amplitude-MSE degradation is
\begin{equation}
  \Delta\mathcal{B}_{A,\max}^{(1)}(\bm{\epsilon})
  \le
  \frac{1}{\mathcal{I}_{\mathrm{eff}}^{*\,2}}
  \sum_{j} |S_j|\,\epsilon_j
  + 2\,|\beta_A^*|\sum_{j} |s_{\beta,j}|\,\epsilon_j,
  \label{eq:autonum:R20}
\end{equation}
and the corresponding normalized robustness metric is
\begin{equation}
  \rho_A(\bm{\epsilon})
  \triangleq
  \frac{\Delta\mathcal{B}_{A,\max}^{(1)}(\bm{\epsilon})}{\mathcal{B}_A^*}.
  \label{eq:autonum:R21}
\end{equation}
As a practical engineering rule, the design can be considered safely robust when $\rho_A(\bm{\epsilon}) \ll 1$ over the intended operating range; for example, $\rho_A < 0.1$ corresponds to a first-order degradation below $10\%$ of the design-point MSE floor.

\textbf{Step SE-R3: Fragility ranking and diagnosis.}
To identify which physical uncertainty dominates the robustness loss, define the per-parameter contribution scores
\begin{equation}
  c_j^{(\mathrm{F})}
  \triangleq
  \frac{|S_j|\,\epsilon_j}{\mathcal{I}_{\mathrm{eff}}^{*\,2}},
  \qquad
  c_j^{(\mathrm{B})}
  \triangleq
  2\,|\beta_A^*|\,|s_{\beta,j}|\,\epsilon_j,
  \label{eq:autonum:R22}
\end{equation}
and rank parameters by $c_j^{(\mathrm{F})} + c_j^{(\mathrm{B})}$.  For amplitude-oriented bi-exponential designs, the dominant fragility directions are typically $\tau_r$, $\tau_d$, and $b$, while $t_0$ is usually negligible because $S_{t_0} = 0$ by time-translation invariance.

\textbf{Step SE-R4: Robust redesign heuristics.}
If the robustness metric is too large, modify the ladder according to the dominant fragility direction and rerun Steps~SE-A2--SE-A9:
\begin{enumerate}
  \item If $\tau_r$ dominates, reduce over-concentration on the earliest rising edge by spreading the lowest thresholds slightly upward or enlarging their spacing.
  \item If $\tau_d$ dominates, reduce reliance on tail-sensitive crossings by pulling the upper thresholds away from the long-decay region and toward the mid-slope or near-peak region.
  \item If $b$ dominates, raise the lowest threshold or enlarge the gap above baseline so that baseline drift perturbs fewer active crossings.
  \item If $A$ dominates across the expected operating range, replace the single-event design by the multi-event recipe of Section~\ref{sec:recipe_multi}, which explicitly averages over amplitude variability.
\end{enumerate}
This redesign step is intentionally heuristic: it uses the sign-free fragility ranking to decide \emph{where} the design is fragile, then reapplies the original optimal-threshold solver to recover the best ladder under the modified initialization or constraints.

\textbf{Step SE-R5: Second-order confirmation for marginal cases.}
When two candidate ladders have similar design-point performance or when $\rho_A(\bm{\epsilon})$ is not clearly small, evaluate the second-order perturbation estimate Eq.~\eqref{eq:biexp_mse_second}.  The preferred design is the one with the smaller predicted worst-case degradation over the anticipated uncertainty region, not necessarily the one with the smallest nominal $\mathcal{B}_A^*$.

\subsection{Multi-Event Design Recipe}
\label{sec:recipe_multi}

The preceding subsection optimized thresholds for a single design-point pulse (e.g., the 511~keV photopeak).  In practice, the detector encounters a continuous energy spectrum, and low-energy events may trigger only a subset of the hardware thresholds.  This subsection distills the multi-event optimization framework of Section~\ref{sec:biexp_partial} into a self-contained, step-by-step engineering procedure that extends the single-event recipe of Section~\ref{sec:recipe}.  All required formulas are restated explicitly.

\emph{Assumption} (Well-specified multi-event model): The recipe below uses the well-specified simplification of the multi-event stationarity system.  That is, the active-set bias terms in Eq.~\eqref{eq:biexp_multi_stationarity} are neglected, and the threshold update is governed by the bias-free form Eq.~\eqref{eq:multi_event_bias_free}.  This simplification is exact only when the model misspecification is sufficiently small over the representative event set.

\emph{Perturbative multi-event error analysis.}
Let $\mathbf{V} = [V_1,\dots,V_N]^T$ denote the common threshold vector, and define the idealized and misspecified multi-event objectives
\begin{align}
  \mathcal{C}_0(\mathbf{V})
   & \triangleq
  \sum_{p=1}^{P} w_p\,
  \bigl[\Delta\mathcal{I}_{\mathrm{eff}}^{(A),\mathrm{act},(p)}(\mathbf{V})\bigr]^{-1},
  \label{eq:autonum:ME0a}   \\
  \mathcal{C}_{\varepsilon}(\mathbf{V})
   & \triangleq
  \mathcal{C}_0(\mathbf{V})
  + \sum_{p=1}^{P} w_p\,
  \bigl[\beta_{A,\mathrm{act}}^{(p)}(\mathbf{V};\varepsilon)\bigr]^2,
  \label{eq:autonum:ME0b}
\end{align}
with minimizers
\(\mathbf{V}_{\mathrm{multi}}^{(0)} = \arg\min \mathcal{C}_0\)
and
\(\mathbf{V}_{\mathrm{multi}}^{(\varepsilon)} = \arg\min \mathcal{C}_{\varepsilon}\).
For compactness, write the misspecification penalty as
\begin{equation}
  \begin{aligned}
    \mathcal{D}_{\varepsilon}(\mathbf{V})
     & \triangleq
    \sum_{p=1}^{P} w_p\,
    \bigl[\beta_{A,\mathrm{act}}^{(p)}(\mathbf{V};\varepsilon)\bigr]^2,     \\
    \mathcal{C}_{\varepsilon}(\mathbf{V})
     & = \mathcal{C}_0(\mathbf{V}) + \mathcal{D}_{\varepsilon}(\mathbf{V}).
  \end{aligned}
  \label{eq:autonum:ME0bprime}
\end{equation}

Evaluate the waveform mismatch at the active crossings induced by the ideal thresholds $\mathbf{V}_{\mathrm{multi}}^{(0)}$.  For event $p$, let $t_k^{(p,0)}$ denote its $2K_p$ active crossing times at $\mathbf{V}_{\mathrm{multi}}^{(0)}$, and define the active-set mean and maximum absolute mismatch amplitudes
\begin{equation}
  \bar{\varepsilon}^{(p)}
  \triangleq
  \frac{1}{2K_p}\sum_{k=1}^{2K_p}
  \bigl|\varepsilon(t_k^{(p,0)})\bigr|,
  \qquad
  \varepsilon_{\max}^{(p)}
  \triangleq
  \max_{1\le k\le 2K_p}
  \bigl|\varepsilon(t_k^{(p,0)})\bigr|.
  \label{eq:autonum:ME0c}
\end{equation}
The corresponding direct active-set amplitude-bias projection is the restricted Gauss--Newton projection (cf.\ Eq.~\eqref{eq:bias_active}):
\begin{equation}
  \begin{aligned}
    \beta_{A,\mathrm{act}}^{(p)}
     & \triangleq
    \Bigl[
      \bigl((\mathbf{J}_{\mathrm{act}}^{(p)})^T\mathbf{W}_{\mathrm{act}}^{(p)}\mathbf{J}_{\mathrm{act}}^{(p)}\bigr)^{-1} \\
      & \qquad \cdot
      (\mathbf{J}_{\mathrm{act}}^{(p)})^T\mathbf{W}_{\mathrm{act}}^{(p)}
      \Delta\mathbf{t}_{\mathrm{bias}}^{\mathrm{act},(p)}
      \Bigr]_A
    \Bigg|_{\mathbf{V}=\mathbf{V}_{\mathrm{multi}}^{(0)}},
  \end{aligned}
  \label{eq:autonum:ME0d}
\end{equation}
where $\Delta\mathbf{t}_{\mathrm{bias}}^{\mathrm{act},(p)}$ is formed from the active crossings of event $p$.  Defining
\begin{equation}
  \begin{aligned}
    m_{\min}^{(p)}
     & \triangleq
    \min_{1\le k\le 2K_p}
    \bigl|\partial_t f(t_k^{(p,0)};\bm{\theta}^{(p)})\bigr|, \\
    C_{\beta}^{(p)}
     & \triangleq
    \left\|\mathbf{e}_A^T
    \bigl((\mathbf{J}_{\mathrm{act}}^{(p)})^T\mathbf{W}_{\mathrm{act}}^{(p)}\mathbf{J}_{\mathrm{act}}^{(p)}\bigr)^{-1}
    (\mathbf{J}_{\mathrm{act}}^{(p)})^T\mathbf{W}_{\mathrm{act}}^{(p)}\right\|_2,
  \end{aligned}
  \label{eq:autonum:ME0e}
\end{equation}
the same argument used for Eq.~\eqref{eq:autonum:R6f} yields
\begin{equation}
  \begin{aligned}
    \bigl|\beta_{A,\mathrm{act}}^{(p)}\bigr|
     & \le
    C_{\beta}^{(p)}
    \frac{\sqrt{2K_p\,\bar{\varepsilon}^{(p)}\,\varepsilon_{\max}^{(p)}}}
    {m_{\min}^{(p)}},   \\
    \bigl(\beta_{A,\mathrm{act}}^{(p)}\bigr)^2
     & = O\!\bigl(\bar{\varepsilon}^{(p)}\,\varepsilon_{\max}^{(p)}\bigr).
  \end{aligned}
  \label{eq:autonum:ME0f}
\end{equation}

Define the weighted aggregate mismatch scale
\begin{equation}
  \delta_{\mathrm{multi}}
  \triangleq
  \sum_{p=1}^{P} w_p\,\bar{\varepsilon}^{(p)}\,\varepsilon_{\max}^{(p)}.
  \label{eq:autonum:ME0g}
\end{equation}
Assume furthermore that the ideal multi-event design has a nondegenerate local minimum, with Hessian
\begin{equation}
  \mathbf{H}_{0,\mathrm{multi}}
  \triangleq
  \nabla_{\mathbf{V}}^2 \mathcal{C}_0\bigl(\mathbf{V}_{\mathrm{multi}}^{(0)}\bigr)
  \succ 0.
  \label{eq:autonum:ME0h}
\end{equation}
Then the misspecified stationarity system Eq.~\eqref{eq:biexp_multi_stationarity} yields, by a first-order implicit-function expansion around $\mathbf{V}_{\mathrm{multi}}^{(0)}$,
\begin{equation}
  \begin{aligned}
    \mathbf{V}_{\mathrm{multi}}^{(\varepsilon)}
    - \mathbf{V}_{\mathrm{multi}}^{(0)}
     & = -\mathbf{H}_{0,\mathrm{multi}}^{-1}
    \nabla_{\mathbf{V}}
    \mathcal{D}_{\varepsilon}(\mathbf{V})
    \Big|_{\mathbf{V}=\mathbf{V}_{\mathrm{multi}}^{(0)}} \\
     & \quad + O\!\bigl(\delta_{\mathrm{multi}}^2\bigr),
  \end{aligned}
  \label{eq:autonum:ME0i}
\end{equation}
so the threshold perturbation is second-order in the waveform-mismatch amplitudes.  Consequently,
\begin{equation}
  \mathcal{C}_{\varepsilon}\bigl(\mathbf{V}_{\mathrm{multi}}^{(\varepsilon)}\bigr)
  - \mathcal{C}_0\bigl(\mathbf{V}_{\mathrm{multi}}^{(0)}\bigr)
  = \mathcal{D}_{\varepsilon}\bigl(\mathbf{V}_{\mathrm{multi}}^{(0)}\bigr)
  + O\!\bigl(\delta_{\mathrm{multi}}^2\bigr).
  \label{eq:autonum:ME0j}
\end{equation}
Equation~\eqref{eq:autonum:ME0j} is the multi-event analogue of Eq.~\eqref{eq:autonum:R6i}: the dominant gap between the true misspecified multi-event amplitude cost and the idealized cost is determined by the weighted sum of active-set bias projections, while threshold re-optimization changes the optimum only at higher order.

The resulting aggregate and per-event diagnostics are
\begin{equation}
  \begin{aligned}
    \rho_{\mathrm{bias,multi}}^{(A)}
     & \triangleq
    \frac{\mathcal{D}_{\varepsilon}\bigl(\mathbf{V}_{\mathrm{multi}}^{(0)}\bigr)}
    {\mathcal{C}_0\bigl(\mathbf{V}_{\mathrm{multi}}^{(0)}\bigr)}, \\
    \rho_{\mathrm{bias}}^{(A),(p)}
     & \triangleq
    \frac{\bigl(\beta_{A,\mathrm{act}}^{(p)}\bigr)^2}
    {\bigl[\Delta\mathcal{I}_{\mathrm{eff}}^{(A),\mathrm{act},(p)}\bigr]^{-1}}.
  \end{aligned}
  \label{eq:autonum:ME0k}
\end{equation}
As in the single-event case, these ratios provide the engineering validity test for the well-specified simplification: if $\rho_{\mathrm{bias,multi}}^{(A)} \ll 1$ and the per-event tail ratios $\rho_{\mathrm{bias}}^{(A),(p)}$ remain small (e.g., in weighted median/upper-percentile sense), then the bias-free multi-event stationarity Eq.~\eqref{eq:multi_event_bias_free} is adequate.  Otherwise, the present recipe should be interpreted only as a \emph{well-specified baseline/initialization}, and a misspecification-aware re-optimization (or model enrichment / robust fitting) is required before final deployment.

\subsubsection{Additional Required Inputs}
\label{sec:recipe_multi_inputs}

In addition to all inputs listed in Section~\ref{sec:recipe_inputs}, the engineer must supply:

{
  \begin{enumerate}
    \item \emph{Energy spectrum discretization}:
          \begin{itemize}
            \item $P$: number of representative energy bins (events);
            \item $\{A^{(p)}\}_{p=1}^{P}$: bi-exponential waveform scale amplitude for each energy bin $[\mathrm{V}]$;
            \item $\{b^{(p)}\}_{p=1}^{P}$: baseline voltage for each energy bin $[\mathrm{V}]$ (often $b^{(p)} = b$ for all $p$);
            \item $\{w_p\}_{p=1}^{P}$: spectral weight (relative occurrence frequency) for each energy bin, with $\sum_p w_p = 1$.
          \end{itemize}
    \item \emph{Shape parameters} (if energy-dependent):
          \begin{itemize}
            \item $\{\tau_d^{(p)},\, \tau_r^{(p)}\}_{p=1}^{P}$: effective macroscopic decay and rise time constants for each energy bin $[\mathrm{s}]$.  In the common case of shared shape parameters, $\tau_d^{(p)} = \tau_d$ and $\tau_r^{(p)} = \tau_r$ for all $p$.
          \end{itemize}
  \end{enumerate}
}

\emph{Practical guidance}: The spectral weights $\{w_p\}$ should reflect the detector's operational energy distribution.  For PET applications that aim to recover inter-crystal scattering events, a typical choice is to discretize the 100--700~keV energy window into $P = 5$--$20$ bins and assign $w_p$ proportional to the measured or simulated energy histogram counts in each bin~\cite{shao1994triple,sossi1995comparison,yoshida2008doi}.  Higher-energy bins near the photopeak may be assigned larger weights when amplitude fidelity around the dominant operating energy is especially important.

\subsubsection{Preliminary Computations}
\label{sec:recipe_multi_prelim}

Compute all single-event preliminary quantities (Steps~SE-P1--SE-P3 of Section~\ref{sec:recipe_prelim}) for the photopeak event to obtain the initial threshold set.  Additionally, compute the following for each energy bin.

\textbf{Step ME-P1: Per-event peak voltage.}
For each event $p = 1, \ldots, P$, compute (using Eqs.~\eqref{eq:autonum:R3}--\eqref{eq:autonum:R4} with event-specific parameters):
\begin{equation}
  V_{\mathrm{peak}}^{(p)} = A^{(p)} \cdot h(x_p^{(p)}) + b^{(p)},
  \label{eq:autonum:ME1}
\end{equation}
where $x_p^{(p)} = \tau_r^{(p)}\tau_d^{(p)}/(\tau_d^{(p)} - \tau_r^{(p)})\cdot\ln(\tau_d^{(p)}/\tau_r^{(p)})$ is the peak time for event $p$.  In the shared-shape case, $x_p^{(p)} = x_p$ for all $p$.

\textbf{Step ME-P2: Active threshold set.}
For each event $p$ and the current threshold set $\{V_n\}$, determine the active threshold set (Eq.~\eqref{eq:biexp_active_set}):
\begin{equation}
  \mathcal{A}_p = \bigl\{n \in \{1,\ldots,N\} : V_n < V_{\mathrm{peak}}^{(p)}\bigr\},
  \label{eq:autonum:ME2}
\end{equation}
with cardinality $K_p = |\mathcal{A}_p|$.  For the five-parameter amplitude-target model, every event included in the formal multi-event design should satisfy $K_p \ge 3$, so that $2K_p \ge 6$ active timestamps are available and the active amplitude information $\Delta\mathcal{I}_{\mathrm{eff}}^{(A),\mathrm{act},(p)}$ can be finite as required by Section~\ref{sec:biexp_partial}.  The weaker condition $K_p \ge 2$ guarantees only nuisance-block invertibility and is not sufficient for the full amplitude-target CRLB analysis.

\emph{Diagnostic}: If any event $p$ has $K_p < 3$, this indicates that the lowest thresholds are set too high for the corresponding energy bin.  Either lower the lowest thresholds so that at least three threshold pairs remain active, or exclude that energy bin from the optimization (effectively truncating the energy window).

\subsubsection{Multi-Event Threshold Algorithm}
\label{sec:recipe_multi_algorithm}

The following alternating fixed-point iteration computes the spectrally optimal threshold set $\{V_1^*, \ldots, V_N^*\}$ that minimizes the weighted-average amplitude MSE across all energy bins.  It implements the multi-event framework of Section~\ref{sec:biexp_partial} under the well-specified model assumption (Section~\ref{sec:recipe_algorithm}).

\textbf{Step ME-A1: Initialization.}
Use the single-event optimal thresholds $\{V_n^{(0)}\}$ from the photopeak design point (Section~\ref{sec:recipe_algorithm}) as the starting configuration:
\begin{equation}
  V_n^{(0)} = V_n^{*,\mathrm{single}}, \quad n = 1, \ldots, N.
  \label{eq:autonum:ME3}
\end{equation}
This initialization leverages the fact that photopeak events trigger all thresholds, providing a well-conditioned starting point.

\textbf{Step ME-A2: Active set computation.}
For each event $p = 1, \ldots, P$, compute the active threshold set $\mathcal{A}_p$ from the current $\{V_n\}$ via Eq.~\eqref{eq:autonum:ME2}.  Record the active cardinality $K_p = |\mathcal{A}_p|$ for each event.

\textbf{Step ME-A3: Per-event crossing roots and sensitivities.}
For each event $p$ and each active threshold $n \in \mathcal{A}_p$, compute:
\begin{enumerate}
  \item The two crossing roots $x_{r}^{(n,p)}$ and $x_{f}^{(n,p)}$ by solving Eq.~\eqref{eq:autonum:R8} with $A^{(p)}$ and $b^{(p)}$:
        \begin{equation}
          h(x) = \frac{V_n - b^{(p)}}{A^{(p)}};
          \label{eq:autonum:ME4}
        \end{equation}
  \item The five-component waveform sensitivity vectors $\mathbf{g}_k^{(n,p)}$ via Eq.~\eqref{eq:autonum:R9}, evaluated at the event-specific parameters $\bm{\theta}^{(p)}$;
  \item The composite noise denominators $D_k^{(n,p)}$ via Eq.~\eqref{eq:autonum:R10}, evaluated at $A^{(p)}$, $b^{(p)}$.
\end{enumerate}

\textbf{Step ME-A4: Per-event FIM block assembly.}
For each event $p$, construct the active nuisance--nuisance and nuisance--target FIM blocks by summing only over active thresholds (cf.\ Eq.~\eqref{eq:autonum:R11}):
\begin{equation}
  \begin{aligned}
    \bm{\mathcal{I}}_{\bm{\eta}'\bm{\eta}'}^{\mathrm{act},(p)} & = \sum_{n \in \mathcal{A}_p}\; \bm{\mathcal{I}}_{\bm{\eta}'\bm{\eta}'}^{(n)}(\bm{\theta}^{(p)}), \\
    \bm{\mathcal{I}}_{\bm{\eta}' A}^{\mathrm{act},(p)}         & = \sum_{n \in \mathcal{A}_p}\; \bm{\mathcal{I}}_{\bm{\eta}' A}^{(n)}(\bm{\theta}^{(p)}),
  \end{aligned}
  \label{eq:autonum:ME5}
\end{equation}
where each per-threshold block $\bm{\mathcal{I}}^{(n)}(\bm{\theta}^{(p)})$ is assembled from Step~ME-A3 using Eq.~\eqref{eq:autonum:R11} with event-specific quantities.

\textbf{Step ME-A5: Per-event nuisance coupling update ($\bm{\gamma}_p$-step).}
For each event $p$, compute the event-specific nuisance coupling vector (Eq.~\eqref{eq:biexp_lambda_p}):
\begin{equation}
  \bm{\gamma}_p^{\mathrm{act}} = \bigl[\bm{\mathcal{I}}_{\bm{\eta}'\bm{\eta}'}^{\mathrm{act},(p)}\bigr]^{-1}\; \bm{\mathcal{I}}_{\bm{\eta}' A}^{\mathrm{act},(p)} \;\in\; \mathbb{R}^4.
  \label{eq:autonum:ME6}
\end{equation}
Note that $\bm{\gamma}_p^{\mathrm{act}}$ differs across events both because the FIM values depend on $\bm{\theta}^{(p)}$ and because the summation range $\mathcal{A}_p$ varies with event amplitude.

\textbf{Step ME-A6: Per-event projected Fisher and effective information.}
For each event $p$ and each active threshold $n \in \mathcal{A}_p$, compute the nuisance-projected sensitivities (cf.\ Eq.~\eqref{eq:autonum:R13}):
\begin{equation}
  \tilde{g}_k^{(n,p)} = g_{A,k}^{(n,p)} - [\bm{\gamma}_p^{\mathrm{act}}]^T\,\mathbf{g}_{\bm{\eta}',k}^{(n,p)}, \quad k \in \{r,f\},
  \label{eq:autonum:ME7}
\end{equation}
and the per-event effective information for amplitude:
\begin{equation}
  \Delta\mathcal{I}_{\mathrm{eff}}^{(A),\mathrm{act},(p)} = \sum_{n \in \mathcal{A}_p} \left[\frac{[\tilde{g}_r^{(n,p)}]^2}{D_r^{(n,p)}} + \frac{[\tilde{g}_f^{(n,p)}]^2}{D_f^{(n,p)}}\right].
  \label{eq:autonum:ME8}
\end{equation}

\textbf{Step ME-A7: Multi-event threshold update ($V$-step).}
Update each threshold $V_n$ ($n = 1, \ldots, N$) by solving a coordinate-ascent subproblem derived from the multi-event stationarity system (Eq.~\eqref{eq:biexp_multi_stationarity}), with the current active sets $\{\mathcal{A}_p\}$, event-specific couplings $\{\bm{\gamma}_p^{\mathrm{act}}\}$, and relevance weights $w_p/[\Delta\mathcal{I}_{\mathrm{eff}}^{(A),\mathrm{act},(p)}]^2$ frozen at the current iterate.  For each threshold $n$, maximize the weighted objective:
\begin{equation}
  \begin{split}
    V_n^* & = \arg\max_{V \in (b_{\min},\, V_{\mathrm{peak}}^{\max})}                                                                                                                                       \\
          & \quad \sum_{p:\,n \in \mathcal{A}_p} \frac{w_p}{\bigl[\Delta\mathcal{I}_{\mathrm{eff}}^{(A),\mathrm{act},(p)}\bigr]^2}\; \Phi_{A}^{(n)}(V;\,\bm{\gamma}_p^{\mathrm{act}},\, \bm{\theta}^{(p)}),
  \end{split}
  \label{eq:autonum:ME9}
\end{equation}
with the search interval clipped by the chosen same-branch spacing floor: the diagonal-covariance diagnostic $\Delta V_{\min}$ (Eq.~\eqref{eq:autonum:R5}) if independent crossings are enforced, or an empirical/coverage floor otherwise.  Here $b_{\min} = \min_p b^{(p)}$ and $V_{\mathrm{peak}}^{\max} = \max_p V_{\mathrm{peak}}^{(p)}$.  When diagonal single-pair weights are retained, this interval must also be intersected with the event-wise pair-gap admissible set, i.e., the set of voltages for which $t_f(V;\bm{\theta}^{(p)})-t_r(V;\bm{\theta}^{(p)}) \ge \tau_c$ for every currently active event $p$ with $n \in \mathcal{A}_p$.  If this condition is intentionally relaxed for some active event, that event should instead be evaluated with its full within-pair covariance block rather than the diagonal form Eq.~\eqref{eq:autonum:ME8}.

With active sets, neighboring spacing bounds, and the frozen relevance weights fixed, each update is a one-dimensional optimization solvable by golden-section search or Brent's method.  The key difference from the single-event case (Eq.~\eqref{eq:autonum:R15}) is the spectral weighting: each event's contribution is weighted by $w_p / [\Delta\mathcal{I}_{\mathrm{eff}}^{(A),\mathrm{act},(p)}]^2$, which gives greater influence to events with \emph{lower} effective information (typically low-energy events), ensuring that the thresholds serve the worst-case energy bins.

\textbf{Step ME-A8: Active set monitoring and convergence check.}
After updating $\{V_n\}$, recompute all active sets $\mathcal{A}_p$ (Step~ME-A2).  If any active set changes discretely (i.e., a threshold crosses an event's peak voltage), restart the $\bm{\gamma}_p$-update from Step~ME-A5 before continuing.

Repeat Steps~ME-A2--ME-A7 until the threshold values converge:
\begin{equation}
  \max_{1 \le n \le N} \left|V_n^{(\ell+1)} - V_n^{(\ell)}\right| < \varepsilon_V.
  \label{eq:autonum:ME10}
\end{equation}
A typical tolerance is $\varepsilon_V = 0.1~\mathrm{mV}$.

\emph{Convergence note}: Discrete changes in the active sets $\mathcal{A}_p$ may cause non-smooth behavior in the objective.  In practice, such transitions are rare after a few initial iterations (since the single-event initialization already places thresholds below the photopeak), and the algorithm converges within 10--30 iterations for typical detector configurations.

\textbf{Step ME-A9: Performance evaluation.}
At convergence, the weighted-average amplitude MSE across the energy spectrum is:
\begin{equation}
  \bar{\mathcal{B}}_{A}^{*} = \sum_{p=1}^{P} w_p \cdot \bigl[\Delta\mathcal{I}_{\mathrm{eff}}^{(A),\mathrm{act},(p)}\bigr]^{-1}.
  \label{eq:autonum:ME11}
\end{equation}
If upstream amplitude variances $\sigma_{A,\mathrm{front}}^{2,(p)}$ are available for the representative bins, a convenient spectrally averaged relative amplitude resolution is:
\begin{equation}
  \begin{aligned}
    \overline{\mathcal{R}}_{A,\mathrm{FWHM}}
     & = 2\sqrt{2\ln 2} \cdot \sqrt{\sum_{p=1}^{P} w_p\,
                                \frac{\mathcal{B}_{A}^{(p)} + \sigma_{A,\mathrm{front}}^{2,(p)}}{\bigl(A^{(p)}\bigr)^2}}.
  \end{aligned}
  \label{eq:autonum:ME12}
\end{equation}

For diagnostic purposes, one may also inspect the per-event amplitude MSE $\mathcal{B}_{A}^{(p)} = [\Delta\mathcal{I}_{\mathrm{eff}}^{(A),\mathrm{act},(p)}]^{-1}$ to verify that no individual energy bin is disproportionately degraded relative to the single-event optimum.

\subsubsection{Summary}

\begin{enumerate}
  \item \textbf{Characterize} your detector and energy spectrum: in addition to the single-event parameters (Section~\ref{sec:recipe_inputs}), determine the representative energy bins $\{A^{(p)}, b^{(p)}, w_p\}_{p=1}^{P}$ from measured or simulated energy histograms.
  \item \textbf{Initialize} with the single-event optimal thresholds from the photopeak design point (Section~\ref{sec:recipe_algorithm}).
  \item \textbf{Compute active sets}: for each energy bin, determine which thresholds lie below the event's peak voltage (Eq.~\eqref{eq:autonum:ME2}).  Verify $K_p \ge 3$ for all bins that are to remain inside the formal five-parameter amplitude-target design regime.
  \item \textbf{Iterate} the multi-event fixed-point loop (Steps~ME-A2--ME-A7 above): for each event, compute its active FIM blocks and nuisance coupling; then update each threshold to maximize the spectrally weighted projected Fisher contribution, giving greater weight to information-poor (low-energy) events.
  \item \textbf{Monitor} active set transitions, verify convergence (Eq.~\eqref{eq:autonum:ME10}), and check that the final ladder satisfies both the chosen same-branch spacing floor and the active-event pair-gap condition wherever diagonal single-pair weights are used.
  \item \textbf{Evaluate} spectrally averaged performance via Eqs.~\eqref{eq:autonum:ME11}--\eqref{eq:autonum:ME12} and inspect per-event diagnostics.
\end{enumerate}

\emph{Reduction check}: Setting $P = 1$ with $A^{(1)} = A_{\mathrm{design}}$, $b^{(1)} = b$, and $w_1 = 1$ recovers the single-event recipe of Section~\ref{sec:recipe} exactly, as $\mathcal{A}_1 = \{1,\ldots,N\}$ and $\bm{\gamma}_1^{\mathrm{act}} = \bm{\gamma}^*$.

\subsection{Model-Mismatch Diagnostics}
\label{sec:delta_validation}

The threshold optimization algorithm of Section~\ref{sec:recipe_algorithm} is derived under the simplifying assumption that the bi-exponential model is well-specified, i.e., $\varepsilon(t) \approx 0$.  If this assumption is violated, the experimental validation is still fully executable: the objective shifts from confirming $\varepsilon(t) \approx 0$ to quantifying misspecification impact and deciding whether correction/re-optimization is required.  Following the perturbative analysis in Eqs.~(R.6a)--(R.6j), we therefore evaluate the amplitude-target misspecification level by directly estimating the bias projection $\beta_{A}^{\mathrm{direct}}$ and the normalized diagnostic ratio
\begin{equation}
  \rho_{\mathrm{bias}}^{(A)}
  \triangleq
  \frac{(\beta_{A}^{\mathrm{direct}})^2}{\mathcal{B}_0^{(A)}},
\end{equation}
where $\mathcal{B}_0^{(A)} \equiv \mathcal{B}_0^{(A)}(\mathbf{V}^{(0)}) = [\Delta\mathcal{I}_{\mathrm{eff}}^{(A),\mathrm{tot}}]^{-1}$ is the idealized amplitude variance floor at the design thresholds.  If $\rho_{\mathrm{bias}}^{(A)} \ll 1$, then Eq.~(R.6i) implies that the misspecification-induced gap in amplitude MSE is dominated by $(\beta_{A}^{\mathrm{direct}})^2$, while threshold re-optimization contributes only higher-order corrections.  We now apply this test to a single-channel 50~GS/s dataset containing 10 WFM files (1000 pulses/file; 10{,}000 pulses total).  Here $y_{\mathrm{data}}(t)$ denotes the linearly interpolated recorded waveform of an individual pulse, so it can be evaluated at fitted crossing times.

\textbf{Methodology.}
For each of the 10{,}000 recorded pulses, we:
\begin{enumerate}
  \item fit the five-parameter bi-exponential model $\hat{\bm{\theta}} = [\hat{A},\, \hat{t}_0,\, \hat{\tau}_d,\, \hat{\tau}_r,\, \hat{b}]^T$ to the full waveform (all 50{,}000 oscilloscope sampling points) via nonlinear least squares;
  \item set $N = 8$ equally spaced thresholds in $(b,\, V_{\mathrm{peak}})$;
  \item for each threshold $n$, solve for the rising- and falling-edge crossing times $t_{r,n}$ and $t_{f,n}$ on the fitted curve, then assemble the ordered sequence $\{t_k\}_{k=1}^{K}$ via Eq.~\eqref{eq:crossing_index_map}, giving $K = 2N = 16$ crossings;
  \item evaluate the crossing-point mismatch samples $\varepsilon_k \triangleq y_{\mathrm{data}}(t_k) - f(t_k;\hat{\bm{\theta}})$ at the fitted crossings by linear interpolation of the recorded waveform; because each $t_k$ satisfies $f(t_k;\hat{\bm{\theta}})=V_{\ell(k)}$ on the fitted curve, this equals the voltage mismatch against the corresponding threshold level;
  \item compute the amplitude variance floor $\mathcal{B}_0^{(A)} = [(\mathbf{J}^T\mathbf{W}\mathbf{J})^{-1}]_{A,A}$ and the direct Fisher bias projection $\beta_{A}^{\mathrm{direct}}$ (Eq.~\eqref{eq:beta_direct}), along with the empirical mismatch diagnostics $\bar{\varepsilon}$ and $\varepsilon_{\max}$ (Eq.~R.6c).
\end{enumerate}

The noise parameters are estimated from the data: the thermal noise $\sigma_{\mathrm{th}} = 9.51~\mathrm{mV}$ (from the pre-pulse baseline standard deviation), the TDC jitter $\sigma_{\mathrm{TDC}} = \Delta t_{\mathrm{samp}}/\sqrt{12} = 5.77~\mathrm{ps}$ (from the 20~ps sampling interval), and a conservative single-photoelectron amplitude $\kappa = 1~\mathrm{mV}$.

\emph{Assumption} (Stationary empirical noise calibration): These three noise constants are treated as event-independent over the acquisition run and threshold-independent after conditioning on the crossing voltage through the affine term $\kappa(V-b)$.  This is reasonable for a short calibration dataset with stable oscilloscope settings and a fixed detector bias.  If the baseline variance, gain, or timing jitter drifts during acquisition, the diagonal weight matrix $\mathbf{W}$ should be recalibrated in time blocks or augmented with nuisance parameters.

To quantify the impact of model misspecification, we evaluate the bias-to-variance ratio using the \emph{direct Fisher projection} $\beta_{A}^{\mathrm{direct}}$ (Eq.~R.6d), which projects the crossing-point mismatch onto the amplitude coordinate through the Fisher metric.  Consistent with Eqs.~(R.6d)--(R.6j), this projection uses the full Jacobian and weight matrix and therefore measures the physically relevant component of waveform mismatch for amplitude estimation, rather than a raw pointwise voltage error.  The resulting ratio $\rho_{\mathrm{bias}}^{(A)} = (\beta_{A}^{\mathrm{direct}})^2/\mathcal{B}_0^{(A)}$ is the operational diagnostic used below.

\raggedbottom

\begin{figure}[H]
  \centering
  \includegraphics[width=0.98\columnwidth]{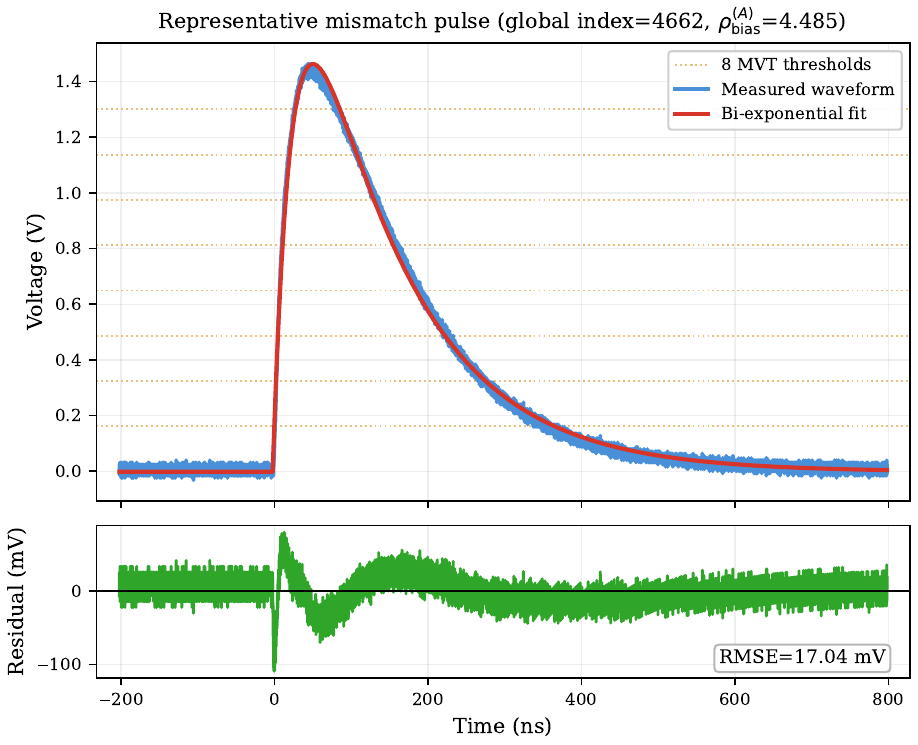}
  \caption{Representative single-channel pulse illustrating model misspecification in the concatenated 10,000-pulse dataset (global index 4662): the measured waveform (blue) is overlaid with the bi-exponential fit (red), eight MVT thresholds are superimposed as horizontal dotted lines, and the residual trace is shown in the lower panel.  For this pulse, $\rho_{\mathrm{bias}}^{(A)} = 4.485$, so the structured residual pattern is not negligible for amplitude estimation under the current bi-exponential model.}
  \label{fig:delta_waveform_mismatch}
\end{figure}

Figure~\ref{fig:delta_waveform_mismatch} provides a representative example of the observed mismatch by overlaying a measured pulse and its bi-exponential fit, together with the residual trace.

\textbf{Results.}
All 10{,}000 fits converge successfully.  Table~\ref{tab:delta_results} summarizes the key diagnostic statistics for the single-channel, amplitude-target analysis.

\begin{table}[H]
  \centering
  \caption{Empirical $\varepsilon(t) \approx 0$ diagnostic statistics for amplitude-target analysis over 10{,}000 single-channel pulses (50~GS/s).}
  \label{tab:delta_results}
  \begin{tabular}{l c c c}
    \hline
    Quantity                                                                             & Median                & Mean                 & 95th pctl \\
    \hline
    $\bar{\varepsilon}$ (mV)                                                             & 9.53                  & 18.8                 & 28.6      \\
    $\varepsilon_{\max}$ (mV)                                                            & 25.5                  & 65.8                 & 89.3      \\
    $\bar{\varepsilon}/\sigma_{\mathrm{th}}$                                             & 1.00                  & 1.98                 & 3.00      \\
    $|\beta_{A}^{\mathrm{direct}}|$ (V)                                                  & 0.255                 & $7.30 \times 10^{4}$ & 0.566     \\
    $\sqrt{\mathcal{B}_0^{(A)}}$ (V)                                                     & 0.315                 & $2.46 \times 10^{4}$ & 0.544     \\[4pt]
    $\rho_{\mathrm{bias}}^{(A)} = (\beta_{A}^{\mathrm{direct}})^2 / \mathcal{B}_0^{(A)}$ & $5.98 \times 10^{-1}$ & $2.21 \times 10^{2}$ & 4.49      \\
    \hline
  \end{tabular}
\end{table}

\begin{figure}[!t]
  \centering
  \includegraphics[width=0.98\columnwidth]{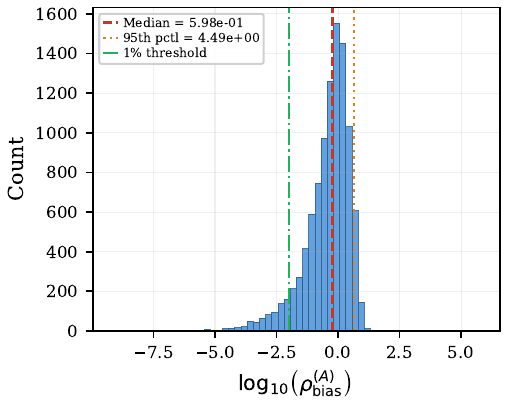}
  \caption{Distribution of the direct amplitude bias-to-variance ratio $\rho_{\mathrm{bias}}^{(A)} = (\beta_A^{\mathrm{direct}})^2 / \mathcal{B}_0^{(A)}$ over 10{,}000 single-channel pulses. The median ratio is $5.98 \times 10^{-1}$ (95th percentile $= 4.49$). The green dash-dotted line marks the 1\% threshold.}
  \label{fig:delta_validation}
\end{figure}

Figure~\ref{fig:delta_validation} visualizes the per-pulse ratio $\rho_{\mathrm{bias}}^{(A)} = (\beta_A^{\mathrm{direct}})^2 / \mathcal{B}_0^{(A)}$.

\begin{figure}[!t]
  \centering
  \includegraphics[width=0.98\columnwidth]{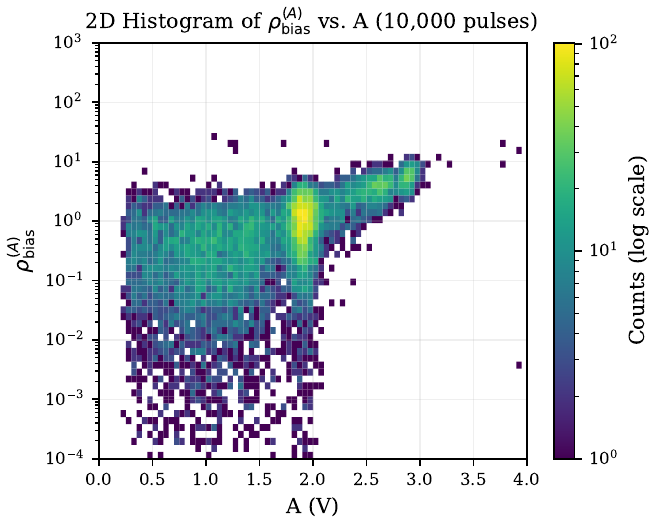}
  \caption{Two-dimensional histogram of the direct amplitude bias-to-variance ratio $\rho_{\mathrm{bias}}^{(A)}$ versus fitted amplitude $A$ over 10{,}000 single-channel pulses. The horizontal axis is restricted to $0$--$4~\mathrm{V}$ to capture a broader portion of the operational amplitude range while excluding only the most extreme outliers.}
  \label{fig:rhoA_vs_A_hist2d}
\end{figure}

Figure~\ref{fig:rhoA_vs_A_hist2d} shows the joint distribution of $\rho_{\mathrm{bias}}^{(A)}$ and the fitted amplitude parameter $A$, with the horizontal axis displayed over the practically relevant range $0$--$4~\mathrm{V}$.

Three conclusions emerge:
\begin{enumerate}
  \item \emph{The amplitude-target ratio is not in the asymptotic small-bias regime.}
        The direct projection gives median $\rho_{\mathrm{bias}}^{(A)} = 5.98 \times 10^{-1}$ with 95th percentile $4.49$, so $(\beta_A^{\mathrm{direct}})^2$ is not negligible relative to $\mathcal{B}_0^{(A)}$ for this dataset.  Therefore, the condition $\rho_{\mathrm{bias}}^{(A)} \ll 1$ is generally not satisfied here.

  \item \emph{Crossing mismatch remains near the voltage-noise scale, but amplitude sensitivity to mismatch is stronger.}
        The median $\bar{\varepsilon}/\sigma_{\mathrm{th}} = 1.00$ indicates that crossing-point mismatch is still baseline-noise scale in voltage space.  However, after projection through the amplitude Fisher geometry, the induced amplitude-bias contribution is substantial for many pulses, indicating higher sensitivity of the amplitude coordinate to model mismatch in this operating condition.

  \item \emph{Bound-based means are dominated by heavy tails, so robust statistics are the preferred engineering diagnostic.}
        As shown in Table~\ref{tab:delta_results} and Fig.~\ref{fig:delta_validation}, arithmetic means of $|\beta_A^{\mathrm{direct}}|$, $\sqrt{\mathcal{B}_0^{(A)}}$, and ratio metrics are strongly inflated by a small number of ill-conditioned pulses, whereas medians and upper percentiles remain stable and interpretable.
\end{enumerate}

\textbf{Summary.}
For the single-channel 50~GS/s dataset (10,000 pulses), amplitude-target validation shows that model-mismatch bias is non-negligible relative to the ideal amplitude variance floor: the direct median ratio is $\rho_{\mathrm{bias}}^{(A)} \approx 0.60$, not $\ll 1$.  Hence, for $A$-oriented threshold design in this detector/readout setting, the strict $\varepsilon(t) \approx 0$ simplification is not uniformly valid at the same confidence level as the timing-target case.  This does not block the validation pipeline; instead, it indicates that misspecification-aware correction (via the direct bias projection term), or model enrichment / robust fitting, should be applied before finalizing closed-form threshold settings.

\flushbottom

% =========================================================================
%  X. EXPERIMENTAL VALIDATION OF THRESHOLD DESIGN
% =========================================================================
\section{Experimental Validation for Single-Event Design}
\label{sec:exp_validation}

The preceding section developed detailed engineering recipes for MVT threshold optimization and assessed, through the misspecification diagnostic $\rho_{\mathrm{bias}}^{(A)}$, whether the well-specified model assumption is empirically justified.  The present section complements that analysis by directly evaluating the \emph{end-to-end} performance of the theory-guided threshold design on physical detector data.  Using the same single-channel 50~GS/s dataset described in Section~\ref{sec:exp_setup} (10{,}000 $^{22}$Na scintillation pulses, each fitted to the five-parameter bi-exponential model), we first extract the energy spectrum from the fitted amplitudes and identify the 511~keV photopeak (Section~\ref{sec:energy_spectrum}), which defines the design-point operating condition.  Subsequent subsections apply the optimal-threshold recipe of Section~\ref{sec:recipe} and compare the resulting reconstruction performance against conventional uniform threshold placements.

\subsection{Photopeak Selection}
\label{sec:energy_spectrum}

To define the design-point population for threshold optimization, we construct a pulse-height spectrum using the directly observable peak voltage $V_{\mathrm{peak}}$ of each waveform.  For each of the 10{,}000 recorded pulses, the peak voltage is extracted by scanning the full oscilloscope trace (50{,}000 samples at 50~GS/s) and recording the maximum instantaneous voltage:
\begin{equation}
  V_{\mathrm{peak},i} \;=\; \max_{1 \le k \le m}\, v_{i,k},
  \qquad i = 1,\ldots,10{,}000,
  \label{eq:vpeak_def}
\end{equation}
where $v_{i,k}$ denotes the $k$-th voltage sample of the $i$-th pulse.

\emph{Assumption} (Peak-voltage energy proxy): The raw peak voltage $V_{\mathrm{peak}}$ is used as a model-free proxy for deposited energy.  This requires the scintillator light yield and SiPM gain to remain approximately linear over the selected energy window, and it assumes that saturation and pile-up do not reorder the pulse-height spectrum.  The assumption is reasonable for photopeak windowing in the present calibration data because it avoids the amplitude--shape coupling of the fitted coefficient $A$; it should not be interpreted as an absolute energy calibration without an independent calibration curve.

\begin{figure}[!t]
  \centering
  \includegraphics[width=\columnwidth]{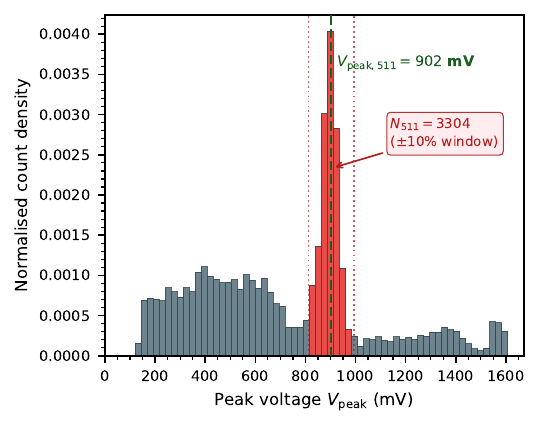}
  \caption{Normalised peak-voltage histogram (pulse-height spectrum) of 10{,}000 $^{22}$Na scintillation pulses recorded at 50~GS/s.  Each $V_{\mathrm{peak}}$ is the maximum voltage sample observed in the raw oscilloscope waveform, providing a model-free energy proxy.  The bin width is 24~mV ($3\times$ the 8~mV ADC quantisation step).  The spectrum shows the Compton continuum (grey) and the 511~keV photopeak at $V_{\mathrm{peak},511} = 902~\mathrm{mV}$ (dashed green line).  The red-shaded region marks the $\pm 10\%$ energy window $[812,\, 993]~\mathrm{mV}$, which selects $N_{511} = 3{,}304$ photopeak events for design-point threshold optimisation.}
  \label{fig:energy_spectrum}
\end{figure}

Figure~\ref{fig:energy_spectrum} displays the normalised $V_{\mathrm{peak}}$ histogram.  The spectrum exhibits two characteristic features of a $^{22}$Na source interacting in an inorganic scintillator:
\begin{enumerate}
  \item A broad \emph{Compton continuum} at lower voltages, arising from gamma photons that undergo one or more Compton scatters and deposit only a fraction of their energy in the crystal;
  \item A prominent \emph{photopeak} near the upper end of the continuum, corresponding to events in which the 511~keV annihilation gamma photon deposits its full energy via photoelectric absorption.
\end{enumerate}

To locate the photopeak, we identify the histogram mode in the upper half of the $V_{\mathrm{peak}}$ range (i.e., above the median), thereby excluding the Compton continuum from the search.  This yields $V_{\mathrm{peak},511} = 902~\mathrm{mV}$.

\textbf{Energy windowing.}
We define the photopeak energy window as $[0.9\,V_{\mathrm{peak},511},\; 1.1\,V_{\mathrm{peak},511}] = [812,\; 993]~\mathrm{mV}$, corresponding to a $\pm 10\%$ peak-voltage acceptance band centred at the photopeak.  Of the 10{,}000 recorded pulses, $N_{511} = 3{,}304$ events (33.0\%) fall within this window (highlighted in red in Fig.~\ref{fig:energy_spectrum}).  These photopeak-selected pulses collectively define the design-point population for threshold optimization.

\subsection{Uniform-Threshold Baseline}
\label{sec:uniform_mvt}

Having identified the $N_{511} = 3{,}304$ photopeak pulses, we now evaluate the amplitude-reconstruction accuracy of an MVT digitiser equipped with $N = 8$ uniformly spaced thresholds.

\textbf{Threshold placement.}
For each photopeak pulse, the raw waveform peak voltage was already determined via Eq.~\eqref{eq:vpeak_def}.  To guarantee that every threshold is crossed by all pulses in the population, we set the threshold upper bound to $V_{\mathrm{peak,min}} = \min_i V_{\mathrm{peak},i} = 816~\mathrm{mV}$ (the smallest observed peak voltage within the energy window) and the lower bound to the median baseline $b = 3.6~\mathrm{mV}$ (obtained from the full-waveform fits).  The $N = 8$ thresholds are then placed at equally spaced voltages strictly between $b$ and $V_{\mathrm{peak,min}}$:
\begin{equation}
  V_n = b + \frac{n}{N+1}\,(V_{\mathrm{peak,min}} - b),
  \qquad n = 1,\ldots,8,
  \label{eq:uniform_thresholds}
\end{equation}
yielding the threshold set $\{93.9,\; 184.1,\; 274.4,\; 364.7,\; 454.9,\;$
$545.2,\; 635.5,\; 725.7\}~\mathrm{mV}$.

\textbf{MVT digitisation and reconstruction.}
For each photopeak pulse, we simulate MVT digitisation exactly as in Section~\ref{sec:exp_setup}: we construct the piecewise-linear interpolant of the recorded waveform, scan all interpolated threshold crossings, retain the earliest rising timestamp, and assign the falling timestamp by the latest post-peak crossing rule of Eq.~\eqref{eq:offline_falling_timestamp_rule}.  The resulting $2N = 16$ simulated MVT timestamps (8~rising, 8~falling) are then fed into the Gauss--Newton solver (Section~\ref{sec:gauss_newton}) with the full-waveform fit as the initial estimate.  Of the 3{,}304 photopeak pulses, 3{,}300 (99.9\%) yield a successful MVT reconstruction.

\emph{Assumption} (Common initialization for validation): The full-waveform fit is used only as a common initializer for the MVT Gauss--Newton solve, so that the comparison isolates the information content of different threshold sets rather than the performance of a separate real-time initialization algorithm.  A deployed MVT system would require a hardware-feasible initializer or a global search safeguard; the present validation therefore measures threshold-design performance under controlled initialization.

\textbf{Reference standard.}
The reference amplitude $A_{\mathrm{ref}}$ for each pulse is obtained by fitting the same five-parameter bi-exponential model (Eq.~\eqref{eq:biexp}) to the full waveform (all 50{,}000 sampling points) via nonlinear least-squares.

\emph{Assumption} (Full-waveform fit as reconstruction reference): The full-waveform fit is used as the empirical reference against which MVT reconstructions are compared.  This is reasonable for an end-to-end threshold-placement study because the reference fit uses roughly three orders of magnitude more samples than the 16 MVT timestamps and all threshold layouts are compared against the same reference.  It is not an absolute ground truth: any structural bias shared by the bi-exponential full-waveform model remains in $A_{\mathrm{ref}}$, so the reported errors quantify loss relative to the high-sample reference model rather than absolute deposited-energy error.

\begin{figure}[!t]
  \centering
  \includegraphics[width=\columnwidth]{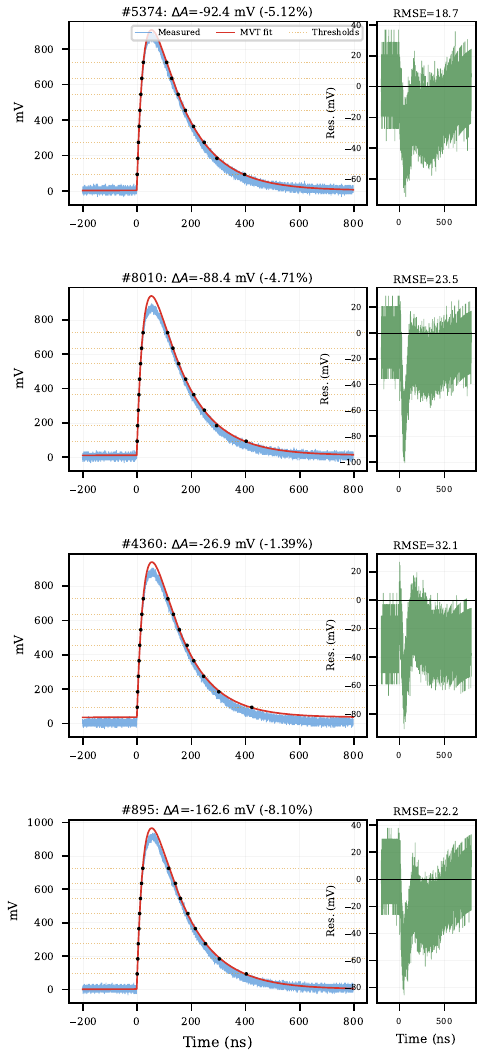}
  \caption{Four representative photopeak pulses reconstructed with $N=8$ uniform thresholds.  Left: measured waveform (blue), MVT-reconstructed fit (red), threshold levels (orange dashed), and simulated MVT timestamps (black dots).  Right: fit residual.  The per-pulse amplitude deviation $\Delta A = A_{\mathrm{uniform}} - A_{\mathrm{ref}}$ and its relative value are indicated in each panel title.}
  \label{fig:uniform_mvt_waveforms}
\end{figure}

\textbf{Representative waveforms.}
Figure~\ref{fig:uniform_mvt_waveforms} displays four representative photopeak pulses spanning the amplitude range.  Each panel shows the recorded waveform, the 8~uniform threshold levels, the $2N = 16$ simulated MVT timestamps (black dots), and the MVT-reconstructed fit (red curve), along with the fit residual (right column).

\begin{figure}[!t]
  \centering
  \includegraphics[width=\columnwidth]{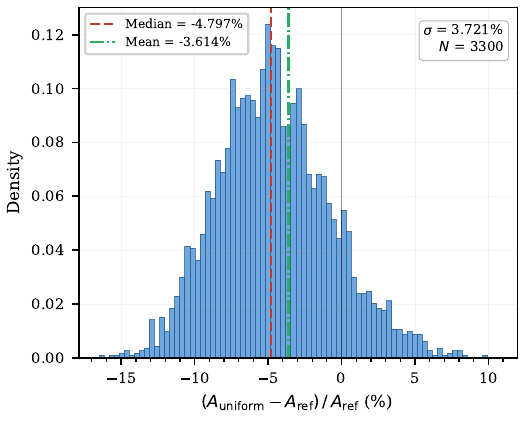}
  \caption{Distribution of the relative amplitude error $(A_{\mathrm{uniform}} - A_{\mathrm{ref}})/A_{\mathrm{ref}}$ for the $N=8$ uniform-threshold MVT reconstruction over 3{,}300 photopeak pulses.  The dashed red line marks the median ($-4.8\%$) and the dot-dashed green line the mean ($-3.6\%$).  The systematic negative bias and the $3.7\%$ spread (std, outlier-excluded)~indicate that uniform threshold spacing is suboptimal, motivating the Fisher-information-optimised placement of Section~\ref{sec:recipe_algorithm}.}
  \label{fig:uniform_mvt_dA}
\end{figure}

\textbf{Amplitude reconstruction accuracy.}
Figure~\ref{fig:uniform_mvt_dA} presents the distribution of the relative amplitude error $\Delta A / A_{\mathrm{ref}} = (A_{\mathrm{uniform}} - A_{\mathrm{ref}}) / A_{\mathrm{ref}}$ over the 3{,}300 successfully reconstructed pulses.  The distribution is centred near $-4.8\%$ (median) with a standard deviation of $3.7\%$ (after excluding outliers beyond the $1.5\times\mathrm{IQR}$ fences).  The key statistics are:
\begin{itemize}
  \item Median bias: $-4.8\%$ (systematic underestimation);
  \item Mean bias: $-3.6\%$;
  \item 51.3\% of pulses satisfy $|\Delta A / A_{\mathrm{ref}}| < 5\%$.
\end{itemize}
The reported standard deviation is computed after removing reconstruction outliers via the standard $1.5\times\mathrm{IQR}$ fence rule: denoting $Q_1$ and $Q_3$ the 25th and 75th percentiles of $\Delta A/A_{\mathrm{ref}}$, all events with $\Delta A/A_{\mathrm{ref}} \notin [Q_1 - 1.5\,\mathrm{IQR},\; Q_3 + 1.5\,\mathrm{IQR}]$ (where $\mathrm{IQR} = Q_3 - Q_1$) are excluded from the standard-deviation calculation.  For the uniform layout this removes 38 of 3{,}300 events (1.2\%), predominantly cases in which the Gauss--Newton solver converged to a secondary local minimum with a grossly incorrect amplitude (relative errors exceeding $50\%$).  The same procedure is applied to the D-optimal results below (41 of 3{,}296 events, 1.2\%).  Median, mean, and pass-rate statistics are always computed on the full population without any exclusion.

The systematic negative bias reflects the inherent information loss when reducing 50{,}000~waveform samples to 16~crossing timestamps with a na\"ive uniform-spacing strategy: the thresholds cluster near the baseline where the signal-to-noise ratio is poorest, while undersampling the high-voltage region near the pulse peak where the amplitude sensitivity is greatest.  This motivates the Fisher-information-based threshold optimisation developed in the preceding sections.

\subsection{D-Optimal Threshold Design}
\label{sec:doptimal_mvt}

We now compare the uniform-threshold baseline of Section~\ref{sec:uniform_mvt} against thresholds selected by maximising the Fisher information content, keeping all other aspects of the pipeline---energy-window selection, the piecewise-linear earliest-rise / latest-fall MVT digitisation of Section~\ref{sec:exp_setup}, the Gauss--Newton solver, and the reference standard---identical.

\subsubsection{Motivation}

The engineering recipe of Section~\ref{sec:recipe_algorithm} maximises the nuisance-projected effective information $\Phi_A(V)$ (Eq.~\eqref{eq:projected_fisher}) for the amplitude parameter $A$ via the Schur complement decomposition (Section~\ref{sec:master_eq}).  Under the regularity assumptions of Section~\ref{sec:opt_derivation}, the resulting threshold set minimises the marginal Cram\'{e}r--Rao lower bound for $A$, i.e.\ the $(A,A)$ entry of the inverse Fisher matrix $[\mathbf{I}^{-1}]_{AA}$---which equals the reciprocal of the effective information $[\Delta\mathcal{I}_{\mathrm{eff}}^{(A)}]^{-1}$ when all nuisance parameters are jointly profiled (Section~\ref{sec:master_eq})---among all feasible $N$-threshold configurations.  One might therefore expect this method to deliver the best possible amplitude reconstruction in practice.  This subsection explains why it does not, and motivates the alternative D-optimal criterion adopted below.

\paragraph{The implicit assumption.}
The CRLB analysis presupposes that the full parameter vector $\bm{\theta} = [A,\,t_0,\,\tau_d,\,\tau_r,\,b]^T$ can be jointly estimated from the $2N = 16$ crossing timestamps.  Formally, the CRLB $[\mathbf{I}^{-1}]_{AA}$ is defined only when the Fisher information matrix $\mathbf{I}(\bm{\theta};\,\mathbf{V})$ is non-singular.  The single-parameter objective $\Phi_A(V)$ accounts for nuisance parameters through the Schur complement, but it does \emph{not} penalise configurations in which $\mathbf{I}$ is near-singular.  In other words, $\Phi_A$ can remain large even when the full FIM has one or more eigenvalues close to zero---a situation where the bound is formally valid but practically useless, because no estimator can jointly resolve all five parameters from the data.

\paragraph{The clustering mechanism.}
For the bi-exponential model with $M = 5$ parameters, the sensitivity gradient $\mathbf{g}_{n,e} \in \mathbb{R}^5$ (Eq.~\eqref{eq:biexp_sensitivity}) varies continuously with the threshold voltage $V_n$.  The nuisance-projected score $\Phi_A(V_n)$ (Eq.~\eqref{eq:projected_fisher}) rewards thresholds where the amplitude gradient $g_A = h(x_c)$ is large \emph{and} maximally orthogonal to the nuisance gradient subspace spanned by $g_{t_0},\,g_{\tau_d},\,g_{\tau_r},\,g_b$ (Eq.~\eqref{eq:nuisance_coupling_vector}).  Classical optimal-design theory~\cite{atkinson1995d,atkinson1989construction} shows that for a $p$-parameter model, the single-parameter-optimal continuous design measure concentrates its mass on at most $\lceil p/2 \rceil$ support voltages.  With $p = 5$, this means the optimizer drives all $N = 8$ thresholds toward approximately three distinct voltage levels.  In the amplitude-target geometry, these support points cluster near the pulse peak (where $g_A = h(x_c) \to h_{\mathrm{peak}}$) and near the baseline (where $g_b = 1$ provides orthogonal baseline information), leaving the mid-range of the pulse---which carries the shape information for $\tau_d$ and $\tau_r$---essentially unsampled.

\paragraph{Numerical consequences for the Gauss--Newton solver.}
The threshold clustering described above has a direct and destructive consequence for parameter reconstruction.  With all thresholds concentrated near three voltages, the $16 \times 5$ Jacobian matrix $\mathbf{J}$ (Eq.~\eqref{eq:jacobian}) consists of near-duplicate rows: crossings at similar voltages produce similar delay offsets $x_c$ and therefore nearly identical sensitivity vectors $\mathbf{g}_{n,e}$.  The normal matrix $\mathbf{J}^T\mathbf{W}\mathbf{J}$ (Eq.~\eqref{eq:gauss_newton}) inherits this rank deficiency: its smallest singular values correspond to the shape-parameter directions $(\tau_d,\,\tau_r)$ that are inadequately sampled.  Concretely, the condition number $\kappa(\mathbf{J}^T\mathbf{W}\mathbf{J})$ can exceed $10^{10}$ under the clustered configuration, compared with $\sim 10^{4}$ under uniform spacing.

The practical effect is that the Gauss--Newton iteration (Eq.~\eqref{eq:gauss_newton}) becomes numerically unstable.  At each step, the update $\delta\bm{\theta} = (\mathbf{J}^T\mathbf{W}\mathbf{J})^{-1}\mathbf{J}^T\mathbf{W}\Delta\mathbf{t}$ amplifies small residuals in the $\tau_d$--$\tau_r$ directions by factors of $10^6$ or more.  The solver either diverges outright, or converges to a spurious local minimum with physically unreasonable shape parameters---which in turn corrupts the amplitude estimate $\hat{A}$ despite the nominally optimal marginal CRLB $[\mathbf{I}^{-1}]_{AA}$ for $A$.

\paragraph{The design--reconstruction gap.}
This failure exposes a fundamental gap between the \emph{design objective} and the \emph{reconstruction objective}:
\begin{itemize}
  \item The \textbf{design objective} (single-parameter CRLB minimisation) asks: ``Given that all parameters \emph{can} be estimated, which thresholds give the best $A$-precision?''
  \item The \textbf{reconstruction objective} (Gauss--Newton convergence) requires: ``Can all parameters \emph{actually} be estimated jointly from these thresholds?''
\end{itemize}
The single-parameter optimizer addresses the first question without enforcing the precondition of the second.  It optimises the \emph{bound} on $\mathrm{Var}(\hat{A})$ but inadvertently destroys the \emph{feasibility} of the estimator that is supposed to attain it.

\paragraph{Engineering regularization: D-optimal design.}
We therefore use D-optimality as an \emph{engineering regularization} of the reconstruction problem, not as a claim that the marginal Schur-complement optimum is mathematically wrong. The Schur-complement objective optimizes the profiled $A$-information under the premise that the nuisance directions are already identifiable; in finite-threshold reconstruction, however, the same objective can drive the threshold sensitivities toward near-collinearity, making the Jacobian and the Gauss--Newton normal matrix nearly singular. The D-optimal criterion---maximising $\log\det\widetilde{\mathbf{I}}(\bm{\theta};\,\mathbf{V})$ after the fixed normalization described below---penalises this collapse by rewarding joint identifiability of all parameter directions. Through the determinant factorization in Eq.~\eqref{eq:det_schur_factorisation}, it may be viewed as adding a nuisance-identifiability regularizer $\log\det(\mathbf{I}_{\bm{\eta}\bm{\eta}})$ to the effective $A$-information term. Although the resulting $A$-bound $[\mathbf{I}^{-1}]_{AA}$ is generically larger than the Schur-complement-optimal bound, the full estimator is better conditioned and can deliver stable amplitude estimates in regimes where the clustered single-parameter design is unusable.

\paragraph{Dimensional normalization.}
Because a determinant of a Fisher matrix depends on parameter units, all D-optimal determinant evaluations are performed in dimensionless local coordinates. Let $\mathbf{D}_{\theta}=\operatorname{diag}(s_A,s_{t_0},s_{\tau_d},s_{\tau_r},s_b)$ collect fixed design-point parameter scales (for example $s_A=A_{\mathrm{design}}$, $s_{\tau_d}=\tau_{d,\mathrm{design}}$, $s_{\tau_r}=\tau_{r,\mathrm{design}}$, with analogous acquisition-window or amplitude scales for $t_0$ and $b$). The determinant objective uses
\begin{equation}
  \widetilde{\mathbf{I}}(\bm{\theta};\mathbf{V})
  = \mathbf{D}_{\theta}^{T}\mathbf{I}(\bm{\theta};\mathbf{V})\mathbf{D}_{\theta},
  \label{eq:dimensionless_fisher}
\end{equation}
and $\log\det\widetilde{\mathbf{I}}$ is therefore dimensionless. Since $\mathbf{D}_{\theta}$ is fixed during threshold optimization, this normalization changes the determinant by a constant offset and does not alter the optimizer; it only makes the criterion physically well-defined.

\paragraph{Formal relationship between D-optimal and Schur-complement objectives.}
The connection is made precise by the Schur complement factorisation of the
determinant.  Partitioning~$\mathbf{I}$ into the
$(M{-}1)\times(M{-}1)$ nuisance block~$\mathbf{I}_{\bm{\eta}\bm{\eta}}$
and the scalar target entry~$\mathcal{I}_{AA}$
(Eqs.~\eqref{eq:I_eta_eta}--\eqref{eq:I_tau_tau}):
The same identities apply to the normalized matrix $\widetilde{\mathbf{I}}$; for compact notation, this paragraph writes $\mathbf{I}$ for the Fisher matrix in the chosen fixed local coordinates.
\begin{equation}
  \det\mathbf{I}
  \;=\;
  \det(\mathbf{I}_{\bm{\eta}\bm{\eta}})
  \;\cdot\;
  \Delta\mathcal{I}_{\mathrm{eff}}^{(A)}.
  \label{eq:det_schur_factorisation}
\end{equation}
Equivalently, the amplitude CRLB admits a purely determinantal
representation via the cofactor expansion:
\begin{equation}
  [\mathbf{I}^{-1}]_{AA}
  \;=\;
  \frac{\det(\mathbf{I}_{\bm{\eta}\bm{\eta}})}{\det(\mathbf{I})}.
  \label{eq:crlb_det_ratio}
\end{equation}
Taking logarithms, the D-optimal objective decomposes additively:
\begin{equation}
  \log\det\mathbf{I}
  \;=\;
  \underbrace{\log\det(\mathbf{I}_{\bm{\eta}\bm{\eta}})}_{\text{nuisance identifiability}}
  \;+\;
  \underbrace{\log\Delta\mathcal{I}_{\mathrm{eff}}^{(A)}}_{\text{effective $A$-information}}.
  \label{eq:dopt_additive_decomposition}
\end{equation}
The Schur-complement objective maximises only the second
term; D-optimality optimises their sum.  The first term
$\log\det(\mathbf{I}_{\bm{\eta}\bm{\eta}})$ acts as a
\emph{nuisance-identifiability regulariser} that penalises threshold
configurations rendering the shape and baseline parameters
$(\tau_d,\tau_r,b)$ unresolvable---precisely the pathology
responsible for the threshold clustering described above.

The regularising effect extends to eigenvalue control.
Let $\xi_1 \le \cdots \le \xi_M$ denote the eigenvalues
of~$\mathbf{I}$.  Because $\det\mathbf{I} = \prod_j\xi_j$,
maximising the determinant under a bounded trace
$\mathrm{tr}(\mathbf{I}) = \sum_j\xi_j$ is equivalent,
by the AM--GM inequality, to equalising the eigenvalues;
any near-zero eigenvalue forces the product toward zero.
Quantitatively,
\begin{equation}
  \xi_{\min}(\mathbf{I})
  \;\ge\;
  \frac{\det\mathbf{I}}
  {\bigl[\mathrm{tr}(\mathbf{I})/(M{-}1)\bigr]^{M-1}},
  \label{eq:lambda_min_bound}
\end{equation}
so D-optimal design provides a trace-conditional lower bound on the
smallest eigenvalue of the Fisher matrix.  The Schur-complement
objective, by contrast, imposes no such constraint, and the
resulting threshold clustering can drive $\xi_{\min}$ below
$10^{-6}\,\xi_{\max}$ (as documented empirically below).

Finally, applying the Hadamard inequality
$\det(\mathbf{I}_{\bm{\eta}\bm{\eta}}) \le
\prod_{j\neq A}\mathbf{I}_{jj}$ to
Eq.~\eqref{eq:crlb_det_ratio} yields a conditional upper bound on
the amplitude CRLB:
\begin{equation}
  [\mathbf{I}^{-1}]_{AA}
  \;\le\;
  \frac{\prod_{j\neq A}\mathbf{I}_{jj}}{\det(\mathbf{I})}.
  \label{eq:hadamard_crlb_bound}
\end{equation}
Thus, when the nuisance diagonal entries are controlled over the
feasible voltage interval, increasing $\det\mathbf{I}$ tightens this
Hadamard upper envelope while also improving joint conditioning.  This
is the operative reason D-optimality is useful here: it discourages
nearly singular designs without pretending that it is identical to the
single-parameter Schur-complement optimum for $A$.

\subsubsection{D-Optimal Criterion}

To ensure that \emph{all} five model parameters $\bm{\theta} = [A,\,t_0,\,\tau_d,\,\tau_r,\,b]^T$ (Eq.~\eqref{eq:biexp_theta}) are jointly well-resolved from the $2N = 16$~crossing timestamps, we replace the single-parameter objective (Eq.~\eqref{eq:projected_fisher}) with the \emph{D-optimal} criterion from classical optimal experimental design: we maximise the log-determinant of the normalized $5 \times 5$ Fisher information matrix,
\begin{equation}
  \mathbf{V}^{*} = \arg\max_{\mathbf{V}} \,\log\det \widetilde{\mathbf{I}}(\bm{\theta};\,\mathbf{V}),
  \label{eq:doptimal_criterion}
\end{equation}
subject to the constraint that $\mathbf{V} = [V_1,\ldots,V_N]^T$ lies within the feasible voltage interval $(b,\,V_{\mathrm{peak,min}})$ with a minimum inter-threshold spacing (defined below).  Here $\widetilde{\mathbf{I}}$ is obtained from the same Fisher information matrix as in Eq.~\eqref{eq:biexp_pair_fim}, summed over all $N$~threshold pairs and normalized by Eq.~\eqref{eq:dimensionless_fisher}.

Maximising $\det\widetilde{\mathbf{I}}$ is equivalent to minimising the volume of the joint confidence ellipsoid in the normalized coordinates.  This simultaneously regularises all nuisance directions and keeps the amplitude bound $[\mathbf{I}^{-1}]_{AA}$ well defined.  It should not be read as a strict monotonic guarantee that $[\mathbf{I}^{-1}]_{AA}$ decreases whenever $\det\widetilde{\mathbf{I}}$ increases, because the nuisance determinant in Eq.~\eqref{eq:crlb_det_ratio} also changes with the threshold set.  The reason for using D-optimality here is joint conditioning: it prevents the near-singular Fisher matrices produced by the single-parameter Schur-complement objective while retaining substantial amplitude information.

\emph{Relationship to the preceding theory.}  It is important to emphasise that D-optimal design does \emph{not} introduce new theoretical machinery.  Every building block---the paired-crossing observation geometry (Section~\ref{sec:paired}), the bi-exponential sensitivity gradients $\mathbf{g}_{n,e}$ (Eq.~\eqref{eq:biexp_sensitivity}), the Poisson-consistent composite noise denominator $D_{n,e}$ (Section~\ref{sec:noise}, Appendix~\ref{app:poisson}), and the rank-one Fisher assembly $\mathbf{I} = \sum \mathbf{g}\mathbf{g}^T/D$ (Eq.~\eqref{eq:biexp_pair_fim})---is inherited verbatim from Sections~\ref{sec:deterministic}--\ref{sec:optimal}.  The \emph{only} modification is the scalar objective function: $\Phi_A(V)$ (Schur complement, Eq.~\eqref{eq:projected_fisher}) is replaced by $\log\det\widetilde{\mathbf{I}}(\bm{\theta};\,\mathbf{V})$.  This replacement turns the single-parameter bound objective into a joint-identifiability objective while preserving the full information-geometric content of the framework.  The coordinate-descent solver described below is likewise a direct adaptation of the fixed-point iteration in Section~\ref{sec:recipe_algorithm}, with the per-threshold sub-problem objective changed from $\Phi_A$ to $\log\det\widetilde{\mathbf{I}}$.

\subsubsection{Implementation Details}

\textbf{Step~1: Design-point parameters.}
The Fisher matrix depends on the unknown true parameters $\bm{\theta}$.  Following standard locally optimal design practice, we evaluate it at the \emph{median} full-waveform fit over the $N_{511} = 3{,}304$ photopeak pulses.  Since the Fisher matrix for the bi-exponential model depends only on the shape parameters through the delay variable $x_c = t_{n,e} - t_0$ (Section~\ref{sec:paired}), it is invariant to $t_0$ itself.  The required design-point quantities are therefore the four shape--amplitude parameters
\begin{equation}
  \bm{\theta}_{\mathrm{design}}
  = \bigl[\tilde{A},\;\tilde{\tau}_d,\;\tilde{\tau}_r,\;\tilde{b}\bigr]^T
  = \mathrm{median}\bigl\{\hat{\bm{\theta}}_i\bigr\}_{i=1}^{N_{511}},
\end{equation}
each computed as the element-wise median over the successfully fitted photopeak pulses.

\emph{Assumption} (Locally representative photopeak design point): The element-wise median full-waveform fit is taken as a representative photopeak pulse for locally optimal threshold design.  This is reasonable for the narrow $\pm10\%$ photopeak window because pulse amplitudes and effective shape parameters are tightly clustered.  It is not a claim of global optimality over the entire energy spectrum; parameter variation is handled by the robustness and multi-event analyses rather than by this single design point.

\textbf{Step~2: Fisher information matrix.}
For a threshold set $\mathbf{V} = [V_1,\ldots,V_N]^T$, each threshold $V_n$ that satisfies $b < V_n < V_{\mathrm{peak,model}}$ generates a rising-edge crossing at $t_{n,r}$ and a falling-edge crossing at $t_{n,f}$ (Section~\ref{sec:paired}), where $V_{\mathrm{peak,model}} = \tilde{A}\,h(x_{\mathrm{peak}}) + \tilde{b}$ is the peak voltage computed from the design-point parameters via $x_{\mathrm{peak}}$ (Eq.~\eqref{eq:biexp_xp}).  Since $V_{\mathrm{peak,min}} < V_{\mathrm{peak,model}}$, all thresholds within the search interval $(b,\,V_{\mathrm{peak,min}})$ are guaranteed to produce valid crossing pairs.

At each crossing, the five-component sensitivity gradient (Eq.~\eqref{eq:pair_sensitivity}) specialises to the bi-exponential partials already derived in Eq.~\eqref{eq:biexp_sensitivity}:
\begin{equation}
  \mathbf{g}_{n,e}
  = \bigl[g_{A},\;g_{t_0},\;g_{\tau_d},\;g_{\tau_r},\;g_{b}\bigr]^T_{\!t = t_{n,e}}
  = \begin{bmatrix}
    h(x_c)                              \\
    -A\,h'(x_c)                         \\
    A\,x_c\,e^{-x_c/\tau_d}\!/\tau_d^2  \\
    -A\,x_c\,e^{-x_c/\tau_r}\!/\tau_r^2 \\
    1
  \end{bmatrix}\!,
  \label{eq:gradient_vec}
\end{equation}
where $e \in \{r,f\}$, $x_c = t_{n,e} - t_0$ is the crossing-time offset, and $h(x) = e^{-x/\tau_d} - e^{-x/\tau_r}$ is the unit-amplitude shape function
(Eq.~\eqref{eq:biexp}).

The composite noise denominator at each crossing follows the same structure as the pair-level denominators $D_r, D_f$ defined in Section~\ref{sec:paired}, combining the thermal floor $\sigma_{\mathrm{th}}^2$, the Poisson shot-noise term $\kappa(V_n - b)$ (Eq.~\eqref{eq:poisson_voltage}), and the slope-projected TDC jitter (Section~\ref{sec:bandwidth}):
\begin{equation}
  D_{n,e} = \sigma_{\mathrm{th}}^2 + \kappa\,(V_n - b)
  + \sigma_{\mathrm{TDC}}^2\!\left[g_{t_0,n,e}\right]^2,
  \label{eq:noise_weight}
\end{equation}
with $\sigma_{\mathrm{th}} = 9.51~\mathrm{mV}$, $\kappa = 1~\mathrm{mV}$, and $\sigma_{\mathrm{TDC}} = \Delta t_{\mathrm{samp}}/\sqrt{12} = 5.77~\mathrm{ps}$.

These denominators inherit the surrogate status of Section~\ref{sec:stochastic}: they are the diagonal, pseudo-Gaussian, first-order weights used to construct an analytically tractable Fisher objective. Any residual colored noise, crossing correlation, comparator/TDC nonideality, or waveform nonlinearity not represented by Eq.~\eqref{eq:noise_weight} is interpreted as part of the practical misspecification budget and is assessed by the diagnostic $\rho_{\mathrm{bias}}^{(A)}$ in Section~\ref{sec:delta_validation}.

The full $5 \times 5$ Fisher information matrix is the sum of all pair contributions (Eq.~\eqref{eq:biexp_pair_fim}):
\begin{equation}
  \mathbf{I}(\bm{\theta};\,\mathbf{V})
  = \sum_{n=1}^{N}\sum_{e \in \{r,f\}}
  \frac{\mathbf{g}_{n,e}\,\mathbf{g}_{n,e}^T}{D_{n,e}}.
  \label{eq:full_fisher}
\end{equation}

\textbf{Step~3: Dynamic-range coverage constraint.}
Unconstrained D-optimal design for a five-parameter model concentrates the $N$~thresholds at approximately $\lceil 5/2 \rceil = 3$ support voltages (and their permutations across rising/falling edges), leaving the mid-range unsampled.  To prevent this, we impose a minimum inter-threshold spacing
\begin{equation}
  \Delta V_{\min}^{\mathrm{cov}} = \frac{V_{\mathrm{peak,min}} - b}{2N},
  \label{eq:dv_min_coverage}
\end{equation}
which guarantees that no two adjacent thresholds are closer than half the average spacing, forcing coverage of the full dynamic range.

\emph{Remark} (Relation to the bandwidth diagnostic).
This coverage spacing $\Delta V_{\min}^{\mathrm{cov}}$ is an engineering heuristic for dynamic-range regularity and is conceptually distinct from the physical bandwidth diagnostic $\Delta V_{\min}^{\mathrm{bw}} = A(\tau_r^{-1} - \tau_d^{-1})/(2B)$ (Eq.~\eqref{eq:autonum:R5}, Section~\ref{sec:bandwidth}), which uses the global rising-edge slew and the proxy correlation scale $\tau_c \approx 1/(2B)$.  For the nominal fast-rise parameters $A \sim 1~\mathrm{V}$, $\tau_r = 0.5~\mathrm{ns}$, $\tau_d = 40~\mathrm{ns}$, and the experimental bandwidth $B = 16~\mathrm{GHz}$, this worst-case diagnostic gives $\Delta V_{\min}^{\mathrm{bw}} \approx 62~\mathrm{mV}$, which is of the same order as $\Delta V_{\min}^{\mathrm{cov}} \approx 50~\mathrm{mV}$ for $N = 8$.  However, the $62~\mathrm{mV}$ figure is an upper-envelope estimate built from the singular-limit slope as $x \to 0^+$; over most of the admissible rising-edge threshold range, the actual local slew is smaller and the corresponding local decorrelation spacing is therefore also smaller.  We therefore treat Eq.~\eqref{eq:autonum:R5} as a conservative physical diagnostic rather than a hard spacing constraint; $\Delta V_{\min}^{\mathrm{cov}}$ is still used as the explicit coverage/conditioning constraint, while any residual correlation between closely spaced rising-edge crossings is regarded as part of the practical model mismatch discussed below.  If one instead wishes to work with a smaller empirically determined \emph{effective} noise bandwidth, that alternative $B$ should be stated explicitly.

\textbf{Step~4: Coordinate-descent optimisation.}
Starting from the uniform initialisation $V_n^{(0)} = b + n(V_{\mathrm{peak,min}} - b)/(N+1)$, clipped to $[b + \Delta V_{\min}^{\mathrm{cov}},\; V_{\mathrm{peak,min}} - \Delta V_{\min}^{\mathrm{cov}}]$, and recording the best solution $\mathbf{V}^{\mathrm{best}} \leftarrow \mathbf{V}^{(0)}$, we iterate:
\begin{enumerate}
  \item \emph{Coordinate sweep.}  For $n = 1,\ldots,N$ sequentially: the search bounds use the already-updated lower neighbour $L_n = \max(b + \Delta V_{\min}^{\mathrm{cov}},\; V_{n-1}^{\mathrm{new}} + \Delta V_{\min}^{\mathrm{cov}})$ and the not-yet-updated upper neighbour $U_n = \min(V_{\mathrm{peak,min}} - \Delta V_{\min}^{\mathrm{cov}},\; V_{n+1}^{\mathrm{old}} - \Delta V_{\min}^{\mathrm{cov}})$.  Within these bounds, the objective function is evaluated by substituting $V_n$ into the \emph{current outer-iteration} vector $\mathbf{V}^{\mathrm{old}}$ (i.e.\ only $V_n$ varies; the remaining entries retain their pre-sweep values):
        \begin{equation}
          V_n^{*} = \arg\max_{V_n \in [L_n,\, U_n]} \log\det \widetilde{\mathbf{I}}\!\bigl(\bm{\theta}_{\mathrm{design}};\,\mathbf{V}^{\mathrm{old}}|_{V_n}\bigr).
          \label{eq:coord_step}
        \end{equation}
        Each scalar sub-problem is solved by Brent's bounded method (SciPy \texttt{minimize\_scalar}, \texttt{method='bounded'}) with absolute tolerance $\texttt{xatol} = 10^{-6}~\mathrm{V}$.  If the solver fails or returns a non-finite value, the threshold retains its previous position.
  \item \emph{Damped update.}  $\mathbf{V}^{\mathrm{new}} \leftarrow \mathbf{V}^{\mathrm{old}} + \alpha\,(\mathbf{V}^{*} - \mathbf{V}^{\mathrm{old}})$ with damping factor $\alpha = 0.5$.
  \item \emph{Projection.}  Sort $\mathbf{V}^{\mathrm{new}}$ in ascending order, clip each element to $(b + \Delta V_{\min}^{\mathrm{cov}},\; V_{\mathrm{peak,min}} - \Delta V_{\min}^{\mathrm{cov}})$, and then sweep upward: if $V_{n}^{\mathrm{new}} - V_{n-1}^{\mathrm{new}} < \Delta V_{\min}^{\mathrm{cov}}$, set $V_{n}^{\mathrm{new}} \leftarrow V_{n-1}^{\mathrm{new}} + \Delta V_{\min}^{\mathrm{cov}}$.
  \item \emph{Best-iterate tracking.}  If $\log\det\widetilde{\mathbf{I}}(\bm{\theta}_{\mathrm{design}};\,\mathbf{V}^{\mathrm{new}}) > \log\det\widetilde{\mathbf{I}}(\bm{\theta}_{\mathrm{design}};\,\mathbf{V}^{\mathrm{best}})$, update $\mathbf{V}^{\mathrm{best}} \leftarrow \mathbf{V}^{\mathrm{new}}$.
  \item \emph{Termination.}  Stop when $\max_n |V_n^{\mathrm{new}} - V_n^{\mathrm{old}}| < 0.1~\mathrm{mV}$, or after a maximum of 80 outer iterations.
\end{enumerate}
The final output is $\mathbf{V}^{\mathrm{best}}$, the iterate with the highest $\log\det\widetilde{\mathbf{I}}$ observed during the entire run.

\emph{Remark (Local optimality).}
Because $\log\det\widetilde{\mathbf{I}}(\bm{\theta};\,\mathbf{V})$ is generally non-concave in $\mathbf{V}$ for the bi-exponential model, coordinate descent guarantees convergence only to a \emph{local} maximum.  Global optimality would require exhaustive search or a branch-and-bound procedure, which is computationally prohibitive for $N = 8$ continuous variables.  In practice, we mitigate this by (i)~initializing from a uniform grid that already covers the full dynamic range, (ii)~applying damped updates ($\alpha = 0.5$) to avoid oscillations, and (iii)~tracking the best iterate across the entire run.  Repeated runs with perturbed initializations (uniform $\pm 10\%$ random jitter) converge to the same threshold set within $<1~\mathrm{mV}$, suggesting that the basin of attraction is broad and the reported solution is at least a strong local optimum.

\subsubsection{Results and Comparison}

\textbf{Optimised thresholds.}
The algorithm converges in $\sim$30 iterations.  The resulting threshold set is
\begin{equation}
  \mathbf{V}^{*} = \{54,\; 105,\; 213,\; 264,\; 315,\; 664,\; 714,\; 765\}~\mathrm{mV}.
  \label{eq:dopt_thresholds}
\end{equation}
This exhibits a characteristic two-cluster pattern: five thresholds in the rising-edge region (54--315~mV), where the waveform slope carries timing and shape information, and three near the peak (664--765~mV), where the amplitude sensitivity $\partial V/\partial A = h(x_c)$ is largest.  Compared with the uniform layout, the D-optimal design deliberately leaves the low-information mid-range (315--664~mV) unsampled, reallocating those measurement resources to the high-information regions.

\emph{Remark (Amplitude scalability).}
An important practical question is whether the D-optimal threshold set rescales with pulse amplitude: if $A$ doubles, can one simply double the thresholds?  To analyze this, define normalized thresholds $\eta_n \triangleq (V_n - b)/A$.  The crossing positions $x_{n,e}$ depend only on $(\eta_n, \tau_d, \tau_r)$, so the sensitivity gradients $\mathbf{g}_{n,e}$ (Eq.~\eqref{eq:biexp_sensitivity}) scale predictably with $A$.  However, the noise denominator (Eq.~\eqref{eq:noise_weight})
$D_{n,e} = \sigma_{\mathrm{th}}^2 + \kappa A\eta_n + \sigma_{\mathrm{TDC}}^2\,[A h'(x_c)]^2$
is \emph{not} homogeneous in $A$: it mixes terms of order $O(1)$, $O(A)$, and $O(A^2)$.  Because the Fisher matrix depends on $\mathbf{g}\mathbf{g}^T/D$, the optimal $\{\eta_n^*\}$ shift as the relative balance of thermal, Poisson, and TDC noise changes with amplitude.

In the Poisson-dominated regime ($\kappa A\eta \gg \sigma_{\mathrm{th}}^2$ and the TDC term is subdominant), $D_{n,e} \approx \kappa A\eta_n$ and the ratio $\mathbf{g}\mathbf{g}^T/D$ becomes independent of $A$ in the normalized coordinate $\eta$, yielding an approximate affine scaling law $V_n^*(A') \approx b + (A'/A)(V_n^* - b)$.  For the present dataset ($A_{\mathrm{design}} \approx 0.85~\mathrm{V}$, $\sigma_{\mathrm{th}} = 9.51~\mathrm{mV}$, $\kappa = 1~\mathrm{mV}$), the Poisson term $\kappa A\eta$ ranges from $\sim$50~mV$^2$ (lowest threshold) to $\sim$650~mV$^2$ (highest), while $\sigma_{\mathrm{th}}^2 \approx 90~\mathrm{mV}^2$.  Thus the low-threshold region is only marginally Poisson-dominated, and exact amplitude rescaling is not justified---the optimizer should be re-run at each new design-point amplitude, using the coordinate-descent procedure of Step~4.

\begin{figure}[!t]
  \centering
  \includegraphics[width=\columnwidth]{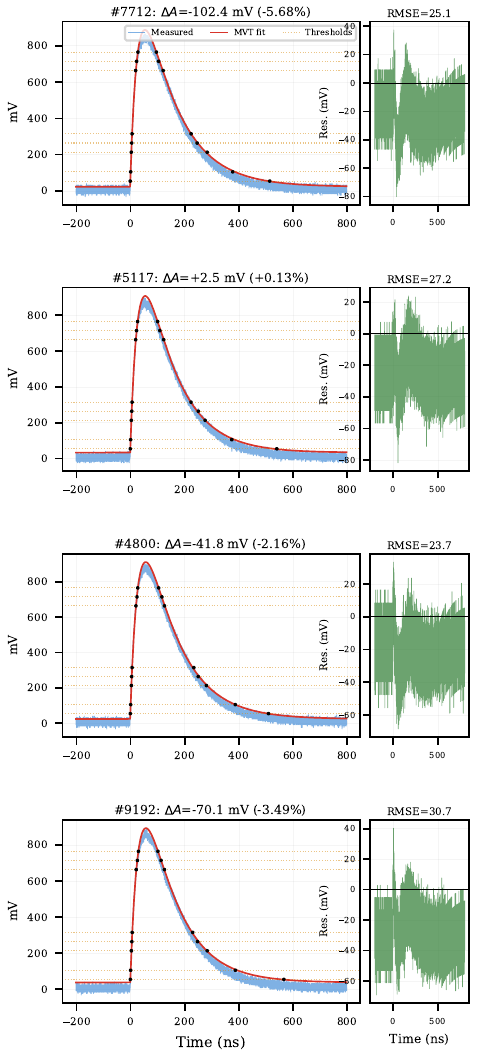}
  \caption{Four representative photopeak pulses reconstructed with the $N=8$ D-optimal thresholds.  Same format as Fig.~\ref{fig:uniform_mvt_waveforms}.  The non-uniform threshold spacing, denser near the rising edge and the peak, yields markedly smaller amplitude deviations $\Delta A$.}
  \label{fig:doptimal_waveforms}
\end{figure}

\textbf{Representative waveforms.}
Figure~\ref{fig:doptimal_waveforms} shows four photopeak pulses reconstructed with the D-optimal thresholds.  The simulated MVT timestamps, fits, and residuals demonstrate visibly tighter reconstruction than the uniform case.

\begin{figure}[!t]
  \centering
  \includegraphics[width=\columnwidth]{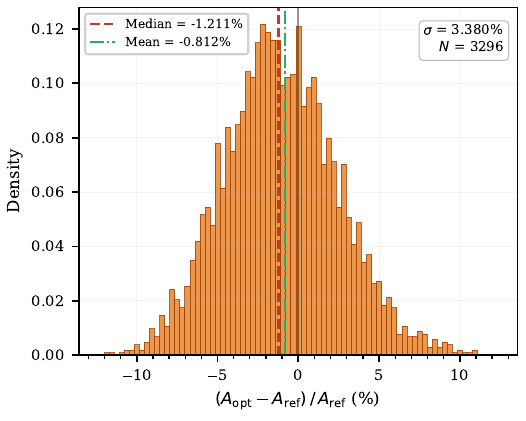}
  \caption{Distribution of the relative amplitude error for the D-optimal ($N=8$) threshold reconstruction (cf.~Fig.~\ref{fig:uniform_mvt_dA} for the uniform layout).  The median bias improves from $-4.8\%$ to $-1.2\%$ and the fraction within $\pm 5\%$ increases from 51\% to 82\%, confirming that Fisher-information-guided threshold placement substantially improves amplitude estimation.}
  \label{fig:doptimal_dA}
\end{figure}

\textbf{Amplitude reconstruction accuracy.}
Table~\ref{tab:uniform_vs_dopt} compares the two threshold designs.  The D-optimal thresholds reduce the median amplitude bias from $-4.8\%$ (uniform) to $-1.2\%$ and increase the fraction of pulses satisfying $|\Delta A/A_{\mathrm{ref}}| < 5\%$ from 51.3\% to 82.0\%.  Figure~\ref{fig:doptimal_dA} shows the full error distribution, which is substantially narrower and more symmetric than the uniform case (Fig.~\ref{fig:uniform_mvt_dA}), confirming that Fisher-information-guided threshold placement yields a measurable improvement in amplitude estimation even under realistic model-mismatch conditions.

\begin{table}[!t]
  \centering
  \caption{Amplitude reconstruction accuracy: uniform vs.\ D-optimal thresholds ($N=8$).}
  \label{tab:uniform_vs_dopt}
  \begin{tabular}{lcc}
    \hline\hline
    Metric                                & Uniform          & D-optimal        \\
    \hline
    Successful reconstructions            & 3{,}300 (99.9\%) & 3{,}296 (99.8\%) \\
    Median $\Delta A / A_{\mathrm{ref}}$  & $-4.8\%$         & $-1.2\%$         \\
    Mean $\Delta A / A_{\mathrm{ref}}$    & $-3.6\%$         & $-0.8\%$         \\
    $\sigma(\Delta A / A_{\mathrm{ref}})$ & $3.7\%$          & $3.4\%$          \\
    $|\Delta A / A_{\mathrm{ref}}| < 1\%$ & $9.6\%$          & $20.7\%$         \\
    $|\Delta A / A_{\mathrm{ref}}| < 5\%$ & $51.3\%$         & $82.0\%$         \\
    \hline\hline
  \end{tabular}
\end{table}

% =========================================================================
%  XI. EXPERIMENTAL VALIDATION FOR MULTI-EVENT DESIGN
% =========================================================================
\section{Experimental Validation for Multi-Event Design}
\label{sec:exp_validation_multi}

The preceding section validated the single-event threshold design on a narrow photopeak population ($V_{\mathrm{peak}} \in [812,\; 993]~\mathrm{mV}$), where the accepted peak voltages---and therefore the fitted scale amplitudes---cluster tightly around the 511~keV design point.  In realistic PET detector operation, however, a substantial fraction of recorded events arise from inter-crystal scattering (ICS) or Compton interactions in which only a portion of the incident gamma-ray energy is deposited in a given crystal~\cite{fu2016recovery,lee2020recovery}.  Recovering these low-energy events is essential for improving spatial resolution and sensitivity in high-resolution detector arrays, yet their reduced pulse height means that the upper hardware thresholds may not be triggered---precisely the partial-triggering regime addressed by the multi-event theory of Sections~\ref{sec:partial_trigger} and~\ref{sec:biexp_partial}.  This section therefore extends the experimental validation to a broad energy window that spans the Compton continuum and the photopeak, testing whether the D-optimal threshold design derived from a photopeak-centred Fisher analysis retains its advantage when applied to a heterogeneous event population with widely varying amplitudes and active threshold sets.

\subsection{Extended Energy Window Selection}
\label{sec:multi_energy_window}

The photopeak selection of Section~\ref{sec:energy_spectrum} restricted attention to a narrow $\pm 10\%$ band around the 511~keV peak ($V_{\mathrm{peak}} \in [812,\; 993]~\mathrm{mV}$), yielding $N_{511} = 3{,}304$ events whose peak voltages cluster tightly around the design point.  To evaluate reconstruction performance across the broader energy range encountered in practice, we now extend the pulse-height acceptance window to include the Compton continuum.

\textbf{Energy windowing.}
We define the extended energy window as $[200,\; 993]~\mathrm{mV}$:
\begin{equation}
  V_{\mathrm{peak}} \;\in\; [200,\; 993]~\mathrm{mV}.
  \label{eq:multi_energy_window}
\end{equation}
The lower bound $V_{\mathrm{peak,lo}} = 200~\mathrm{mV}$ is chosen to retain events whose deposited energy is substantially below the photopeak while excluding the lowest-amplitude triggers that are dominated by electronic noise.  The upper bound $V_{\mathrm{peak,hi}} = 993~\mathrm{mV}$ is identical to the photopeak ceiling of Section~\ref{sec:energy_spectrum}.

Using the same model-free peak-voltage extraction procedure (Eq.~\eqref{eq:vpeak_def}), we select all pulses whose maximum instantaneous voltage falls within this window.  Of the 10{,}000 recorded pulses, $N_{\mathrm{ext}} = 8{,}085$ events (80.8\%) satisfy the criterion.  This population spans a continuous range of deposited energies---from low-energy Compton-scattered events near $200~\mathrm{mV}$ through the Compton edge and up to the 511~keV photopeak---and therefore exhibits substantial amplitude heterogeneity: the ratio of the highest to the lowest accepted peak voltage exceeds $4{:}1$.

\begin{figure}[!t]
  \centering
  \includegraphics[width=\columnwidth]{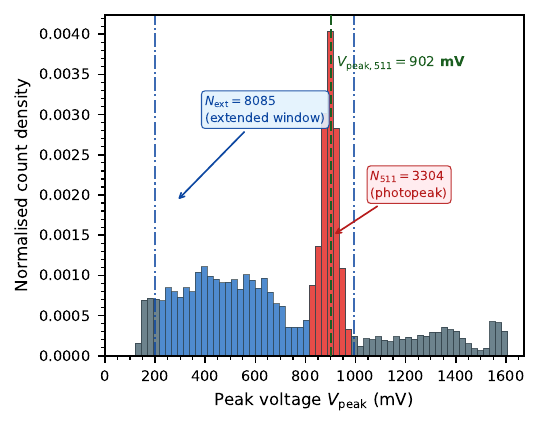}
  \caption{Normalised peak-voltage histogram of 10{,}000 $^{22}$Na scintillation pulses (same dataset as Fig.~\ref{fig:energy_spectrum}).  The blue-shaded region marks the extended energy window $[200,\; 993]~\mathrm{mV}$ used for multi-event validation, selecting $N_{\mathrm{ext}} = 8{,}085$ events (80.8\%).  The red sub-region identifies the $\pm 10\%$ photopeak window ($N_{511} = 3{,}304$).  The extended population spans the Compton continuum and the photopeak, with a peak-voltage ratio exceeding $4{:}1$.}
  \label{fig:energy_spectrum_multi}
\end{figure}

Figure~\ref{fig:energy_spectrum_multi} displays the extended energy window overlaid on the $V_{\mathrm{peak}}$ histogram.  The blue-shaded region marks the full $[200,\; 993]~\mathrm{mV}$ acceptance band, while the red-shaded sub-region identifies the photopeak events ($N_{511} = 3{,}304$) that served as the design-point population in the preceding section.

\subsection{Uniform-Threshold Baseline}
\label{sec:uniform_mvt_multi}

Having identified the $N_{\mathrm{ext}} = 8{,}085$ extended-window pulses, we now evaluate the amplitude-reconstruction accuracy of the same $N = 8$ uniformly spaced MVT digitiser used in the single-event analysis, applied to this broader population.

\textbf{Threshold placement.}
The threshold set is identical to Section~\ref{sec:uniform_mvt}: the $N = 8$ thresholds are placed at equally spaced voltages in $(b,\; V_{\mathrm{peak,min}})$ (Eq.~\eqref{eq:uniform_thresholds}), giving
\begin{equation*}
  \begin{gathered}
    \{93.9,\; 184.1,\; 274.4,\; 364.7,\; \\
    454.9,\; 545.2,\; 635.5,\; 725.7\}~\mathrm{mV}.
  \end{gathered}
\end{equation*}
This set was designed for the photopeak population, where all events trigger all 8~thresholds.  In the extended window, however, many events have $V_{\mathrm{peak}} < 725.7~\mathrm{mV}$ and therefore activate only a subset of the threshold ladder.

\textbf{Active threshold assignment.}
For each pulse in the extended population, we determine the active threshold set $\mathcal{A} = \{n : V_n < V_{\mathrm{peak},i}\}$, where $V_{\mathrm{peak},i}$ is the measured model-free peak-voltage proxy of Eq.~\eqref{eq:vpeak_def}.  In the formal theory of Eqs.~\eqref{eq:active_set} and \eqref{eq:biexp_active_set}, the active set is defined by the unknown model peak $p(\bm{\theta}_i)=A^{(i)}h_{\mathrm{peak}}^{(i)}+b^{(i)}$; in the experimental pipeline we therefore use the observed $V_{\mathrm{peak},i}$ as its practical surrogate when deciding which thresholds were actually crossed by the recorded waveform.  Each active threshold contributes one rising-edge timestamp and one falling-edge timestamp selected by Eq.~\eqref{eq:offline_falling_timestamp_rule}; inactive thresholds are simply absent from the observation vector.  Consistent with Section~\ref{sec:biexp_partial}, the formal five-parameter amplitude-target regime begins at $|\mathcal{A}| \ge 3$ (i.e., at least 6~crossing timestamps); events below that threshold fall outside the finite-amplitude-bound theory and are not used in the reported active-count statistics.

\emph{Assumption} (Measured peak as active-set surrogate): The sampled peak voltage is assumed to classify active thresholds reliably, i.e., if $V_n < V_{\mathrm{peak},i}$ then the recorded waveform contains a physically meaningful crossing pair for that threshold.  This is reasonable when thresholds are separated from the peak by more than the local voltage noise and when the oscilloscope bandwidth is high enough to resolve the maximum.  Events with thresholds nearly tangent to the noisy peak may be misclassified or yield ill-conditioned pair gaps; these cases contribute to the practical reconstruction failures and are part of the reported end-to-end performance.

\textbf{MVT digitisation and reconstruction.}
For each pulse, we simulate MVT digitisation using only the active thresholds, again following Section~\ref{sec:exp_setup}: all interpolated threshold crossings of the piecewise-linear waveform are scanned, the earliest rising timestamp is retained, and the falling timestamp is selected by Eq.~\eqref{eq:offline_falling_timestamp_rule}.  The resulting $2|\mathcal{A}|$ simulated MVT timestamps are then fed into the Gauss--Newton solver (Section~\ref{sec:gauss_newton}) with the full-waveform fit as the initial estimate.  Of the 8{,}085 extended-window pulses, 7{,}342 (90.8\%) yield a successful MVT reconstruction.  The lower success rate compared with the photopeak population (99.9\%, Section~\ref{sec:uniform_mvt}) reflects the reduced number of crossing timestamps available for low-amplitude events, which makes the Gauss--Newton normal matrix more ill-conditioned.

\textbf{Active threshold statistics.}
Table~\ref{tab:active_thr_dist_uniform} reports the distribution of the active threshold count $K_{\mathrm{act}} = |\mathcal{A}|$ over the 7{,}342 successfully reconstructed pulses.  Approximately half (49.3\%) of the events trigger all 8~thresholds (these are predominantly photopeak events), while the remaining 50.7\% operate in the partial-triggering regime with $K_{\mathrm{act}} \in \{3,\ldots,7\}$.

\begin{table}[!t]
  \centering
  \caption{Active threshold distribution for the uniform-threshold MVT reconstruction over the extended energy window ($N_{\mathrm{ext}} = 8{,}085$ pulses, 7{,}342 successful).}
  \label{tab:active_thr_dist_uniform}
  \begin{tabular}{c c c}
    \hline\hline
    $K_{\mathrm{act}}$ & Count   & Fraction \\
    \hline
    3                  & 495     & 6.7\%    \\
    4                  & 897     & 12.2\%   \\
    5                  & 870     & 11.8\%   \\
    6                  & 803     & 10.9\%   \\
    7                  & 657     & 8.9\%    \\
    8                  & 3{,}620 & 49.3\%   \\
    \hline
    Total              & 7{,}342 & 100\%    \\
    \hline\hline
  \end{tabular}
\end{table}

\textbf{Reference standard.}
The reference amplitude $A_{\mathrm{ref}}$ for each pulse is obtained by fitting the same five-parameter bi-exponential model (Eq.~\eqref{eq:biexp}) to the full waveform (all 50{,}000 sampling points) via nonlinear least-squares, identical to Section~\ref{sec:uniform_mvt}.

\begin{figure}[!t]
  \centering
  \includegraphics[width=\columnwidth]{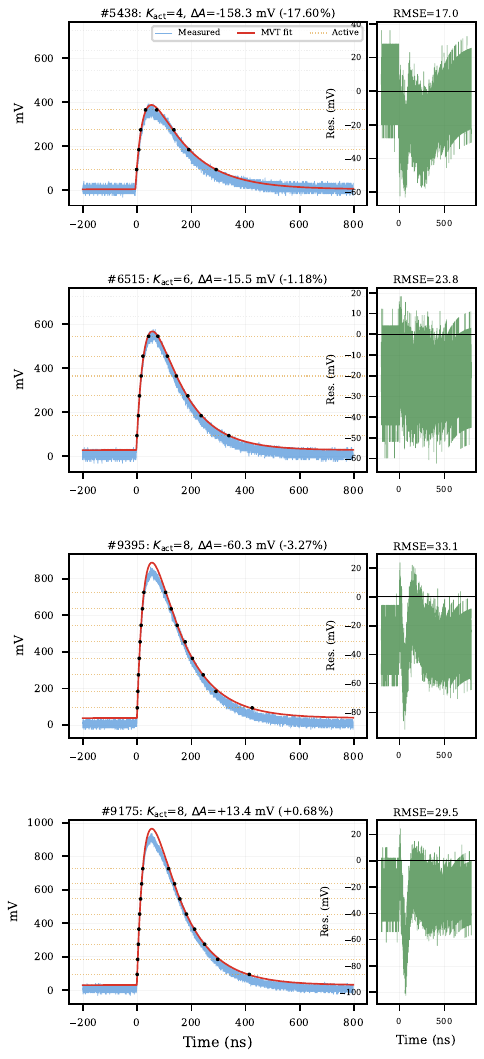}
  \caption{Four representative pulses from the extended energy window $[200,\; 993]~\mathrm{mV}$, reconstructed with $N = 8$ uniform thresholds.  Left: measured waveform (blue), MVT-reconstructed fit (red), active threshold levels (orange dashed), inactive thresholds (grey), and simulated MVT timestamps (black dots).  Right: fit residual.  The number of active thresholds $K_{\mathrm{act}}$ and the per-pulse amplitude deviation $\Delta A$ are indicated in each panel title.  Low-energy events with fewer active thresholds show larger amplitude errors, reflecting the reduced information content of partial triggering.}
  \label{fig:uniform_mvt_multi_waveforms}
\end{figure}

\textbf{Representative waveforms.}
Figure~\ref{fig:uniform_mvt_multi_waveforms} displays four representative pulses spanning the extended amplitude range, from a low-energy Compton event ($K_{\mathrm{act}} < 8$) to a photopeak event ($K_{\mathrm{act}} = 8$).  Each panel shows the recorded waveform, the active and inactive threshold levels, the simulated MVT timestamps (black dots), the MVT-reconstructed fit (red curve), and the fit residual.

\begin{figure}[!t]
  \centering
  \includegraphics[width=\columnwidth]{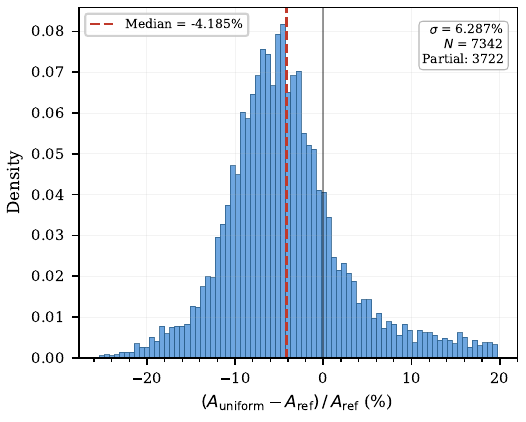}
  \caption{Distribution of the relative amplitude error $(A_{\mathrm{uniform}} - A_{\mathrm{ref}})/A_{\mathrm{ref}}$ for the $N = 8$ uniform-threshold MVT reconstruction over 7{,}342 extended-window pulses.  The dashed red line marks the median ($-4.2\%$).  The mean ($+20.6\%$) is dominated by a small fraction of partial-triggering events with grossly incorrect reconstructions and is therefore not displayed.  The wider spread compared with the photopeak-only distribution (Fig.~\ref{fig:uniform_mvt_dA}) is due to partial-triggering events with $K_{\mathrm{act}} < 8$.}
  \label{fig:uniform_mvt_multi_dA}
\end{figure}

\textbf{Amplitude reconstruction accuracy.}
Figure~\ref{fig:uniform_mvt_multi_dA} presents the distribution of the relative amplitude error $\Delta A / A_{\mathrm{ref}} = (A_{\mathrm{uniform}} - A_{\mathrm{ref}}) / A_{\mathrm{ref}}$ over the 7{,}342 successfully reconstructed pulses.  The key statistics are:
\begin{itemize}
  \item Median bias: $-4.2\%$ (systematic underestimation);
  \item Outlier-excluded standard deviation: $6.3\%$;
  \item 36.7\% of pulses satisfy $|\Delta A / A_{\mathrm{ref}}| < 5\%$.
\end{itemize}
Compared with the photopeak-only results of Section~\ref{sec:uniform_mvt} (median bias $-4.8\%$, $\sigma = 3.7\%$, 51.3\% within $\pm 5\%$), the extended-window population shows a comparable median bias but substantially wider spread and lower pass rate.  The degradation is driven by the low-energy events that trigger only 3--5 thresholds: these partial-triggering events have fewer crossing timestamps, a less well-conditioned Jacobian $\mathbf{J}$, and therefore larger amplitude estimation variance---precisely as predicted by the effective information analysis of Section~\ref{sec:partial_trigger}.  The outlier-excluded standard deviation increases from 3.7\% to 6.3\%, and the fraction of reconstruction outliers (events excluded by the $1.5 \times \mathrm{IQR}$ fence rule of Section~\ref{sec:uniform_mvt}) rises to 964 of 7{,}342 events (13.1\%), compared with 1.2\% in the photopeak-only case.  As before, median, mean, and pass-rate statistics are computed on the full population without any exclusion; only the reported standard deviation uses the outlier-excluded subset.

\subsection{D-Optimal Threshold Design}
\label{sec:doptimal_mvt_multi}

We now apply the D-optimal framework of Section~\ref{sec:doptimal_mvt} to the extended energy window, keeping the Section~\ref{sec:exp_setup} piecewise-linear earliest-rise / latest-fall MVT digitisation, the partial-triggering logic, the Gauss--Newton solver, the outlier diagnostics, and the reference standard identical to the uniform-threshold baseline of Section~\ref{sec:uniform_mvt_multi}.

\textbf{Multi-event D-optimal criterion.}
The single-event D-optimal design of Section~\ref{sec:doptimal_mvt} maximises $\log\det\widetilde{\mathbf{I}}(\bm{\theta}_{\mathrm{design}};\,\mathbf{V})$ at a single photopeak design point.  For the extended energy window, where the amplitude population spans a $4{:}1$ peak-voltage ratio, we extend this to a spectrally weighted objective.  The energy window is discretised into $P = 10$ representative amplitude bins at the 5th through 95th percentiles of the fitted amplitude distribution.  In this validation subsection we deliberately assign equal bin weights $w_p = 1/P$ to isolate the geometric effect of threshold redistribution across the amplitude range; this is a controlled study choice rather than the deployment prescription of Section~\ref{sec:recipe_multi_inputs}, where $w_p$ should reflect the operational energy distribution.  For each event $p$, the Fisher information matrix is assembled from the active thresholds $\mathcal{A}_p$ only (Eq.~\eqref{eq:autonum:ME2}), normalized by the same fixed parameter-scale convention, and the threshold set is obtained by maximising
\begin{equation}
  \mathbf{V}_{\mathrm{multi\text{-}D}}^{*} = \arg\max_{\mathbf{V}} \;\sum_{p=1}^{P} w_p\,\log\det \widetilde{\mathbf{I}}^{\mathrm{act},(p)}(\bm{\theta}^{(p)};\,\mathbf{V}),
  \label{eq:multi_doptimal}
\end{equation}
subject to the coverage constraint $\Delta V_{\min}^{\mathrm{cov}} = (V_{\mathrm{peak,max}} - V_{\mathrm{floor}}) / (2N)$, where $V_{\mathrm{peak,max}}$ is the upper admitted peak voltage in the extended window and $V_{\mathrm{floor}}$ is the lower feasible threshold floor used by the solver (cf.\ Eq.~\eqref{eq:dv_min_coverage}).  This formulation is the direct multi-event extension of the single-event D-optimal criterion: it maximises joint parameter identifiability (via $\det\widetilde{\mathbf{I}}$) for each event class while balancing the threshold budget across the amplitude range through the chosen bin weights $w_p$.

\emph{Assumption} (Representative-bin population surrogate): The percentile bins and equal weights are used as a controlled surrogate for the extended event population, not as an operational spectrum model.  This choice is reasonable for isolating how D-optimal threshold geometry behaves across a wide amplitude range.  For deployment, the bins and weights should be replaced by measured or simulated event rates, detection priorities, and any application-specific utility assigned to Compton or inter-crystal-scatter recovery.

\textbf{Threshold configuration.}
The coordinate-descent solver converges in 25~iterations to
\begin{equation*}
  \begin{aligned}
    \mathbf{V}_{\mathrm{multi\text{-}D}}
      = &[76.6,\; 119.9,\; 230.4,\; 333.6,\; 425.4, \\
        &530.7,\; 609.3,\; 682.5]~\mathrm{mV}.
  \end{aligned}
\end{equation*}
In contrast to the single-event D-optimal set (Eq.~\eqref{eq:dopt_thresholds}), which exhibits a distinctive two-cluster pattern (five thresholds $\le 315~\mathrm{mV}$, three $\ge 664~\mathrm{mV}$), the multi-event D-optimal thresholds redistribute more of the threshold budget into the low- and mid-amplitude region.  This reflects the influence of the low-energy events in the objective: the equal-bin weighting across the amplitude range lowers the first two thresholds and pulls the upper cluster downward, increasing the number of events with feasible active sets while reducing the sampling density near the photopeak peak-region.

\textbf{Active threshold statistics.}
Among the $N_{\mathrm{ext}} = 8{,}085$ extended-window pulses, 7{,}665 (94.8\%) are successfully reconstructed.  The active threshold counts are: $K_{\mathrm{act}} = 8$ for 3{,}871 events (50.5\%), with partial-triggering events distributed across $K_{\mathrm{act}} = 3$--$7$ (49.5\%).

\begin{figure}[!t]
  \centering
  \includegraphics[width=\columnwidth]{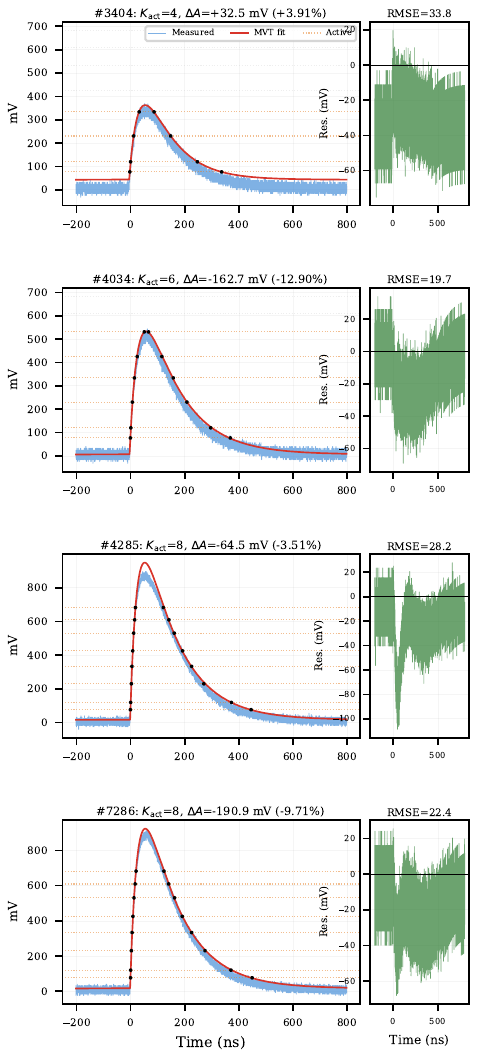}
  \caption{Four representative pulses from the extended energy window $[200,\; 993]~\mathrm{mV}$, reconstructed with the $N = 8$ multi-event D-optimal thresholds (Eq.~\eqref{eq:multi_doptimal}).  Same format as Fig.~\ref{fig:uniform_mvt_multi_waveforms}.  The redistributed low- and mid-amplitude threshold spacing contrasts with the two-cluster pattern of the single-event D-optimal design (Fig.~\ref{fig:doptimal_waveforms}).}
  \label{fig:doptimal_mvt_multi_waveforms}
\end{figure}

\textbf{Representative waveforms.}
Figure~\ref{fig:doptimal_mvt_multi_waveforms} shows four exemplary reconstructions spanning the amplitude range.

\textbf{Amplitude reconstruction accuracy.}
Table~\ref{tab:multi_uniform_vs_dopt} compares the uniform and D-optimal threshold designs on the extended-window population.

\begin{table}[!t]
  \centering
  \caption{Amplitude reconstruction accuracy on the extended energy window: uniform vs.\ multi-event D-optimal thresholds ($N=8$).  The $1.5 \times \mathrm{IQR}$ fence rule of Section~\ref{sec:uniform_mvt} is applied identically to both configurations.}
  \label{tab:multi_uniform_vs_dopt}
  \begin{tabular}{lcc}
    \hline\hline
    Metric                                  & Uniform          & D-optimal        \\
    \hline
    Successful reconstructions              & 7{,}342 (90.8\%) & 7{,}665 (94.8\%) \\
    Median $\Delta A / A_{\mathrm{ref}}$    & $-4.2\%$         & $-9.8\%$         \\
    $\sigma({\Delta A / A_{\mathrm{ref}}})$ & $6.3\%$          & $6.7\%$          \\
    $|\Delta A / A_{\mathrm{ref}}| < 5\%$   & $36.7\%$         & $17.9\%$         \\
    \hline\hline
  \end{tabular}
\end{table}

The multi-event D-optimal design increases the reconstruction success rate (94.8\% vs.\ 90.8\%), consistent with its lower first thresholds and broader active-set coverage.  However, the median amplitude bias worsens from $-4.2\%$ to $-9.8\%$, the inlier spread is slightly larger ($\sigma = 6.7\%$ vs.\ $6.3\%$), and the $\pm 5\%$ pass rate decreases substantially.  This outcome contrasts sharply with the single-event D-optimal result (Table~\ref{tab:uniform_vs_dopt}), where the same criterion improved all reported amplitude metrics simultaneously.  Two structural factors explain why the D-optimal criterion succeeds in the single-event regime but not in the multi-event amplitude-reconstruction regime:
\begin{enumerate}
  \item \emph{Thresholds-per-event bottleneck}: with only $N = 8$ thresholds spanning a $4{:}1$ amplitude ratio, each low-energy event triggers only $K_{\mathrm{act}} = 3$--$5$ thresholds.  For the five-parameter model, this yields at most $2K_{\mathrm{act}} = 6$--$10$ crossing observations to estimate 5 unknowns---marginal measurement redundancy.  In the single-event case, $K_{\mathrm{act}} = N = 8$ for every pulse, providing 16 observations for 5 parameters and ample redundancy for the D-optimal criterion to leverage.
  \item \emph{Objective--estimator mismatch}: the multi-event D-optimal objective maximises joint parameter identifiability for representative bins, not the empirical amplitude error after nonlinear reconstruction under waveform mismatch.  Lowering the first thresholds improves convergence and active-set feasibility, but the redistributed threshold geometry still leaves the amplitude coordinate sensitive to model residuals and nuisance-parameter coupling.
\end{enumerate}

These findings suggest that with a fixed budget of $N = 8$ thresholds, per-event coverage and threshold positioning interact in a non-trivial way.  The multi-event D-optimal layout improves feasibility and convergence, but the uniform layout remains more robust in amplitude accuracy for this dataset.  Section~\ref{sec:discuss_multi_gap} provides a detailed analysis showing that these structural factors are ultimately manifestations of a single underlying cause: model misspecification amplified by the D-optimal threshold redistribution.

\begin{figure}[!t]
  \centering
  \includegraphics[width=\columnwidth]{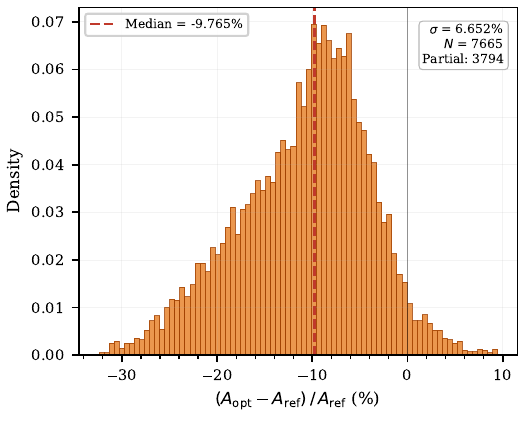}
  \caption{Distribution of the relative amplitude error $(A_{\mathrm{opt}} - A_{\mathrm{ref}})/A_{\mathrm{ref}}$ for the $N = 8$ multi-event D-optimal MVT reconstruction over 7{,}665 extended-window pulses (cf.~Fig.~\ref{fig:uniform_mvt_multi_dA} for the uniform layout).  The dashed red line marks the median ($-9.8\%$).  The larger systematic bias and lower $\pm 5\%$ pass rate confirm that the D-optimal criterion does not translate into amplitude-reconstruction improvement when partial triggering and model mismatch limit the per-event measurement redundancy.}
  \label{fig:doptimal_mvt_multi_dA}
\end{figure}

Figure~\ref{fig:doptimal_mvt_multi_dA} presents the full error distribution.

  % =========================================================================
  %  XII. DISCUSSION
  % =========================================================================
  \section{Discussion}
  \label{sec:discussion}

  \subsection{Scope and Positioning of the Framework}

  The central contribution of this work is not a single threshold-placement rule, but the formalization of prior-based MVT as a structured inverse problem. Once the problem is stated in that form, identifiability, stochastic error propagation, model mismatch, and threshold optimization are no longer disconnected engineering issues; they become coupled aspects of the same paired-crossing geometry. That shift in viewpoint is the main conceptual contribution of the paper: MVT is treated not as an empirical threshold heuristic, but as a mathematically defined inference problem.

  At the general level, the theory is deliberately \emph{model-agnostic}. Any strictly unimodal pulse family satisfying the standing regularity conditions can be inserted into the same machinery for deterministic recovery, error analysis, and threshold design. The bi-exponential model of Section~\ref{sec:optimal} is therefore not the conceptual endpoint of the paper, but the first complete instantiation of the formalism.

  That instantiation closes the loop between formalization and experiment. It turns the general theory into explicit reconstruction formulas and executable design rules, and it makes the main empirical result interpretable: Fisher-guided threshold design succeeds in the photopeak regime, but loses predictive power over a broad energy window once partial triggering and model mismatch become dominant. In that sense, the experiments do more than benchmark a threshold rule; they reveal both the reach and the regime boundary of the theory.

  \subsection{Assumptions and Their Practical Implications}

  The closed-form results are deliberately explicit about their assumptions.  Those assumptions are not incidental technicalities; they mark the regime in which the theory should be read literally and the regime in which it should be read as a structured approximation.

  \paragraph{High-SNR Gaussian approximation.}
  The Poisson-to-Gaussian bridge invoked in Section~\ref{sec:noise} requires large effective photoelectron counts.  For photopeak events in LYSO this is a reasonable approximation, which is why the Gaussianized likelihood is useful here.  At sufficiently low signal levels, however, the crossing errors retain a genuinely non-Gaussian structure, and a Poisson-native inference theory would be more faithful than the high-SNR limit adopted in the manuscript.

  \paragraph{Single-photoelectron linearity and stationarity.}
  Section~\ref{sec:noise} represents the macroscopic voltage as a sum of shifted copies of a common single-photoelectron impulse response $h_e(t)$.  For direct-readout pulses over a moderate dynamic range, this is a sensible first-order model.  At high occupancy, however, SiPM saturation, recharge dynamics, afterpulsing, or front-end nonlinearity can violate it.  The present framework does not ignore those departures; it relocates them into the mismatch term $\varepsilon(t)$ and therefore makes them part of the boundary of the well-specified regime.

  \paragraph{Slowly varying intensity in the Campbell reduction.}
  The Campbell-theorem reduction assumes that the photon intensity $\nu(t)$ varies slowly across one $h_e$ window.  This is well satisfied on the pulse tail and remains a useful surrogate near the peak, but it is weakest on the fastest rising edge where the effective rise time is not asymptotically larger than the single-photoelectron response width.  The manuscript therefore uses Eq.~\eqref{eq:poisson_voltage} as a controlled first-order variance surrogate, not as an exact microscopic law.

  \paragraph{Diagonal covariance and bandwidth diagnostic.}
  Statistical independence across crossings is diagnosed through the same-branch spacing condition $|V_i - V_j| \ge L_{ij}\tau_c$ together with the within-pair separation condition $t_f(V_n)-t_r(V_n) \ge \tau_c$ when a single threshold pair is treated diagonally.  This is not a cosmetic regularizer; it is the point at which the information model meets the analog bandwidth of the readout chain.  If these conditions fail, the deterministic MVT observations remain valid, but the stochastic calculation should retain off-diagonal covariance terms or adopt an empirically calibrated spacing rule.

  \paragraph{Static threshold calibration.}
  The theory treats the hardware thresholds $\{V_k\}$ as deterministic, event-independent reference voltages.  For calibrated laboratory electronics over short acquisition intervals, this is a reasonable approximation.  If threshold hysteresis, offset drift, or comparator metastability becomes appreciable, however, the effective crossing condition is no longer governed by a single fixed ladder.  In that regime, the thresholds themselves should be promoted to nuisance parameters or controlled by tighter calibration.

  \paragraph{Pulse-model adequacy and misspecification.}
  The five-parameter bi-exponential pulse $f(t;\bm{\theta})$ captures the dominant scintillation morphology but is not exact.  Real detector waveforms contain secondary components, sensor nonlinearities, and baseline distortions, all of which enter the mismatch field $\varepsilon(t)$.  Section~\ref{sec:delta_validation} shows that this mismatch is not negligible in the amplitude-target geometry (median $\rho_{\mathrm{bias}}^{(A)} = 5.98 \times 10^{-1}$, 95th percentile $4.49$).  This observation also explains the experimental split of the paper: in the single-event regime, Fisher-guided threshold redistribution is still useful because it improves conditioning without reducing coverage; in the multi-event regime, the same mismatch becomes decisive because threshold redistribution both concentrates crossings in high-mismatch regions and reduces per-event redundancy.  A richer pulse model would therefore do more than tighten fit residuals; it would enlarge the regime in which Fisher-optimal designs are quantitatively reliable.

  \paragraph{Experimental surrogates.}
  The validation also relies on three practical surrogates: raw peak voltage is used as an energy and active-set proxy, the full-waveform bi-exponential fit is used as the reconstruction reference, and offline linear interpolation together with the earliest-rise/latest-fall rule is used to emulate an MVT digitizer.  These choices are appropriate for controlled comparison because they are applied identically to all threshold designs and because the oscilloscope sampling rate is much faster than the fitted pulse dynamics.  They nevertheless delimit interpretation: the experiments compare threshold layouts relative to a high-sample reference model and a specified timestamping convention, not against an independent calorimetric ground truth or a fully characterized comparator ASIC.

  \paragraph{Local design and population weighting.}
  The D-optimal thresholds are locally optimal with respect to the chosen design point or representative-bin population.  The photopeak design assumes that the median full-waveform fit represents the narrow 511~keV window, while the multi-event design uses percentile amplitude bins and deliberately equal weights to expose geometric effects across the extended energy range.  Those choices are useful for analysis and diagnosis, but a deployed design should replace them with the operational spectrum, channel calibration, and application-specific weights.

  \paragraph{Stable baseline and no pile-up.}
  Finally, the framework assumes a deterministic DC offset $b$ and one dominant pulse per acquisition window.  For isolated scintillation events at modest count rate, this is reasonable.  In practice, baseline wander, pile-up, or strong afterpulsing can add low-frequency structure or even produce additional local maxima.  The present formalism can absorb such effects only indirectly, through enlarged nuisance models, preprocessing, or a future multimodal extension of the observation geometry.

  \subsection{Interpretation of Threshold Geometry}

  Four geometric lessons are worth emphasizing.

  First, threshold utility is global rather than local.  Because the design objective is nuisance-profiled, the best thresholds are not simply those with the steepest slope or the lowest local noise.  They are the thresholds that shape the \emph{joint} information geometry most favorably after nuisance coupling has been accounted for.  A threshold that is individually weak can still be valuable if it rotates the Fisher geometry enough to decorrelate nuisance directions from the target parameter.

  Second, once partial triggering is admitted, ``optimal'' becomes population-dependent.  A threshold set that is optimal at one design point need not remain optimal over an amplitude range in which different events activate different subsets of the ladder.  The multi-event theory shows that threshold design is fundamentally an active-set design problem, not merely a single-waveform optimization problem.  The resulting spectral partition of the ladder---lower thresholds serving the full population, upper thresholds refining high-amplitude events---is therefore a structural consequence of the model, not an ad hoc engineering rule.

  Third, robustness is part of the design problem rather than a postscript to it.  The optimal thresholds are local solutions whose quality depends on parameter perturbations and model error.  A threshold set is therefore useful not only if it is stationary at a nominal point, but also if that stationarity is not excessively fragile under realistic uncertainty.  In practice, this means that any candidate ladder should be accompanied by a fragility audit based on the sensitivity vector $\mathbf{S}$ and the normalized robustness metric $\rho(\bm{\epsilon})$.

  Fourth, the experiments expose a genuine \emph{design--reconstruction gap}.  The single-parameter Schur-complement objective can drive thresholds toward a small number of support voltages, producing an appealing scalar CRLB while leaving the full Fisher matrix nearly singular.  In that regime the bound is mathematically small but operationally useless, because the Gauss--Newton solver cannot reliably recover all parameters.  Replacing the single-parameter objective with the D-optimal criterion $\max \log\det \mathbf{I}$ closes that gap in the single-event regime by forcing all Fisher eigenvalues to remain substantial.  The resulting improvement in conditioning is precisely what turns the information-theoretic design from a formal bound into a usable reconstruction strategy.

  \subsection{Single-Event vs.\ Multi-Event Optimization: The CRB--MSE Gap}
  \label{sec:discuss_multi_gap}

  The most informative experimental result of the manuscript is the asymmetry between single-event and multi-event D-optimal design.  On the photopeak population, D-optimal thresholds improve all reported amplitude metrics (Table~\ref{tab:uniform_vs_dopt}).  On the broad energy window, they do not (Table~\ref{tab:multi_uniform_vs_dopt}).  The issue is not that the derivation is mathematically incorrect; it is that the CRB-based design objective and the practical estimator stop targeting the same error once model mismatch and partial triggering become large.

  The first gap is between variance optimality and MSE optimality.  The multi-event criterion minimizes a Fisher-based variance proxy, but the experimentally dominant error over the broad energy window is systematic amplitude bias, not variance.  Once bias contributes at the same scale as the CRB floor, reducing variance alone is no longer enough; threshold moves that improve the Fisher objective can still worsen the practical MSE.

  The second gap is measurement redundancy.  In the single-event regime, all $N=8$ thresholds are active for every pulse, so the estimator works with 16 crossing timestamps to recover 5 parameters.  In the multi-event regime, low-energy events often trigger only $K_{\mathrm{act}} = 3$--$5$ thresholds, yielding only 6--10 timestamps.  That is a marginal observation budget for a five-parameter model, so the Gauss--Newton solver cannot be expected to approach the CRB uniformly even if the Fisher design itself is internally consistent.

  The third gap is the interaction between conditioning, active-set coverage, and bias projection.  Lowering the first thresholds can increase the number of feasible reconstructions, as observed in Table~\ref{tab:multi_uniform_vs_dopt}, but those thresholds are then no longer available to refine other portions of the waveform.  With a fixed threshold budget, the same ladder must serve low-energy Compton events, mid-range events, and photopeak events; a placement that improves joint Fisher conditioning or convergence for one part of the population can still project waveform mismatch unfavorably onto the amplitude coordinate elsewhere.

  These gaps have a single unifying explanation: model mismatch is already too large for the well-specified simplification to be read as quantitatively accurate over the full event population.  Section~\ref{sec:delta_validation} shows that the amplitude bias-to-variance ratio has median $\rho_{\mathrm{bias}}^{(A)} = 5.98 \times 10^{-1}$ and 95th percentile $4.49$.  The broad-window design problem therefore operates outside the regime $\rho_{\mathrm{bias}}^{(A)} \ll 1$ in which a Fisher-only objective is expected to predict practical MSE ordering.  Under these conditions, threshold placement directly controls how waveform mismatch is projected into amplitude bias.

  This is also why the uniform baseline remains competitive.  Uniform spacing spreads crossings across the waveform in a simple amplitude-agnostic way, which balances exposure to the mismatch profile $\varepsilon(t)$ even though it is not Fisher-optimal.  In the single-event regime, D-optimality can improve conditioning without changing the active set, so the Fisher objective remains useful in practice.  In the multi-event regime, the redistributed ladder improves convergence but changes how residual waveform error is sampled and projected.  The resulting bias amplification overwhelms the feasibility gain for amplitude reconstruction.

  The consequence is not that the framework fails, but that it cleanly identifies its own regime boundary.  The multi-event experiments should be read as an empirical validation of the diagnostic role of $\rho_{\mathrm{bias}}^{(A)}$: once that ratio is no longer small, Fisher-optimal thresholds need not be MSE-optimal thresholds.  Three routes follow naturally from this diagnosis: optimize a misspecification-aware objective such as $\mathcal{C}_{\varepsilon}(\mathbf{V})$, replace the pulse family with a model that materially reduces $\varepsilon(t)$, or move from CRB-based criteria to minimax MSE criteria that penalize both variance and worst-case bias.

  \subsection{Limitations and Directions for Future Work}

  The main limitations of the present study point directly to the next stage of the program.

  \emph{Unimodality constraint.}  The embedding analysis relies on strict unimodality, which guarantees exactly two crossings per threshold.  Extending the theory to multi-modal waveforms---for example, under strong afterpulsing, pile-up, or more complicated detector responses---would require a combinatorial generalization of the observation cone $\Omega_K$.

  \emph{Pulse-model improvement.}  The single-event experiments show that Fisher-guided threshold design is already useful under the current bi-exponential model, but the residual amplitude error is still dominated by waveform mismatch rather than stochastic noise.  A richer pulse family that captures additional scintillation or sensor dynamics would not merely improve fits; it would enlarge the experimentally validated domain of Fisher-optimal design.

  \emph{Timing-target validation.}  The present experiments validate the framework only in the amplitude-target geometry.  The theory itself is target-agnostic, so the natural next validation is for $\vartheta = t_0$, where the optimal ladder is expected to shift toward the steep rising edge.  That experiment would test whether the same framework delivers equally actionable gains for coincidence-timing reconstruction.

  \emph{Misspecification-aware multi-event design.}  The broad-window failure of the CRB-based multi-event objective suggests that future multi-event work should optimize beyond the Fisher proxy alone.  The clearest options are to optimize $\mathcal{C}_{\varepsilon}(\mathbf{V})$ directly when calibration data permit estimating $\varepsilon(t)$, to switch to a minimax MSE criterion, or to reduce mismatch enough that the Fisher objective becomes reliable again.

  \emph{Hardware and systems co-design.}  The manuscript optimizes threshold voltages for a fixed architecture.  A natural extension is joint optimization over the number of comparators, timestamp resolution, and threshold positions $(N, \mathrm{LSB}, \{V_n\})$, so that the present information analysis is tied directly to circuit-level power, area, and implementation limits.  At a larger scale, the same ideas should be extended to detector arrays in which channel-to-channel variation, transport heterogeneity, and calibration uncertainty add new nuisance dimensions.

  \emph{Adaptive and non-Gaussian regimes.}  The current recipes are offline and tied to a nominal operating point.  Systems with drifting gain, changing scintillation yield, or evolving temperature suggest adaptive threshold updates driven by the robustness quantities of Section~\ref{sec:robustness}.  Likewise, low-light regimes with small $N_{pe}$ call for a Poisson-native extension of the estimation theory rather than the Gaussian approximation used here.

  % =========================================================================
  %  XI. CONCLUSION
  % =========================================================================
  \section{Conclusion}
  \label{sec:conclusion}

  Prior-based Multi-Voltage Threshold sampling reconstructs pulse parameters from sparse threshold-crossing times rather than full waveforms, making parameter recovery inherently a model-dependent inverse problem. This paper addressed the missing formal mathematical statement for prior-based MVT by formalizing it for strictly unimodal pulse families as a structured inverse problem.

  On that foundation, we developed the first unified theory of prior-based MVT, comprising deterministic identifiability, a stochastic timing-error model with leading-order mismatch bias, and a nuisance-profiled threshold-design theory centered on effective information, including robust single-event and partial-trigger multi-event operation. The bi-exponential instantiation showed how this theory yields explicit reconstruction formulas, executable design rules, and testable predictions.

  Experimental validation on a 10{,}000-pulse $^{22}$Na/LYSO/SiPM dataset confirmed the utility of Fisher-guided design in the photopeak regime while also revealing the regime boundary at which partial triggering and model mismatch limit its predictive power. The experiments therefore validate both the reach and the limits of the theory.

  The main contribution is therefore not another threshold heuristic, but the first unified mathematical foundation for prior-based MVT. It recasts prior-based MVT from an empirical design problem as a principled inferential framework and provides a model-agnostic basis for richer pulse families and future MVT theories.

  % =========================================================================
  %  APPENDICES
  % =========================================================================
  \appendices
  \numberwithin{equation}{section}
  \renewcommand{\theequation}{\Alph{section}.\arabic{equation}}

  % ----- Appendix A -----
  \section{Jacobian Derivation via Geometric Projection}
  \label{app:jacobian}
  \emph{(Derivation of Eq.~\eqref{eq:jacobian} in Section~\ref{sec:model}.)}

  \noindent\textbf{Setup.}
  Consider the $k$-th ordered crossing in the observation vector, with associated threshold $V_{\ell(k)}$ from Eq.~\eqref{eq:crossing_index_map}. The continuous macroscopic pulse $f(t;\bm{\theta})$ intersects this fixed hardware voltage level at a time instant $t_k$. This crossing is defined implicitly by the level-set constraint function:
  \begin{equation}
    F_k(t_k, \bm{\theta}) \;\triangleq\; f(t_k;\,\bm{\theta}) \;-\; V_{\ell(k)} \;=\; 0.
    \label{eq:autonum:A1}
  \end{equation}

  \noindent\textbf{Transversality condition.}
  Because the parameter domain is restricted to $\bm{\theta} \in \Theta_V$ (so $p(\bm{\theta}) > V_{\max} \ge V_{\ell(k)}$), the crossing point $t_k$ is guaranteed to lie on a strictly monotone branch of $f$ (either rising or falling), away from the stationary peak where $\partial_t f = 0$.
  Consequently, the partial derivative of $F_k$ with respect to $t_k$ is non-vanishing:
  \begin{equation}
    \frac{\partial F_k}{\partial t_k}
    \;=\; \frac{\partial f(t_k;\,\bm{\theta})}{\partial t}
    \;\neq\; 0
    \qquad \forall\; \bm{\theta} \in \Theta_V.
    \label{eq:autonum:A2}
  \end{equation}

  \noindent\textbf{Total differential.}
  Now treat $t_k$ as an implicit function of the parameter vector $\bm{\theta}$, i.e., $t_k = t_k(\bm{\theta})$. The constraint $F_k(t_k(\bm{\theta}),\,\bm{\theta}) = 0$ must hold identically for all admissible $\bm{\theta}$. Taking the total differential of both sides with respect to all variables yields:
  \begin{equation}
    dF_k
    \;=\;
    \frac{\partial F_k}{\partial t_k}\,dt_k
    \;+\;
    \sum_{j=1}^{M} \frac{\partial F_k}{\partial \theta_j}\,d\theta_j
    \;=\; 0.
    \label{eq:autonum:A3}
  \end{equation}
  Substituting the definition $F_k = f - V_{\ell(k)}$ (noting that $V_{\ell(k)}$ is a constant with $dV_{\ell(k)} = 0$), each partial derivative expands as:
  \begin{equation}
    \frac{\partial F_k}{\partial t_k} = \frac{\partial f(t_k;\,\bm{\theta})}{\partial t},
    \qquad
    \frac{\partial F_k}{\partial \theta_j} = \frac{\partial f(t_k;\,\bm{\theta})}{\partial \theta_j}.
    \label{eq:autonum:A4}
  \end{equation}
  Inserting \eqref{eq:autonum:A4} into \eqref{eq:autonum:A3} produces:
  \begin{equation}
    \frac{\partial f(t_k;\,\bm{\theta})}{\partial t}\,dt_k
    \;+\;
    \sum_{j=1}^{M}
    \frac{\partial f(t_k;\,\bm{\theta})}{\partial \theta_j}\,d\theta_j
    \;=\; 0.
    \label{eq:autonum:A5}
  \end{equation}

  \noindent\textbf{Isolating $dt_k$.}
  Because the transversality condition \eqref{eq:autonum:A2} guarantees $\partial_t f \neq 0$, we may divide both sides of \eqref{eq:autonum:A5} by $\partial_t f(t_k;\bm{\theta})$ and rearrange:
  \begin{equation}
    dt_k
    \;=\;
    -\sum_{j=1}^{M}
    \frac{\partial f(t_k;\,\bm{\theta})/\partial \theta_j}
    {\partial f(t_k;\,\bm{\theta})/\partial t}
    \;d\theta_j.
    \label{eq:autonum:A6}
  \end{equation}
  Reading off the coefficient of $d\theta_j$, we identify the $(k,j)$-th element of the Jacobian matrix $\mathbf{J} \in \mathbb{R}^{K \times M}$:
  \begin{equation}
    J_{kj}
    \;\equiv\;
    \frac{\partial t_k}{\partial \theta_j}
    \;=\;
    -\,\frac{\partial f(t_k;\,\bm{\theta})\,/\,\partial \theta_j}
    {\partial f(t_k;\,\bm{\theta})\,/\,\partial t}.
    \label{eq:autonum:A7}
  \end{equation}
  This reproduces Eq.~\eqref{eq:jacobian} in the main text.

  \noindent\textbf{Implicit Function Theorem guarantee.}
  Equation~\eqref{eq:autonum:A7} is not merely a formal manipulation. The classical Implicit Function Theorem states: if $F_k \in C^1$ and $\partial F_k / \partial t_k \neq 0$ at a solution point $(t_k^*,\,\bm{\theta}^*)$, then there exists an open neighborhood of $\bm{\theta}^*$ in which $t_k(\bm{\theta})$ is a uniquely defined $C^1$-smooth function. Because $f$ is assumed $C^2$, the constraint $F_k$ is $C^2$, and consecutively $t_k(\bm{\theta})$ is $C^2$-smooth, ensuring $J_{kj}$ is itself $C^1$-continuous across $\Theta_V$.

  \noindent\textbf{Physical dimensional check.}
  Let $[\theta_j]$ denote the physical dimension of the $j$-th parameter. Then $[\partial f / \partial \theta_j] = [\mathrm{V}] / [\theta_j]$ and $[\partial f / \partial t] = [\mathrm{V/s}]$. Hence $[J_{kj}] = ([\mathrm{V}]/[\theta_j]) / ([\mathrm{V/s}]) = [\mathrm{s}]/[\theta_j]$, which is dimensionally consistent with the definition $J_{kj} = \partial t_k / \partial \theta_j$.
  \hfill$\square$

  % ----- Appendix B -----
  \section{Poisson Shot Noise Projection via Campbell's Theorem}
  \label{app:poisson}
  \emph{(Derivation of Eq.~\eqref{eq:poisson_voltage} in Section~\ref{sec:noise}.)}

  \subsection*{B.1\quad Physical model of the photocurrent}
  The scintillation crystal emits optical photons whose detection times form an inhomogeneous Poisson point process $\{s_i\}$ with time-varying intensity $\nu(t)$ $[\mathrm{photons/s}]$. Each detected photon $s_i$ triggers a single-photoelectron avalanche in the photosensor. Let $h_e(t)$ $[\mathrm{V}]$ denote the effective output-voltage impulse response produced by one detected photoelectron after the sensor and front-end chain (i.e., a photoelectron created at time $s_i$ contributes a voltage $h_e(t - s_i)$ to the readout output).

  The total macroscopic output voltage is the superposition of all SPE responses:
  \begin{equation}
    y(t) = \sum_{i} h_e(t - s_i).
    \label{eq:autonum:B1}
  \end{equation}

  \subsection*{B.2\quad Campbell's theorem: mean}
  Campbell's theorem (first moment) states that for an inhomogeneous Poisson process with intensity $\nu(t)$, the expected value of the filtered sum \eqref{eq:autonum:B1} is:
  \begin{equation}
    \mu_y(t) \;\triangleq\; \mathbb{E}[y(t)]
    = \int_{-\infty}^{+\infty} \nu(s)\,h_e(t - s)\,ds
    = (\nu * h_e)(t),
    \label{eq:autonum:B2}
  \end{equation}
  where $*$ denotes temporal convolution. This integral is the macroscopic expected voltage that we identify with the ideal pulse model: $\mu_y(t) = f(t;\bm{\theta})$.

  \subsection*{B.3\quad Campbell's theorem: variance}
  The second-moment form of Campbell's theorem gives the instantaneous variance of the shot-noise process:
  \begin{equation}
    \sigma_p^2(t) \;\triangleq\; \mathrm{Var}[y(t)]
    = \int_{-\infty}^{+\infty} \nu(s)\,h_e^2(t - s)\,ds
    = (\nu * h_e^2)(t).
    \label{eq:autonum:B3}
  \end{equation}
  The key difference from \eqref{eq:autonum:B2} is that the kernel is $h_e^2$ instead of $h_e$.

  \subsection*{B.4\quad Slowly-varying envelope approximation}
  For fast scintillators, the effective SPE voltage response $h_e(t)$ is exceedingly narrow compared to the macroscopic photon rate $\nu(t)$. Specifically, let $\tau_e$ be the effective width of $h_e(t)$. If $\nu(t)$ varies negligibly over any interval of width $\tau_e$, we may pull $\nu$ outside the convolution integral.

  As a representative order-of-magnitude example for the fast LYSO/SiPM regime considered in the main text, one may take $\tau_e \approx 0.5$--$1\,\mathrm{ns}$ for the effective SPE/front-end response, $\tau_r \approx 0.5\,\mathrm{ns}$ for a sharp macroscopic rising edge, and $\tau_d \approx 40\,\mathrm{ns}$ for the decay scale~\cite{lecoq2016scintillation,gundacker2020silicon}. Thus $\tau_e/\tau_d \approx 0.01$--$0.03 \ll 1$ on the decay branch, whereas $\tau_e/\tau_r \approx 1$--$2$ on the fastest rising segment; the approximation is therefore strong on the decay branch and only first-order on the rising edge.

  Applying this to equation \eqref{eq:autonum:B2}:
  \begin{align}
    \mu_y(t) & = \int \nu(s)\,h_e(t-s)\,ds \nonumber              \\
             & \approx \nu(t) \int_{-\infty}^{+\infty} h_e(u)\,du
    \;\triangleq\; \nu(t)\,\mathcal{H}_1,
    \label{eq:autonum:B4}
  \end{align}
  where the substitution $u = t - s$ was made, and $\mathcal{H}_1 \triangleq \int h_e(u)\,du$ $[\mathrm{V \cdot s}]$ is the total area (charge-to-voltage gain integral) of the SPE pulse.

  Applying the same approximation to equation \eqref{eq:autonum:B3}:
  \begin{equation}
    \sigma_p^2(t) \approx \nu(t) \int_{-\infty}^{+\infty} h_e^2(u)\,du
    \;\triangleq\; \nu(t)\,\mathcal{H}_2,
    \label{eq:autonum:B5}
  \end{equation}
  where $\mathcal{H}_2 \triangleq \int h_e^2(u)\,du$ $[\mathrm{V}^2\!\cdot\!\mathrm{s}]$.

  \subsection*{B.5\quad Eliminating $\nu(t)$ in favor of $\mu_y(t)$}
  From the approximation \eqref{eq:autonum:B4}, $\nu(t) \approx \mu_y(t) / \mathcal{H}_1$. Substituting into \eqref{eq:autonum:B5} gives the first-order relation:
  \begin{equation}
    \sigma_p^2(t) \approx \frac{\mu_y(t)}{\mathcal{H}_1}\,\mathcal{H}_2
    = \mu_y(t)\,\frac{\mathcal{H}_2}{\mathcal{H}_1}
    \;\triangleq\; \kappa\,\mu_y(t).
    \label{eq:autonum:B6}
  \end{equation}
  The proportionality constant $\kappa$ is:
  \begin{equation}
    \kappa \;\triangleq\; \frac{\mathcal{H}_2}{\mathcal{H}_1}
    = \frac{\displaystyle\int h_e^2(u)\,du}{\displaystyle\int h_e(u)\,du}.
    \label{eq:autonum:B7}
  \end{equation}

  \noindent\textbf{Dimensional check:} $[\mathcal{H}_2] = [\mathrm{V}^2\!\cdot\!\mathrm{s}]$ and $[\mathcal{H}_1] = [\mathrm{V}\!\cdot\!\mathrm{s}]$, so $[\kappa] = [\mathrm{V}]$. Then $[\kappa\,\mu_y] = [\mathrm{V}]\cdot[\mathrm{V}] = [\mathrm{V}^2]$, which is the correct dimension for a voltage variance.

  Physically, $\kappa$ represents the effective single-photoelectron voltage amplitude: the voltage-squared contribution per detected photon per unit mean voltage.

  \subsection*{B.6\quad Evaluation at the threshold crossing}
  At a deterministic mean crossing instant $t_c$ associated with threshold level $V$, the mean voltage equals that threshold by construction:
  \begin{equation}
    \mu_y(t_c) = V.
    \label{eq:autonum:B8}
  \end{equation}
  Substituting \eqref{eq:autonum:B8} into the Poisson variance \eqref{eq:autonum:B6} and adding the independent, signal-independent thermal noise floor $\sigma_{\mathrm{th}}^2$ $[\mathrm{V}^2]$ yields the first-order affine analog voltage model used in the main text:
  \begin{equation}
    \sigma_V^2(V) = \sigma_{\mathrm{th}}^2 + \kappa\,V.
    \label{eq:autonum:B9}
  \end{equation}
  This reproduces Eq.~\eqref{eq:poisson_voltage}.

  \noindent\textbf{Physical interpretation:} The variance is an affine function of the threshold level. The constant term $\sigma_{\mathrm{th}}^2$ represents signal-independent electronic noise (Johnson--Nyquist, $1/f$, etc.), while the linear term $\kappa V$ captures the signal-dependent quantum shot noise---higher thresholds correspond to detecting more photoelectrons per unit time, hence greater shot-noise power.
  \hfill$\square$

  % ----- Appendix C -----
  \section{Slepian-Bangs Formula and the High-SNR Limit}
  \label{app:slepian}
  \emph{(Justification of the approximation $\mathcal{I}(\bm{\theta}) \approx \mathbf{J}^T\mathbf{W}\mathbf{J}$ used in Sections~\ref{sec:mle}, \ref{sec:misspecification}, and elsewhere.)}

  \subsection*{C.1\quad The exact Slepian-Bangs formula}
  Consider a Gaussian observation model $\mathbf{t} \sim \mathcal{N}\bigl(\bm{\mu}(\bm{\theta}),\;\bm{\Sigma}(\bm{\theta})\bigr)$, where both the mean vector $\bm{\mu}(\bm{\theta}) \in \mathbb{R}^K$ and the covariance matrix $\bm{\Sigma}(\bm{\theta}) \in \mathbb{R}^{K \times K}$ depend on the parameter $\bm{\theta}$. The Fisher Information Matrix (FIM) is defined as:
  \begin{equation}
    \mathcal{I}_{ij}
    = -\mathbb{E}\!\left[\frac{\partial^2 \ln p(\mathbf{t};\bm{\theta})}{\partial\theta_i\,\partial\theta_j}\right].
    \label{eq:autonum:C1}
  \end{equation}
  The Slepian-Bangs formula provides a closed-form evaluation:
  \begin{equation}
    \mathcal{I}_{ij}
    = \underbrace{
      \left(\frac{\partial\bm{\mu}}{\partial\theta_i}\right)^{\!T}
      \bm{\Sigma}^{-1}
      \left(\frac{\partial\bm{\mu}}{\partial\theta_j}\right)
    }_{\displaystyle \mathcal{I}_{ij}^{(\mathrm{mean})}}
    + \underbrace{
      \frac{1}{2}\,\mathrm{Tr}\!\left[
        \bm{\Sigma}^{-1}\frac{\partial\bm{\Sigma}}{\partial\theta_i}\;
        \bm{\Sigma}^{-1}\frac{\partial\bm{\Sigma}}{\partial\theta_j}
        \right]
    }_{\displaystyle \mathcal{I}_{ij}^{(\mathrm{cov})}}.
    \label{eq:autonum:C2}
  \end{equation}
  The first term $\mathcal{I}^{(\mathrm{mean})}$ quantifies information carried by shifts in the mean observation; the second term $\mathcal{I}^{(\mathrm{cov})}$ captures information encoded in changes of the noise covariance structure itself.

  \subsection*{C.2\quad Identification with the MVT model}
  In the MVT setting:
  \begin{itemize}
  \item The mean vector is $\bm{\mu}(\bm{\theta}) = \mathbf{t}(\bm{\theta}) = S(\bm{\theta})$, so $\partial\bm{\mu}/\partial\theta_j$ is the $j$-th column of the Jacobian $\mathbf{J}$.
  \item The covariance is diagonal: $\bm{\Sigma} = \mathrm{diag}(\sigma_1^2, \dots, \sigma_K^2)$, with $\sigma_k^2$ given by Eq.~\eqref{eq:variance}.
  \end{itemize}
  The mean-shift term becomes:
  \begin{equation}
    \mathcal{I}_{ij}^{(\mathrm{mean})}
    = \sum_{k=1}^{K} \frac{J_{ki}\,J_{kj}}{\sigma_k^2}
    = (\mathbf{J}^T \mathbf{W} \mathbf{J})_{ij},
    \label{eq:autonum:C3}
  \end{equation}
  which is exactly the geometric Fisher matrix used throughout the main text.

  \subsection*{C.3\quad Order-of-magnitude analysis of $\mathcal{I}^{(\mathrm{cov})}$}
  To compare the two terms, introduce a dimensionless noise-level parameter $\epsilon$ that uniformly scales all noise sources. Specifically, write $\sigma_{\mathrm{th}}^2 = \epsilon^2\,\bar{\sigma}_{\mathrm{th}}^2$, $\kappa = \epsilon^2\,\bar{\kappa}$, and $\sigma_{\mathrm{TDC}}^2 = \epsilon^2\,\bar{\sigma}_{\mathrm{TDC}}^2$, where the barred quantities are $\mathcal{O}(1)$. Then:
  \begin{equation}
    \sigma_k^2 = \epsilon^2\,\bar{\sigma}_k^2 \implies \bm{\Sigma} = \epsilon^2\,\bar{\bm{\Sigma}},
    \label{eq:autonum:C4}
  \end{equation}
  where $\bar{\bm{\Sigma}} = \mathrm{diag}(\bar{\sigma}_1^2, \dots, \bar{\sigma}_K^2)$ is $\mathcal{O}(1)$.

  \noindent\textbf{Scaling of $\mathcal{I}^{(\mathrm{mean})}$:}
  \begin{align}
     & \bm{\Sigma}^{-1} = \epsilon^{-2}\,\bar{\bm{\Sigma}}^{-1}
    \quad\implies \nonumber                                                                          \\
     & \mathcal{I}^{(\mathrm{mean})} = \mathbf{J}^T (\epsilon^{-2}\bar{\bm{\Sigma}}^{-1}) \mathbf{J}
    = \epsilon^{-2}\,\mathbf{J}^T \bar{\bm{\Sigma}}^{-1} \mathbf{J}
    \sim \mathcal{O}(\epsilon^{-2}).
    \label{eq:autonum:C5}
  \end{align}

  \noindent\textbf{Scaling of $\partial_{\theta_i}\bm{\Sigma}$:}
  Since $\sigma_k^2$ depends on $\bm{\theta}$ through the local slew rate $\partial_t f(t_k;\bm{\theta})$ in the projected analog-variance term $(\sigma_{\mathrm{th}}^2 + \kappa V_{\ell(k)})/[\partial_t f(t_k;\bm{\theta})]^2$, and equivalently through any model-specific rewriting of that same slope dependence (e.g., Eq.~\eqref{eq:biexp_variance}), we have $\partial_{\theta_i}\bm{\Sigma} = \epsilon^2\,\partial_{\theta_i}\bar{\bm{\Sigma}} \sim \mathcal{O}(\epsilon^2)$. The additive TDC term $\sigma_{\mathrm{TDC}}^2$ in Eq.~\eqref{eq:variance} is parameter-independent.

  \noindent\textbf{Scaling of $\mathcal{I}^{(\mathrm{cov})}$:}
  Substituting the scalings into the trace formula:
  \begin{align}
    \mathcal{I}_{ij}^{(\mathrm{cov})}
     & = \frac{1}{2}\,\mathrm{Tr}\!\left[
                                     (\epsilon^{-2}\bar{\bm{\Sigma}}^{-1})
                                     (\epsilon^2\,\partial_{\theta_i}\bar{\bm{\Sigma}})
                                     (\epsilon^{-2}\bar{\bm{\Sigma}}^{-1})
                                     (\epsilon^2\,\partial_{\theta_j}\bar{\bm{\Sigma}})
                                     \right] \nonumber \\
     & = \frac{1}{2}\,\epsilon^{-2+2-2+2}\,
    \mathrm{Tr}\!\left[
                   \bar{\bm{\Sigma}}^{-1}\partial_{\theta_i}\bar{\bm{\Sigma}}\;
                   \bar{\bm{\Sigma}}^{-1}\partial_{\theta_j}\bar{\bm{\Sigma}}
                   \right] \nonumber         \\
     & = \frac{1}{2}\,\epsilon^{0}\,
    \mathrm{Tr}\!\left[\cdots\right]
    \sim \mathcal{O}(1).
    \label{eq:autonum:C6}
  \end{align}
  Here the exponents cancel step by step: $\epsilon^{-2} \cdot \epsilon^{+2} \cdot \epsilon^{-2} \cdot \epsilon^{+2} = \epsilon^0 = 1$.

  \subsection*{C.4\quad Conclusion}
  Comparing the two contributions:
  \begin{equation}
    \frac{\mathcal{I}^{(\mathrm{cov})}}{\mathcal{I}^{(\mathrm{mean})}}
    \sim \frac{\mathcal{O}(1)}{\mathcal{O}(\epsilon^{-2})}
    = \mathcal{O}(\epsilon^2) \;\to\; 0
    \quad\text{as}\quad \epsilon \to 0.
    \label{eq:autonum:C7}
  \end{equation}
  In the high-SNR regime ($\epsilon \ll 1$, corresponding to high-photon-count scintillation events typical in PET), the covariance-derivative term is negligible relative to the mean-shift term. Therefore:
  \begin{equation}
    \mathcal{I}_{ij}
    \approx \mathcal{I}_{ij}^{(\mathrm{mean})}
    = (\mathbf{J}^T \mathbf{W} \mathbf{J})_{ij}.
    \label{eq:autonum:C8}
  \end{equation}
  This justifies dropping $\mathcal{I}^{(\mathrm{cov})}$ and using the simpler geometric FIM throughout the paper.
  \hfill$\square$

  % ----- Appendix D -----
  \section{Derivation of Systematic Bias via QMLE Score Equation}
  \label{app:mismatch_proof}
  \emph{(Derivation of Eq.~\eqref{eq:dt_bias_scalar} and Eq.~\eqref{eq:bias_vector} in Section~\ref{sec:misspecification}.)}

  \subsection*{D.1\quad Problem setup}
  Let the true physical waveform be:
  \begin{equation}
    y_{\mathrm{true}}(t) = f(t;\,\bm{\theta}_0) + \varepsilon(t),
    \label{eq:autonum:D1}
  \end{equation}
  where $\bm{\theta}_0$ is the true parameter, and $\varepsilon(t)$ is a deterministic (non-random) model misspecification function. The estimator, however, assumes a perfect model $f(t;\bm{\theta})$ and seeks $\hat{\bm{\theta}}$ that best fits the observed data.
  Throughout this appendix, expectations are taken over zero-mean stochastic timing noise only; $\varepsilon(t)$ is held fixed.

  \subsection*{D.2\quad Temporal bias from voltage mismatch}
  At the $k$-th ordered crossing associated with threshold $V_{\ell(k)}$, the misspecified physical crossing time $t_k^{(\varepsilon)}$ satisfies $y_{\mathrm{true}}(t_k^{(\varepsilon)}) = V_{\ell(k)}$, i.e.:
  \begin{equation}
    f(t_k^{(\varepsilon)};\,\bm{\theta}_0) + \varepsilon(t_k^{(\varepsilon)}) = V_{\ell(k)}.
    \label{eq:autonum:D2}
  \end{equation}
  Meanwhile, the ideal (unperturbed) crossing $t_k^{(0)}$ satisfies $f(t_k^{(0)};\,\bm{\theta}_0) = V_{\ell(k)}$. Define the temporal perturbation $\Delta t_{\mathrm{bias},k} \triangleq t_k^{(\varepsilon)} - t_k^{(0)}$.

  Expanding both $f(t_k^{(\varepsilon)};\,\bm{\theta}_0)$ and $\varepsilon(t_k^{(\varepsilon)})$ around $t_k^{(0)}$ to first order in $\Delta t_{\mathrm{bias},k}$:
  \begin{equation}
    \begin{aligned}
      f(t_k^{(0)};\,\bm{\theta}_0)
       & + \bigl[\partial_t f(t_k^{(0)};\,\bm{\theta}_0)
        + \partial_t \varepsilon(t_k^{(0)})\bigr]\,\Delta t_{\mathrm{bias},k} \\
       & + \varepsilon(t_k^{(0)}) \;\approx\; V_{\ell(k)}.
    \end{aligned}
    \label{eq:autonum:D3}
  \end{equation}
  If the mismatch is locally flat near the crossing, in the sense $|\partial_t \varepsilon(t_k^{(0)})| \ll |\partial_t f(t_k^{(0)};\,\bm{\theta}_0)|$, the correction involving $\partial_t\varepsilon$ is higher order. Using $f(t_k^{(0)};\,\bm{\theta}_0) = V_{\ell(k)}$ then gives:
  \begin{equation}
    \Delta t_{\mathrm{bias},k} \;\approx\; -\frac{\varepsilon(t_k^{(0)})}{\partial_t f(t_k^{(0)};\,\bm{\theta}_0)}.
    \label{eq:autonum:D4}
  \end{equation}
  In vector form, $\Delta\mathbf{t}_{\mathrm{bias}} = [\Delta t_{\mathrm{bias},1}, \dots, \Delta t_{\mathrm{bias},K}]^T$, each element given by \eqref{eq:autonum:D4}. This is a deterministic shift that persists under ensemble averaging.

  \subsection*{D.3\quad Approximate QMLE score equation}
  The Quasi-Maximum Likelihood Estimator (QMLE) assumes the model $f$ is correct and minimizes the weighted residual cost (Eq.~\eqref{eq:cost}). Under the same high-SNR locally constant-covariance approximation used in Section~\ref{sec:mle} and Appendix~\ref{app:slepian}, the first-order optimality condition reduces to the approximate score equation:
  \begin{equation}
    \mathbf{s}(\hat{\bm{\theta}})
    \;\triangleq\;
    \nabla_{\bm{\theta}} \ln L\big|_{\hat{\bm{\theta}}}
    = \mathbf{J}^T(\hat{\bm{\theta}})\,\mathbf{W}\,
    \bigl[\mathbf{t}_{\mathrm{obs}} - \mathbf{t}_{\mathrm{model}}(\hat{\bm{\theta}})\bigr]
    = \mathbf{0}.
    \label{eq:autonum:D5}
  \end{equation}

  \subsection*{D.4\quad Linearization around $\bm{\theta}_0$}
  Taylor-expand the model predictions around the true parameter:
  \begin{equation}
    \mathbf{t}_{\mathrm{model}}(\hat{\bm{\theta}})
    \;\approx\;
    \mathbf{t}_{\mathrm{model}}(\bm{\theta}_0)
    + \mathbf{J}_0\,(\hat{\bm{\theta}} - \bm{\theta}_0),
    \label{eq:autonum:D6}
  \end{equation}
  where $\mathbf{J}_0 \triangleq \mathbf{J}(\bm{\theta}_0)$. The observed timestamps are:
  \begin{equation}
    \mathbf{t}_{\mathrm{obs}}
    = \mathbf{t}_{\mathrm{model}}(\bm{\theta}_0)
    + \Delta\mathbf{t}_{\mathrm{bias}}
    + \bm{\epsilon},
    \label{eq:autonum:D7}
  \end{equation}
  where $\bm{\epsilon}$ is the zero-mean stochastic noise ($\mathbb{E}[\bm{\epsilon}] = \mathbf{0}$).

  \subsection*{D.5\quad Substitution and expectation}
  Insert \eqref{eq:autonum:D6} and \eqref{eq:autonum:D7} into \eqref{eq:autonum:D5}:
  \begin{align}
    \mathbf{0}
     & = \mathbf{J}_0^T\,\mathbf{W}\,
    \bigl[
      \mathbf{t}_{\mathrm{model}}(\bm{\theta}_0)
      + \Delta\mathbf{t}_{\mathrm{bias}} + \bm{\epsilon} \nonumber\\
      &\quad  - \mathbf{t}_{\mathrm{model}}(\bm{\theta}_0)
      - \mathbf{J}_0\,(\hat{\bm{\theta}} - \bm{\theta}_0)
      \bigr] \nonumber \\
     & = \mathbf{J}_0^T\,\mathbf{W}\,
    \bigl[
      \Delta\mathbf{t}_{\mathrm{bias}} + \bm{\epsilon}
      - \mathbf{J}_0\,(\hat{\bm{\theta}} - \bm{\theta}_0)
      \bigr].
    \label{eq:autonum:D8}
  \end{align}
  Taking the ensemble expectation of both sides (using $\mathbb{E}[\bm{\epsilon}] = \mathbf{0}$):
  \begin{align}
    \mathbf{J}_0^T\,\mathbf{W}\,
    \bigl[
      \Delta\mathbf{t}_{\mathrm{bias}}
      &- \mathbf{J}_0\,
      \underbrace{\mathbb{E}[\hat{\bm{\theta}} - \bm{\theta}_0]}_{\bm{\beta}}
      \bigr]
    = \mathbf{0}.
    \label{eq:autonum:D9}
  \end{align}

  \subsection*{D.6\quad Solving for the bias vector}
  Rearranging \eqref{eq:autonum:D9}:
  \begin{equation}
    \mathbf{J}_0^T\,\mathbf{W}\,\mathbf{J}_0\;\bm{\beta}
    = \mathbf{J}_0^T\,\mathbf{W}\,\Delta\mathbf{t}_{\mathrm{bias}}.
    \label{eq:autonum:D10}
  \end{equation}
  The matrix $\mathbf{J}_0^T\mathbf{W}\mathbf{J}_0$ is the Fisher Information Matrix $\mathcal{I}(\bm{\theta}_0)$, which is invertible by the global immersion condition ($\operatorname{rank}(\mathbf{J}_0) = M$). Left-multiplying both sides by $(\mathbf{J}_0^T\mathbf{W}\mathbf{J}_0)^{-1}$ yields the leading-order bias projection:
  \begin{equation}
    \bm{\beta}
    \approx \bigl(\mathbf{J}_0^T\,\mathbf{W}\,\mathbf{J}_0\bigr)^{-1}\,
    \mathbf{J}_0^T\,\mathbf{W}\,\Delta\mathbf{t}_{\mathrm{bias}}.
    \label{eq:autonum:D11}
  \end{equation}
  Equation~\eqref{eq:autonum:D4} reproduces Eq.~\eqref{eq:dt_bias_scalar}, and Eq.~\eqref{eq:autonum:D11} reproduces Eq.~\eqref{eq:bias_vector} in the main text.

  \noindent\textbf{Geometric interpretation:} The expression $(\mathbf{J}_0^T\mathbf{W}\mathbf{J}_0)^{-1}\mathbf{J}_0^T\mathbf{W}$ is the Moore--Penrose weighted pseudoinverse of $\mathbf{J}_0$. It projects the $K$-dimensional temporal bias vector $\Delta\mathbf{t}_{\mathrm{bias}}$ onto the $M$-dimensional tangent space of the parameter manifold, using the metric tensor $\mathbf{W}$. In other words, the estimator absorbs the model mismatch by finding the closest point on the model manifold in the weighted least-squares sense, resulting in a systematic parameter offset.
  \hfill$\square$

  % ----- Appendix E -----
  \section{Single-Pair FIM Derivation and Bias-Aware Effective MSE}
  \label{app:schur}
  \emph{(Derivation of Eq.~\eqref{eq:pair_fim} and Eq.~\eqref{eq:schur_general}, plus interpretation of the single-pair profiled score in Sections~\ref{sec:pair_general}--\ref{sec:master_eq}.)}

  Throughout this appendix we adopt the same diagonal single-pair approximation as Section~\ref{sec:pair_general}: the threshold is admissible ($b < V < p(\bm{\theta})$), the pair gap satisfies $t_f(V;\bm{\theta}) - t_r(V;\bm{\theta}) \ge \tau_c$, and otherwise the diagonal precision block below should be replaced by the full $2 \times 2$ within-pair precision matrix.

  \subsection*{E.1\quad Single-pair Jacobian and weight matrix}

  For a single threshold $V$ generating a rising-edge crossing at $t_r$ and a falling-edge crossing at $t_f$, the local observation is the $2$-vector $\mathbf{t}_{\mathrm{pair}} = [t_r,\; t_f]^T$. From Eq.~\eqref{eq:jacobian}, the $2 \times M$ Jacobian is:
  \begin{align}
    \mathbf{J}_{\mathrm{pair}}
     & =
    \begin{bmatrix}
      \partial t_r / \partial\theta_1 & \cdots & \partial t_r / \partial\theta_M \\[2pt]
      \partial t_f / \partial\theta_1 & \cdots & \partial t_f / \partial\theta_M
    \end{bmatrix} \nonumber \\
     & =
    \begin{bmatrix}
      -g_{1,r}/\dot{f}_r & \cdots & -g_{M,r}/\dot{f}_r \\[2pt]
      -g_{1,f}/\dot{f}_f & \cdots & -g_{M,f}/\dot{f}_f
    \end{bmatrix},
    \label{eq:autonum:E1}
  \end{align}
  where we use the shorthand $\dot{f}_r \triangleq \partial_t f(t_r;\bm{\theta})$, $\dot{f}_f \triangleq \partial_t f(t_f;\bm{\theta})$, and the waveform sensitivity $g_{j,r} \triangleq \partial f(t_r;\bm{\theta})/\partial\theta_j$ as defined in Eq.~\eqref{eq:pair_sensitivity}.

  The $2 \times 2$ diagonal precision matrix is:
  \begin{equation}
    \mathbf{W}_{\mathrm{pair}}
    = \mathrm{diag}\!\left(\frac{1}{\sigma_r^2},\;\frac{1}{\sigma_f^2}\right),
    \label{eq:autonum:E2}
  \end{equation}
  where from Eq.~\eqref{eq:variance}:
  \begin{equation}
    \begin{split}
      \sigma_r^2 & = \sigma_{\mathrm{TDC}}^2 + \frac{\sigma_{\mathrm{th}}^2 + \kappa (V-b)}{\dot{f}_r^2}, \\
      \sigma_f^2 & = \sigma_{\mathrm{TDC}}^2 + \frac{\sigma_{\mathrm{th}}^2 + \kappa (V-b)}{\dot{f}_f^2}.
    \end{split}
    \label{eq:autonum:E3}
  \end{equation}
  We define the composite noise denominators (referenced in the main text as $D_r$, $D_f$):
  \begin{align}
    D_r & \triangleq \sigma_r^2\,\dot{f}_r^2 = \sigma_{\mathrm{th}}^2 + \kappa (V-b) + \sigma_{\mathrm{TDC}}^2\,\dot{f}_r^2, \nonumber \\
    D_f & \triangleq \sigma_f^2\,\dot{f}_f^2 = \sigma_{\mathrm{th}}^2 + \kappa (V-b) + \sigma_{\mathrm{TDC}}^2\,\dot{f}_f^2.
    \label{eq:pair_denominators}
  \end{align}
  Here again $b$ denotes any deterministic baseline offset, with $b=0$ in the baseline-free general convention.
  Hence $1/\sigma_r^2 = \dot{f}_r^2 / D_r$ and $1/\sigma_f^2 = \dot{f}_f^2 / D_f$.

  \subsection*{E.2\quad Computing $\mathcal{I}^{(\mathrm{pair})} = \mathbf{J}_{\mathrm{pair}}^T \mathbf{W}_{\mathrm{pair}} \mathbf{J}_{\mathrm{pair}}$}

  The $(j, l)$ element of the $M \times M$ Fisher matrix is:
  \begin{align}
    \mathcal{I}_{jl}^{(\mathrm{pair})}
     & = \sum_{k \in \{r,f\}} \frac{1}{\sigma_k^2}\,
    \frac{\partial t_k}{\partial\theta_j}\,
    \frac{\partial t_k}{\partial\theta_l}.
    \label{eq:autonum:E5}
  \end{align}
  Substituting $\partial t_k/\partial\theta_j = -g_{j,k}/\dot{f}_k$ from \eqref{eq:autonum:E1}:
  \begin{align}
    \mathcal{I}_{jl}^{(\mathrm{pair})}
     & = \frac{1}{\sigma_r^2}\cdot\frac{g_{j,r}}{\dot{f}_r}\cdot\frac{g_{l,r}}{\dot{f}_r}
    + \frac{1}{\sigma_f^2}\cdot\frac{g_{j,f}}{\dot{f}_f}\cdot\frac{g_{l,f}}{\dot{f}_f} \nonumber \\
     & = \frac{g_{j,r}\,g_{l,r}}{\sigma_r^2\,\dot{f}_r^2}
    + \frac{g_{j,f}\,g_{l,f}}{\sigma_f^2\,\dot{f}_f^2}.
    \label{eq:autonum:E6}
  \end{align}
  Recognizing from \eqref{eq:pair_denominators} that $\sigma_k^2\,\dot{f}_k^2 = D_k$:
  \begin{equation}
    \mathcal{I}_{jl}^{(\mathrm{pair})}
    = \frac{g_{j,r}\,g_{l,r}}{D_r}
    + \frac{g_{j,f}\,g_{l,f}}{D_f}.
    \label{eq:autonum:E7}
  \end{equation}
  This reproduces Eq.~\eqref{eq:pair_fim}. Note how the temporal slopes $\dot{f}_r$ and $\dot{f}_f$ have completely cancelled between the Jacobian and the weight matrix, leaving only the waveform sensitivities $g_{j,k}$ and the composite denominators $D_k$.

  \subsection*{E.3\quad Generalized Schur complement for effective information}

  For a single threshold pair, partition the pairwise FIM as
  \begin{equation}
    \mathcal{I}^{(\mathrm{pair})}
    =
    \begin{bmatrix}
      \bm{\mathcal{I}}_{\bm{\eta}\bm{\eta}}   & \bm{\mathcal{I}}_{\bm{\eta}\vartheta} \\[3pt]
      \bm{\mathcal{I}}_{\bm{\eta}\vartheta}^T & \mathcal{I}_{\vartheta\vartheta}
    \end{bmatrix}
    = \begin{bmatrix} \mathbf{A} & \mathbf{u} \\ \mathbf{u}^T & c \end{bmatrix},
    \label{eq:autonum:E8}
  \end{equation}
  where $\mathbf{A} = \bm{\mathcal{I}}_{\bm{\eta}\bm{\eta}}$, $\mathbf{u} = \bm{\mathcal{I}}_{\bm{\eta}\vartheta}$, and $c = \mathcal{I}_{\vartheta\vartheta}$. Because Eq.~\eqref{eq:pair_fim} expresses the pairwise FIM as the sum of two weighted rank-one outer products, the nuisance--target block is always a linear combination of the nuisance sensitivity vectors:
  \begin{equation}
    \mathbf{u}
    = \frac{g_{\vartheta,r}\,\mathbf{g}_{\bm{\eta},r}}{D_r}
    + \frac{g_{\vartheta,f}\,\mathbf{g}_{\bm{\eta},f}}{D_f}
    \in \operatorname{span}\{\mathbf{g}_{\bm{\eta},r},\mathbf{g}_{\bm{\eta},f}\}
    = \operatorname{range}(\mathbf{A}).
    \label{eq:autonum:E9}
  \end{equation}

  Define the quadratic form
  \begin{equation}
    \psi(\bm{\gamma})
    \triangleq
    c - 2\,\bm{\gamma}^{T}\mathbf{u} + \bm{\gamma}^{T}\mathbf{A}\bm{\gamma}.
    \label{eq:autonum:E10}
  \end{equation}
  Since $\mathbf{u} \in \operatorname{range}(\mathbf{A})$, we have $\mathbf{A}\mathbf{A}^{+}\mathbf{u} = \mathbf{u}$. Completing the square gives
  \begin{align}
    \psi(\bm{\gamma})
     & = c - \mathbf{u}^{T}\mathbf{A}^{+}\mathbf{u}
                                         + (\bm{\gamma} - \mathbf{A}^{+}\mathbf{u})^{T}
    \mathbf{A}
    (\bm{\gamma} - \mathbf{A}^{+}\mathbf{u}).
    \label{eq:autonum:E11}
  \end{align}
  Because $\mathbf{A} \succeq 0$, the second term is non-negative, so the minimum of $\psi$ is attained on the affine set
  \begin{equation}
    \bm{\gamma}
    \in \mathbf{A}^{+}\mathbf{u} + \ker(\mathbf{A}),
    \label{eq:autonum:E12}
  \end{equation}
  and the minimum value is
  \begin{equation}
    \Delta\mathcal{I}_{\mathrm{eff}}^{(\vartheta)}(V)
    \triangleq
    \min_{\bm{\gamma}} \psi(\bm{\gamma})
    = c - \mathbf{u}^{T}\mathbf{A}^{+}\mathbf{u}
    = \mathcal{I}_{\vartheta\vartheta}
    - \bm{\mathcal{I}}_{\bm{\eta}\vartheta}^{T}
    \bm{\mathcal{I}}_{\bm{\eta}\bm{\eta}}^{+}
    \bm{\mathcal{I}}_{\bm{\eta}\vartheta}.
    \label{eq:autonum:E13}
  \end{equation}
  This is precisely the generalized Schur complement used in Eq.~\eqref{eq:schur_general}. If $\mathbf{A}$ is invertible, Eq.~\eqref{eq:autonum:E13} reduces to the ordinary Schur complement.

  To recover the ordinary profiled CRLB in the nonsingular case, choose an orthogonal basis $\mathbf{Q} = [\mathbf{Q}_r\;\mathbf{Q}_0]$ such that $\operatorname{range}(\mathbf{A}) = \operatorname{span}(\mathbf{Q}_r)$ and $\ker(\mathbf{A}) = \operatorname{span}(\mathbf{Q}_0)$. When $\mathbf{A}$ is invertible, the nullspace block is absent and
  \begin{equation}
    \mathbf{Q}^{T}\mathbf{A}\mathbf{Q}
    = \bm{\Lambda},
    \qquad
    \mathbf{Q}^{T}\mathbf{u} = \tilde{\mathbf{u}},
    \label{eq:autonum:E14}
  \end{equation}
  with $\bm{\Lambda} \succ 0$. Under the orthogonal similarity transform $\operatorname{diag}(\mathbf{Q}^{T},1)$, the pairwise FIM becomes
  \begin{equation}
    \widetilde{\mathcal{I}}^{(\mathrm{pair})}
    = \begin{bmatrix}
      \bm{\Lambda}           & \tilde{\mathbf{u}} \\[2pt]
      \tilde{\mathbf{u}}^{T} & c
    \end{bmatrix}.
    \label{eq:autonum:E15}
  \end{equation}
  Ordinary block inversion gives the $(\vartheta,\vartheta)$ entry of $\bigl(\widetilde{\mathcal{I}}^{(\mathrm{pair})}\bigr)^{-1}$ as
  \begin{equation}
    \left[c - \tilde{\mathbf{u}}^{T}\bm{\Lambda}^{-1}\tilde{\mathbf{u}}\right]^{-1}
    = \left[c - \mathbf{u}^{T}\mathbf{A}^{+}\mathbf{u}\right]^{-1}
    = \left[\Delta\mathcal{I}_{\mathrm{eff}}^{(\vartheta)}(V)\right]^{-1}.
    \label{eq:autonum:E16}
  \end{equation}
  Thus, when $\mathbf{A}$ is invertible, the profiled CRLB coincides with the reciprocal of the Schur-complement value in Eq.~\eqref{eq:autonum:E13}.

  \subsection*{E.4\quad Interpretation of the singular-pair case}

  Equation~\eqref{eq:autonum:E13} is the exact generalized Schur complement and variational value associated with a single threshold pair. When $\mathbf{A}$ is invertible, Eq.~\eqref{eq:autonum:E13} reduces to the ordinary Schur complement, and its reciprocal is the usual profiled CRLB for the target parameter.

  In the singular case, however, this reciprocal-CRLB interpretation need not survive. Because a single pair supplies only two scalar edge observations, if $\dim(\bm{\eta}) \ge 2$ and the two nuisance sensitivity vectors are linearly independent, one can solve
  \begin{equation}
    \bm{\gamma}^{T}\mathbf{g}_{\bm{\eta},r} = g_{\vartheta,r},
    \qquad
    \bm{\gamma}^{T}\mathbf{g}_{\bm{\eta},f} = g_{\vartheta,f},
    \label{eq:autonum:E17}
  \end{equation}
  which makes the two projected edge sensitivities vanish simultaneously and therefore yields $\Delta\mathcal{I}_{\mathrm{eff}}^{(\vartheta)}(V)=0$. The Moore--Penrose inverse of the full pairwise FIM may nevertheless have a finite $(\vartheta,\vartheta)$ entry, so in general
  \begin{equation}
    \left[(\mathcal{I}^{(\mathrm{pair})})^{+}\right]_{\vartheta\vartheta}
    \neq
    \left[\Delta\mathcal{I}_{\mathrm{eff}}^{(\vartheta)}(V)\right]^{-1}
    \qquad\text{for singular } \mathbf{A}.
    \label{eq:autonum:E18}
  \end{equation}

  Accordingly, in the general $M$-parameter framework Eq.~\eqref{eq:autonum:E13} is interpreted as the nuisance-profiled single-pair information score entering the joint variational analysis of Appendix~\ref{app:synergy}, not as a standalone finite MSE bound. A finite model-based bound is recovered only after sufficiently many threshold pairs are aggregated so that $\bm{\mathcal{I}}_{\bm{\eta}\bm{\eta}}^{\mathrm{tot}} \succ 0$, as in Eq.~\eqref{eq:synergy_bound}. \hfill$\square$

  % ----- Appendix F -----
  \section{Proof of Systematic Synergy via Variational Generalized Schur Complement}
  \label{app:synergy}
  \emph{(Derivation of the superadditivity bound Eq.~\eqref{eq:synergy_bound} in Section~\ref{sec:synergy}.)}

  \subsection*{F.1\quad Notation}
  Let $N$ independent threshold pairs be indexed by $n = 1, \dots, N$. Each pair contributes to the full FIM additively (by independence). Partitioning $\bm{\theta} = [\bm{\eta}^T,\,\vartheta]^T$ as before, the $n$-th pair's FIM block components are:
  \begin{align}
    \bm{A}_n     & \;\triangleq\; \bm{\mathcal{I}}_{\bm{\eta}\bm{\eta}}^{(n)} \;\succeq\; 0,
                 &
    \mathbf{u}_n & \;\triangleq\; \bm{\mathcal{I}}_{\bm{\eta}\vartheta}^{(n)} \;\in\; \mathbb{R}^{M-1}, \nonumber \\
    c_n          & \;\triangleq\; \mathcal{I}_{\vartheta\vartheta}^{(n)} \;>\; 0.
    \label{eq:autonum:F1}
  \end{align}
  Because each pairwise FIM is a sum of two weighted rank-one outer products, $\mathbf{u}_n \in \operatorname{range}(\bm{A}_n)$ for every $n$ (Appendix~E). The total nuisance block is
  \begin{equation}
    \bm{A}_{\mathrm{tot}} = \sum_{n=1}^{N} \bm{A}_n,
    \qquad
    \mathbf{u}_{\mathrm{tot}} = \sum_{n=1}^{N} \mathbf{u}_n,
    \qquad
    c_{\mathrm{tot}} = \sum_{n=1}^{N} c_n,
    \label{eq:autonum:F2}
  \end{equation}
  and the global nuisance identifiability assumption of the main text implies $\bm{A}_{\mathrm{tot}} \succ 0$.

  \subsection*{F.2\quad Per-pair variational characterization}
  For the $n$-th pair, define
  \begin{equation}
    \psi_n(\bm{\gamma})
    \triangleq
    c_n - 2\,\bm{\gamma}^{T}\mathbf{u}_n + \bm{\gamma}^{T}\bm{A}_n\bm{\gamma}.
    \label{eq:autonum:F3}
  \end{equation}
  By Appendix~E, the generalized per-pair effective information is
  \begin{equation}
    \Delta\mathcal{I}_{\mathrm{eff}}^{(\vartheta),(n)}
    = \min_{\bm{\gamma}} \psi_n(\bm{\gamma})
    = c_n - \mathbf{u}_n^T\bm{A}_n^{+}\mathbf{u}_n,
    \label{eq:autonum:F4}
  \end{equation}
  and the minimizer set is
  \begin{equation}
    \bm{\gamma}
    \in \bm{A}_n^{+}\mathbf{u}_n + \ker(\bm{A}_n).
    \label{eq:autonum:F5}
  \end{equation}
  Since $\bm{A}_n \succeq 0$, each $\psi_n$ is a convex quadratic function, so
  \begin{equation}
    \psi_n(\bm{\gamma}) \ge \Delta\mathcal{I}_{\mathrm{eff}}^{(\vartheta),(n)}
    \qquad \forall\; \bm{\gamma} \in \mathbb{R}^{M-1}.
    \label{eq:autonum:F6}
  \end{equation}

  \subsection*{F.3\quad Global variational characterization}
  Summing Eq.~\eqref{eq:autonum:F3} over all pairs gives
  \begin{align}
    \Psi(\bm{\gamma})
     & \triangleq \sum_{n=1}^{N} \psi_n(\bm{\gamma}) \nonumber          \\
     & = c_{\mathrm{tot}} - 2\,\bm{\gamma}^{T}\mathbf{u}_{\mathrm{tot}}
    + \bm{\gamma}^{T}\bm{A}_{\mathrm{tot}}\bm{\gamma}.
    \label{eq:autonum:F7}
  \end{align}
  Because $\bm{A}_{\mathrm{tot}} \succ 0$, the minimizer is unique:
  \begin{equation}
    \bm{\gamma}^{*} = \bm{A}_{\mathrm{tot}}^{-1}\mathbf{u}_{\mathrm{tot}},
    \label{eq:autonum:F8}
  \end{equation}
  and the minimum value is the total effective information:
  \begin{equation}
    \Delta\mathcal{I}_{\mathrm{eff}}^{(\vartheta),\mathrm{tot}}
    = \min_{\bm{\gamma}} \Psi(\bm{\gamma})
    = c_{\mathrm{tot}} - \mathbf{u}_{\mathrm{tot}}^{T}\bm{A}_{\mathrm{tot}}^{-1}\mathbf{u}_{\mathrm{tot}}.
    \label{eq:autonum:F9}
  \end{equation}

  \subsection*{F.4\quad Superadditivity of effective information}
  Applying Eq.~\eqref{eq:autonum:F6} to the common choice $\bm{\gamma} = \bm{\gamma}^{*}$ and summing over $n$ yields
  \begin{align}
    \Delta\mathcal{I}_{\mathrm{eff}}^{(\vartheta),\mathrm{tot}}
     & = \Psi(\bm{\gamma}^{*}) \tag{by \eqref{eq:autonum:F9}} \nonumber                       \\
     & = \sum_{n=1}^{N} \psi_n(\bm{\gamma}^{*}) \tag{by \eqref{eq:autonum:F7}} \nonumber      \\
     & \ge \sum_{n=1}^{N} \Delta\mathcal{I}_{\mathrm{eff}}^{(\vartheta),(n)}.
    \label{eq:autonum:F10}
  \end{align}
  This is exactly the desired superadditivity inequality. No per-pair invertibility assumption is required: singular pairs simply have non-unique minimizer sets in Eq.~\eqref{eq:autonum:F5}, while the minimum values remain uniquely defined. Equality in Eq.~\eqref{eq:autonum:F10} holds if and only if the global minimizer $\bm{\gamma}^{*}$ belongs to every per-pair minimizer set,
  \begin{equation}
    \bm{\gamma}^{*}
    \in
    \bigcap_{n=1}^{N}\left(\bm{A}_n^{+}\mathbf{u}_n + \ker(\bm{A}_n)\right).
    \label{eq:autonum:F10a}
  \end{equation}
  When all $\bm{A}_n$ are invertible, this reduces to the familiar condition $\bm{A}_n^{-1}\mathbf{u}_n=\bm{\gamma}^{*}$ for every $n$; when some $\bm{A}_n$ are singular, mismatch between particular pseudoinverse representatives is not by itself decisive because the null directions in $\ker(\bm{A}_n)$ also belong to the minimizer set.

  \subsection*{F.5\quad Translation to the MSE bound}
  Since $\Delta\mathcal{I}_{\mathrm{eff}}^{(\vartheta),\mathrm{tot}} \ge \sum_n \Delta\mathcal{I}_{\mathrm{eff}}^{(\vartheta),(n)}$ and the function $x \mapsto x^{-1}$ is monotonically decreasing for $x > 0$:
  \begin{equation}
    \bigl(\Delta\mathcal{I}_{\mathrm{eff}}^{(\vartheta),\mathrm{tot}}\bigr)^{-1}
    \;\le\;
    \left(\sum_{n=1}^N \Delta\mathcal{I}_{\mathrm{eff}}^{(\vartheta),(n)}\right)^{\!-1}.
    \label{eq:autonum:F11}
  \end{equation}
  Adding the total squared bias $\beta_{\vartheta,\mathrm{tot}}^2$ to both sides preserves the inequality:
  \begin{equation}
    \underbrace{
      \bigl(\Delta\mathcal{I}_{\mathrm{eff}}^{(\vartheta),\mathrm{tot}}\bigr)^{-1} + \beta_{\vartheta,\mathrm{tot}}^2
    }_{\displaystyle \mathcal{B}_{\vartheta,\min}^{\mathrm{tot}}}
    \;\le\;
    \left(\sum_{n=1}^N \Delta\mathcal{I}_{\mathrm{eff}}^{(\vartheta),(n)}\right)^{\!-1}
    + \beta_{\vartheta,\mathrm{tot}}^2.
    \label{eq:autonum:F12}
  \end{equation}
  This reproduces Eq.~\eqref{eq:synergy_bound}. The inequality is weak in general and becomes strict precisely when the intersection condition in Eq.~\eqref{eq:autonum:F10a} fails.
  \hfill$\square$

  % ----- Appendix G -----
  \section{Bi-Exponential Peak Time and Peak Value}
  \label{app:peak_time}
  \emph{(Derivation of the peak time $x_p$ and peak value $h_{\mathrm{peak}}$ referenced in Section~\ref{sec:biexp_model}.)}

  \subsection*{G.1\quad Setting up the optimization}
  The normalized shape function is $h(x) = e^{-\lambda_d x} - e^{-\lambda_r x}$ for $x > 0$, with $\lambda_d = 1/\tau_d$, $\lambda_r = 1/\tau_r$, and $0 < \lambda_d < \lambda_r$.

  The peak occurs at $h'(x_p) = 0$. Computing:
  \begin{equation}
    h'(x) = -\lambda_d\,e^{-\lambda_d x} + \lambda_r\,e^{-\lambda_r x}.
    \label{eq:autonum:G1}
  \end{equation}
  Setting $h'(x_p) = 0$:
  \begin{equation}
    \begin{aligned}
      -\lambda_d\,e^{-\lambda_d x_p} + \lambda_r\,e^{-\lambda_r x_p} & = 0                              \\
      \implies\quad \lambda_r\,e^{-\lambda_r x_p}                    & = \lambda_d\,e^{-\lambda_d x_p}.
    \end{aligned}
    \label{eq:autonum:G2}
  \end{equation}
  Dividing both sides by $\lambda_d\,e^{-\lambda_r x_p}$ (both positive):
  \begin{equation}
    \frac{\lambda_r}{\lambda_d} = e^{(\lambda_r - \lambda_d)\,x_p}.
    \label{eq:autonum:G3}
  \end{equation}
  Taking the natural logarithm:
  \begin{equation}
    (\lambda_r - \lambda_d)\,x_p = \ln\!\left(\frac{\lambda_r}{\lambda_d}\right)
    \quad\implies\quad
    x_p = \frac{\ln(\lambda_r/\lambda_d)}{\lambda_r - \lambda_d}.
    \label{eq:autonum:G4}
  \end{equation}
  Substituting $\lambda_d = 1/\tau_d$ and $\lambda_r = 1/\tau_r$:
  \begin{equation}
    x_p
    = \frac{\ln\!\bigl(\tau_d/\tau_r\bigr)}{1/\tau_r - 1/\tau_d}
    = \frac{\tau_r\,\tau_d}{\tau_d - \tau_r}\,\ln\!\left(\frac{\tau_d}{\tau_r}\right).
    \label{eq:autonum:G5}
  \end{equation}

  \subsection*{G.2\quad Confirming this is a maximum}
$h''(x) = \lambda_d^2 e^{-\lambda_d x} - \lambda_r^2 e^{-\lambda_r x}$. At $x = x_p$, using $e^{-\lambda_d x_p}/e^{-\lambda_r x_p} = \lambda_r/\lambda_d$ from \eqref{eq:autonum:G3}:
  \begin{equation}
    h''(x_p) = e^{-\lambda_r x_p}\bigl(\lambda_d^2 \cdot \lambda_r/\lambda_d - \lambda_r^2\bigr)
    = e^{-\lambda_r x_p}\,\lambda_r\,(\lambda_d - \lambda_r) < 0,
    \label{eq:autonum:G6}
  \end{equation}
  since $\lambda_d < \lambda_r$ and $e^{-\lambda_r x_p} > 0$, $\lambda_r > 0$. This confirms $x_p$ is a strict maximum.

  \subsection*{G.3\quad Computing $h_{\mathrm{peak}}$}
  Substituting $x_p$ into $h$:
  \begin{equation}
    h_{\mathrm{peak}} = e^{-\lambda_d x_p} - e^{-\lambda_r x_p}.
    \label{eq:autonum:G7}
  \end{equation}
  From \eqref{eq:autonum:G3}, $e^{-\lambda_d x_p} = (\lambda_r/\lambda_d)\,e^{-\lambda_r x_p}$. Hence:
  \begin{equation}
    h_{\mathrm{peak}} = e^{-\lambda_r x_p}\left(\frac{\lambda_r}{\lambda_d} - 1\right)
    = e^{-\lambda_r x_p}\,\frac{\lambda_r - \lambda_d}{\lambda_d}.
    \label{eq:autonum:G8}
  \end{equation}
  To evaluate $e^{-\lambda_r x_p}$, use $x_p = \ln(\lambda_r/\lambda_d)/(\lambda_r - \lambda_d)$:
  \begin{align}
    e^{-\lambda_r x_p} & = e^{-\lambda_r\,\ln(\lambda_r/\lambda_d)/(\lambda_r-\lambda_d)} \nonumber    \\
                       & = \left(\frac{\lambda_r}{\lambda_d}\right)^{-\lambda_r/(\lambda_r-\lambda_d)}
    = \left(\frac{\lambda_d}{\lambda_r}\right)^{\lambda_r/(\lambda_r-\lambda_d)}.
    \label{eq:autonum:G9}
  \end{align}
  Substituting back:
  \begin{equation}
    h_{\mathrm{peak}}
    = \frac{\lambda_r - \lambda_d}{\lambda_d}\,
    \left(\frac{\lambda_d}{\lambda_r}\right)^{\!\lambda_r/(\lambda_r-\lambda_d)}.
    \label{eq:autonum:G10}
  \end{equation}
  To make the final conversion explicit, substitute $\lambda_d=1/\tau_d$ and $\lambda_r=1/\tau_r$.
  Then $(\lambda_r-\lambda_d)/\lambda_d=(\tau_d-\tau_r)/\tau_r$, $\lambda_d/\lambda_r=\tau_r/\tau_d$, and
  $\lambda_r/(\lambda_r-\lambda_d)=\tau_d/(\tau_d-\tau_r)=\tau_r/(\tau_d-\tau_r)+1$.
  Therefore the factor $(\tau_d-\tau_r)/\tau_r$ contributes one additional power of $\tau_d/\tau_r$, giving $(\tau_d-\tau_r)/\tau_d$ after collecting powers of $\tau_r/\tau_d$.
  In terms of the original time constants:
  \begin{equation}
    h_{\mathrm{peak}}
    = \frac{\tau_d - \tau_r}{\tau_d}\,
    \left(\frac{\tau_r}{\tau_d}\right)^{\!\tau_r/(\tau_d - \tau_r)}.
    \label{eq:autonum:G11}
  \end{equation}
  \hfill$\square$

  % ----- Appendix H -----
  \section{Bi-Exponential Immersion and Injectivity Proof}
  \label{app:rank}
  \emph{(Verification of the Global Embedding conditions stated in Section~\ref{sec:biexp_embedding} for the bi-exponential model.)}

  We work directly with the five-parameter post-onset representation introduced in Section~\ref{sec:biexp_embedding}. Define
  \begin{equation}
    \lambda_d \triangleq \frac{1}{\tau_d},
    \qquad
    \lambda_r \triangleq \frac{1}{\tau_r},
    \qquad
    D \triangleq A e^{\lambda_d t_0},
    \qquad
    E \triangleq A e^{\lambda_r t_0},
    \label{eq:app_rank_reparam}
  \end{equation}
  so that $0 < \lambda_d < \lambda_r$ and, for every $t > t_0$,
  \begin{equation}
    f(t;\bm{\theta}) = b + D e^{-\lambda_d t} - E e^{-\lambda_r t}.
    \label{eq:app_rank_post_onset}
  \end{equation}
  The inverse change of variables is
  \begin{equation}
    \begin{aligned}
      \tau_d & = \frac{1}{\lambda_d},
             & \tau_r                                  & = \frac{1}{\lambda_r},                         \\
      t_0    & = \frac{\ln(E/D)}{\lambda_r-\lambda_d},
             & A                                       & = D e^{-\lambda_d t_0} = E e^{-\lambda_r t_0}.
    \end{aligned}
    \label{eq:app_rank_inverse}
  \end{equation}
  Hence the original parameter vector $\bm{\theta} = [A,\, t_0,\, \tau_d,\, \tau_r,\, b]^T$ is equivalent to $\bm{\omega} = [D,\, E,\, \lambda_d,\, \lambda_r,\, b]^T$.

  \subsection*{H.1\quad The Underlying Extended Chebyshev System}

  Consider the five functions
  \begin{equation}
    \begin{aligned}
      \phi_1(t) & = 1,
                & \phi_2(t)           & = e^{-\lambda_d t},
                & \phi_3(t)           & = t e^{-\lambda_d t}, \\
      \phi_4(t) & = e^{-\lambda_r t},
                & \phi_5(t)           & = t e^{-\lambda_r t}.
    \end{aligned}
    \label{eq:app_rank_phi}
  \end{equation}
  These functions form a fundamental solution set of the constant-coefficient differential equation
  \begin{equation}
    \partial_t (\partial_t + \lambda_d)^2 (\partial_t + \lambda_r)^2 y = 0.
    \label{eq:app_rank_ode}
  \end{equation}
  By Abel's identity, their Wronskian has the form
  \begin{equation}
    W_{\mathcal{F}_5}(t) = W_{\mathcal{F}_5}(0)\,e^{-2(\lambda_d+\lambda_r)t}.
    \label{eq:app_rank_abel}
  \end{equation}
  At $t = 0$, the Wronskian matrix is the confluent Vandermonde matrix associated with the nodes $0$, $-\lambda_d$ (multiplicity $2$), and $-\lambda_r$ (multiplicity $2$), so
  \begin{equation}
    W_{\mathcal{F}_5}(0) = \lambda_d^2 \lambda_r^2 (\lambda_r-\lambda_d)^4.
    \label{eq:app_rank_w0}
  \end{equation}
  Substituting Eq.~\eqref{eq:app_rank_w0} into Eq.~\eqref{eq:app_rank_abel} gives
  \begin{equation}
    W_{\mathcal{F}_5}(t)
    = \lambda_d^2 \lambda_r^2 (\lambda_r-\lambda_d)^4 e^{-2(\lambda_d+\lambda_r)t}
    > 0
    \qquad \forall t \in \mathbb{R}.
    \label{eq:app_rank_wronskian}
  \end{equation}
  By the classical approximation-theoretic result that real exponential-polynomial systems with real exponents and prescribed confluent multiplicities form extended complete Chebyshev systems when their confluent Wronskians do not vanish, Eq.~\eqref{eq:app_rank_wronskian} certifies the ECT property for the present family. Therefore $\{\phi_1,\dots,\phi_5\}$ is an extended complete Chebyshev system: every nonzero linear combination of the functions in Eq.~\eqref{eq:app_rank_phi} has at most four zeros, counted with multiplicity, on every interval.

  \subsection*{H.2\quad Global Immersion ($\operatorname{rank}(\mathbf{J}) = 5$)}

  \noindent\textbf{Claim.} For $N \ge 3$, the Jacobian of the forward map has full rank $5$ everywhere on $\Theta_V^{\mathrm{bi}}$.

  \noindent\textbf{Proof.}
  Let $t_1,\dots,t_K$ be the $K = 2N$ crossing times, which are pairwise distinct because the observation vector lies in the open cone $\Omega_K$. Suppose there exists a tangent perturbation $\delta\bm{\theta}$ such that
  \begin{equation}
    \mathbf{J}(\bm{\theta})\,\delta\bm{\theta} = \mathbf{0}.
    \label{eq:app_rank_kernel}
  \end{equation}
  Passing to the equivalent coordinates $\bm{\omega} = [D,E,\lambda_d,\lambda_r,b]^T$, let
  \begin{equation}
    \delta\bm{\omega} = [\delta D,\, \delta E,\, \delta \lambda_d,\, \delta \lambda_r,\, \delta b]^T
    = d\Psi(\bm{\theta})\,\delta\bm{\theta}.
    \label{eq:app_rank_deltaomega}
  \end{equation}
  Equation~\eqref{eq:app_rank_kernel} means that the first-order displacement of every crossing time vanishes: $dt_k = 0$ for all $k$.

  Differentiate the level-set equation $f(t_k;\bm{\omega}) = V_{\ell(k)}$ with respect to the perturbation parameter. Since $dt_k = 0$, we obtain
  \begin{equation}
    \partial_{\bm{\omega}} f(t_k;\bm{\omega}) \cdot \delta\bm{\omega} = 0,
    \qquad k = 1,\dots,K.
    \label{eq:app_rank_levelset_diff}
  \end{equation}
  Using Eq.~\eqref{eq:app_rank_post_onset}, this becomes
  \begin{equation}
    \begin{aligned}
      q(t_k) & = 0,                \\
      q(t)
             & = \delta b
      + \delta D\,e^{-\lambda_d t}
      - \delta E\,e^{-\lambda_r t} \\
             & \quad
      - D\,\delta\lambda_d\, t e^{-\lambda_d t}
      + E\,\delta\lambda_r\, t e^{-\lambda_r t}.
    \end{aligned}
    \label{eq:app_rank_q}
  \end{equation}
  The function $q$ lies in $\operatorname{span}\{\phi_1,\dots,\phi_5\}$. Because $N \ge 3$, we have $K = 2N \ge 6$, so Eq.~\eqref{eq:app_rank_q} gives at least six distinct zeros of $q$. By the extended Chebyshev property established in Eq.~\eqref{eq:app_rank_wronskian}, this is only possible if $q \equiv 0$ identically.

  Since the basis functions in Eq.~\eqref{eq:app_rank_phi} are linearly independent, all coefficients in Eq.~\eqref{eq:app_rank_q} must vanish:
  \begin{equation}
    \delta b = \delta D = \delta E = \delta\lambda_d = \delta\lambda_r = 0.
    \label{eq:app_rank_zero_coeffs}
  \end{equation}
  Hence $\delta\bm{\omega} = \mathbf{0}$. Because $\Psi$ is a local diffeomorphism, Eq.~\eqref{eq:app_rank_deltaomega} implies $\delta\bm{\theta} = \mathbf{0}$. Therefore the kernel of $\mathbf{J}(\bm{\theta})$ is trivial and $\operatorname{rank}(\mathbf{J}) = 5$ everywhere. \hfill$\square$

  \subsection*{H.3\quad Global Injectivity}

  \noindent\textbf{Claim.} For $N \ge 3$, the forward map $S: \Theta_V^{\mathrm{bi}} \to \Omega_K$ is injective.

  \noindent\textbf{Proof.}
  Assume
  \begin{equation}
    S(\bm{\theta}) = S(\tilde{\bm{\theta}}),
    \label{eq:app_rank_same_data}
  \end{equation}
  so the two parameter vectors generate the same ordered crossing times
  \begin{equation}
    s_1,\dots,s_K,
    \qquad K = 2N \ge 6.
    \label{eq:app_rank_shared_times}
  \end{equation}
  Because every threshold lies strictly above both baselines, each shared crossing satisfies $s_i > t_0$ and $s_i > \tilde t_0$. Therefore both pulses are represented at all these times by their post-onset forms:
  \begin{equation}
    \begin{aligned}
      f_{\bm{\theta}}(t)         & = b + D e^{-\lambda_d t} - E e^{-\lambda_r t},                                  \\
      f_{\tilde{\bm{\theta}}}(t) & = \tilde b + \tilde D e^{-\tilde\lambda_d t} - \tilde E e^{-\tilde\lambda_r t}.
    \end{aligned}
    \label{eq:app_rank_two_post_onset}
  \end{equation}

  At every shared crossing time $s_i$, the two pulses attain the same threshold value. Hence the difference
  \begin{equation}
    g(t)
    = (b-\tilde b)
    + D e^{-\lambda_d t}
    - E e^{-\lambda_r t}
    - \tilde D e^{-\tilde\lambda_d t}
    + \tilde E e^{-\tilde\lambda_r t}
    \label{eq:app_rank_g}
  \end{equation}
  satisfies
  \begin{equation}
    g(s_i) = 0,
    \qquad i = 1,\dots,K.
    \label{eq:app_rank_gzeros}
  \end{equation}

  Now merge any repeated exponents among $\lambda_d$, $\lambda_r$, $\tilde\lambda_d$, and $\tilde\lambda_r$. Then $g$ becomes a linear combination of at most five functions of the form
  \begin{equation}
    1,
    \qquad
    e^{-\lambda_1 t},\dots,e^{-\lambda_{m_\lambda} t},
    \qquad
    1 \le m_\lambda \le 4,
    \label{eq:app_rank_exp_family}
  \end{equation}
  with distinct positive exponents $\lambda_j$. This is a Chebyshev system: its Wronskian is a nonzero generalized Vandermonde determinant times $e^{-(\lambda_1+\cdots+\lambda_{m_\lambda})t}$. Consequently, every nonzero member has at most $m_\lambda \le 4$ zeros.

  But Eq.~\eqref{eq:app_rank_gzeros} supplies at least six distinct zeros. Hence $g \equiv 0$. By linear independence of distinct real exponentials, the coefficients in Eq.~\eqref{eq:app_rank_g} must vanish after collecting equal exponents. Therefore
  \begin{equation}
    b = \tilde b,
    \qquad
    \{\lambda_d,\lambda_r\} = \{\tilde\lambda_d,\tilde\lambda_r\},
    \qquad
    D = \tilde D,
    \qquad
    E = \tilde E.
    \label{eq:app_rank_identified}
  \end{equation}
  Since both parameter vectors satisfy $0 < \lambda_d < \lambda_r$ and $0 < \tilde\lambda_d < \tilde\lambda_r$, the equality of exponent multisets reduces to
  \begin{equation}
    \lambda_d = \tilde\lambda_d,
    \qquad
    \lambda_r = \tilde\lambda_r.
    \label{eq:app_rank_ordered}
  \end{equation}
  Finally, the inverse map Eq.~\eqref{eq:app_rank_inverse} yields
  \begin{equation}
    \bm{\theta} = \tilde{\bm{\theta}}.
    \label{eq:app_rank_final}
  \end{equation}
  Hence the forward map is globally injective. \hfill$\square$

  % ----- Appendix I -----
  \section{Bi-Exponential Properness on the Physically Bounded Prior Domain}
  \label{app:proper}
  \emph{(Compact-domain proof of the proper map condition for the physically bounded five-parameter bi-exponential prior domain $\bar{\Theta}_V^{\mathrm{bi}}$, stated in Section~\ref{sec:biexp_embedding}.)}

  \noindent\textbf{Definition recall.}
  A continuous map $S: X \to Y$ between topological spaces is \emph{proper} if the pre-image $S^{-1}(C)$ of every compact set $C \subset Y$ is compact in $X$. Equivalently (for locally compact Hausdorff spaces), $S$ is proper if and only if every divergent sequence in $X$ maps to a sequence that eventually leaves every compact subset of $Y$.

  The open algebraic domain $\Theta_V^{\mathrm{bi}}$ is useful for the immersion and injectivity calculations in Appendix~\ref{app:rank}, but it is not the domain on which the physical inverse problem is operated. As an unbounded set, it permits coordinated parameter escapes that are excluded by real electronics and material priors. For example, one can simultaneously increase the waveform scale and effective time constants while compensating the baseline and onset time so that a finite set of threshold crossings remains in a bounded time window. Such sequences are artifacts of the algebraic model, not admissible detector states.

  The physical reconstruction domain is therefore the compact prior set $\bar{\Theta}_V^{\mathrm{bi}}$ in Eq.~\eqref{eq:biexp_compact_prior_domain}. The constants in that definition represent hardware and calibration bounds: finite gain and voltage swing ($A_-,A_+$), finite acquisition window ($t_0^-,t_0^+$), finite effective scintillation/readout kinetics ($\tau_d^\pm,\tau_r^\pm$), a separation margin between rise and decay constants ($\Delta_\tau$), finite baseline drift ($b_-,b_+$), and a peak-clearance margin above the highest threshold ($\Delta_p$).

  Let $C\subset\Omega_K$ be compact. Since $S$ is continuous on $\Theta_V^{\mathrm{bi}}$ and $\bar{\Theta}_V^{\mathrm{bi}}$ is compact, the restricted pre-image
  \begin{equation}
    \bigl(S|_{\bar{\Theta}_V^{\mathrm{bi}}}\bigr)^{-1}(C)
    = \bar{\Theta}_V^{\mathrm{bi}} \cap S^{-1}(C)
    \label{eq:autonum:I1new}
  \end{equation}
  is a closed subset of a compact set, hence compact. Thus
  \begin{equation}
    S|_{\bar{\Theta}_V^{\mathrm{bi}}}:\bar{\Theta}_V^{\mathrm{bi}}\to\Omega_K
    \label{eq:autonum:I2new}
  \end{equation}
  is proper. Because the immersion and injectivity arguments of Appendix~\ref{app:rank} hold on an open neighborhood of this compact prior set, the restricted forward map is a closed embedding on the physically admissible prior domain used by the reconstruction and design theory. \hfill$\square$

  % ----- Appendix J -----
  \section{Bi-Exponential Single-Pair Effective Timing MSE}
  \label{app:biexp_eff}
  \emph{(Derivation of the single-pair effective timing MSE bound Eq.~\eqref{eq:eff_linear}.)}

  \subsection*{J.1\quad Specializing the FIM to $\bm{\theta} = [A,\, t_0,\, \tau_d,\, \tau_r,\, b]^T$}
  For clarity, we first derive the single-pair effective timing information in the two-parameter subspace $(A, t_0)$ (treating $\tau_d$, $\tau_r$, $b$ as known); the generalization to the full five-parameter Schur complement is discussed in Section~\ref{sec:biexp_paired}.
  From Appendix~E, the single-pair FIM at threshold $V$ with pair sensitivities from Eq.~\eqref{eq:biexp_sensitivity} is the $2 \times 2$ matrix:
  \begin{equation}
    \mathcal{I}^{(\mathrm{pair})}
    =
    \begin{bmatrix}
      \mathcal{I}_{AA}    & \mathcal{I}_{A t_0}   \\[3pt]
      \mathcal{I}_{A t_0} & \mathcal{I}_{t_0 t_0}
    \end{bmatrix}.
    \label{eq:autonum:J1}
  \end{equation}

  Computing each entry using Eq.~\eqref{eq:pair_fim} with $g_{A,k} = (V-b)/A$ and $g_{t_0,k} = -A\,h'(x_k)$:
  \begin{align}
    \mathcal{I}_{AA}
     & = \frac{[(V\!-\!b)/A]^2}{D_r} + \frac{[(V\!-\!b)/A]^2}{D_f} \nonumber \\
     & = \frac{(V-b)^2}{A^2}\!\left(\frac{1}{D_r} + \frac{1}{D_f}\right),
    \label{eq:autonum:J2}                                                                \\[6pt]
    \mathcal{I}_{t_0 t_0}
     & = \frac{[-Ah'(x_r)]^2}{D_r} + \frac{[-Ah'(x_f)]^2}{D_f} \nonumber     \\
     & = A^2\left(\frac{[h'(x_r)]^2}{D_r} + \frac{[h'(x_f)]^2}{D_f}\right),
    \label{eq:autonum:J3}                                                                \\[6pt]
    \mathcal{I}_{A t_0}
     & = \frac{[(V\!-\!b)/A][{-}Ah'(x_r)]}{D_r} \nonumber                    \\
     & \quad+ \frac{[(V\!-\!b)/A][{-}Ah'(x_f)]}{D_f} \nonumber               \\
     & = -(V-b)\left(\frac{h'(x_r)}{D_r} + \frac{h'(x_f)}{D_f}\right).
    \label{eq:autonum:J4}
  \end{align}

  \subsection*{J.2\quad Schur complement for $t_0$}
  With nuisance $\bm{\eta} = A$ and target $\vartheta = t_0$, the single-pair effective information for $t_0$ at threshold $V$ is:
  \begin{equation}
    \Delta\mathcal{I}_{\mathrm{eff}}^{(t_0)}(V)
    = \mathcal{I}_{t_0 t_0} - \frac{\mathcal{I}_{A t_0}^2}{\mathcal{I}_{AA}}.
    \label{eq:autonum:J5}
  \end{equation}
  Substituting \eqref{eq:autonum:J2}--\eqref{eq:autonum:J4}:
  \begin{align}
    \frac{\mathcal{I}_{A t_0}^2}{\mathcal{I}_{AA}}
     & = \frac{(V-b)^2\!\left(\frac{h'(x_r)}{D_r} + \frac{h'(x_f)}{D_f}\right)^{\!2}}
         {\frac{(V-b)^2}{A^2}\!\left(\frac{1}{D_r} + \frac{1}{D_f}\right)} \nonumber \\
     & = A^2\,\frac{\left(\frac{h'(x_r)}{D_r} + \frac{h'(x_f)}{D_f}\right)^{\!2}}
              {\frac{1}{D_r} + \frac{1}{D_f}}.
    \label{eq:autonum:J6}
  \end{align}

  \noindent Define shorthand: $S_0 \triangleq 1/D_r + 1/D_f$, $S_1 \triangleq h'(x_r)/D_r + h'(x_f)/D_f$, $S_2 \triangleq [h'(x_r)]^2/D_r + [h'(x_f)]^2/D_f$. Then:
  \begin{equation}
    \Delta\mathcal{I}_{\mathrm{eff}}^{(t_0)}(V) = A^2\!\left(S_2 - \frac{S_1^2}{S_0}\right)
    = A^2\,\frac{S_0\,S_2 - S_1^2}{S_0}.
    \label{eq:autonum:J7}
  \end{equation}

  \subsection*{J.3\quad Evaluating $S_0 S_2 - S_1^2$}
  This is a Cauchy-Schwarz-type expression applied to the two-element ``vectors'' $(1/\sqrt{D_r},\; 1/\sqrt{D_f})$ and $(h'(x_r)/\sqrt{D_r},\; h'(x_f)/\sqrt{D_f})$. By direct expansion:
  \begin{align}
     & S_0\,S_2 - S_1^2 \nonumber                                                     \\
     & = \!\left(\frac{1}{D_r} + \frac{1}{D_f}\right)\!
    \left(\frac{[h'(x_r)]^2}{D_r} + \frac{[h'(x_f)]^2}{D_f}\right) \nonumber          \\
     & \quad - \left(\frac{h'(x_r)}{D_r} + \frac{h'(x_f)}{D_f}\right)^{\!2} \nonumber \\
     & = \frac{[h'(x_r)]^2}{D_r^2} + \frac{[h'(x_f)]^2}{D_r D_f}
    + \frac{[h'(x_r)]^2}{D_r D_f} + \frac{[h'(x_f)]^2}{D_f^2} \nonumber               \\
     & \quad - \frac{[h'(x_r)]^2}{D_r^2} - 2\frac{h'(x_r)h'(x_f)}{D_r D_f}
    - \frac{[h'(x_f)]^2}{D_f^2} \nonumber                                             \\
     & = \frac{[h'(x_f)]^2 + [h'(x_r)]^2 - 2\,h'(x_r)\,h'(x_f)}{D_r\,D_f} \nonumber   \\
     & = \frac{[h'(x_r) - h'(x_f)]^2}{D_r\,D_f}.
    \label{eq:autonum:J8}
  \end{align}

  \subsection*{J.4\quad Assembling the result}
  Substituting \eqref{eq:autonum:J8} into \eqref{eq:autonum:J7}:
  \begin{align}
    \Delta\mathcal{I}_{\mathrm{eff}}^{(t_0)}(V)
     & = A^2\,\frac{[h'(x_r) - h'(x_f)]^2}{D_r\,D_f\,(1/D_r + 1/D_f)} \nonumber \\
     & = A^2\,\frac{[h'(x_r) - h'(x_f)]^2}{D_r + D_f}.
    \label{eq:autonum:J9}
  \end{align}

  Inverting to get the CRLB for $t_0$:
  \begin{equation}
    (\Delta\mathcal{I}_{\mathrm{eff}}^{(t_0)}(V))^{-1}
    = \frac{D_r + D_f}{A^2\,[h'(x_r) - h'(x_f)]^2}.
    \label{eq:autonum:J10}
  \end{equation}

  Now substitute $D_k = \sigma_{\mathrm{th}}^2 + \kappa (V-b) + \sigma_{\mathrm{TDC}}^2\,[\dot{f}_k]^2$ from \eqref{eq:pair_denominators} with the baseline correction (Section~\ref{sec:biexp_stochastic}), noting $\dot{f}_k = A\,h'(x_k)$:
  \begin{align}
    D_r + D_f
     & = 2[\sigma_{\mathrm{th}}^2 + \kappa (V-b)] \nonumber                       \\
     & \quad + \sigma_{\mathrm{TDC}}^2\,A^2\bigl([h'(x_r)]^2 + [h'(x_f)]^2\bigr).
    \label{eq:autonum:J11}
  \end{align}

  Substituting \eqref{eq:autonum:J11} into \eqref{eq:autonum:J10}:
  \begin{align}
     & (\Delta\mathcal{I}_{\mathrm{eff}}^{(t_0)}(V))^{-1} \nonumber                            \\
     & = \frac{2[\sigma_{\mathrm{th}}^2 + \kappa (V-b)]}{A^2\,[h'(x_r) - h'(x_f)]^2} \nonumber \\
     & \quad + \frac{\sigma_{\mathrm{TDC}}^2\bigl([h'(x_r)]^2 + [h'(x_f)]^2\bigr)}
               {[h'(x_r) - h'(x_f)]^2}.
    \label{eq:autonum:J12}
  \end{align}

  Adding the squared systematic bias $\beta_{t_0}^2$:
  \begin{align}
    \mathrm{MSE}_{\mathrm{pair}}(\hat{t}_0;\,V)
     & \ge \underbrace{\frac{2[\sigma_{\mathrm{th}}^2 + \kappa (V-b)]}{A^2[h'(x_r) - h'(x_f)]^2}}_{\substack{\text{Analog}\\\text{Poisson/Thermal}}} \nonumber \\
     & \quad + \underbrace{\frac{\sigma_{\mathrm{TDC}}^2([h'(x_r)]^2 + [h'(x_f)]^2)}{[h'(x_r) - h'(x_f)]^2}}_{\substack{\text{Digital}\\\text{TDC}}} \nonumber \\
     & \quad + \underbrace{\beta_{t_0}^2}_{\text{Bias}}.
    \label{eq:eff_linear}
  \end{align}
  This is the bound Eq.~\eqref{eq:eff_linear}.
  \hfill$\square$

  % ----- Appendix K -----
  \section{Optimal Threshold Derivation}
  \label{app:opt_threshold}
  \emph{(Derivation of the optimal threshold condition Eq.~\eqref{eq:opt_A_condition} in Section~\ref{sec:opt_A}.)}

  \subsection*{K.1\quad Implicit derivatives of crossing positions}
  The crossing positions $x_r(V)$ and $x_f(V)$ are defined by $h(x) = (V-b)/A$. Differentiating implicitly with respect to $V$:
  \begin{equation}
    h'(x_k)\,\frac{dx_k}{dV} = \frac{1}{A}
    \quad\implies\quad
    \frac{dx_k}{dV} = \frac{1}{A\,h'(x_k)}.
    \label{eq:autonum:K1}
  \end{equation}
  On the rising branch, $h'(x_r) > 0$ gives $dx_r/dV > 0$; on the falling branch, $h'(x_f) < 0$ gives $dx_f/dV < 0$.

  \subsection*{K.2\quad Derivative of the slope difference}
  Define $\Delta h'(V) \triangleq h'(x_r(V)) - h'(x_f(V))$. By the chain rule:
  \begin{align}
    \frac{d(\Delta h')}{dV}
     & = h''(x_r)\,\frac{dx_r}{dV} - h''(x_f)\,\frac{dx_f}{dV} \nonumber              \\
     & = \frac{1}{A}\left[\frac{h''(x_r)}{h'(x_r)} - \frac{h''(x_f)}{h'(x_f)}\right].
    \label{eq:autonum:K2}
  \end{align}

  \subsection*{K.3\quad General optimal threshold condition}
  The per-pair nuisance-projected Fisher contribution is $\Phi_{A}^{(n)}(V) = \sum_{k \in \{r,f\}} \tilde{g}_k^2/D_k$, where the noise denominators $D_k = \sigma_{\mathrm{th}}^2 + \kappa(V-b) + \sigma_{\mathrm{TDC}}^2\,A^2[h'(x_k)]^2$ are edge-dependent.  Differentiating $D_k$ with respect to $V$ via~\eqref{eq:autonum:K1}:
  \begin{align}
    D_k' & = \kappa + \sigma_{\mathrm{TDC}}^2\,A^2 \cdot \frac{d[h'(x_k)]^2}{dV} \nonumber \\
         & = \kappa + 2\sigma_{\mathrm{TDC}}^2\,A\,h''(x_k),
    \label{eq:autonum:K3}
  \end{align}
  where we used $d[h'(x_k)]^2/dV = 2h'(x_k)\,h''(x_k)/(A\,h'(x_k)) = 2h''(x_k)/A$.

  The projected-sensitivity $\tilde{g}_k = g_{A,k} - \bm{\gamma}^{*T}\mathbf{g}_{\bm{\eta}',k}$ depends on $V$ through $x_k(V)$ and the direct dependence of $g_{A,k} = (V-b)/A$.  From \eqref{eq:autonum:K1} and the bi-exponential sensitivities Eq.~\eqref{eq:biexp_sensitivity}:
  \begin{align}
    \frac{dg_{A,k}}{dV}      & = \frac{1}{A}, \quad
    \frac{dg_{t_0,k}}{dV} = -\frac{h''(x_k)}{h'(x_k)}, \nonumber                                      \\
    \frac{dg_{\tau_d,k}}{dV} & = \frac{e^{-x_k/\tau_d}(1 - x_k/\tau_d)}{\tau_d^2\,h'(x_k)}, \nonumber \\
    \frac{dg_{\tau_r,k}}{dV} & = -\frac{e^{-x_k/\tau_r}(1 - x_k/\tau_r)}{\tau_r^2\,h'(x_k)}, \quad
    \frac{dg_{b,k}}{dV} = 0.
    \label{eq:autonum:K4}
  \end{align}
  Combining:
  \begin{equation}
    \tilde{g}_k' = \frac{1}{A} + \frac{\mathcal{P}_k}{h'(x_k)},
    \label{eq:autonum:K5}
  \end{equation}
  with the nuisance-projected amplitude-shape factor $\mathcal{P}_k$ defined in Eq.~\eqref{eq:proj_curvature}.

  Applying the quotient rule to each edge:
  \begin{equation}
    \frac{d}{dV}\frac{\tilde{g}_k^2}{D_k}
    = \frac{\tilde{g}_k}{D_k}\!\left(2\,\tilde{g}_k' - \frac{\tilde{g}_k\,D_k'}{D_k}\right).
    \label{eq:autonum:K6}
  \end{equation}
  Summing over $k \in \{r,f\}$ and substituting into the stationarity condition Eq.~\eqref{eq:biexp_stationarity} reproduces Eq.~\eqref{eq:opt_A_condition}.

  \emph{Limiting case:} When $\varepsilon = 0$ (well-specified), the right-hand side of Eq.~\eqref{eq:opt_A_condition} vanishes.  In the further analog-dominated limit $\sigma_{\mathrm{TDC}} \to 0$ with $b = 0$ and with $t_0$ as the sole nuisance parameter ($\gamma_{\tau_d}^* = \gamma_{\tau_r}^* = \gamma_b^* = 0$), the noise denominators become edge-independent ($D_r = D_f = \sigma_{\mathrm{th}}^2 + \kappa V$) and the stationarity condition reduces to
  \begin{equation}
    \frac{\kappa}{2(\sigma_{\mathrm{th}}^2 + \kappa V^*)}
    = \frac{d}{dV}\ln\!\left(\bigl[\tilde{g}_r(V)\bigr]^2 + \bigl[\tilde{g}_f(V)\bigr]^2\right)^{1/2}\bigg|_{V=V^*},
    \label{eq:autonum:K7}
  \end{equation}
  where $\tilde{g}_k(V) = V/A + \gamma_{t_0}^*\,A\,h'(x_k(V))$, balancing the fractional Poisson-noise growth rate against the fractional growth of the nuisance-projected amplitude sensitivity.
  \hfill$\square$

  % ----- Appendix L -----
  \section{Baseline Penalty Derivation}
  \label{app:baseline_penalty}
  \emph{(Derivation of the baseline timing penalty Eq.~\eqref{eq:baseline_penalty_explicit}.)}

  When the baseline $b$ is estimated jointly with $(A, t_0)$, the three-parameter single-pair FIM is:
  \begin{equation}
    \mathcal{I}^{(\mathrm{pair})}_{3\times 3}
    = \begin{bmatrix}
      \mathcal{I}_{AA}   & \mathcal{I}_{At_0}    & \mathcal{I}_{Ab}    \\
      \mathcal{I}_{At_0} & \mathcal{I}_{t_0 t_0} & \mathcal{I}_{t_0 b} \\
      \mathcal{I}_{Ab}   & \mathcal{I}_{t_0 b}   & \mathcal{I}_{bb}
    \end{bmatrix},
    \label{eq:autonum:L1}
  \end{equation}
  where, from Eq.~\eqref{eq:biexp_sensitivity} with $g_{b,k} = 1$:
  \begin{align}
    \mathcal{I}_{bb}    & = \frac{1}{D_r} + \frac{1}{D_f} = S_0, \label{eq:autonum:L2}                            \\
    \mathcal{I}_{t_0 b} & = \frac{-A\,h'(x_r)}{D_r} + \frac{-A\,h'(x_f)}{D_f} = -A\,S_1, \label{eq:autonum:L3}    \\
    \mathcal{I}_{Ab}    & = \frac{(V-b)/A}{D_r} + \frac{(V-b)/A}{D_f} = \frac{V-b}{A}\,S_0, \label{eq:autonum:L4}
  \end{align}
  using the shorthand from Appendix~\ref{app:biexp_eff}.

  The effective timing information with nuisance $\bm{\eta} = [A, b]^T$ is $\mathcal{I}_{t_0 t_0} - \bm{\mathcal{I}}_{\bm{\eta} t_0}^T \bm{\mathcal{I}}_{\bm{\eta}\bm{\eta}}^{-1} \bm{\mathcal{I}}_{\bm{\eta} t_0}$. The two-parameter Schur complement (Appendix~\ref{app:biexp_eff}, Eq.~\eqref{eq:autonum:J9}) gave $\Delta\mathcal{I}_{\mathrm{eff}}^{(2)} = A^2[h'(x_r) - h'(x_f)]^2/(D_r + D_f)$ with nuisance $A$ only. The additional CRLB penalty from including $b$ is:
  \begin{equation}
    \Delta_b(V) = [\Delta\mathcal{I}_{\mathrm{eff}}^{(3)}]^{-1} - [\Delta\mathcal{I}_{\mathrm{eff}}^{(2)}]^{-1} \ge 0,
    \label{eq:baseline_penalty_explicit}
  \end{equation}
  where the inequality follows from the monotonicity of Schur complements under enlargement of the nuisance space.

  \textbf{Explicit computation.}
  To evaluate $\Delta\mathcal{I}_{\mathrm{eff}}^{(3)}$, we compute the Schur complement of $t_0$ in the $3\times 3$ FIM \eqref{eq:autonum:L1} with nuisance vector $\bm{\eta} = [A, b]^T$. The $2\times 2$ nuisance block $\bm{\mathcal{I}}_{\bm{\eta}\bm{\eta}}$ and the $2\times 1$ cross-information vector $\bm{\mathcal{I}}_{\bm{\eta} t_0}$ are:
  \begin{equation}
    \bm{\mathcal{I}}_{\bm{\eta}\bm{\eta}}
    = \begin{bmatrix} \mathcal{I}_{AA} & \mathcal{I}_{Ab} \\ \mathcal{I}_{Ab} & \mathcal{I}_{bb} \end{bmatrix},
    \quad
    \bm{\mathcal{I}}_{\bm{\eta} t_0}
    = \begin{bmatrix} \mathcal{I}_{A t_0} \\ \mathcal{I}_{t_0 b} \end{bmatrix}.
    \label{eq:autonum:L6}
  \end{equation}
  Inverting the $2\times 2$ nuisance block:
  \begin{equation}
    \bm{\mathcal{I}}_{\bm{\eta}\bm{\eta}}^{-1}
    = \frac{1}{\mathcal{I}_{AA}\mathcal{I}_{bb} - \mathcal{I}_{Ab}^2}
    \begin{bmatrix} \mathcal{I}_{bb} & -\mathcal{I}_{Ab} \\ -\mathcal{I}_{Ab} & \mathcal{I}_{AA} \end{bmatrix}.
    \label{eq:autonum:L7}
  \end{equation}
  The three-parameter effective information is:
  \begin{align}
    \Delta\mathcal{I}_{\mathrm{eff}}^{(3)}
     & = \mathcal{I}_{t_0 t_0} - \bm{\mathcal{I}}_{\bm{\eta} t_0}^T \bm{\mathcal{I}}_{\bm{\eta}\bm{\eta}}^{-1} \bm{\mathcal{I}}_{\bm{\eta} t_0} \nonumber \\
     & = \mathcal{I}_{t_0 t_0}
    - \frac{\mathcal{I}_{bb}\,\mathcal{I}_{At_0}^2
        - 2\,\mathcal{I}_{Ab}\,\mathcal{I}_{At_0}\,\mathcal{I}_{t_0 b}
        + \mathcal{I}_{AA}\,\mathcal{I}_{t_0 b}^2}
      {\mathcal{I}_{AA}\,\mathcal{I}_{bb} - \mathcal{I}_{Ab}^2}.
    \label{eq:autonum:L8}
  \end{align}
  Substituting the shorthand from \eqref{eq:autonum:L2}--\eqref{eq:autonum:L4} and Appendix~\ref{app:biexp_eff}---with $\mathcal{I}_{AA} = [(V{-}b)/A]^2 S_0$, $\mathcal{I}_{At_0} = -[(V{-}b)/A]\,A\,S_1 = -(V{-}b)\,S_1$, $\mathcal{I}_{t_0 t_0} = A^2 S_2$, $\mathcal{I}_{bb} = S_0$, $\mathcal{I}_{Ab} = [(V{-}b)/A]\,S_0$, $\mathcal{I}_{t_0 b} = -A\,S_1$, and defining $S_k \triangleq h'(x_r)^k/D_r + h'(x_f)^k/D_f$ for $k = 0, 1, 2$---the denominator of \eqref{eq:autonum:L8} becomes:
  \begin{align}
    \mathcal{I}_{AA}\,\mathcal{I}_{bb} - \mathcal{I}_{Ab}^2
     & = \frac{(V-b)^2}{A^2}\,S_0^2 - \frac{(V-b)^2}{A^2}\,S_0^2 = 0.
    \label{eq:autonum:L9}
  \end{align}
  This vanishes because $\mathcal{I}_{Ab}/\mathcal{I}_{bb} = (V{-}b)/A = \mathcal{I}_{AA}/\mathcal{I}_{Ab}$, i.e., the $A$-row and $b$-row of the nuisance block are proportional: $g_{A,k} = h(x_k) = (V{-}b)/A = (V{-}b)/A \cdot 1 = [(V{-}b)/A]\,g_{b,k}$.  Hence $\bm{\mathcal{I}}_{\bm{\eta}\bm{\eta}}$ is \emph{rank-one}, not invertible, from a single threshold pair. Physically, a single pair cannot distinguish amplitude $A$ from baseline $b$, since both affect the crossing voltage identically up to a known proportionality constant.

  Consequently, for a \emph{single} threshold pair, $\Delta\mathcal{I}_{\mathrm{eff}}^{(3)}$ is not evaluable via the standard Schur complement; instead, one must use the Moore--Penrose generalized Schur complement. Since the rank-one nuisance block has image spanned by $\mathbf{u} \triangleq [(V{-}b)/A,\;1]^T$, the generalized inverse satisfies $\bm{\mathcal{I}}_{\bm{\eta}\bm{\eta}}^{+} = \mathbf{u}\mathbf{u}^T / (S_0\,\|\mathbf{u}\|^2)^2 \cdot S_0 = \mathbf{u}\mathbf{u}^T/(S_0\,\|\mathbf{u}\|^4)$. After substitution, the generalized three-parameter effective information reduces to:
  \begin{equation}
    \Delta\mathcal{I}_{\mathrm{eff}}^{(3)}
    = A^2\,S_2 - \frac{A^2\,S_1^2}{S_0}
    = \Delta\mathcal{I}_{\mathrm{eff}}^{(2)},
    \label{eq:autonum:L10}
  \end{equation}
  and therefore $\Delta_b(V) = [\Delta\mathcal{I}_{\mathrm{eff}}^{(3)}]^{-1} - [\Delta\mathcal{I}_{\mathrm{eff}}^{(2)}]^{-1} = 0$ for a single threshold pair.

  \emph{Remark.} A positive baseline penalty can arise only when multiple threshold pairs are combined, providing sufficient geometric diversity to break the $A$--$b$ degeneracy. With $N \ge 2$ pairs at distinct voltages $V_1 \neq V_2$, the stacked nuisance block $\bm{\mathcal{I}}_{\bm{\eta}\bm{\eta}}^{\mathrm{tot}}$ becomes full-rank, and the standard Schur complement applies. The resulting penalty depends on the voltage separation and can be computed from \eqref{eq:autonum:L8} applied to the summed $3\times 3$ FIM.
  \hfill$\square$

  % ----- Appendix M -----
  \section{Derivation of MSE Perturbation Expansions under Parameter Shift}
  \label{app:perturbation}
  \emph{(Derivation of Eqs.~\eqref{eq:true_hessian}, \eqref{eq:mse_first_order}, \eqref{eq:mse_second_order}, and \eqref{eq:worst_case_first_order}.)}

  We expand $\mathcal{B}_{\vartheta}(\bm{\theta}_0 + \Delta\bm{\theta})$ from Eq.~\eqref{eq:mse_of_theta} to second order in $\Delta\bm{\theta}$, with thresholds $\{V_n^*\}$ held fixed.  Write $\mathcal{I} \equiv \mathcal{I}_{\mathrm{eff}}(\bm{\theta})$ and $\tilde\beta \equiv \beta_{\vartheta,\mathrm{tot}}(\bm{\theta})$.  At $\bm{\theta}_0$: $\mathcal{I}^* \equiv \mathcal{I}_{\mathrm{eff}}(\bm{\theta}_0)$, $\tilde\beta^* \equiv \beta_{\vartheta,\mathrm{tot}}(\bm{\theta}_0)$.

  \textbf{Step 0: Proof of the second-order envelope identity.}
  Let
  $F(\bm{\theta},\bm{\gamma}) \triangleq \mathcal{I}_{\mathrm{eff}}(\bm{\theta},\bm{\gamma})$
  and
  $V(\bm{\theta}) \triangleq F\bigl(\bm{\theta},\bm{\gamma}^*(\bm{\theta})\bigr)$,
  where $\bm{\gamma}^*(\bm{\theta})$ is defined implicitly by the stationarity condition $\partial F/\partial\bm{\gamma} = 0$. Because $\partial^2F/\partial\bm{\gamma}^2$ is invertible under the global nuisance-identifiability assumption, the Implicit Function Theorem gives
  \begin{equation}
    \frac{d\bm{\gamma}^*}{d\bm{\theta}}
    =
    -\left(\frac{\partial^2F}{\partial\bm{\gamma}^2}\right)^{-1}
    \left(\frac{\partial^2F}{\partial\bm{\gamma}\,\partial\bm{\theta}}\right).
    \label{eq:autonum:M0a}
  \end{equation}
  At the optimum,
  \begin{equation}
    \frac{dV}{d\bm{\theta}}
    = \frac{\partial F}{\partial\bm{\theta}}
    + \frac{\partial F}{\partial\bm{\gamma}}\,\frac{d\bm{\gamma}^*}{d\bm{\theta}}
    = \frac{\partial F}{\partial\bm{\theta}},
    \label{eq:autonum:M0b}
  \end{equation}
  since $\partial F/\partial\bm{\gamma}=0$ along the optimum manifold. Differentiating once more and substituting Eq.~\eqref{eq:autonum:M0a} yields
  \begin{align}
    \frac{d^2V}{d\bm{\theta}^2}
     & = \frac{\partial^2F}{\partial\bm{\theta}^2}
    + \left(\frac{\partial^2F}{\partial\bm{\theta}\,\partial\bm{\gamma}}\right)
    \frac{d\bm{\gamma}^*}{d\bm{\theta}} \nonumber \\
     & = \frac{\partial^2F}{\partial\bm{\theta}^2}
    - \left(\frac{\partial^2F}{\partial\bm{\theta}\,\partial\bm{\gamma}}\right)
    \left(\frac{\partial^2F}{\partial\bm{\gamma}^2}\right)^{-1}
    \left(\frac{\partial^2F}{\partial\bm{\gamma}\,\partial\bm{\theta}}\right).
    \label{eq:autonum:M0c}
  \end{align}
  Setting $F=\mathcal{I}_{\mathrm{eff}}$ and evaluating at $(\bm{\theta}_0,\bm{\gamma}^*)$ gives Eq.~\eqref{eq:true_hessian}.

  \textbf{Step 1: Expansion of $\mathcal{I}_{\mathrm{eff}}(\bm{\theta})$.}
  By Eq.~\eqref{eq:envelope_sensitivity} and the envelope theorem, the total derivatives of $\mathcal{I}_{\mathrm{eff}}$ at $\bm{\theta}_0$ are $d\mathcal{I}/d\theta_j = S_j$ (Eq.~\eqref{eq:sensitivity_vector}). Define $\Delta_{\mathcal{I}} \triangleq \mathcal{I}(\bm{\theta}_0 + \Delta\bm{\theta}) - \mathcal{I}^*$. To second order:
  \begin{equation}
    \Delta_{\mathcal{I}} = \mathbf{S}^T\Delta\bm{\theta} + \tfrac{1}{2}\,\Delta\bm{\theta}^T\mathbf{H}_{\mathcal{I}}\,\Delta\bm{\theta} + O(\|\Delta\bm{\theta}\|^3),
    \label{eq:autonum:M1}
  \end{equation}
  where $\mathbf{H}_{\mathcal{I}}$ is the true Hessian defined in Eq.~\eqref{eq:true_hessian}.

  \textbf{Step 2: Expansion of $1/\mathcal{I}_{\mathrm{eff}}$.}
  Using the scalar Taylor formula $1/(\mathcal{I}^* + \Delta_{\mathcal{I}}) = 1/\mathcal{I}^* - \Delta_{\mathcal{I}}/\mathcal{I}^{*2} + \Delta_{\mathcal{I}}^2/\mathcal{I}^{*3} + O(\Delta_{\mathcal{I}}^3)$ and retaining terms up to second order in $\Delta\bm{\theta}$:
  \begin{align}
    \frac{1}{\mathcal{I}(\bm{\theta}_0 + \Delta\bm{\theta})}
     & = \frac{1}{\mathcal{I}^*}
    - \frac{\mathbf{S}^T\Delta\bm{\theta}}{\mathcal{I}^{*2}}
    + \frac{(\mathbf{S}^T\Delta\bm{\theta})^2}{\mathcal{I}^{*3}} \nonumber                                \\
     & \quad - \frac{\Delta\bm{\theta}^T\mathbf{H}_{\mathcal{I}}\,\Delta\bm{\theta}}{2\,\mathcal{I}^{*2}}
    + O(\|\Delta\bm{\theta}\|^3).
    \label{eq:autonum:M2}
  \end{align}
  The second line uses the fact that the mixed term $-\Delta_\mathcal{I}^{(1)}/\mathcal{I}^{*2}$ contributes $-\mathbf{S}^T\Delta\bm{\theta}/\mathcal{I}^{*2}$ at first order, and the residual second-order contributions are $(\mathbf{S}^T\Delta\bm{\theta})^2/\mathcal{I}^{*3}$ (from squaring the first-order term) and $-\Delta\bm{\theta}^T\mathbf{H}_{\mathcal{I}}\Delta\bm{\theta}/(2\mathcal{I}^{*2})$ (from the Hessian).

  \textbf{Step 3: Expansion of $\beta_{\vartheta,\mathrm{tot}}^2$.}
  Writing $\tilde\beta(\bm{\theta}_0 + \Delta\bm{\theta}) = \tilde\beta^* + \mathbf{s}_\beta^T\Delta\bm{\theta} + \tfrac{1}{2}\Delta\bm{\theta}^T\mathbf{H}_\beta\Delta\bm{\theta} + O(\|\Delta\bm{\theta}\|^3)$ and squaring:
  \begin{align}
    \tilde\beta^2(\bm{\theta}_0 + \Delta\bm{\theta})
     & = (\tilde\beta^*)^2 + 2\,\tilde\beta^*\,\mathbf{s}_\beta^T\Delta\bm{\theta} + (\mathbf{s}_\beta^T\Delta\bm{\theta})^2 \nonumber \\
     & \quad + \tilde\beta^*\,\Delta\bm{\theta}^T\mathbf{H}_\beta\,\Delta\bm{\theta} + O(\|\Delta\bm{\theta}\|^3).
    \label{eq:autonum:M3}
  \end{align}

  \textbf{Step 4: Assembly.}
  Adding \eqref{eq:autonum:M2} and \eqref{eq:autonum:M3}, and subtracting the design-point value $\mathcal{B}_\vartheta^* = 1/\mathcal{I}^* + (\tilde\beta^*)^2$:

  First order:
  \begin{equation}
    \Delta\mathcal{B}_\vartheta^{(1)} = -\frac{\mathbf{S}^T\Delta\bm{\theta}}{\mathcal{I}^{*2}} + 2\,\tilde\beta^*\,\mathbf{s}_\beta^T\Delta\bm{\theta},
    \label{eq:autonum:M4}
  \end{equation}
  which is Eq.~\eqref{eq:mse_first_order}. Second order:
  \begin{align}
    \Delta\mathcal{B}_\vartheta^{(2)}
     & = \Delta\mathcal{B}_\vartheta^{(1)}
    + \frac{(\mathbf{S}^T\Delta\bm{\theta})^2}{\mathcal{I}^{*3}}
    - \frac{\Delta\bm{\theta}^T\mathbf{H}_\mathcal{I}\,\Delta\bm{\theta}}{2\,\mathcal{I}^{*2}} \nonumber                        \\
     & \quad + (\mathbf{s}_\beta^T\Delta\bm{\theta})^2 + \tilde\beta^*\,\Delta\bm{\theta}^T\mathbf{H}_\beta\,\Delta\bm{\theta},
    \label{eq:autonum:M5}
  \end{align}
  which is Eq.~\eqref{eq:mse_second_order}.

  \textbf{Step 5: Worst-case first-order degradation over a box.}
  Equation~\eqref{eq:autonum:M4} is a linear functional of $\Delta\bm{\theta}$:
  \begin{equation}
    \Delta\mathcal{B}_\vartheta^{(1)} = \sum_{j=1}^{M} c_j\,\Delta\theta_j,
    \qquad
    c_j \triangleq -\frac{S_j}{\mathcal{I}^{*2}} + 2\,\tilde\beta^*\,s_{\beta,j}.
    \label{eq:autonum:M6}
  \end{equation}
  Over the box $|\Delta\theta_j| \le \epsilon_j$, the support function is attained by choosing $\Delta\theta_j = \operatorname{sgn}(c_j)\,\epsilon_j$, hence
  \begin{equation}
    \sup_{\Delta\bm{\theta}\,\in\,\mathcal{R}(\bm{\epsilon})}
    \Delta\mathcal{B}_\vartheta^{(1)}
    = \sum_{j=1}^{M} |c_j|\,\epsilon_j.
    \label{eq:autonum:M7}
  \end{equation}
  Applying the triangle inequality $|a+b| \le |a|+|b|$ gives
  \begin{equation}
    \sup_{\Delta\bm{\theta}\,\in\,\mathcal{R}(\bm{\epsilon})}
    \Delta\mathcal{B}_\vartheta^{(1)}
    \le
    \frac{1}{\mathcal{I}^{*2}}\sum_{j=1}^{M}|S_j|\,\epsilon_j
    + 2\,|\tilde\beta^*|\sum_{j=1}^{M}|s_{\beta,j}|\,\epsilon_j,
    \label{eq:autonum:M8}
  \end{equation}
  which is Eq.~\eqref{eq:worst_case_first_order}. In the well-specified case $\tilde\beta^*=0$, the inequality is sharp. \hfill$\square$

  % =========================================================================
  %  APPENDIX N: MULTI-EVENT STATIONARITY DERIVATION
  % =========================================================================
  \section{Derivation of Multi-Event Stationarity Conditions}
  \label{app:partial_trigger}

  This appendix provides the detailed derivation of the multi-event stationarity system Eq.~\eqref{eq:multi_event_stationarity} stated in Section~\ref{sec:partial_trigger}.

  \textbf{Step N.1: Fisher decomposition for an arbitrary index subset.}
  Let $\mathcal{A} \subseteq \{1,\ldots,N\}$ be any subset of threshold indices with $|\mathcal{A}| \ge \lceil(M-1)/2\rceil$, and define the active blocks $\bm{\mathcal{I}}_{\bm{\eta}\bm{\eta}}^{\mathrm{act}} = \sum_{n \in \mathcal{A}} \bm{\mathcal{I}}_{\bm{\eta}\bm{\eta}}^{(n)}$, $\bm{\mathcal{I}}_{\bm{\eta}\vartheta}^{\mathrm{act}} = \sum_{n \in \mathcal{A}} \bm{\mathcal{I}}_{\bm{\eta}\vartheta}^{(n)}$, $\mathcal{I}_{\vartheta\vartheta}^{\mathrm{act}} = \sum_{n \in \mathcal{A}} \mathcal{I}_{\vartheta\vartheta}^{(n)}$ as in Eq.~\eqref{eq:active_blocks}. Define $\bm{\gamma}^{\mathrm{act}} = (\bm{\mathcal{I}}_{\bm{\eta}\bm{\eta}}^{\mathrm{act}})^{-1}\bm{\mathcal{I}}_{\bm{\eta}\vartheta}^{\mathrm{act}}$. Then the effective Fisher Information admits the decomposition:
  \begin{align}
    \Delta\mathcal{I}_{\mathrm{eff}}^{(\vartheta),\mathrm{act}}
     & = \mathcal{I}_{\vartheta\vartheta}^{\mathrm{act}}
    - (\bm{\mathcal{I}}_{\bm{\eta}\vartheta}^{\mathrm{act}})^T
                                                               (\bm{\mathcal{I}}_{\bm{\eta}\bm{\eta}}^{\mathrm{act}})^{-1}
    \bm{\mathcal{I}}_{\bm{\eta}\vartheta}^{\mathrm{act}} \nonumber                                           \\
     & = \sum_{n \in \mathcal{A}}
    \left[\mathcal{I}_{\vartheta\vartheta}^{(n)}
      - 2(\bm{\gamma}^{\mathrm{act}})^T \bm{\mathcal{I}}_{\bm{\eta}\vartheta}^{(n)}
      + (\bm{\gamma}^{\mathrm{act}})^T \bm{\mathcal{I}}_{\bm{\eta}\bm{\eta}}^{(n)} \bm{\gamma}^{\mathrm{act}}
      \right] \nonumber \\
     & = \sum_{n \in \mathcal{A}}
    \Phi_{\vartheta}^{(n)}(V_n;\,\bm{\gamma}^{\mathrm{act}}).
    \label{eq:autonum:N1}
  \end{align}
  The second equality follows by substituting the definitions of the summed blocks and expanding the quadratic form, exactly as in the proof of the Fisher decomposition (Section~\ref{sec:opt_derivation}), with the summation restricted to $\mathcal{A}$. The algebraic identity depends only on the block structure, not on the cardinality of the index set. \hfill$\square$

  \textbf{Step N.2: Differentiation of the multi-event cost.}
  The multi-event cost function is $\mathcal{C}(\{V_n\}) = \sum_{p=1}^P w_p \,\mathcal{B}_{\vartheta,\min}(\{V_n\};\bm{\theta}^{(p)})$. For a given threshold $V_n$, consider the contribution of the $p$-th event:
  \begin{equation}
    \mathcal{B}_{\vartheta,\min}^{(p)}
    = \left[\sum_{m \in \mathcal{A}_p} \Phi_{\vartheta}^{(m)}(V_m;\,\bm{\gamma}_p^{\mathrm{act}})\right]^{-1}
    + \beta_{\vartheta,\mathrm{act}}^{(p)\,2}.
    \label{eq:autonum:N2}
  \end{equation}
  If $n \notin \mathcal{A}_p$ and one remains on the same active-set stratum $V_n > p(\bm{\theta}^{(p)})$, then neither the sum in the Fisher term nor the bias term depends on $V_n$ (the $n$-th threshold produces no crossings for event $p$), so $\partial\mathcal{B}_{\vartheta,\min}^{(p)}/\partial V_n = 0$.

  If $n \in \mathcal{A}_p$, the derivative with respect to $V_n$ involves two channels: (i)~the direct dependence of $\Phi_{\vartheta}^{(n)}$ on $V_n$, and (ii)~the implicit dependence of $\bm{\gamma}_p^{\mathrm{act}}$ on $V_n$ through the FIM blocks. We handle the implicit channel via the envelope theorem in Step~N.3.

  \emph{Boundary case}: When the pulse peak $p(\bm{\theta}^{(p)})$ exactly equals a threshold $V_n$, the active set $\mathcal{A}_p$ changes discontinuously as $V_n$ varies. In this degenerate case, the cost $\mathcal{C}$ has directional derivatives but is not differentiable in the classical sense at $V_n = p(\bm{\theta}^{(p)})$. By the assumption that no pulse peak coincides with any optimal threshold (stated in Section~\ref{sec:partial_trigger}), this boundary case is excluded at the optimum.

  \textbf{Step N.3: Envelope theorem for the active coupling vector.}
  For a fixed event $p$ with $n \in \mathcal{A}_p$, the active Fisher decomposition \eqref{eq:autonum:N1} expresses $\Delta\mathcal{I}_{\mathrm{eff}}^{(\vartheta),\mathrm{act},(p)}$ as a function of $V_n$ and $\bm{\gamma}_p^{\mathrm{act}}$.  The coupling vector $\bm{\gamma}_p^{\mathrm{act}}$ satisfies the stationarity condition $\partial \Delta\mathcal{I}_{\mathrm{eff}}^{(\vartheta),\mathrm{act},(p)}/\partial\bm{\gamma} = 0$ at $\bm{\gamma} = \bm{\gamma}_p^{\mathrm{act}}$ (this is the defining property of the Schur complement, identical to the argument in Section~\ref{sec:robustness}, Eq.~\eqref{eq:envelope_sensitivity}). Therefore, by the envelope theorem:
  \begin{align}
    \frac{d\,\Delta\mathcal{I}_{\mathrm{eff}}^{(\vartheta),\mathrm{act},(p)}}{dV_n}
     & = \frac{\partial\,\Delta\mathcal{I}_{\mathrm{eff}}^{(\vartheta),\mathrm{act},(p)}}{\partial V_n}\bigg|_{\bm{\gamma}_p^{\mathrm{act}}} \nonumber \\
     & = \frac{d\,\Phi_{\vartheta}^{(n)}(V;\,\bm{\gamma}_p^{\mathrm{act}})}{dV}\bigg|_{V=V_n},
    \label{eq:autonum:N3}
  \end{align}
  since only the $n$-th term in the sum \eqref{eq:autonum:N1} depends on $V_n$.

  Combining with the chain rule through $\mathcal{B}_{\vartheta,\min}^{(p)} = [\Delta\mathcal{I}_{\mathrm{eff}}^{(\vartheta),\mathrm{act},(p)}]^{-1} + \beta_{\vartheta,\mathrm{act}}^{(p)\,2}$:
  \begin{align}
    \frac{\partial\,\mathcal{B}_{\vartheta,\min}^{(p)}}{\partial V_n}
     & = -\frac{1}{[\Delta\mathcal{I}_{\mathrm{eff}}^{(\vartheta),\mathrm{act},(p)}]^{2}}\,
    \frac{d\,\Phi_{\vartheta}^{(n)}(V;\,\bm{\gamma}_p^{\mathrm{act}})}{dV}\bigg|_{V_n} \nonumber \\
     & \quad + 2\,\beta_{\vartheta,\mathrm{act}}^{(p)}\,
    \frac{\partial\,\beta_{\vartheta,\mathrm{act}}^{(p)}}{\partial V_n}.
    \label{eq:autonum:N4}
  \end{align}

  \textbf{Step N.4: Assembly of the stationarity system.}
  Substituting \eqref{eq:autonum:N4} into the gradient $\partial\mathcal{C}/\partial V_n = \sum_{p:\,n\in\mathcal{A}_p} w_p\,\partial\mathcal{B}_{\vartheta,\min}^{(p)}/\partial V_n$ and setting to zero yields directly the multi-event stationarity system Eq.~\eqref{eq:multi_event_stationarity}. \hfill$\square$

  % =========================================================================
  %  APPENDIX O: BI-EXPONENTIAL ROBUSTNESS DERIVATION
  % =========================================================================
  \section{Bi-Exponential Robustness: Sensitivity Vector and Hessian Derivation}
  \label{app:biexp_robustness}

  This appendix provides the detailed derivation of the Fisher sensitivity vector $\mathbf{S}$ and the Hessian $\mathbf{H}_{\mathcal{I}}$ for the bi-exponential model, as stated in Section~\ref{sec:biexp_robustness}.

  \textbf{Step O.1: Implicit differentiation of crossing positions.}
  The crossing positions $x_r(\bm{\theta})$ and $x_f(\bm{\theta})$ at threshold $V_n$ are defined implicitly by $A\,h(x) + b = V_n$, i.e.,
  \begin{equation}
    h(x_k) = \frac{V_n - b}{A}, \quad k \in \{r, f\}.
    \label{eq:autonum:O1}
  \end{equation}
  Total differentiation with respect to each parameter $\theta_j$, holding $V_n$ fixed:
  \begin{equation}
    h'(x_k)\,\frac{\partial x_k}{\partial\theta_j}
    = \frac{\partial}{\partial\theta_j}\!\left[\frac{V_n - b}{A}\right]
    - \frac{\partial h}{\partial\theta_j}\bigg|_{x_k},
    \label{eq:autonum:O2}
  \end{equation}
  where $\partial h/\partial\theta_j|_{x_k}$ accounts for the explicit dependence of $h$ on $\tau_d$ and $\tau_r$ (through the exponential terms). Solving:
  \begin{align}
    \frac{\partial x_k}{\partial A}
                                      & = \frac{1}{h'(x_k)}\!\left[-\frac{V_n - b}{A^2}\right]
    = -\frac{h(x_k)}{A\,h'(x_k)}, \label{eq:autonum:O3a}                                                                    \\
    \frac{\partial x_k}{\partial t_0} & = 0, \label{eq:autonum:O3b}                                                         \\
    \frac{\partial x_k}{\partial\tau_d}
                                      & = -\frac{1}{h'(x_k)}\,\frac{x_k\,e^{-x_k/\tau_d}}{\tau_d^2}, \label{eq:autonum:O3c} \\
    \frac{\partial x_k}{\partial\tau_r}
                                      & = \frac{1}{h'(x_k)}\,\frac{x_k\,e^{-x_k/\tau_r}}{\tau_r^2}, \label{eq:autonum:O3d}  \\
    \frac{\partial x_k}{\partial b}
                                      & = -\frac{1}{A\,h'(x_k)}, \label{eq:autonum:O3e}
  \end{align}
  confirming Eq.~\eqref{eq:crossing_derivatives}. Note that \eqref{eq:autonum:O3b} follows because the shape variable $x_k$ is geometrically determined by the level-set equation $h(x_k) = (V_n - b)/A$ and is therefore invariant under absolute time translations of $t_0$.

  \textbf{Step O.2: Differentiation of projected sensitivities.}
  The projected sensitivity at edge $k$ of the $n$-th pair is (Eq.~\eqref{eq:biexp_proj_sens}):
  \begin{equation}
    \tilde{g}_k^{(n)} = g_{A,k} - \bm{\gamma}^{*T}\mathbf{g}_{\bm{\eta}',k},
    \label{eq:autonum:O4}
  \end{equation}
  where $g_{A,k} = h(x_k)$ and $\mathbf{g}_{\bm{\eta}',k} = [-A\,h'(x_k),\, Ax_ke^{-x_k/\tau_d}/\tau_d^2,\, -Ax_ke^{-x_k/\tau_r}/\tau_r^2,\, 1]^T$ (Eq.~\eqref{eq:biexp_sensitivity}). Since $\bm{\gamma}^*$ is held fixed (by the envelope theorem), differentiating with respect to $\theta_j$:
  \begin{equation}
    \frac{\partial\tilde{g}_k^{(n)}}{\partial\theta_j}
    = \frac{\partial g_{A,k}}{\partial\theta_j}
    - \bm{\gamma}^{*T}\,\frac{\partial\mathbf{g}_{\bm{\eta}',k}}{\partial\theta_j}.
    \label{eq:autonum:O5}
  \end{equation}
  Each derivative $\partial g_{A,k}/\partial\theta_j$ and $\partial\mathbf{g}_{\bm{\eta}',k}/\partial\theta_j$ is computed via the chain rule through $x_k(\bm{\theta})$ using the crossing derivatives \eqref{eq:autonum:O3a}--\eqref{eq:autonum:O3e}. For example:
  \begin{align}
    \frac{\partial g_{A,k}}{\partial A}
     & = h'(x_k)\,\frac{\partial x_k}{\partial A} \nonumber \\
     & = -\frac{h(x_k)}{A}.
    \label{eq:autonum:O6}
  \end{align}
  Using the level-set identity $g_{A,k}=h(x_k)=(V_n-b)/A$, the full target-sensitivity derivative list is
  \begin{equation}
    \begin{aligned}
      \frac{\partial g_{A,k}}{\partial A}      & = -\frac{h(x_k)}{A},
      & \frac{\partial g_{A,k}}{\partial t_0}   & = 0, \\
      \frac{\partial g_{A,k}}{\partial \tau_d} & = 0,
      & \frac{\partial g_{A,k}}{\partial \tau_r} & = 0,
      & \frac{\partial g_{A,k}}{\partial b}     & = -\frac{1}{A}.
    \end{aligned}
    \label{eq:autonum:O6b}
  \end{equation}
  For the nuisance components, using Kronecker deltas $\delta_{jA}$, $\delta_{j\tau_d}$, and $\delta_{j\tau_r}$ to select the corresponding parameter derivative, the chain rule gives
  \begin{equation}
    \frac{\partial g_{t_0,k}}{\partial\theta_j}
    = -\delta_{jA}\,h'(x_k)
    - A\,h''(x_k)\,\frac{\partial x_k}{\partial\theta_j},
    \label{eq:autonum:O6c}
  \end{equation}
  \begin{equation}
    \begin{aligned}
      \frac{\partial g_{\tau_d,k}}{\partial\theta_j}
       ={}& \delta_{jA}\,\frac{x_k e^{-x_k/\tau_d}}{\tau_d^2}
      + A\left[
        \frac{e^{-x_k/\tau_d}(1-x_k/\tau_d)}{\tau_d^2}\,\frac{\partial x_k}{\partial\theta_j}
      \right. \\
      &\left.
        + \delta_{j\tau_d}\,\frac{x_k e^{-x_k/\tau_d}(x_k-2\tau_d)}{\tau_d^4}
      \right],
    \end{aligned}
    \label{eq:autonum:O6d}
  \end{equation}
  \begin{equation}
    \begin{aligned}
      \frac{\partial g_{\tau_r,k}}{\partial\theta_j}
       ={}& -\delta_{jA}\,\frac{x_k e^{-x_k/\tau_r}}{\tau_r^2}
      - A\left[
        \frac{e^{-x_k/\tau_r}(1-x_k/\tau_r)}{\tau_r^2}\,\frac{\partial x_k}{\partial\theta_j}
      \right. \\
      &\left.
        + \delta_{j\tau_r}\,\frac{x_k e^{-x_k/\tau_r}(x_k-2\tau_r)}{\tau_r^4}
      \right],
    \end{aligned}
    \label{eq:autonum:O6e}
  \end{equation}
  together with $\partial g_{b,k}/\partial\theta_j = 0$. Substituting Eqs.~\eqref{eq:autonum:O6b}--\eqref{eq:autonum:O6e} into Eq.~\eqref{eq:autonum:O5} yields every component of $\partial\tilde g_k^{(n)}/\partial\theta_j$ explicitly.

  \textbf{Step O.3: Differentiation of noise denominators.}
  The composite noise denominator is $D_k = \sigma_{\mathrm{th}}^2 + \kappa(V_n - b) + \sigma_{\mathrm{TDC}}^2\,A^2[h'(x_k)]^2$ (Eq.~\eqref{eq:biexp_phi}). At fixed $V_n$:
  \begin{align}
    \frac{\partial D_k}{\partial A}
     & = 2\sigma_{\mathrm{TDC}}^2\,A\,[h'(x_k)]^2 \notag \\
     &\quad + 2\sigma_{\mathrm{TDC}}^2\,A^2\,h'(x_k)\,h''(x_k)\,\frac{\partial x_k}{\partial A}, \label{eq:autonum:O7a} \\
    \frac{\partial D_k}{\partial t_0}
     & = 0, \label{eq:autonum:O7b}                                                                              \\
    \frac{\partial D_k}{\partial\tau_d}
     & = 2\sigma_{\mathrm{TDC}}^2\,A^2\,h'(x_k) \notag \\
     &\quad \left[
       h''(x_k)\,\frac{\partial x_k}{\partial\tau_d}
     \right. \notag \\
     &\qquad \left.
       + \frac{e^{-x_k/\tau_d}(\tau_d-x_k)}{\tau_d^3}
     \right], \label{eq:autonum:O7c} \\
    \frac{\partial D_k}{\partial\tau_r}
     & = 2\sigma_{\mathrm{TDC}}^2\,A^2\,h'(x_k) \notag \\
     &\quad \left[
       h''(x_k)\,\frac{\partial x_k}{\partial\tau_r}
     \right. \notag \\
     &\qquad \left.
       + \frac{e^{-x_k/\tau_r}(x_k-\tau_r)}{\tau_r^3}
     \right], \label{eq:autonum:O7d} \\
    \frac{\partial D_k}{\partial b}
     & = -\kappa \notag \\
     &\quad + 2\sigma_{\mathrm{TDC}}^2\,A^2\,h'(x_k)\,h''(x_k) \notag \\
     &\qquad \frac{\partial x_k}{\partial b}. \label{eq:autonum:O7e}
  \end{align}
  Equations~\eqref{eq:autonum:O7a}--\eqref{eq:autonum:O7e} therefore provide the complete five-parameter derivative list for the composite denominators.

  \textbf{Step O.4: Assembly of $S_j$ components.}
  Substituting Steps~O.2--O.3 into Eq.~\eqref{eq:biexp_S_expanded}:
  \begin{align}
    S_j & = \sum_{n=1}^{N}\sum_{k \in \{r,f\}}
    \frac{1}{D_k}\!\left[
                     2\,\tilde{g}_k^{(n)}\,\frac{\partial\tilde{g}_k^{(n)}}{\partial\theta_j}
                     - \frac{[\tilde{g}_k^{(n)}]^2}{D_k}\,\frac{\partial D_k}{\partial\theta_j}
                     \right], \nonumber \\
        & \quad j = A, t_0, \tau_d, \tau_r, b.
    \label{eq:autonum:O8}
  \end{align}
  Each term in the double sum is expressed purely in terms of the bi-exponential shape function $h$, its derivatives $h'$, $h''$, the crossing positions $x_r^{(n)}$, $x_f^{(n)}$, the coupling vector $\bm{\gamma}^*$, the noise parameters $(\sigma_{\mathrm{th}}^2, \kappa, \sigma_{\mathrm{TDC}}^2)$, and the physical parameters $\bm{\theta}_0$. \hfill$\square$

  \textbf{Step O.5: Bias sensitivity vector.}
  The total bias $\beta_{A,\mathrm{tot}} = [(\mathbf{J}^T\mathbf{W}\mathbf{J})^{-1}\mathbf{J}^T\mathbf{W}\,\Delta\mathbf{t}_{\mathrm{bias}}]_A$ (Eq.~\eqref{eq:b_tau_tot}) depends on $\bm{\theta}$ through $\mathbf{J}$, $\mathbf{W}$, and $\Delta\mathbf{t}_{\mathrm{bias}}$. Denoting $\mathbf{M} \triangleq (\mathbf{J}^T\mathbf{W}\mathbf{J})^{-1}\mathbf{J}^T\mathbf{W}$ and using the identity $d(\mathbf{A}^{-1}) = -\mathbf{A}^{-1}(d\mathbf{A})\mathbf{A}^{-1}$:
  \begin{align}
    \frac{\partial\,\beta_{A,\mathrm{tot}}}{\partial\theta_j}
    & = \mathbf{e}_A^T\left[
                             \frac{\partial\mathbf{M}}{\partial\theta_j}\,\Delta\mathbf{t}_{\mathrm{bias}}
                             + \mathbf{M}\,\frac{\partial\,\Delta\mathbf{t}_{\mathrm{bias}}}{\partial\theta_j}
                             \right],
    \label{eq:autonum:O9}
  \end{align}
  where $\partial\mathbf{M}/\partial\theta_j$ involves the derivatives of both the Jacobian $\mathbf{J}$ and the weight matrix $\mathbf{W}$ with respect to $\theta_j$, and $\partial\Delta\mathbf{t}_{\mathrm{bias}}/\partial\theta_j$ involves the derivative of the misspecification-induced temporal bias at each crossing. \hfill$\square$

  \textbf{Steps O.6--O.7: Hessian computation.}
  The cross-derivative $\partial^2\mathcal{I}_{\mathrm{eff}}/\partial\theta_j\,\partial\gamma_k$ is obtained by differentiating the expanded Fisher decomposition Eq.~\eqref{eq:decomp_expanded} with respect to $\theta_j$:
  \begin{align}
     & \frac{\partial^2\mathcal{I}_{\mathrm{eff}}}{\partial\theta_j\,\partial\gamma_k}
    = \sum_{n=1}^{N}\left[
                      -2\,\frac{\partial[\bm{\mathcal{I}}_{\bm{\eta}\vartheta}^{(n)}]_k}{\partial\theta_j}\right.\nonumber\\
                      &\quad\left.+ 2\sum_{l=1}^{M-1}
                      \frac{\partial[\bm{\mathcal{I}}_{\bm{\eta}\bm{\eta}}^{(n)}]_{kl}}{\partial\theta_j}\,\gamma_l^*
                      \right].
    \label{eq:autonum:O10}
  \end{align}
  Note that $\bm{\gamma}$ is an independent variable in $\mathcal{I}_{\mathrm{eff}}(\bm{\theta},\bm{\gamma})$, so $\partial\gamma_l/\partial\theta_j \equiv 0$ by definition; the implicit response $d\bm{\gamma}^*/d\bm{\theta}$ enters only through the second-order envelope correction in Eq.~\eqref{eq:true_hessian}. The na\"ive Hessian $\partial^2\mathcal{I}_{\mathrm{eff}}/\partial\theta_j\partial\theta_k|_{\bm{\gamma}^*}$ is computed by directly differentiating \eqref{eq:autonum:O8} with respect to $\theta_k$. The full Hessian Eq.~\eqref{eq:biexp_hessian} is then assembled as (Eq.~\eqref{eq:true_hessian}):
  \begin{equation}
    [\mathbf{H}_{\mathcal{I}}]_{jk}
    = \frac{\partial^2\mathcal{I}_{\mathrm{eff}}}{\partial\theta_j\,\partial\theta_k}\bigg|_{\bm{\gamma}^*}
    - \frac{1}{2}\sum_{a,b=1}^{M-1}
    \frac{\partial^2\mathcal{I}_{\mathrm{eff}}}{\partial\theta_j\,\partial\gamma_a}\,
    [\bm{\mathcal{I}}_{\bm{\eta}\bm{\eta}}^{\mathrm{tot}}]^{-1}_{ab}\,
    \frac{\partial^2\mathcal{I}_{\mathrm{eff}}}{\partial\gamma_b\,\partial\theta_k}.
    \label{eq:autonum:O11}
  \end{equation}
  This is a $5 \times 5$ symmetric matrix; the correction involves a $(M{-}1) \times (M{-}1) = 4 \times 4$ solve with the nuisance block. \hfill$\square$

  % =========================================================================
  %  APPENDIX P: TDC QUANTIZATION ERROR VARIANCE
  % =========================================================================
  \section{TDC Quantization Error Variance}
  \label{app:quantization_error}

  This appendix provides a rigorous self-contained proof that midpoint assignment for timeline threshold crossings introduces a uniform quantization error with variance $\Delta t_{\mathrm{bin}}^2/12$ and hence standard deviation $\Delta t_{\mathrm{bin}}/\sqrt{12}$. Identifying the bin width $\Delta t_{\mathrm{bin}}$ with the TDC least-significant bit, $\Delta t_{\mathrm{bin}} = \mathrm{LSB}$, yields the main-text relation $\sigma_{\mathrm{TDC}}^2 = \mathrm{LSB}^2/12$.

  Consider a continuous physical time variable $t_{\mathrm{true}}$ denoting the exact instance a signal crosses a predetermined voltage threshold. In a digital sampling system with bin width $\Delta t_{\mathrm{bin}}$, the time axis is partitioned into discrete bins of width $\Delta t_{\mathrm{bin}}$. Let the $i$-th bin span the interval $[i\Delta t_{\mathrm{bin}}, (i+1)\Delta t_{\mathrm{bin}})$.

  When the digitizer records that a crossing event occurred within the $i$-th bin (i.e., $t_{\mathrm{true}} \in [i\Delta t_{\mathrm{bin}}, (i+1)\Delta t_{\mathrm{bin}})$), the exact temporal position within the bin remains unknown. In the absence of any prior information that biases the sub-bin arrival time within that interval, the true crossing time $t_{\mathrm{true}}$ is modeled as being uniformly distributed across the width of the bin.

  The midpoint assignment method estimates the crossing time as the center of the bin:
  \begin{equation}
    t_{\mathrm{est}} = \left(i + \frac{1}{2}\right)\Delta t_{\mathrm{bin}}.
  \end{equation}

  The quantization error $\epsilon_{\mathrm{q}}$ is the difference between the estimated time and the true time:
  \begin{equation}
    \epsilon_{\mathrm{q}} = t_{\mathrm{est}} - t_{\mathrm{true}}.
  \end{equation}
  Since $t_{\mathrm{true}} \sim \mathcal{U}(i\Delta t_{\mathrm{bin}}, (i+1)\Delta t_{\mathrm{bin}})$, where $\mathcal{U}$ denotes the continuous uniform distribution, a linear shift of the variable implies that the error $\epsilon_{\mathrm{q}}$ is uniformly distributed over the interval $[-\Delta t_{\mathrm{bin}}/2, \Delta t_{\mathrm{bin}}/2]$. Thus, its probability density function $p(\epsilon_{\mathrm{q}})$ is given by:
  \begin{equation}
    p(\epsilon_{\mathrm{q}}) = \begin{cases}
      \frac{1}{\Delta t_{\mathrm{bin}}}, & \text{for } -\frac{\Delta t_{\mathrm{bin}}}{2} \le \epsilon_{\mathrm{q}} \le \frac{\Delta t_{\mathrm{bin}}}{2} \\
      0,                  & \text{otherwise}.
    \end{cases}
  \end{equation}

  Since the distribution is symmetric around zero, the expected value (mean) of the error is:
  \begin{equation}
    \mathbb{E}[\epsilon_{\mathrm{q}}] = \int_{-\Delta t_{\mathrm{bin}}/2}^{\Delta t_{\mathrm{bin}}/2} \epsilon_{\mathrm{q}} \cdot \frac{1}{\Delta t_{\mathrm{bin}}} \, d\epsilon_{\mathrm{q}} = 0.
  \end{equation}

  The variance of the quantization error is therefore $\sigma_{\mathrm{TDC}}^2 = \mathbb{E}[\epsilon_{\mathrm{q}}^2] - (\mathbb{E}[\epsilon_{\mathrm{q}}])^2$. Calculating the second moment yields:
  \begin{align}
    \sigma_{\mathrm{TDC}}^2 & = \int_{-\Delta t_{\mathrm{bin}}/2}^{\Delta t_{\mathrm{bin}}/2} \epsilon_{\mathrm{q}}^2 \cdot \frac{1}{\Delta t_{\mathrm{bin}}} \, d\epsilon_{\mathrm{q}} \nonumber             \\
         & = \frac{1}{\Delta t_{\mathrm{bin}}} \left[ \frac{\epsilon_{\mathrm{q}}^3}{3} \right]_{-\Delta t_{\mathrm{bin}}/2}^{\Delta t_{\mathrm{bin}}/2} \nonumber            \\
         & = \frac{1}{3\Delta t_{\mathrm{bin}}} \left( \frac{\Delta t_{\mathrm{bin}}^3}{8} - \left(-\frac{\Delta t_{\mathrm{bin}}^3}{8}\right) \right) \nonumber \\
         & = \frac{1}{3\Delta t_{\mathrm{bin}}} \left( \frac{2\Delta t_{\mathrm{bin}}^3}{8} \right) = \frac{\Delta t_{\mathrm{bin}}^2}{12}.
  \end{align}

  Taking the square root gives the standard deviation of the uniform quantization error:
  \begin{equation}
    \sqrt{\sigma_{\mathrm{TDC}}^2} = \frac{\Delta t_{\mathrm{bin}}}{\sqrt{12}}.
  \end{equation}
  This establishes the standard $\Delta t_{\mathrm{bin}}/\sqrt{12}$ intrinsic timing precision limit imposed by the discrete bin width. \hfill$\square$

  % =========================================================================
  %  NOTATION TABLE
  % =========================================================================
  \clearpage
  \onecolumn
  \section*{Notation}
  \label{sec:notation}
  \renewcommand{\arraystretch}{1.0}
  \setlength{\LTleft}{0pt}
  \setlength{\LTright}{0pt}
  \setlength{\LTpre}{0pt}
  \setlength{\LTpost}{0pt}
  \scriptsize
  \begin{longtable}{@{}p{3.2cm}p{13.0cm}@{}}
  \caption{Summary of notation and diagnostics used throughout this paper.}\label{tab:notation}\\
  \hline
  \textbf{Symbol} & \textbf{Description} \\
  \hline
  \endfirsthead
  \multicolumn{2}{@{}l@{}}{\tablename~\thetable\ (\textit{Continued}): Summary of notation and diagnostics.}\\
  \hline
  \textbf{Symbol} & \textbf{Description} \\
  \hline
  \endhead
  \hline
  \multicolumn{2}{r@{}}{\textit{Continued on next page}}\\
  \endfoot
  \hline
  \endlastfoot
  \multicolumn{2}{@{}l}{\emph{Pulse model parameters}} \\
$f(t;\bm{\theta})$ & Continuous ideal pulse model (Sec.~\ref{sec:deterministic}) \\
$\Theta \subset \mathbb{R}^M$ & Admissible physical parameter domain for $\bm{\theta}$ (Sec.~\ref{sec:model}) \\
$\Theta_{\mathrm{adm}}$ & Physically admissible domain used in the global embedding theorem, with $\Theta_{\mathrm{adm}} \subseteq \Theta_V$ \\
$\bm{\theta}=[\theta_1,\ldots,\theta_M]^T \in \mathbb{R}^M$ & Physical parameter vector; for the bi-exponential model $\bm{\theta} = [A,\,t_0,\,\tau_d,\,\tau_r,\,b]^T$, $M=5$ \\
$M$ & Dimension of the parameter vector $\bm{\theta}$ \\
$t_p(\bm{\theta})$, $p(\bm{\theta})$ & Peak time and peak amplitude of the pulse model \\
$A$ $[\mathrm{V}]$ & Waveform scale amplitude, proportional to deposited energy; the peak voltage is $A h_{\mathrm{peak}}+b$ \\
$A_{\mathrm{ref}}$ $[\mathrm{V}]$ & Full-waveform-fit reference scale amplitude used for the experimental amplitude-error metrics in Sections~\ref{sec:uniform_mvt} and~\ref{sec:uniform_mvt_multi} \\
$A_{\mathrm{design}}$, $A^{(p)}$ $[\mathrm{V}]$ & Single-event design-point and representative-event waveform scale amplitudes; both are scale amplitudes, not observable peak voltages \\
$t_0$ $[\mathrm{s}]$ & Absolute photon arrival time \\
$\tau_d$ $[\mathrm{s}]$ & Effective macroscopic decay time constant of the fitted voltage waveform \\
$\tau_r$, $\tau_{r,\mathrm{eff}}$, $\tau_{r,\mathrm{scint}}$ $[\mathrm{s}]$ & Rise-time constants; $\tau_r$ in the fitted bi-exponential waveform denotes the effective macroscopic value, while $\tau_{r,\mathrm{scint}}$ denotes the intrinsic scintillation value \\
$\tau_d^{(p)}$, $\tau_r^{(p)}$, $x_p^{(p)}$ & Representative-event decay/rise time constants and relative peak time used in the multi-event recipe \\
$\lambda_d \triangleq 1/\tau_d$ $[\mathrm{s}^{-1}]$ & Inverse decay rate constant \\
$\lambda_r \triangleq 1/\tau_r$ $[\mathrm{s}^{-1}]$ & Inverse rise rate constant ($\lambda_r > \lambda_d$) \\
$D$, $E$ $[\mathrm{V}]$ & Reparameterized exponential amplitudes $A e^{\lambda_d t_0}$ and $A e^{\lambda_r t_0}$ used in the bi-exponential immersion/injectivity proof (Sec.~\ref{sec:biexp_embedding}, Appendix~\ref{app:rank}) \\
$\Psi$, $\bm{\omega}$, $\delta\bm{\omega}$ & Bi-exponential reparameterization, its coordinate vector $\bm{\omega}=[D,E,\lambda_d,\lambda_r,b]^T$, and tangent perturbation used in Appendix~\ref{app:rank} \\
$b$ $[\mathrm{V}]$ & DC voltage baseline offset \\
$h(x)$, $h'(x)$, $h''(x)$ & Normalized shape function $e^{-x/\tau_d} - e^{-x/\tau_r}$ and its first two derivatives (Sec.~\ref{sec:biexp_model}) \\
$u(\cdot)$ & Heaviside unit step enforcing causality in the bi-exponential pulse model \\
$h_{\mathrm{peak}}$, $x_p$ & Maximum value and normalized peak time of $h(x)$ \\
$\mathcal{F}_5(\lambda_d,\lambda_r)$ & Five-function extended Chebyshev family $\{1, e^{-\lambda_d t}, t e^{-\lambda_d t}, e^{-\lambda_r t}, t e^{-\lambda_r t}\}$ used in the rank proof (Sec.~\ref{sec:biexp_embedding}, Appendix~\ref{app:rank}) \\
$\phi_i(t)$, $i=1,\ldots,5$ & Basis functions of the extended Chebyshev system in Appendix~\ref{app:rank} \\
$W_{\mathcal{F}_5}(t)$ & Wronskian of $\mathcal{F}_5(\lambda_d,\lambda_r)$ certifying the extended Chebyshev property (Sec.~\ref{sec:biexp_embedding}, Appendix~\ref{app:rank}) \\
$q(t)$, $g(t)$ & Auxiliary exponential combinations used in Appendix~\ref{app:rank} \\
$\bar{\Theta}_V^{\mathrm{bi}}$ & Physically bounded compact prior domain for the bi-exponential embedding claim (Eq.~\eqref{eq:biexp_compact_prior_domain}, Appendix~\ref{app:proper}) \\
$A_-$, $A_+$, $t_0^-$, $t_0^+$, $\tau_d^-$, $\tau_d^+$, $\tau_r^-$, $\tau_r^+$, $b_-$, $b_+$ & Hardware/calibration bounds defining $\bar{\Theta}_V^{\mathrm{bi}}$ \\
$\Delta_\tau$, $\Delta_p$ & Compact-prior margins for rise--decay separation and peak clearance above the highest threshold \\
$\bm{\theta}_0$ & True (or design-point) physical parameter vector (Sec.~\ref{sec:misspecification}, \ref{sec:robustness}) \\
$\bm{\theta}_{\mathrm{design}}$, $\tilde{A}$, $\tilde{\tau}_d$, $\tilde{\tau}_r$, $\tilde{b}$ & Median photopeak design-point vector and its fitted amplitude/shape/baseline components used in local D-optimal validation \\
$\varepsilon(t)$ & Deterministic model misspecification residual: $y_{\mathrm{true}}(t) = f(t;\bm{\theta}_0) + \varepsilon(t)$ (Sec.~\ref{sec:misspecification}) \\
  \hline
  \multicolumn{2}{@{}l}{\emph{Hardware and observation}} \\
$V_n$, $V_k$ $[\mathrm{V}]$ & Generic hardware threshold voltage; use $V_{\ell(k)}$ when an ordered-crossing index $k$ is already in play \\
$\ell(k)$ & Map from ordered-crossing index $k \in \{1,\dots,K\}$ to the associated physical threshold index in Eq.~\eqref{eq:crossing_index_map} \\
$V_{\max}$ $[\mathrm{V}]$ & Highest hardware threshold, i.e., $V_N$ \\
$V_{\mathrm{HW,max}}$ $[\mathrm{V}]$ & Finite hardware/DAC upper bound imposed on the highest threshold in the multi-event design problem \\
$V_{\mathrm{peak}}$, $V_{\mathrm{peak}}^{(p)}$ $[\mathrm{V}]$ & Peak voltage of the single-event design pulse / of representative event $p$ in the multi-event setting \\
$V_{\mathrm{peak},i}$, $V_{\mathrm{peak},511}$ $[\mathrm{V}]$ & Empirical maximum-sample peak voltage of pulse $i$ and photopeak location in the pulse-height spectrum \\
$V_{\mathrm{peak,lo}}$, $V_{\mathrm{peak,hi}}$, $V_{\mathrm{peak,max}}$, $V_{\mathrm{floor}}$ $[\mathrm{V}]$ & Extended-window lower/upper peak-voltage bounds, upper admitted peak-voltage limit, and lower feasible threshold floor used in multi-event validation \\
$N_{511}$, $N_{\mathrm{ext}}$ & Numbers of events selected by the photopeak and extended energy windows in experimental validation \\
$V_{\mathrm{peak,min}}$, $V_{\mathrm{peak,model}}$ $[\mathrm{V}]$ & Minimum observed peak voltage in the selected population / model-predicted design-point peak voltage used to delimit feasible thresholds in Section~IX \\
$N$ & Number of threshold levels \\
$K = 2N$ & Total number of observations per event \\
$\mathbf{t} \in \mathbb{R}^K$ & Ordered observation vector of threshold-crossing times \\
$v_i$, $v_{i,k}$ $[\mathrm{V}]$ & Oscilloscope voltage samples in a single waveform and the $k$-th sample of pulse $i$ used in empirical peak extraction \\
$\Delta t_{\mathrm{samp}}$ $[\mathrm{s}]$ & Oscilloscope sampling interval used in the offline MVT emulation and validation noise calibration \\
$m$ & Number of oscilloscope samples in a waveform when defining empirical peak voltages \\
$t_{r,k}$, $t_{f,k}$ $[\mathrm{s}]$ & Rising and falling crossing times at $V_k$ \\
$x_k \triangleq t_k - t_0$ $[\mathrm{s}]$ & Generic relative crossing time; in threshold-specific notation the two roots are $x_{r,k}$ and $x_{f,k}$ (Sec.~\ref{sec:biexp_model}) \\
$x_r(V)$, $x_f(V)$ $[\mathrm{s}]$ & Normalized rising/falling crossing times at threshold $V$: solutions of $h(x) = (V-b)/A$ \\
$\eta_n \triangleq (V_n-b)/A$, $x_{n,e}$ & Normalized threshold coordinate and edge-specific crossing-position offset used in the amplitude-scaling analysis of D-optimal thresholds \\
$\tilde v(t)$, $\hat t_p$ & Piecewise-linear interpolant of the recorded oscilloscope waveform and observed time of its maximum used to split pre-peak and post-peak crossings in the offline MVT simulation (Sec.~\ref{sec:exp_setup}) \\
$t_{\mathrm{cross}}$ $[\mathrm{s}]$ & Linearly interpolated threshold-crossing time in the offline MVT emulation \\
$\mathcal{T}_{j}^{(r)}$, $\mathcal{T}_{j}^{(f)}$ & Pre-peak and post-peak threshold-crossing sets of $\tilde v(t)$ at threshold $V_j$ in the offline simulation (Sec.~\ref{sec:exp_setup}) \\
$\hat t_{r,j}$, $\hat t_{f,j}$ $[\mathrm{s}]$ & Final offline-simulated rising/falling timestamps selected as the earliest pre-peak rising crossing and latest post-peak falling crossing, respectively (Sec.~\ref{sec:exp_setup}) \\
$\Omega_K$, $\Omega_{2K_{\mathrm{act}}}$ & Full and active-set observation cones for ordered crossing-time vectors \\
$\Theta_V$, $\Theta_V^{\mathrm{bi}}$, $\bar{\Theta}_V^{\mathrm{bi}}$ & Dynamically feasible parameter domains in the general and bi-exponential theories, and the compact physical prior domain used for the closed-embedding statement \\
$B$ $[\mathrm{Hz}]$ & Analog front-end bandwidth \\
$\tau_c \approx 1/(2B)$ & Noise autocorrelation time \\
$\Delta V_{\min}$, $\Delta V_{\min}^{\mathrm{bw}}$, $\Delta V_{\min}^{\mathrm{cov}}$ & Minimum threshold spacing, its bandwidth-derived form, and its coverage/conditioning form \\
$L_{\mathrm{max}}^{(r)}$ & Bi-exponential rising-edge global slew bound $A(\lambda_r-\lambda_d)$ entering Eq.~\eqref{eq:biexp_bandwidth} \\
  \hline
  \multicolumn{2}{@{}l}{\emph{Noise and variance}} \\
$\nu(t)$ $[\mathrm{s}^{-1}]$ & Instantaneous photon intensity of the inhomogeneous Poisson process \\
$y(t)$, $\mu_y(t)$ $[\mathrm{V}]$ & Instantaneous macroscopic output voltage and its mean waveform $\mu_y(t)=\mathbb{E}[y(t)]$ (Sec.~\ref{sec:noise}, Appendix~\ref{app:poisson}) \\
$\sigma_p^2(t)$ $[\mathrm{V}^2]$ & Intrinsic Poisson shot-noise voltage variance before adding the thermal floor (Appendix~\ref{app:poisson}) \\
$\mathcal{H}_1$, $\mathcal{H}_2$ & Single-photoelectron impulse area integrals $\int h_e(u)\,du$ and $\int h_e^2(u)\,du$ used in the Campbell reduction (Appendix~\ref{app:poisson}) \\
$\sigma_V^2(V)$ $[\mathrm{V}^2]$ & Analog voltage variance as a function of threshold level $V$ \\
$\Delta V$, $\Delta t$ & Local voltage fluctuation and induced timing jitter in the linearized error-propagation argument; these are perturbations, not fixed hardware intervals \\
$\sigma_{\mathrm{th}}^2$ $[\mathrm{V}^2]$ & Stationary thermal noise variance of the analog front-end \\
$\kappa$ $[\mathrm{V}]$ & Effective single-photoelectron voltage, $\int h_e^2(x)\,dx \big/ \int h_e(x)\,dx$ \\
$\mathrm{LSB}$ $[\mathrm{s}]$ & TDC least-significant-bit time resolution entering $\sigma_{\mathrm{TDC}}^2 = \mathrm{LSB}^2/12$ \\
$t_{\mathrm{true}}$, $t_{\mathrm{est}}$, $\Delta t_{\mathrm{bin}}$ $[\mathrm{s}]$ & Exact crossing time, midpoint-assigned digitized time, and TDC quantization bin width used in Appendix~\ref{app:quantization_error} \\
$\epsilon_{\mathrm{q}}$ $[\mathrm{s}]$, $\sigma_{\mathrm{TDC}}^2$ $[\mathrm{s}^2]$ & Uniform TDC quantization error within one bin and its variance (Appendix~\ref{app:quantization_error}) \\
$\mathcal{N}(\cdot,\cdot)$, $\mathcal{U}(\cdot,\cdot)$ & Normal and continuous uniform distributions used in the stochastic timing and quantization models \\
$\sigma_k^2$ $[\mathrm{s}^2]$ & Approximate total temporal variance at the $k$-th ordered crossing, after slope projection and TDC addition \\
$t_{k,\mathrm{true}}$ $[\mathrm{s}]$ & Exact physical $k$-th ordered crossing instant satisfying $f(t;\bm{\theta}) = V_{\ell(k)}$ (Sec.~\ref{sec:bandwidth}) \\
$h_e(t)$ $[\mathrm{V}]$ & Single-photoelectron voltage impulse response (Sec.~\ref{sec:noise}) \\
$\tau_e$ $[\mathrm{s}]$ & Effective duration of the single-photoelectron impulse response $h_e(t)$ (Sec.~\ref{sec:noise}) \\
$N_{\mathrm{pe}}(V)$ & Mean number of photoelectrons effectively contributing near threshold level $V$, used to justify the Gaussian regime (Sec.~\ref{sec:mle}) \\
$D_k$ & Composite noise denominator at edge $k$ \\
$D_r(V)$, $D_f(V)$ & Composite noise denominators at the rising and falling edges of threshold $V$ (Sec.~\ref{sec:pair_general}) \\
$\dot{f}_r$, $\dot{f}_f$ & Rising- and falling-edge temporal slopes $\partial_t f(t_r;\bm{\theta})$ and $\partial_t f(t_f;\bm{\theta})$ used in Appendix~\ref{app:schur} \\
$\sigma_r^2$, $\sigma_f^2$ $[\mathrm{s}^2]$ & Rising- and falling-edge temporal variances for a single threshold pair in Appendix~\ref{app:schur} \\
  \hline
  \multicolumn{2}{@{}l}{\emph{Estimation and information}} \\
$F_k(t_k, \bm{\theta})$ & Level-set constraint function $\triangleq f(t_k;\bm{\theta}) - V_{\ell(k)}$ for the $k$-th ordered crossing (Sec.~\ref{sec:model}, Appendix~\ref{app:jacobian}) \\
$S:\Theta_V \to \Omega_K$ & Forward map (parameters $\to$ crossing times) \\
$\partial_t f(t;\bm{\theta})$ & Temporal derivative (slew rate) of the pulse model \\
$\mathbf{t} \in \mathbb{R}^K$, $\mathbf{t}(\bm{\theta})$, $t_k(\bm{\theta})$ & Observed crossing-time vector, model-predicted crossing-time vector, and the $k$-th ordered crossing component \\
$\mathbf{J} \in \mathbb{R}^{K\times M}$ & Jacobian matrix $J_{kj} = \partial t_k/\partial\theta_j$ \\
$\mathbf{W}$ & Precision (weight) matrix $\bm{\Sigma}^{-1}$ \\
$L(\bm{\theta}\mid\mathbf{t})$, $\ln L(\bm{\theta}\mid\mathbf{t})$ & Gaussian likelihood and log-likelihood under the locally diagonal noise model (Sec.~\ref{sec:mle}) \\
$\mathcal{E}_{\mathrm{WLS}}(\bm{\theta})$ & Weighted non-linear least-squares objective $\triangleq \frac{1}{2}\Delta\mathbf{t}^T\mathbf{W}\Delta\mathbf{t}$ (Sec.~\ref{sec:mle}) \\
$\Delta\mathbf{t}(\bm{\theta})$, $\Delta\mathbf{t}^{(\ell)}$ & Generic residual vector $\mathbf{t}-\mathbf{t}(\bm{\theta})$ and residual vector at iteration $\ell$ \\
$\bm{\theta}^{(\ell)}$, $\Delta\bm{\theta}$ & Current Gauss--Newton iterate and parameter update \\
$\mathbf{H}$ & Gauss-Newton normal matrix / approximate Hessian $\mathbf{J}^T\mathbf{W}\mathbf{J}$ (Sec.~\ref{sec:gauss_newton}) \\
$\bm{\epsilon}$ & Gaussian measurement error vector: $\mathbf{t} = S(\bm{\theta}) + \bm{\epsilon}$ (Sec.~\ref{sec:mle}) \\
$\bm{\Sigma}$ & Measurement covariance matrix; diagonal approximation $\operatorname{diag}(\sigma_1^2,\ldots,\sigma_K^2)$ when crossing errors are independent \\
$\mathcal{I}(\bm{\theta})$, $\mathbf{I}$ & Fisher Information Matrix; under the high-SNR locally constant-covariance approximation, $\mathcal{I}(\bm{\theta}) \approx \mathbf{J}^T \mathbf{W}\,\mathbf{J}$, while later bi-exponential sections use boldface $\mathbf{I}$ for the same object \\
$g_{j,k}$ & Waveform sensitivity $\partial f/\partial\theta_j$ evaluated at crossing $k$ \\
$g_{j,r}(V)$, $g_{j,f}(V)$, $\mathbf{g}_{n,e}$ & Pairwise component sensitivities and the full edge-sensitivity vector used in the bi-exponential Fisher assembly \\
$\hat{\bm{\theta}}$ & Estimated parameter vector returned by the QMLE / Gauss--Newton fit \\
$\mathbf{e}_A$ & Coordinate unit vector selecting the amplitude component $A$ in bias projections \\
$\bm{\beta}$ & Systematic bias vector under model misspecification (Sec.~\ref{sec:misspecification}) \\
$\Delta\mathbf{t}_{\mathrm{bias}}$ & Temporal bias vector assembled from all crossing-time biases \\
$\Delta t_{\mathrm{bias},k}$ & Elementwise temporal bias at ideal ordered crossing $k$: $\approx -\varepsilon(t_k)/\partial_t f(t_k;\bm{\theta}_0)$, with $t_k$ defined by $f(t_k;\bm{\theta}_0)=V_{\ell(k)}$ in Sec.~\ref{sec:misspecification} \\
$\Delta t_{\mathrm{char}}$ $[\mathrm{s}]$ & Characteristic temporal width of the pulse (e.g., FWHM), used as perturbation scale (Sec.~\ref{sec:misspecification}) \\
$\partial_t \varepsilon(t)$ & Temporal derivative of the mismatch field, used in the local-flatness assumption near a crossing \\
$\mathbf{t}_{\mathrm{obs}}$, $\mathbf{t}_{\mathrm{model}}$, $\mathbf{J}_0$ & Observed/model-predicted timestamp vectors and Jacobian evaluated at $\bm{\theta}_0$ in Appendix~\ref{app:mismatch_proof} \\
$\mathbf{s}(\hat{\bm{\theta}})$ & Approximate QMLE score vector used in Appendix~\ref{app:mismatch_proof} \\
  \hline
  \multicolumn{2}{@{}l}{\emph{Misspecification operators and high-SNR decomposition (Sec.~\ref{sec:misspecification}, Appendix~\ref{app:slepian})}} \\
$\mathrm{MSE}(\hat{\bm{\theta}})$ & Mean-squared-error matrix $\mathbb{E}[(\hat{\bm{\theta}}-\bm{\theta}_0)(\hat{\bm{\theta}}-\bm{\theta}_0)^T]$ \\
$\mathrm{Cov}(\hat{\bm{\theta}})$ & Covariance matrix of $\hat{\bm{\theta}}$ about its mean $\mathbb{E}[\hat{\bm{\theta}}]$ \\
$\mathbf{A} \succeq \mathbf{B}$ & Loewner partial order: $\mathbf{A}-\mathbf{B}$ is positive semidefinite \\
$\mathcal{I}_{\mathrm{true}}$ & True score covariance (the middle matrix in the sandwich covariance under model misspecification) \\
$\bm{\mu}(\bm{\theta})$ & Mean observation vector in the Slepian-Bangs formula (Appendix~\ref{app:slepian}) \\
$\mathcal{I}^{(\mathrm{mean})}$, $\mathcal{I}^{(\mathrm{cov})}$ & Mean-shift and covariance-derivative contributions to the Fisher Information Matrix in the Slepian-Bangs decomposition \\
$\epsilon$ & Dimensionless noise-level scaling parameter used in Appendix~\ref{app:slepian} \\
$\bar{\sigma}_{\mathrm{th}}^2$, $\bar{\kappa}$, $\bar{\sigma}_{\mathrm{TDC}}^2$, $\bar{\sigma}_k^2$ & Rescaled $\mathcal{O}(1)$ noise coefficients defined by $\sigma_{\mathrm{th}}^2=\epsilon^2\bar{\sigma}_{\mathrm{th}}^2$, $\kappa=\epsilon^2\bar{\kappa}$, and $\sigma_k^2=\epsilon^2\bar{\sigma}_k^2$ in Appendix~\ref{app:slepian} \\
$\bar{\bm{\Sigma}}$ & Rescaled $\mathcal{O}(1)$ covariance matrix defined by $\bm{\Sigma} = \epsilon^2\bar{\bm{\Sigma}}$ in Appendix~\ref{app:slepian} \\
  \hline
  \multicolumn{2}{@{}l}{\emph{Misspecification diagnostics (Secs.~\ref{sec:misspecification}, \ref{sec:recipe_algorithm}, \ref{sec:delta_validation})}} \\
$y_{\mathrm{true}}(t)$ & True physical waveform in the misspecified relation $y_{\mathrm{true}}(t) = f(t;\bm{\theta}_0) + \varepsilon(t)$ \\
$y_{\mathrm{data}}(t)$ & Linearly interpolated recorded waveform of an individual pulse used for empirical mismatch evaluation \\
$t_k^{(0)}$, $t_k^{(\varepsilon)}$ & Ideal model crossing and misspecified physical crossing at ordered index $k$ in Appendix~\ref{app:mismatch_proof} \\
$\mathbf{V}^{(0)}$, $\mathbf{V}^{(\varepsilon)}$ & Minimizers of the idealized and misspecified single-event amplitude objectives $\mathcal{B}_0^{(A)}(\mathbf{V})$ and $\mathcal{B}_{\varepsilon}^{(A)}(\mathbf{V})$ \\
$\varepsilon_k$ & Crossing-point mismatch sample at the $k$-th crossing; in the single-event perturbative analysis $\varepsilon_k = \varepsilon(t_k^{(0)})$ \\
$\bar{\varepsilon}$, $\varepsilon_{\max}$ & Mean and maximum absolute crossing-point mismatch over the $K = 2N$ crossings of a single pulse/design-point event \\
$m_{\min}$, $C_{\beta}$ & Minimum local slew magnitude and bias-projection coefficient entering the single-event bound Eq.~\eqref{eq:autonum:R6f} \\
$\beta_{A}^{\mathrm{direct}}$ & Direct Gauss--Newton projection of crossing-time bias onto the amplitude coordinate $A$ (Eq.~\eqref{eq:beta_direct}) \\
$\mathcal{B}_0^{(A)}(\mathbf{V})$, $\mathcal{B}_{\varepsilon}^{(A)}(\mathbf{V})$ & Idealized and misspecified amplitude-target design objectives for threshold optimization (Eqs.~\eqref{eq:autonum:R6a}--\eqref{eq:autonum:R6b}) \\
$\mathbf{H}_0$ & Hessian of the ideal single-event amplitude-target objective at $\mathbf{V}^{(0)}$ (Eq.~\eqref{eq:autonum:R6g}) \\
$\rho_{\mathrm{bias}}^{(A)}$ & Direct amplitude bias-to-variance diagnostic ratio $(\beta_{A}^{\mathrm{direct}})^2 / \mathcal{B}_0^{(A)}$ \\
  \hline
  \multicolumn{2}{@{}l}{\emph{Experimental validation statistics}} \\
$Q_1$, $Q_3$, $\mathrm{IQR}$ & First and third quartiles and interquartile range used in the amplitude-error outlier fence \\
  \hline
  \multicolumn{2}{@{}l}{\emph{Target--nuisance decomposition (Sec.~\ref{sec:pair_general})}} \\
$\vartheta$ & Generic scalar target parameter of interest (e.g., arrival time $t_0$ or amplitude $A$) \\
$\bm{\eta} \in \mathbb{R}^{M-1}$ & Nuisance parameter vector; partition $\bm{\theta} = [\bm{\eta}^T,\,\vartheta]^T$ \\
$\bm{\eta}' \in \mathbb{R}^{4}$ & Bi-exponential amplitude-target nuisance vector $[t_0,\,\tau_d,\,\tau_r,\,b]^T$ used in Sec.~\ref{sec:opt_A} and Appendix~\ref{app:biexp_robustness} \\
$\mathbf{g}_{\bm{\eta},r}$, $\mathbf{g}_{\bm{\eta},f}$ & Rising/falling nuisance-edge sensitivity vectors entering the single-pair block decomposition (Sec.~\ref{sec:pair_general}) \\
$\bm{\mathcal{I}}_{\bm{\eta}\bm{\eta}}^{\mathrm{tot}}$, $\bm{\mathcal{I}}_{\bm{\eta}\vartheta}^{\mathrm{tot}}$, $\mathcal{I}_{\vartheta\vartheta}^{\mathrm{tot}}$ & Block decomposition of the total FIM \\
  \hline
  \multicolumn{2}{@{}l}{\emph{Paired information and effective MSE (Secs.~\ref{sec:pair_general}--\ref{sec:general_opt})}} \\
$\mathcal{I}_{jl}^{(\mathrm{pair})}(V)$ & Single-pair FIM element at threshold $V$ \\
$\mathcal{I}^{(\mathrm{pair})}(V)$ & Full single-pair Fisher Information Matrix at threshold $V$ \\
$\bm{\mathcal{I}}_{\bm{\eta}\bm{\eta}}$, $\bm{\mathcal{I}}_{\bm{\eta}\vartheta}$, $\mathcal{I}_{\vartheta\vartheta}$ & Single-pair target--nuisance block decomposition of the Fisher matrix \\
$\Delta\mathcal{I}_{\mathrm{eff}}^{(\vartheta)}(V)$, $\Delta\mathcal{I}_{\mathrm{eff}}^{(\vartheta),(n)}(V_n)$, $\Phi_A(V)$ & Generalized nuisance-profiled single-pair information score, its $n$-th pair instantiation, and the amplitude-specialized nuisance-projected objective \\
$\Delta\mathcal{I}_{\mathrm{eff}}^{(\vartheta),\mathrm{tot}}$ & Total effective Fisher Information for $\vartheta$ from all $N$ pairs jointly \\
$\beta_{\vartheta}$, $\beta_{\vartheta,\mathrm{tot}}$ & Generic scalar systematic bias in the target coordinate $\vartheta$ and, in the paired-crossing sections, its leading-order multi-pair model-based proxy \\
$\mathbf{J}_{\mathrm{tot}}$, $\mathbf{W}_{\mathrm{tot}}$, $\Delta\mathbf{t}_{\mathrm{bias}}^{\mathrm{tot}}$ & Stacked Jacobian, block-diagonal precision matrix, and concatenated temporal-bias vector over all $N$ threshold pairs \\
$\mathcal{B}_{\vartheta,\min}^{\mathrm{tot}}$ & Leading-order model-based MSE proxy for $\vartheta$ from all $N$ pairs \\
$\mathbf{A}$, $\mathbf{u}$, $c$, $\bm{A}_n$, $\mathbf{u}_n$, $c_n$, $\bm{A}_{\mathrm{tot}}$, $\mathbf{u}_{\mathrm{tot}}$, $c_{\mathrm{tot}}$ & Generic single-pair and per-pair/aggregated quadratic blocks used in Appendices~\ref{app:schur} and \ref{app:synergy} \\
$\mathbf{Q}$, $\mathbf{Q}_r$, $\mathbf{Q}_0$, $\bm{\Lambda}$, $\tilde{\mathbf{u}}$ & Orthogonal basis, range/nullspace blocks, diagonalized nuisance block, and transformed coupling vector used in the generalized-Schur proof \\
$S_0$, $S_1$, $S_2$, $S_k$ & Weighted slope-sum shorthands $S_k \triangleq h'(x_r)^k/D_r + h'(x_f)^k/D_f$ for $k=0,1,2$ in Appendices~\ref{app:biexp_eff} and~\ref{app:baseline_penalty} \\
$\Delta_b(V)$ & Additional CRLB penalty from including the baseline parameter $b$ as a nuisance variable in Appendix~\ref{app:baseline_penalty} \\
$\psi(\bm{\gamma})$, $\psi_n(\bm{\gamma})$, $\Psi(\bm{\gamma})$, $\bm{\gamma}_n^{*}$ & Single-pair, per-pair, and total variational quadratics, and the per-pair optimal nuisance-coupling direction when $\bm{A}_n$ is invertible \\
$[\mathbf{I}^{-1}]_{AA}$, $\mathbf{D}_{\theta}$, $\widetilde{\mathbf{I}}$, $\log\det\widetilde{\mathbf{I}}$ & Amplitude CRLB entry of the inverse FIM, fixed parameter-scale normalization matrix, dimensionless normalized FIM, and the D-optimal log-determinant objective \\
$\xi_j$, $\xi_{\min}$, $\xi_{\max}$ & Eigenvalues and extremal eigenvalues of the Fisher matrix used in the D-optimal conditioning discussion \\
$\mathbf{V}^{*}$, $\mathbf{V}^{\mathrm{old}}$, $\mathbf{V}^{\mathrm{new}}$, $\mathbf{V}^{\mathrm{best}}$ & Coordinate-descent threshold candidate, previous/current iterates, and best recorded threshold vector in D-optimal validation \\
$L_n$, $U_n$, $\alpha$, $\mathbf{V}_{\mathrm{multi\text{-}D}}^{*}$ & Coordinate-wise search bounds, damping factor, and multi-event D-optimal threshold vector \\
  \hline
  \multicolumn{2}{@{}l}{\emph{Nuisance-projected Fisher decomposition (Sec.~\ref{sec:opt_derivation})}} \\
$\bm{\gamma}$, $\bm{\gamma}^{*}$ & Nuisance coupling vector and, in the variational formulation, its optimizer/minimizer \\
$\gamma_{t_0}^{*}$, $\gamma_{\tau_d}^{*}$, $\gamma_{\tau_r}^{*}$, $\gamma_b^{*}$ & Components of the bi-exponential amplitude-target nuisance-coupling vector $\bm{\gamma}^{*} = [\gamma_{t_0}^{*},\,\gamma_{\tau_d}^{*},\,\gamma_{\tau_r}^{*},\,\gamma_b^{*}]^T$ \\
$\tilde{g}_{r}^{(n)}$, $\tilde{g}_{f}^{(n)}$ & Nuisance-projected sensitivities at rising/falling edge of $n$-th pair \\
$D_k'$, $\tilde{g}_k'$ & Threshold derivatives $dD_k/dV$ and $d\tilde{g}_k/dV$ entering the bi-exponential stationarity equations (Sec.~\ref{sec:opt_A}, Appendix~\ref{app:opt_threshold}) \\
$\mathcal{P}_k$ & Nuisance-projected waveform curvature at crossing $k$ (Eq.~\eqref{eq:proj_curvature}) \\
$\Phi_{\vartheta}^{(n)}(V_n;\,\bm{\gamma})$ & Nuisance-projected per-pair Fisher contribution from the $n$-th threshold pair \\
$\mathcal{B}_{\vartheta}^{*}$ & Leading-order model-based MSE proxy at optimal thresholds $\{V_n^*\}$ \\
$\mathrm{MSE}_{A,\mathrm{det}}$, $\mathcal{R}_{A,\mathrm{FWHM}}$ & Single-detector amplitude MSE after adding upstream frontend variance and the corresponding relative FWHM amplitude resolution (Sec.~\ref{sec:opt_A}) \\
$\sigma_{A,\mathrm{front}}^2$ & Aggregate upstream amplitude variance added to the digitizer-level bound when reporting full-chain amplitude resolution \\
$\bar{\mathcal{B}}_{A}^{*}$, $\mathcal{B}_{A}^{(p)}$, $\overline{\mathcal{R}}_{A,\mathrm{FWHM}}$, $\sigma_{A,\mathrm{front}}^{2,(p)}$ & Multi-event averaged amplitude MSE proxy, per-event amplitude MSE proxy, averaged relative FWHM amplitude resolution, and per-event upstream amplitude variance \\
  \hline
  \multicolumn{2}{@{}l}{\emph{Robustness analysis (Sec.~\ref{sec:robustness})}} \\
$\mathcal{I}_{\mathrm{eff}}(\bm{\theta})$ & Effective Fisher Information at fixed thresholds $\{V_n^*\}$ evaluated at parameter $\bm{\theta}$ \\
$\mathcal{I}_{\mathrm{eff}}^{*}$ & $\mathcal{I}_{\mathrm{eff}}(\bm{\theta}_0)$: effective Fisher Information at the design point \\
$\mathcal{B}_{\vartheta}(\bm{\theta})$ & Leading-order model-based MSE proxy at fixed design thresholds $\{V_n^*\}$ evaluated at parameter $\bm{\theta}$ \\
$\mathbf{S} \in \mathbb{R}^M$ & Fisher sensitivity vector; $S_j \triangleq d\,\mathcal{I}_{\mathrm{eff}}/d\theta_j\big|_{\bm{\theta}_0}$ \\
$\mathbf{s}_\beta \in \mathbb{R}^M$ & Bias sensitivity vector; $s_{\beta,j} \triangleq \partial\,\beta_{\vartheta,\mathrm{tot}}/\partial\theta_j\big|_{\bm{\theta}_0}$ \\
$\mathbf{H}_{\mathcal{I}}$, $\mathbf{H}_\beta$ & Hessian matrices of $\mathcal{I}_{\mathrm{eff}}$ and $\beta_{\vartheta,\mathrm{tot}}$ w.r.t.\ $\bm{\theta}$ at $\bm{\theta}_0$ \\
$\epsilon_j$ & Componentwise parameter uncertainty bound: $|\Delta\theta_j| \le \epsilon_j$ \\
$\mathcal{R}(\bm{\epsilon})$ & Hyper-rectangular parameter uncertainty region \\
$\rho(\bm{\epsilon})$ & Relative robustness metric (fractional MSE degradation) \\
$\rho_A(\bm{\epsilon})$, $\Delta\mathcal{B}_{A,\max}^{(1)}(\bm{\epsilon})$ & Amplitude-specialized first-order robustness ratio and worst-case first-order amplitude-MSE degradation \\
$c_j^{(\mathrm{F})}$, $c_j^{(\mathrm{B})}$ & Fisher and bias contribution scores used to rank parameter fragility in the amplitude robustness audit \\
$\delta_{jA}$, $\delta_{j\tau_d}$, $\delta_{j\tau_r}$ & Kronecker deltas selecting the amplitude, decay-time, and rise-time components in Appendix~\ref{app:biexp_robustness} \\
$\Delta\mathcal{B}_{\vartheta}^{(1)}$, $\Delta\mathcal{B}_{\vartheta}^{(2)}$ & First- and second-order perturbative changes of the target MSE proxy under parameter displacement \\
  \hline
  \multicolumn{2}{@{}l}{\emph{Partial-triggering and multi-event optimization (Sec.~\ref{sec:partial_trigger})}} \\
$\mathcal{A}(\bm{\theta})$ & Active threshold set: $\{n : V_n < p(\bm{\theta})\}$ \\
$S_{\mathcal{A}}$ & Restricted forward map on a fixed-active-set stratum, using only the crossings generated by the active threshold set $\mathcal{A}$ \\
$K_{\mathrm{act}}(\bm{\theta})$ & Number of active (triggered) threshold pairs for a generic event: $|\mathcal{A}(\bm{\theta})|$ \\
$\mathcal{I}_{\vartheta\vartheta}^{\mathrm{act}}$, $\bm{\mathcal{I}}_{\bm{\eta}\vartheta}^{\mathrm{act}}$, $\bm{\mathcal{I}}_{\bm{\eta}\bm{\eta}}^{\mathrm{act}}$ & Active target--nuisance block decomposition built from triggered threshold pairs only \\
$\bm{\gamma}^{\mathrm{act}}(\bm{\theta})$, $\bm{\gamma}_p^{\mathrm{act}}$ & Nuisance coupling vector restricted to active pairs for a generic event and its representative-event specialization \\
$\Delta\mathcal{I}_{\mathrm{eff}}^{(\vartheta),\mathrm{act}}$, $\Delta\mathcal{I}_{\mathrm{eff}}^{(\vartheta),\mathrm{act},(p)}$ & Effective Fisher Information from active pairs only for a generic event and for the $p$-th representative event \\
$\mathbf{J}_{\mathrm{act}}$, $\mathbf{W}_{\mathrm{act}}$, $\Delta\mathbf{t}_{\mathrm{bias}}^{\mathrm{act}}$ & Jacobian, precision matrix, and temporal-bias vector restricted to the active crossings of a partially triggering event \\
$\beta_{\vartheta,\mathrm{act}}$, $\beta_{\vartheta,\mathrm{act}}^{(p)}$ & Systematic bias from active pairs only for a generic event and for the $p$-th representative event \\
$\mathcal{B}_{\vartheta,\min}(\{V_n\};\,\bm{\theta})$, $\mathcal{B}_{\vartheta,\min}^{(p)}$ & Leading-order model-based MSE proxy for a partial-triggering event and its $p$-th representative-event specialization \\
$P$ & Number of representative events in multi-event optimization \\
$w_p$ & Importance weight of the $p$-th event ($\sum_p w_p = 1$) \\
$\bm{\theta}^{(p)}$ & Parameter vector of the $p$-th representative event \\
$b^{(p)}$ $[\mathrm{V}]$ & Baseline offset of the $p$-th representative event \\
$\mathcal{A}_p$ & Active threshold set for the $p$-th event: $\mathcal{A}(\bm{\theta}^{(p)})$ \\
$K_p$ & Number of active threshold pairs for representative event $p$: $|\mathcal{A}_p|$ \\
$K_{\min}^{(\vartheta)}$ & Target-specific minimum active-pair feasibility floor in the multi-event optimization \\
$\mathcal{C}(\{V_n\})$ & Weighted average leading-order MSE-proxy cost function (Eq.~\eqref{eq:multi_event_opt}) \\
$\mathcal{C}_0(\mathbf{V})$, $\mathcal{C}_{\varepsilon}(\mathbf{V})$, $\mathcal{D}_{\varepsilon}(\mathbf{V})$ & Idealized multi-event cost, misspecified multi-event cost, and their gap in the amplitude-target misspecification extension (Eqs.~\eqref{eq:autonum:ME0a}--\eqref{eq:autonum:ME0bprime}) \\
$\mathbf{H}_{0,\mathrm{multi}}$ & Hessian of the ideal multi-event amplitude-target objective at $\mathbf{V}_{\mathrm{multi}}^{(0)}$ (Eq.~\eqref{eq:autonum:ME0h}) \\
$\mathbf{V}_{\mathrm{multi}}^{(0)}$, $\mathbf{V}_{\mathrm{multi}}^{(\varepsilon)}$ & Minimizers of the idealized and misspecified multi-event amplitude costs $\mathcal{C}_0$ and $\mathcal{C}_{\varepsilon}$ \\
$\beta_{A,\mathrm{act}}^{(p)}$, $\bar{\varepsilon}^{(p)}$, $\varepsilon_{\max}^{(p)}$, $m_{\min}^{(p)}$, $C_{\beta}^{(p)}$ & Active-set amplitude misspecification diagnostics for representative event $p$ (Eqs.~\eqref{eq:autonum:ME0c}--\eqref{eq:autonum:ME0f}) \\
$\delta_{\mathrm{multi}}$, $\rho_{\mathrm{bias,multi}}^{(A)}$, $\rho_{\mathrm{bias}}^{(A),(p)}$ & Weighted aggregate mismatch scale, global multi-event bias-to-cost diagnostic, and per-event active-set bias-to-variance diagnostic in the misspecification audit (Eqs.~\eqref{eq:autonum:ME0g}--\eqref{eq:autonum:ME0k}) \\
$L_{ij}^{\max}$ & Worst-case slew rate over all events: $\max_p L_{ij}(\bm{\theta}^{(p)})$ \\
$L_{ij}(\bm{\theta})$ & Supremum local slew rate between thresholds $i$ and $j$ (Sec.~\ref{sec:bandwidth}) \\
  \end{longtable}
  \normalsize

  \clearpage
  \twocolumn
\bibliographystyle{IEEEtran}
\bibliography{ref}

@article{atkinson1995d,
  title={D-optimum designs for heteroscedastic linear models},
  author={Atkinson, AC and Cook, RD},
  journal={Journal of the American Statistical Association},
  volume={90},
  number={429},
  pages={204--212},
  year={1995},
  publisher={Taylor \& Francis},
  doi={10.1080/01621459.1995.10476503}
}

@article{atkinson1989construction,
  title={The construction of exact D-optimum experimental designs with application to blocking response surface designs},
  author={Atkinson, Anthony C and Donev, Alexander N},
  journal={Biometrika},
  volume={76},
  number={3},
  pages={515--526},
  year={1989},
  publisher={Oxford University Press},
  doi={10.2307/2336117}
}

@article{gundacker2020silicon,
  title={The silicon photomultiplier: fundamentals and applications of a modern solid-state photon detector},
  author={Gundacker, Stefan and Heering, Arjan},
  journal={Physics in Medicine \& Biology},
  volume={65},
  number={17},
  pages={17TR01},
  year={2020},
  publisher={IOP Publishing},
  doi={10.1088/1361-6560/ab7b2d}
}

@inbook{lecoq2016scintillation,
  title={Scintillation and Inorganic Scintillators},
  author={Lecoq, Paul and Gektin, Alexander and Korzhik, Mikhail},
  booktitle={Inorganic Scintillators for Detector Systems},
  pages={1--41},
  year={2016},
  publisher={Springer International Publishing},
  doi={10.1007/978-3-319-45522-8_1}
}

@book{lee2012smooth,
  title={Introduction to Smooth Manifolds},
  author={Lee, John M.},
  edition={2},
  series={Graduate Texts in Mathematics},
  year={2012},
  publisher={Springer New York},
  address={New York, NY},
  doi={10.1007/978-1-4419-9982-5}
}

@incollection{mukherjee2015approximation,
  title={Approximation Theorems and Whitney's Embedding},
  author={Mukherjee, Amiya},
  booktitle={Differential Topology},
  pages={43--67},
  year={2015},
  publisher={Birkh\"auser},
  address={Cham},
  doi={10.1007/978-3-319-19045-7_2}
}

@article{lane1984the,
  title={The central limit theorem for the Poisson shot-noise process},
  author={Lane, J. A},
  journal={Journal of applied probability},
  volume={21},
  number={2},
  pages={287--301},
  year={1984},
  doi={10.1017/s0021900200024682}
}

@article{johnson1928thermal,
  title={Thermal agitation of electricity in conductors},
  author={Johnson, John Bertrand},
  journal={Physical review},
  volume={32},
  number={1},
  pages={97--109},
  year={1928},
  doi={10.1103/PhysRev.32.97}
}

@article{nyquist1928thermal,
  title={Thermal Agitation of Electric Charge in Conductors},
  author={Nyquist, H},
  journal={Physical review},
  volume={32},
  number={1},
  pages={110--113},
  year={1928},
  doi={10.1103/physrev.32.110}
}

@article{seifert2009simulation,
  title={Simulation of silicon photomultiplier signals},
  author={Seifert, Stefan and Van Dam, Herman T and Huizenga, Jan and Vinke, Ruud and Dendooven, Peter and Lohner, Herbert and Schaart, Dennis R},
  journal={IEEE Transactions on Nuclear Science},
  volume={56},
  number={6},
  pages={3726--3733},
  year={2009},
  publisher={IEEE},
  doi={10.1109/TNS.2009.2030728}
}

@inbook{meyer2002noise,
  title={Noise, Gain and Bandwidth in Analog Design},
  author={Meyer, Robert G.},
  booktitle={Trade-Offs in Analog Circuit Design},
  pages={227--256},
  year={2002},
  publisher={Kluwer Academic Publishers},
  doi={10.1007/0-306-47673-8_8}
}

@article{chen1982minimum,
  title={On minimum step response rise time of linear low-pass systems under the constraint of a given noise bandwidth},
  author={Chen, HB},
  journal={Proceedings of the IEEE},
  volume={70},
  number={4},
  pages={404--406},
  year={1982},
  publisher={IEEE},
  doi={10.1109/PROC.1982.12316}
}

@incollection{anthonys2021jitter,
  title={Jitter and Measurement of Jitter},
  author={Anthonys, Gehan},
  booktitle={Timing Jitter in Time-of-Flight Range Imaging Cameras},
  pages={55--73},
  year={2021},
  publisher={Springer},
  doi={10.1007/978-3-030-94159-8_4}
}

@article{hero1991timing,
  title={Timing estimation for a filtered Poisson process in Gaussian noise},
  author={Hero, AO},
  journal={IEEE transactions on information theory},
  volume={37},
  number={1},
  pages={92--106},
  year={1991},
  publisher={IEEE},
  doi={10.1109/18.61107}
}

@inproceedings{thompson2013measurement,
  title={Measurement of energy and timing resolution of very highly pixellated LYSO crystal blocks with multiplexed SiPM readout for use in a small animal PET/MR insert},
  author={Thompson, Christopher J and Goertzen, Andrew L and Kozlowski, Piotr and Reti{\`e}re, Fabrice and Stortz, Greg and Sossi, Vesna and Zhang, Xuezhu},
  booktitle={2013 IEEE Nuclear Science Symposium and Medical Imaging Conference (2013 NSS/MIC)},
  pages={1--5},
  year={2013},
  organization={IEEE},
  doi={10.1109/NSSMIC.2013.6829134}
}

@article{mao2013crystal,
  title={Crystal growth and scintillation properties of LSO and LYSO crystals},
  author={Mao, Rihua and Wu, Chen and Dai, Ling'En and Lu, Sheng},
  journal={Journal of crystal growth},
  volume={368},
  pages={97--100},
  year={2013},
  publisher={Elsevier},
  doi={10.1016/j.jcrysgro.2013.01.038}
}

@article{slepian1958some,
  title={Some comments on the detection of Gaussian signals in Gaussian noise},
  author={Slepian, D.},
  journal={IRE Transactions on Information Theory},
  volume={4},
  number={2},
  pages={65--68},
  year={1958},
  publisher={IEEE},
  doi={10.1109/tit.1958.1057443}
}

@article{abeida2019slepian,
  title={Slepian--Bangs formula and Cram{\'e}r--Rao bound for circular and non-circular complex elliptical symmetric distributions},
  author={Abeida, Habti and Delmas, Jean-Pierre},
  journal={IEEE Signal Processing Letters},
  volume={26},
  number={10},
  pages={1561--1565},
  year={2019},
  publisher={IEEE},
  doi={10.1109/LSP.2019.2939714}
}

@inproceedings{el2024full,
  title={Full {Slepian--Bangs} Formula for Fisher Information on Lie Groups},
  author={El Bouch, Sara and Labsir, Samy and Renaux, Alexandre and Vil{\`a}-Valls, Jordi and Chaumette, Eric},
  booktitle={2024 58th Asilomar Conference on Signals, Systems, and Computers},
  pages={1306--1310},
  year={2024},
  organization={IEEE},
  doi={10.1109/IEEECONF60004.2024.10943026}
}

@article{rosado2015characterization,
  title={Characterization and modeling of crosstalk and afterpulsing in Hamamatsu silicon photomultipliers},
  author={Rosado, Jaime and Hidalgo, Salvador},
  journal={Journal of Instrumentation},
  volume={10},
  number={10},
  pages={P10031--P10031},
  year={2015},
  doi={10.1088/1748-0221/10/10/P10031}
}

@article{gola2014sipm,
  title={SiPM optical crosstalk amplification due to scintillator crystal: effects on timing performance},
  author={Gola, Alberto and Ferri, Alessandro and Tarolli, Alessandro and Zorzi, Nicola and Piemonte, Claudio},
  journal={Physics in medicine and biology},
  volume={59},
  number={13},
  pages={3615--3635},
  year={2014},
  publisher={IOP Publishing},
  doi={10.1088/0031-9155/59/13/3615}
}

@inproceedings{couce2008parametrization,
  title={Parametrization of SiPM dynamic range contribution to energy resolution of scintillation light readout},
  author={Couce, B and Gomez, F and Iglesias, A and Aguiar, P},
  booktitle={2008 IEEE Nuclear Science Symposium Conference Record},
  pages={3973--3974},
  year={2008},
  organization={IEEE},
  doi={10.1109/NSSMIC.2008.4774153}
}

@inproceedings{fu2016recovery,
  title={Recovery of inter-crystal compton scattering events for sensitivity improvement of sub-250 ps TOF-PET detector},
  author={Fu, Geng and Ivan, Adrian and Qian, Hua},
  booktitle={2016 IEEE Nuclear Science Symposium, Medical Imaging Conference and Room-Temperature Semiconductor Detector Workshop (NSS/MIC/RTSD)},
  pages={1--5},
  year={2016},
  organization={IEEE},
  doi={10.1109/NSSMIC.2016.8069520}
}

@article{lee2020recovery,
  title={Recovery of inter-detector and inter-crystal scattering in brain PET based on LSO and GAGG crystals},
  author={Lee, Seungeun and Kim, Kyeong Yun and Lee, Min Sun and Lee, Jae Sung},
  journal={Physics in Medicine \& Biology},
  volume={65},
  number={19},
  pages={195005},
  year={2020},
  publisher={IOP Publishing},
  doi={10.1088/1361-6560/ab9f5c}
}

@article{comanor1996algorithms,
  title={Algorithms to identify detector Compton scatter in PET modules},
  author={Comanor, KA and Virador, PRG and Moses, WW},
  journal={IEEE transactions on nuclear science},
  volume={43},
  number={4},
  pages={2213--2218},
  year={1996},
  publisher={IEEE},
  doi={10.1109/23.531884}
}

@article{ollinger1995detector,
  title={Detector efficiency and Compton scatter in fully 3D PET},
  author={Ollinger, John M},
  journal={IEEE transactions on nuclear science},
  volume={42},
  number={4},
  pages={1168--1173},
  year={1995},
  publisher={IEEE},
  doi={10.1109/23.467731}
}

@inproceedings{kimble2002scintillation,
  title={Scintillation properties of LYSO crystals},
  author={Kimble, Thomas and Chou, Mitch and Chai, Bruce HT},
  booktitle={2002 IEEE Nuclear Science Symposium Conference Record},
  volume={3},
  pages={1434--1437},
  year={2002},
  organization={IEEE},
  doi={10.1109/NSSMIC.2002.1239590}
}

@article{brunner2017bgo,
  title={BGO as a hybrid scintillator / Cherenkov radiator for cost-effective time-of-flight PET},
  author={Brunner, S. E. and Schaart, D. R},
  journal={Physics in Medicine \& Biology},
  volume={62},
  number={11},
  pages={4421--4439},
  year={2017},
  doi={10.1088/1361-6560/aa6a49}
}

@article{okajima1982characteristics,
  title={Characteristics of a gamma-ray detector using a bismuth germanate scintillator},
  author={Okajima, K. and Takami, K. and Ueda, K. and Kawaguchi, F},
  journal={Review of Scientific Instruments},
  volume={53},
  number={8},
  pages={1285--1286},
  year={1982},
  doi={10.1063/1.1137124}
}

@inproceedings{xu2015optimization,
  title={Optimization of dual mode readout of Sensl SiPM for TOF PET detectors},
  author={Xu, T. and Yao, S. and Chen, S. and Wei, Q. and Liu, Y. and Ma, T},
  booktitle={2015 IEEE Nuclear Science Symposium and Medical Imaging Conference (NSS/MIC)},
  pages={1--4},
  year={2015},
  publisher={IEEE},
  doi={10.1109/nssmic.2015.7581741}
}

@article{grodzicka2013energy,
  title={Energy resolution of small scintillation detectors with SiPM light readout},
  author={Grodzicka, M. and Moszyński, M. and Szczęśniak, T. and Kapusta, M. and Szawłowski, M. and Wolski, D},
  journal={Journal of Instrumentation},
  volume={8},
  number={02},
  pages={P02017--P02017},
  year={2013},
  doi={10.1088/1748-0221/8/02/p02017}
}

@article{iwai2017background,
  title={Background identification system in MEG II experiment based on high-rate scintillation detector with SiPM readout},
  author={Iwai, R},
  journal={Journal of Instrumentation},
  volume={12},
  number={02},
  pages={C02023--C02023},
  year={2017},
  doi={10.1088/1748-0221/12/02/c02023}
}

@article{cates2022low,
  title={Low power implementation of high frequency SiPM readout for Cherenkov and scintillation detectors in TOF-PET},
  author={Cates, J. W. and Choong, W. S},
  journal={Physics in Medicine \& Biology},
  volume={67},
  number={19},
  pages={195009},
  year={2022},
  doi={10.1088/1361-6560/ac8963}
}

@inproceedings{mog2004zero,
  title={Zero crossing determination by linear interpolation of sampled sinusoidal signals},
  author={Mog, G. E. and Ribeiro, E. P},
  booktitle={2004 IEEE/PES Transmission and Distribution Conference and Exposition: Latin America (IEEE Cat. No. 04EX956)},
  pages={799--802},
  year={2004},
  organization={IEEE},
  doi={10.1109/tdc.2004.1432484}
}

@book{hadamard1923lectures,
  title={Lectures on Cauchy's Problem in Linear Partial Differential Equations},
  author={Hadamard, Jacques},
  year={1923},
  publisher={Yale University Press}
}

@article{kabanikhin2008definitions,
  title={Definitions and examples of inverse and ill-posed problems},
  author={Kabanikhin, S. I.},
  journal={Journal of Inverse and Ill-posed Problems},
  volume={16},
  number={4},
  pages={317--357},
  year={2008},
  doi={10.1515/JIIP.2008.019}
}

@misc{clason2020regularization,
  title={Regularization of inverse problems},
  author={Clason, Christian},
  year={2020},
  eprint={2001.00617},
  archivePrefix={arXiv},
  primaryClass={math.FA},
  doi={10.48550/arXiv.2001.00617},
  url={https://arxiv.org/abs/2001.00617}
}

@book{bertero2021inverseimaging,
  title={Introduction to Inverse Problems in Imaging},
  author={Bertero, Mario and Boccacci, Patrizia and De Mol, Christine},
  edition={2},
  year={2021},
  publisher={CRC Press},
  address={Boca Raton},
  doi={10.1201/9781003032755}
}

@book{kingman1992poisson,
  title={Poisson Processes},
  author={Kingman, J. F. C.},
  year={1992},
  publisher={Oxford University Press},
  doi={10.1093/oso/9780198536932.001.0001}
}

@article{bennett1948spectra,
  title={Spectra of Quantized Signals},
  author={Bennett, W. R.},
  journal={Bell System Technical Journal},
  volume={27},
  number={3},
  pages={446--472},
  year={1948},
  doi={10.1002/j.1538-7305.1948.tb01340.x}
}

@book{kay1993estimation,
  title={Fundamentals of Statistical Signal Processing, Volume I: Estimation Theory},
  author={Kay, Steven M.},
  year={1993},
  publisher={Prentice Hall}
}

@IEEEtranBSTCTL{IEEEexample:BSTcontrol,
  CTLdash_repeated_names = "no"
}

@article{chu2023singleline,
  title={Single-line Multi-Voltage Threshold Method for Scintillation Detectors},
  author={Chu, Hyeyeun and Yi, Minseok and Shim, Hyeong Seok and Lee, Jae Sung},
  journal={Journal of Instrumentation},
  volume={18},
  number={06},
  pages={P06021},
  year={2023},
  doi={10.1088/1748-0221/18/06/P06021}
}

@inproceedings{deng2013empirical,
  title={Empirical Bayesian Energy Estimation for Multi-Voltage Threshold Digitizer in {PET}},
  author={Deng, Zhenzhou and Xie, Qingguo},
  booktitle={2013 IEEE Nuclear Science Symposium and Medical Imaging Conference ({NSS/MIC})},
  pages={1--5},
  year={2013},
  organization={IEEE},
  doi={10.1109/NSSMIC.2013.6829196}
}

@article{deng2015quadratic,
  title={Quadratic Programming Time Pickoff Method for Multivoltage Threshold Digitizer in {PET}},
  author={Deng, Zhenzhou and Xie, Qingguo},
  journal={IEEE Transactions on Nuclear Science},
  volume={62},
  number={3},
  pages={805--813},
  year={2015},
  doi={10.1109/TNS.2015.2416349}
}

@inproceedings{deng2017threshold,
  title={Threshold Optimization in Multi-Voltage Threshold Digitizers for {TOF PET} Detector},
  author={Deng, Zhenzhou and Wu, Haodi and Xiong, Wei and Zhao, Anjiang and Han, Chunlei and Lu, Jinyuan and Yang, Yi and Duan, Zhiwen and Xiao, Peng and Tang, Jiang and Xie, Qingguo},
  booktitle={2017 IEEE Nuclear Science Symposium and Medical Imaging Conference ({NSS/MIC})},
  pages={1--3},
  year={2017},
  organization={IEEE},
  doi={10.1109/NSSMIC.2017.8533079}
}

@article{xu2020neural,
  title={Neural-Network-Based Energy Calculation for Multivoltage Threshold Sampling},
  author={Xu, Hao and Wang, Hao and Xu, Feng and Cheng, Ran and Zhang, Bo and Fang, Lei and Xie, Qingguo and Xiao, Peng},
  journal={IEEE Transactions on Radiation and Plasma Medical Sciences},
  volume={4},
  number={3},
  pages={311--318},
  year={2020},
  doi={10.1109/TRPMS.2019.2960129}
}

@article{ling2024novel,
  title={A Novel Peak Picking Multi-Voltage Threshold Digitizer for Pulse Sampling},
  author={Ling, Yiqing and Qiu, Ao and Wan, Lin and Wang, Fei and Zhu, Kezhang and Zhang, Yeping and Xie, Qingguo},
  journal={IEEE Transactions on Radiation and Plasma Medical Sciences},
  volume={8},
  number={3},
  pages={248--256},
  year={2024},
  doi={10.1109/TRPMS.2024.3359241}
}

@article{cheng2024dualended,
  title={Dual-Ended Readout {PET} Detector Based on Multivoltage Threshold Sampling Combined With Convolutional Neural Network for Energy Calculation},
  author={Cheng, Ran and Sun, Mingchen and Wang, Fei and Mu, Dengyun and Liu, Yu and Xie, Qingguo and Qiu, Bensheng and Chen, Xun and Xiao, Peng},
  journal={IEEE Transactions on Radiation and Plasma Medical Sciences},
  volume={8},
  number={7},
  pages={709--717},
  year={2024},
  doi={10.1109/TRPMS.2024.3393235}
}

@book{nocedal2006numerical,
  title={Numerical Optimization},
  author={Nocedal, Jorge and Wright, Stephen J.},
  edition={2},
  year={2006},
  publisher={Springer},
  doi={10.1007/978-0-387-40065-5}
}

@article{hero1991a,
  title={A Lower Bound On PET Timing Estimation With Pulse Pileup},
  author={Hero, A. O. and Clinthorne, N. H. and Rogers, W. L.},
  journal={IEEE Transactions on Nuclear Science},
  volume={38},
  number={2},
  pages={709--712},
  year={1991},
  publisher={IEEE},
  doi={10.1109/23.289378}
}

@inproceedings{ruizgonzalez2016joint,
  title={Joint amplitude and timing estimation for scintillation pulses in GPU},
  author={Ruiz-Gonzalez, M. and Caucci, L. and Furenlid, L. R},
  booktitle={2016 IEEE Nuclear Science Symposium, Medical Imaging Conference and Room-Temperature Semiconductor Detector Workshop (NSS/MIC/RTSD)},
  pages={1--3},
  year={2016},
  publisher={IEEE},
  doi={10.1109/nssmic.2016.8069435}
}

@inproceedings{mohammaddjafari2004on,
  title={On the estimation of a parameter with incomplete knowledge on a nuisance parameter},
  author={Mohammad‐Djafari, A. and Mohammadpour, A},
  booktitle={AIP Conference Proceedings},
  volume={735},
  number={1},
  pages={533--540},
  year={2004},
  publisher={American Institute of Physics},
  doi={10.1063/1.1835253}
}

@article{si2015off,
  title={Off-Grid DOA Estimation Using Alternating Block Coordinate Descent in Compressed Sensing},
  author={Si, W. and Qu, X. and Qu, Z},
  journal={Sensors},
  volume={15},
  number={9},
  pages={21099--21113},
  year={2015},
  doi={10.3390/s150921099}
}

@article{kharfati2026block,
  title={Block coordinate descent methods of centres for solving block-constrained optimization problems},
  author={Kharfati, K and Roubi, A},
  journal={Optimization Methods and Software},
  volume={41},
  number={3},
  pages={577--601},
  year={2026},
  publisher={Taylor \& Francis},
  doi={10.1080/10556788.2026.2617639}
}

@article{shao1994triple,
  title={Triple energy window scatter correction technique in PET},
  author={Shao, Lingxiong and Freifelder, R. and Karp, J. S.},
  journal={IEEE Transactions on Medical Imaging},
  volume={13},
  number={4},
  pages={641--648},
  year={1994},
  publisher={IEEE},
  doi={10.1109/42.363104}
}

@inproceedings{sossi1995comparison,
  title={Comparison of energy window choice and parameter implementation in dual energy window scatter correction performance in 3D PET},
  author={Sossi, V and Barney, JS and Oakes, TR and Ruth, TJ},
  booktitle={1995 IEEE Nuclear Science Symposium and Medical Imaging Conference Record},
  volume={2},
  pages={1060--1063},
  year={1995},
  organization={IEEE},
  doi={10.1109/NSSMIC.1995.510447}
}

@article{yoshida2008doi,
  title={A DOI-dependent extended energy window method to control balance of scatter and true events},
  author={Yoshida, Eiji and Kitamura, Keishi and Shibuya, Kengo and Nishikido, Fumihiko and Hasegawa, Tomoyuki and Yamaya, Taiga and Lam, Chihfung and Inadama, Naoko and Murayama, Hideo},
  journal={IEEE Transactions on Nuclear Science},
  volume={55},
  number={5},
  pages={2475--2481},
  year={2008},
  publisher={IEEE},
  doi={10.1109/TNS.2008.2003357}
}

@article{xie_new_2005,
	title = {A new approach for pulse processing in positron emission tomography},
	volume = {52},
	issn = {1558-1578},
	doi = {10.1109/TNS.2005.852966},
	number = {4},
	journal = {IEEE Transactions on Nuclear Science},
	author = {Xie, Qingguo and Kao, Chien-Min and Hsiau, Zekai and Chen, Chin-Tu},
	month = aug,
	year = {2005},
	pages = {988--995},
}

@article{xie_implementation_2013,
	title = {Implementation of {LYSO}/{PSPMT} {Block} {Detector} {With} {All} {Digital} {DAQ} {System}},
	volume = {60},
	issn = {1558-1578},
	doi = {10.1109/TNS.2013.2245771},
	number = {3},
	journal = {IEEE Transactions on Nuclear Science},
	author = {Xie, Qingguo and Chen, Yuanbao and Zhu, Jun and Liu, Jingjing and Wang, Xi and Liu, Wei and Chen, Xin and Niu, Ming and Wu, Zhongyi and Xi, Daoming and Wang, Luyao and Xiao, Peng and Chen, Chin-Tu and Kao, Chien-Min},
	month = jun,
	year = {2013},
	pages = {1487--1494},
}

@article{xi_fpga-only_2013,
	title = {{FPGA}-{Only} {MVT} {Digitizer} for {TOF} {PET}},
	volume = {60},
	issn = {1558-1578},
	doi = {10.1109/TNS.2013.2277855},
	number = {5},
	journal = {IEEE Transactions on Nuclear Science},
	author = {Xi, Daoming and Kao, Chien-Min and Liu, Wei and Zeng, Chen and Liu, Xiang and Xie, Qingguo},
	month = oct,
	year = {2013},
	pages = {3253--3261},
}

@article{fang_development_2024,
	title = {Development and evaluation of a new high-{TOF}-resolution all-digital brain {PET} system},
	volume = {69},
	issn = {0031-9155},
	doi = {10.1088/1361-6560/ad164d},
	language = {en},
	number = {2},
	journal = {Physics in Medicine \& Biology},
	publisher = {IOP Publishing},
	author = {Fang, Lei and Zhang, Bo and Li, Bingxuan and Zhang, Xiangsong and Zhou, Xiaoyun and Yang, Jigang and Li, Ang and Shi, Xinchong and Liu, Yuqing and Kreissl, Michael and D’Ascenzo, Nicola and Xiao, Peng and Xie, Qingguo},
	month = jan,
	year = {2024},
	pages = {025019},
}

@inproceedings{fang2021atc,
  title = {Automatic Threshold Calibration for {MVT}-{Based} All Digital {PET}},
  author = {Fang, Lei and Zhang, Bo and Li, Bingxuan and Zhang, Chaofan and Xie, Qingguo and Xiao, Peng},
  booktitle = {2021 {IEEE} Nuclear Science Symposium and Medical Imaging Conference ({NSS}/{MIC})},
  pages = {1--2},
  year = {2021},
  organization = {IEEE},
  doi = {10.1109/NSS/MIC44867.2021.9875603}
}

@inproceedings{xie2010baseline,
  title = {An Investigation of Baseline Calibration Method for Digitally Sampling Scintillation Pulses in {PET}},
  author = {Xie, Qingguo and Chen, Yuanbao and Wu, Zhongyi and Zhu, Jun and Wang, Xi and Xi, Daoming and Zhao, Jin},
  booktitle = {2010 {IEEE} Nuclear Science Symposium Conference Record ({NSS}/{MIC})},
  pages = {1480--1482},
  year = {2010},
  organization = {IEEE},
  doi = {10.1109/NSSMIC.2010.5874020}
}

@article{zhang_performance_2025,
	title = {Performance {Evaluation} of {New} {PET}/{CT} {DigitMI} 930},
	volume = {9},
	issn = {2469-7303},
	doi = {10.1109/TRPMS.2025.3526659},
	number = {5},
	journal = {IEEE Transactions on Radiation and Plasma Medical Sciences},
	author = {Zhang, Bo and Li, Bingxuan and Fang, Lei and Zhou, Xiaoyun and Li, Ang and Zhang, Xuan and Liu, Yang and Wang, Zhuo and Kao, Chien-Min and Liu, Yuqing and Zhu, Xiaohua and Wan, Lin and Xiao, Peng and Chen, Xun and Iida, Hidehiro and Knuuti, Juhani and Xie, Qingguo},
	month = may,
	year = {2025},
	pages = {578--585},
}

\end{document}